\documentclass[12pt, a4paper, twoside, openright]{report}
\usepackage{amsthm,mathrsfs,mathptmx,cite,textcase,ifthen}
\usepackage{amssymb}
\usepackage{amsmath}
\usepackage{amsfonts}
\usepackage{lscape}
\usepackage{latexsym}
\usepackage{graphicx}
\usepackage{graphics}
\usepackage{epsfig}
\usepackage{tikz}
\usetikzlibrary{shapes,arrows,positioning}
\usepackage{setspace}
\usepackage[T1]{fontenc}
\usepackage{color}
\UseRawInputEncoding

\newcommand{\properpagestyle}{\pagestyle{myheadings}\markboth{}{}\markright{}}

\makeatletter
\def\cleardoublepage{\clearpage\if@twoside\ifodd\c@page\else
\hbox{}
\vspace*{\fill}
\begin{center}

\end{center}
\vspace{\fill}
\thispagestyle{empty}
\newpage
\if@twocolumn\hbox{}\newpage\fi\fi\fi}
\makeatother

\usepackage[outer=.85in, top=.85in, bottom=.85in, inner=1.2in, ]{geometry}
\usepackage[compact]{titlesec}
\titlespacing{\section}{0pt}{12pt}{12pt}
\titlespacing{\subsection}{0pt}{10pt}{10pt}
\usepackage[]{nomencl}
\def\execute{%
\begingroup
\catcode`\%=12
\catcode`\\=12
\executeaux}
\def\executeaux#1{\immediate\write18{#1}\endgroup}
\execute{makeindex Main.nlo -s nomencl.ist -o Main.nls}
\makenomenclature
\renewcommand{\nomname}{List of symbols}

\renewcommand\contentsname{Contents}

\numberwithin{equation}{section}
\numberwithin{section}{chapter}
\numberwithin{figure}{chapter}
\numberwithin{table}{chapter}

\newcommand{\ThesisTitle}{Detection and Classification of Bipartite and Multipartite Entangled States}
\newcommand{\Student}{Anu Kumari}
\newcommand{\Enrollment}{2K18/PHD/AM/05}
\newcommand{\Supervisor}{Dr.\ Satyabrata Adhikari}

\newcommand{\Institute}{Delhi Technological University}
\newcommand{\SubmissionDate}{October, 2022}

\renewcommand{\bibname}{References}

\begin{document}

\newcommand{\TitlePage}{\thispagestyle{empty} \fontfamily{phv}\selectfont
%\textcolor{blue}{
\begin{center} \large
\textbf{\MakeTextUppercase{\ThesisTitle}}
\end{center}
%}
%\vfill

\begin{center}

\emph{A thesis submitted to} \\ \ \\

\textbf{\large\MakeTextUppercase{\Institute}}\\ \ \\
\emph{in partial fulfillment of the requirements for the award of the degree of}\\ \ \\

  \textbf{\large DOCTOR OF PHILOSOPHY }\\ \ \\
\emph{in}\\ \ \\
\textbf{\large MATHEMATICS } \\ \ \\
\emph{By}\\
\vspace{.3cm}
  \textbf{\large\MakeTextUppercase \Student}\\ \ \\
\emph{Under the Supervision of }  \\
\large \textbf{ \Supervisor} \\
 \end{center}

\begin{figure}[h]
  \centering
  \includegraphics[width=4cm]{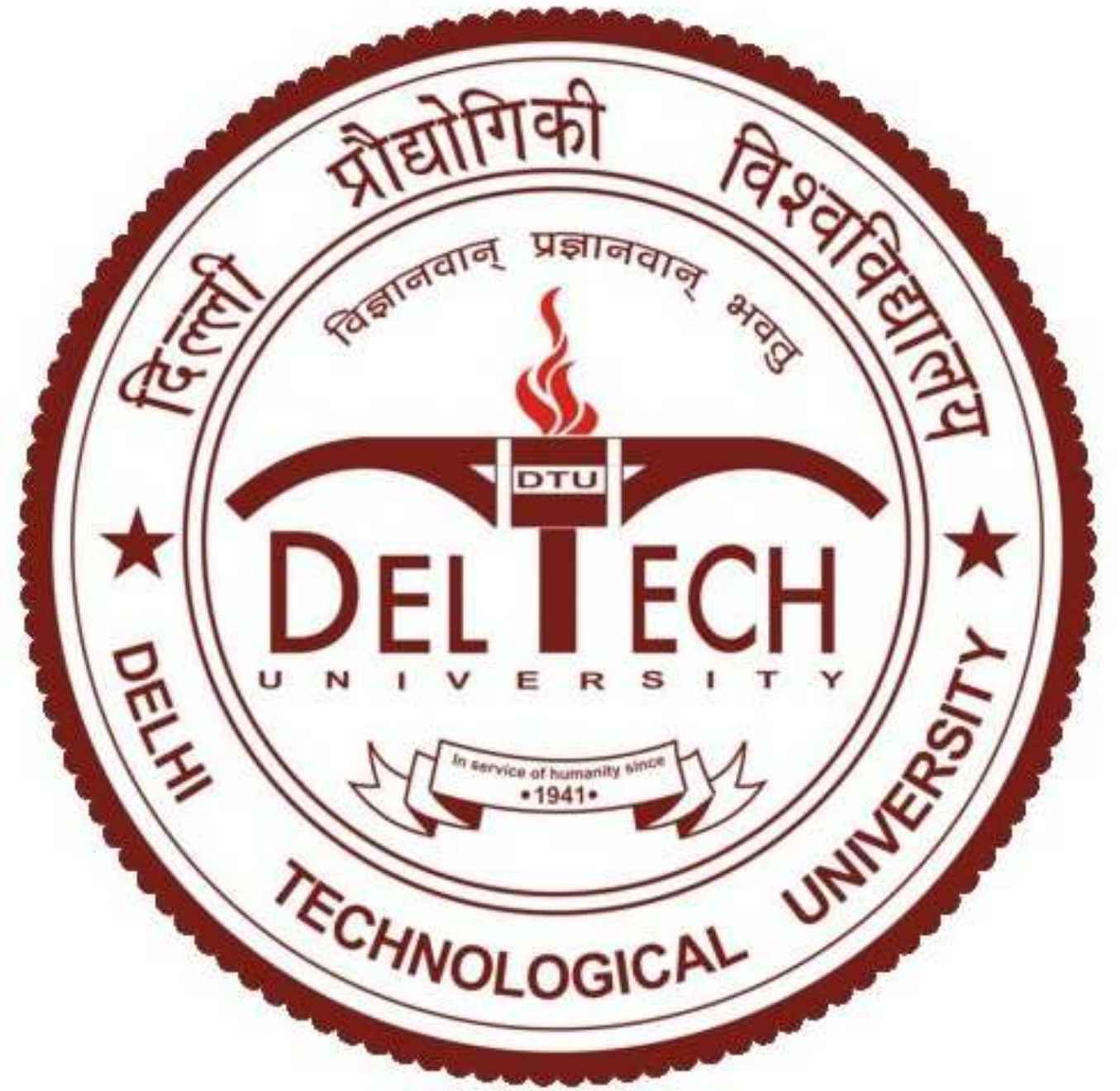}\\
  %\caption{}\label{}
\end{figure}

%\vfill

\begin{center}\large
DEPARTMENT OF APPLIED MATHEMATICS \\
DELHI TECHNOLOGICAL UNIVERSITY\\
(Formerly Delhi College of Engineering)\\
\small BAWANA ROAD, DELHI-110 042, INDIA.
\end{center}
\begin{flushleft}
%\noindent\includegraphics{figure/Sign-Virendra}\\
\textbf{\SubmissionDate\hfill Enroll. No. : \Enrollment}\\
\end{flushleft}
}

\TitlePage
\newpage

\thispagestyle{empty}
\
\newpage

\setcounter{page}{1}

%\TitlePage
%
%\thispagestyle{empty}
%\
%\newpage
%\thispagestyle{empty}\
%
%\ \\

%\fontfamily{rm}\selectfont

\newpage
%\addcontentsline{toc}{chapter}{Title page}
\thispagestyle{empty}
\text{}
    \vspace{8cm}
    \begin{center}
   \large\bf $\copyright$  DELHI TECHNOLOGICAL UNIVERSITY, DELHI, 2022\\ ALL RIGHTS RESERVED.
   \end{center}

\newpage
\
\thispagestyle{empty}
\newpage
\pagenumbering{roman}
\setcounter{page}{1}
\addcontentsline{toc}{chapter}{Declaration page}
%\chapter*{Acknowledgements}
%\thispagestyle{empty}

%\thispagestyle{empty}
\begin{center}
	{ \bf \large  DECLARATION\\ \ \\}
\end{center}

\noindent I declare that the research work reported in this thesis entitled ``\textbf{{\ThesisTitle}}'' for the award of the degree of \emph{Doctor of Philosophy in Mathematics} has been carried out by me under the supervision of \ifthenelse{\isundefined{\SupervisorTwo}}{\emph{\Supervisor}}{{\emph{\Supervisor }} and \emph{\SupervisorTwo}}, Department of Applied Mathematics, Delhi Technological University, Delhi, India.\\

\indent The research work embodied in this thesis, except where otherwise indicated,   is my original research. This thesis has not been submitted by me earlier in part or full to any other University or Institute for the award of any degree or diploma. This thesis does not contain other person's data, graphs, or other information unless specifically acknowledged. \\ \ \\
\\ \ \\
\begin{flushright}
	%\noindent\includegraphics{figure/Sign-Virendra}\\
	{\bf Date : 28.10.2022 \hfill{\bf (\Student)}\\
	}
\end{flushright}

\newpage

\thispagestyle{empty}
\
\newpage
\
\addcontentsline{toc}{chapter}{Certificate page}
\\ \ \\
\begin{center}{ \bf \large
		\underline{CERTIFICATE}\\ \ \\}
\end{center}

\indent On the basis of a declaration submitted by {\bf Ms. \Student}, Ph.D. scholar, I hereby certify that the thesis titled {\bf``\ThesisTitle''} submitted to the Department of Applied Mathematics, Delhi Technological University, Delhi, India for the award of the degree of \emph{Doctor of Philosophy in Mathematics}, is a record of bonafide research work carried out by her under my supervision.

I have read this thesis and, in my opinion, it is fully adequate in scope and quality as a thesis for the degree of Doctor of Philosophy.

To the best of my knowledge, the work reported in this thesis is original and has not been submitted to any other Institution or University in any form for the award of any Degree or Diploma.
\\ \ \\
\begin{flushleft}
	%\noindent\includegraphics{figure/Sign-Virendra}\\
	{\bf (Dr. Satyabrata Adhikari) }\\
	Supervisor and Assistant Professor\\
	Department of Applied Mathematics \\
	Delhi Technological University \\
	Delhi
\end{flushleft}
\vspace{1cm}
\begin{center}
	{\bf (Dr. S. Sivaprasad Kumar)}\\
	Head and Professor\\
	Department of Applied Mathematics\\
	Delhi Technological University \\
	Delhi
\end{center}

\newpage
\thispagestyle{empty}
\
\newpage
%{\fontfamily{sf}\selectfont
\addcontentsline{toc}{chapter}{Acknowledgements}
\begin{center}
	{\bf \large  ACKNOWLEDGEMENTS}
\end{center}
{\textsf{It would not have been possible to write this doctoral thesis without the kind support and help of many individuals around me, to only some of whom it is possible to give a particular mention here. I would like to extend my sincere thanks to all of them.}}

{\textsf{First and foremost, praises and thanks to the GOD, the almighty, for His showers of blessings throughout my research work to complete the research successfully.}}

{\textsf{I would like to express my special gratitude and thanks to my supervisor \Supervisor, Assistant Professor, Department of Applied Mathematics, Delhi Technological University (DTU), Delhi for the continuous support, for his patience, motivation, enthusiasm, and immense knowledge. His guidance helped me in all the time of research and writing of this thesis. It is indeed a great privilege and honor for me to work under his supervision. I am extremely grateful for what he has offered me. Without his guidance and persistent help, this thesis would not have been possible.}}

{\textsf{Besides my supervisor, I would like to thank Prof. S. Sivaprasad Kumar, Head, Department of Applied Mathematics, DTU for providing me the necessary facilities in the Department. }}

{\textsf{My sincere thanks also goes to Prof. Naokant Deo, DRC chairman, Department of Applied Mathematics, DTU for his everlasting support and guidance. I am grateful to Prof. H. C. Taneja, Professor, Department of Applied Mathematics, DTU and all faculty members of the Department of Applied Mathematics, DTU for their constant support and encouragement. }}

{\textsf{I also wish to extend my gratitude to Prof. Tabish Qureshi, Professor and Hony. Director, Centre for Theoretical Physics, Jamia Millia Islamia University, Delhi and Prof. Sibasish Ghosh, Institue of Mathematical Sciences, Chennai for their valuable guidance from the beginning of my Ph.D. research work.}}

{\textsf{I would like to acknowledge the academic and technical support of the Delhi Technological University, Delhi, and its staff, particularly the Academic-PG section for their all kinds of support. The library facilities of the University have been indispensable. I also thank everyone in the Department of Applied Mathematics particularly Mr. Akhil, Ms. Nisha, Mr. Anurag, Mr. Narendra and Ms. Anju for their support and assistance.}}

{\textsf{I am extremely grateful to my family and relatives for their love, prayers, caring, and sacrifices for educating me and preparing me for my future. Special thanks to my parents and husband, they were the pillar of strength and support for me in this entire journey.}}

{\textsf{I place my acknowledge to Ms. Anuma Garg and Ms.Shruti Aggarwal for valuable discussions on the subject matter and for their support. I also place on record, my sense of gratitude to one and all who, directly or indirectly, have lent their helping hand in this journey. }}

{\textsf{My acknowledgements would not be complete without thanking the Council of Scientific and Industrial Research (CSIR), Ministry of Science and Technology, Government of India, for providing fellowship (JRF and SRF) that made my Ph.D. work possible.}}\\\\

\noindent {\bf Date : 28.10.2022} \hfill {\bf (\MakeTextUppercase\Student)}
\\\noindent {\bf Place : DTU, Delhi, India.}
%\newpage\null
%\newpage
%\end{sloppypar}

\newpage
\text{}
\vspace{6cm}
\begin{center}
	\bf{\Huge{ \emph{
				Dedicated}}}\\
	\vspace{1cm}
	\bf{\Huge{ \emph{to}}}\\
	\vspace{1cm}
	\bf{\Huge{ \emph{My Family
			}
		}
	}
\end{center}

{\singlespacing\tableofcontents}
\newpage

%{\fontfamily{sf}\selectfont

\addcontentsline{toc}{chapter}{Preface}
\chapter*{Preface}
%\thispagestyle{empty}
%\noindent\hrulefill
The theory of quantum entanglement is a burgeoning field of quantum information theory. The fact that measurements on spatially separated systems cannot always be described by a locally causal theory is one of the most remarkable features in quantum physics. Although the basic understanding of the bipartite system has been established over the last few decades, still there are many unanswered questions about the entanglement properties of the multipartite system. It is particularly unclear how several concepts from the bipartite case can be generalized to multipartite scenarios in a meaningful way. The detection and classification of multipartite entanglement are discussed in this thesis. The differences and similarities between the bipartite and multipartite situations are discussed, various possible generalizations are presented, and results in several areas are obtained.\\
\noindent The detection and classification of entanglement properties in a multi-qubit system is a topic of great interest. This topic has been extensively studied, and as a result, we discovered various approaches for detecting and classifying multi-qubit, in particular three-qubit entangled states. The emphasis of this work is on a formalism of methods for the detection and classification of bipartite as well as multipartite quantum systems. We have used the method of structural physical approximation of partially transposed matrix (SPA-PT) for the detection of entangled states in arbitrary dimensional bipartite quantum systems. Also, we have proposed criteria for the classification of all possible stochastic local operations and classical communication (SLOCC) inequivalent classes of a pure and mixed three-qubit state using the SPA-PT map. To quantify entanglement, we have defined a new measure of entanglement based on the method of SPA-PT, which we named as "structured negativity". We have shown that this measure can be used to quantify entanglement for negative partial transposed entangled states (NPTES). Since the methods for detection, classification and quantification of entanglement, defined in this thesis are based on SPA-PT, they may be realized in an experiment.\\
\noindent Three-qubit systems have two types of SLOCC inequivalent genuine entangled classes. These classes are known as the GHZ class and the W class. The GHZ class has been proven to be very useful for various quantum information processing tasks such as quantum teleportation, controlled quantum teleportation, and so on. We divide pure three-qubit states from the GHZ class into four distinct subclasses namely $S_{1}$, $S_{2}$, $S_{3}$,
$S_{4}$ and shown that three-qubit states either belong to   $S_{2}$ or $S_{3}$ or
$S_{4}$ may be more efficient than the three-qubit state belonging to $S_{1}$. Moreover, we have constructed various witness operators that can classify the subclasses and demonstrated that the constructed witness operator can be decomposed into Pauli matrices and thus experimentally realized.\\
\noindent Also, we have used coherence to detect and classify the entanglement property of three-qubit states. We have obtained the necessary conditions to discriminate biseparable states from other classes of the three-qubit system. Moreover, we have also obtained the necessary condition that may classify the separable states from other classes of three-qubit entangled states.  Since there are only three types of states in a three-qubit system, so if we found that the detected state is neither a separable nor a biseparable state, we may conclude that the given three-qubit state is a genuine entangled state.\\ \ \\ \ \\

\noindent {\bf Date : 28.10.2022 } \hfill {\bf (\MakeTextUppercase\Student)}
%\hfill\includegraphics{figure/Sign-Virendra}
\\\noindent {\bf Place : Delhi, India}

%}
\newpage
%\
%\thispagestyle{empty}
%list of symbols
%\printnomenclature
%\newpage
%\
%\newpage
%\input{Abreviation/abrvn}
\addcontentsline{toc}{chapter}{List of figures}
{\singlespacing\listoffigures}

\newpage

\addcontentsline{toc}{chapter}{List of tables}
{\singlespacing\listoftables}

\newpage
\
\thispagestyle{empty}
\newpage
\properpagestyle
%\setcounter{page}{1}
%\fontfamily{sf}\selectfont
\pagenumbering{arabic}
\chapter{General Introduction}\label{ch1}
\vspace{1.5cm}
\noindent\hrulefill

\noindent\emph{The introductory chapter consists of basic definitions, basic concepts of linear algebra, and the preliminaries of the results obtained in the literature. We then provide a brief review of the theory of bipartite and multipartite entanglement. In the theory of bipartite entanglement, we have discussed various entanglement detection criteria for the two-qubit system as well as the higher dimensional bipartite quantum system. In the case of multipartite entanglement, we have specifically considered a three-qubit system and studied different schemes for the classification of different classes that may exist in the tripartite entangled system.}\\%Finally, the motivation and plan of the research are discussed.}
\noindent\hrulefill
\newpage
%\noindent \textbf{Introduction}\\

%\title {A Study on Coefficient Estimates, Radius Problems and Differerential Subordination Results For Certain Class of Analytic Functions}
\date{}
\setlength{\parskip}{3pt}
% --------------------------------------------------------------
\noindent The study of data storage and transmission through a noisy channel is known as information theory. It focuses on the quantification of data for communication purposes. It is concerned with determining how much information is contained in a message. Since information is related to uncertainty or randomness, so, to quantify the amount of information, we need to determine the amount of uncertainty. The amount of uncertainty in the value of a random variable or the outcome of a random process can be measured in entropy and thus, entropy can be considered an important metric in information theory. For example, determining the result of a fair coin flip (with two equally likely outcomes) offers less information (lower entropy) than determining the result of a dice roll (with six equally likely outcomes).\\ %There are other useful measures of information by which we can quantify the amount of information are mutual information, channel capacity, error exponents, and relative entropy.\\
\noindent Nowadays, we can split the whole concept of information theory into two parts: (i) classical information theory and (ii) quantum information theory. Classical information theory is the mathematical theory of information-processing tasks utilizing laws of classical physics whereas, quantum information theory is the study of how such tasks can be done using quantum mechanical systems. Quantum Information Theory is concerned with the use of quantum-mechanical features of physical systems to facilitate effective data storage and transfer. Quantum and classical information theories differ significantly due to the underlying features of quantum physics that do not exist in classical physics. For instance, quantum entanglement is one such feature in quantum physics that has no classical analog.\\

\section{Basics of Linear Algebra}
\noindent Let us first recapitulate some basic concepts of linear algebra that would be needed to discuss the topics of quantum information theory. The basic objects of linear algebra are vector spaces. In quantum mechanics, we generally consider the vector space over the field of complex numbers $C$. The elements of vector spaces are known as vectors. They can be represented by column vectors.\\
\textbf{I. Linearly independent and dependent vectors:}
A set of vectors $V=\{v_1,v_2,...,v_n\}$ is said to be linearly independent if there exist scalars $a_1,a_2,....,a_n$ such that 
\begin{eqnarray}
a_1v_1+a_2v_2+...+a_nv_n=0 \implies a_i=0 ~\forall~ i
\end{eqnarray} 
otherwise, the set $V$ is said to be linearly dependent.\\
\textbf{II. Basis:} A basis $\beta$ of a vector space is the spanning set $\beta=\{v_1,v_2,...,v_n\}$ such that every vector in the vector space can be expressed as a linear combination of vectors present in the basis set such that the set $\beta$ is linearly independent. Let $v$ be any vector in the vector space, then $v$ can be expressed as $v=\sum_i{\alpha_iv_i},\alpha \in C$. The number of elements in a basis set is the dimension of the vector space.\\
\textbf{III. Linear space:} A linear space (or vector space) over the field $F$ is a set $X$ which satisfies the following properties:\\
	(i) $x+y \in X$ $\forall$ $x,y \in X$ \\
	(ii) $x+y=y+x$ $\forall$ $x,y \in X$\\
	(iii) $(x+y)+z=x+(y+z)$ $\forall$ $x,y,z \in X$\\
	(iv) There is a zero element (additive identity), denoted by, 0 in $X$ such that\\ $x+0=x,~\forall x\in X$\\
	(v) For every $x\in X, \exists~\text{additive inverse} -x\in X$ such that $x+(-x)=0=-x+x$\\
	(vi) For every $\alpha \in F$ and $x\in X$, $\alpha x\in X$\\
	(vii) For every $\alpha,\beta \in F$ and $x,y\in X$:\\
	(a) $\alpha(\beta x)=(\alpha \beta)x$\\
	(b)$(\alpha+\beta)x=\alpha x+\beta x$\\
	(c)$\alpha(x+y)=\alpha x+\alpha y$\\
	(viii) There is a unit element (multiplicative identity), denoted by 1, in $F$ such that $1.x=x,~\forall x\in X$.\\
	\textbf{IV. Complete linear space:} A normed linear space is said to be complete if every Cauchy sequence converges in the space.\\
	\textbf{V. Norm:} Consider a vector space $X$ defined over a field $F$. A norm $||.||$ on X is defined as a function
	\begin{eqnarray}
	f: X\rightarrow F
	\end{eqnarray}
	which satisfies the following properties:\\
	1. $||\textbf{x}||\geq 0,~\forall~\textbf{x}\in X$.\\
	2. $||\textbf{x}||=0$ $\iff$ $\textbf{x}=0$.\\
	3. $||\alpha \textbf{x}||=|\alpha| ||\textbf{x}||$,~$\forall \textbf{x} \in X,~ \forall \alpha \in F$.\\
	4. $||\textbf{x}+\textbf{y}||\leq ||\textbf{x}||+||\textbf{y}||$, $\forall \textbf{x},\textbf{y} \in X$.\\
	\textbf{VI. Normed linear space:} A normed linear space is a linear vector space over the field of real or complex numbers, on which a norm is defined.\\
	For example, the set $C[a,b]$ of all continuous functions $f(x),~x\in[a,b]$,~is a normed linear space with respect to norm defined as
	\begin{eqnarray}
	||f||_p=[\int_{a}^{b}|f(x)|^p]^{\frac{1}{p}},~~p\geq 1
	\end{eqnarray}
	\textbf{VII. Banach Space:} A complete normed linear space is known as a Banach space.\\
	For instance, The set of all real valued continuous functions $x=x(t)$ defined on the set $[a,b]$, denoted by $C[a, b]$, is a Banach space with respect to the norm defined as
	\begin{eqnarray}
	||x||=\sup_{t\in[a,b]}{|x(t)|},~~x=x(t)\in C[a,b]
	\end{eqnarray}
	%\begin{eqnarray}
	%||f||=\int_{a}^{b}|f(x)|dx
	%\end{eqnarray}
	%where $f(x)$ is a real valued function on $[a,b]$.\\
	\textbf{VIII. Inner Product:} Inner product on a complex linear space $X$ may be defined as a map:
	\begin{eqnarray}
	<.,.>: X \times X\rightarrow C
	\end{eqnarray}
	If we take three arbitrary vectors $\textbf{a},\textbf{b},\textbf{c} \in X$ and two scalars $\alpha, \beta \in C$, then the following properties holds\\
	(a) $<\textbf{a},\alpha \textbf{b}+\beta \textbf{c}>=\alpha^{*}<\textbf{a},\textbf{b}>+\beta^{*}<\textbf{a},\textbf{c}>$, where $\alpha^{*}$ and $\beta^{*}$ denote the complex conjugate of $\alpha$ and $\beta$ respectively.\\
	(b) $<\textbf{b},\textbf{a}>=\overline{<{\textbf{a},\textbf{b}}>}$\\
	(c) $<\textbf{a},\textbf{a}>\geq 0$\\
	(d) $<\textbf{a},\textbf{a}>=0$ if and only if $\textbf{a}=0$.\\
	If $X(F)$ is a vector space with inner product $<.,.>$ defined on $X$. Then, a norm can be defined from inner product as
	\begin{eqnarray}
	||a||=\sqrt{<a,a>},~ \forall a\in X
	\end{eqnarray}
	\textbf{IX. Inner product space:} Let $X$ be a linear space over the field of complex number $C$. Then, X is called an inner product space if for every pair of elements $x,y \in X$, $\exists$ a complex number denoted by $<x,y>$ and is called the inner product of $x$ and $y$. For example, the $n$ dimensional space $R_n$ or $C_n$ forms an inner product space. Let $x=(\xi_1,\xi_2,...,\xi_n)$ and $y=(\nu_1,\nu_2,...,\nu_n)$ be two elements of $C_n$. The inner product of $x$ and $y$
 is denoted by $<x,y>$ and is given by the formula
 \begin{eqnarray}
 <x,y>=\sum_{i=1}^{n}{\xi_i\nu_i^{*}}
 \end{eqnarray}	
 With the help of this inner product , the length of $x=(\xi_1,\xi_2,...,\xi_n)$ can be written as
 \begin{eqnarray}
 ||x||=(\sum_{i=1}^{n}|\xi_i|^2)^{\frac{1}{2}}=\langle x,x\rangle^{\frac{1}{2}}
 \end{eqnarray}
	%For example, the set of $n\times n$ matrices over the field of complex numbers, $M_n(C)$ forms an inner product space.\\
	\textbf{X. Hilbert Space:} Hilbert space is a normed linear space which is complete with respect to the norm derived by the inner product.\\
	(i) A Hilbert space is an inner product space and is complete.\\
	(ii) Inner product spaces are normed linear spaces.\\
	(iii) Hilbert spaces are Banach spaces.\\
	(iv) If for an inner product space, the scalar field is the set of real numbers, then we obtain real inner product space and corresponding real Hilbert space.\\
	(v) In real Hilbert space, we simply have $<x,y>=<y,x>$.\\
	(vi) If for an inner product space, the scalar field is the set of complex numbers, then we obtain complex inner product space and corresponding complex Hilbert space.\\
	\textbf{XI. Orthogonal vectors:} In a vector space, two vectors are said to be orthogonal if their inner product is zero.\\
	\textbf{XII. Unit vector:} If $||a||=1$, then $a$ is known as unit vector.\\
	\textbf{XIII. Orthonormal vector:} A set of vectors $\{v_1,v_2,...,v_n\}$ is said to be orthonormal if each vector $v_i$ in the set is a unit vector and distinct vectors $v_i$ and $v_j$ are orthogonal i.e. $\langle v_i|v_j\rangle = \delta_{ij}$, $i,j\in\{1,2,...n\}$.\\
	\textbf{XIV. Linear operator:} Let $X$ and $Y$ be two vector spaces over the field $F$. An operator $T: X\rightarrow Y$ is said to be linear if
	%An operator T from a vector space $X$ to another vector space $Y$ is said to be linear if 
	\begin{eqnarray}
	T(\sum_i{c_iv_i})=\sum_i{c_iTv_i},~~v_i\in X,~~c_i\in F
	\end{eqnarray}
	A linear operator on a vector space $X$ is a linear transformation from $X$ into $X$. Linear operators may be expressed in terms of matrices.\\
	\textbf{XV. Matrix representation:} %For two vector spaces $X$ and $Y$, a linear operator may be defined as a tranformation $T:X\rightarrow Y$ which maps
	%\begin{eqnarray}
	%T(\sum_i{a_iv_i)}=a_iT(\sum_iv_i)
	%\end{eqnarray}
	%Linear operators may be understood in terms of their matrix representations. 
	Consider two vector spaces $X$ and $Y$ of dimension $m$ and $n$ respectively. Then a linear operator from $X$ to $Y$ may be expressed by a $(m\times n)$ matrix. Let $\beta_1=\{v_1,v_2,...v_m\}$ be basis of $X$ and ${\beta_2}=\{w_1,w_2,...w_n\}$ be basis of $Y$, then 
	\begin{eqnarray}
	T(v_j)=\sum_{i=1}^{m}{a_{ij}w_i}, j\in \{1,2,...,m\}
	\end{eqnarray}
	Then, the matrix of $T$ with respect to the basis ${\beta_1}$ and ${\beta_2}$ can be represented as $A=[a_{ij}]_{m,n}$.\\
	\textbf{XVI. Eigenvalue and eigenvector:} Let $X$ be a vector space over the field of complex numbers $C$ and $T: X\rightarrow X$ be a linear operator. An eigenvector of a linear operator $T$ is a non zero vector $x\in X$ such that $Tx=\lambda x$, $\lambda\in C$ is known as the eigenvalue of the operator $T$ corresponding to eigenvector $x$. Eigenvalues are the solutions of the characteristic equation, which is given by $det|T-\lambda I|=0$ where $det$ denote the determinant and $I$ denote the identity operator.\\
	\textbf{XVII. Transpose of a matrix:} The transpose of a matrix can be defined by interchanging the rows of the matrix into columns or by transforming the columns of the matrix into rows. For a given matrix $A$, the transpose of the matrix $A$ is represented by $A^{T}$.\\
	\textbf{XVIII. Conjugate transpose  of a matrix:} For a linear operator $A$, the conjugate transpose of $A$ is obtained by applying transposition operation on $A$ and applying complex conjugate on each entry. For a given matrix $A$, the conjugate transpose of the matrix $A$ is represented $A^{\dagger}$.\\
	\textbf{XIX. Normal operator:} An operator $A$ is said to be normal if and only if
	\begin{eqnarray}
	AA^{\dagger}=A^{\dagger}A
	\end{eqnarray}
	\textbf{XX. Orthogonal operator:} An operator $A$ is said to be orthogonal if and only if
	\begin{eqnarray}
	AA^{T}=A^{T}A=I
	\end{eqnarray}
	\textbf{XXI. Unitary operator:} An operator $A$ is said to be unitary if and only if
	\begin{eqnarray}
	AA^{\dagger}=A^{\dagger}A=I
	\end{eqnarray}
	\textbf{XXII. Hermitian operator:} An operator $H$ is said to be Hermitian if and only if
	\begin{eqnarray}
	H^{\dagger}=H
	\end{eqnarray}
	The eigenvalues of a Hermitian operators are real numbers.\\
	For example, Pauli Matrices are Hermitian matrices and it can be represented as\\
	\begin{eqnarray}
	\sigma_0 = I =
	\begin{pmatrix}
	1 & 0\\
	0 & 1
	\end{pmatrix}, \sigma_x=
	\begin{pmatrix}
	0 & 1\\
	1 & 0
	\end{pmatrix},
	\sigma_y=
	\begin{pmatrix}
	0 & -i\\
	i & 0
	\end{pmatrix},\sigma_z=
	\begin{pmatrix}
	1 & 0\\
	0 & -1
	\end{pmatrix}
	\label{paulimatrices}
	\end{eqnarray}
	\textbf{XXIII. Properties of Pauli matrices:}\\
	(i) Pauli matrices are Hermitian and unitary.\\
	(ii) $\sigma_x^2=\sigma_y^{2}=\sigma_z^2=I$, where I is the identity matrix.\\
	(iii) det($\sigma_i$)=-1,~~i=x,y,z\\
	(iv) Tr($\sigma_i$)=0,~~i=x,y,z\\
	(v) Eigenvalues of Pauli matrices are $\pm$ 1.\\
	(vi) $[\sigma_x,\sigma_y]=2i\sigma_z$, $[\sigma_y,\sigma_z]=2i\sigma_x$, $[\sigma_z,\sigma_x]=2i\sigma_y$, where $[.]$ denote the commutation relation.\\
	(vii)$\{\sigma_i,\sigma_j\}=0,~~i\neq j,~~i,j\in\{x,y,z\}$, where $\{.\}$ denote anti-commutation relation.\\
		\textbf{XXIV. Partial transpose of a matrix:} Given $A=[A_{i,j}]_{i,j=1}^{n} \in M_n(M_k) $, it's partial transpose $A^{\tau}$ is given by,
	\begin{eqnarray}
	%                       \nonumber % Remove numbering (before each equation)
	A^{\tau}=[{A^{*}_{i,j}}]_{i,j=1}^{n}
	\end{eqnarray}
	Here $M_n(M_k)$ is the set of nxn block matrices with each block in $M_k$.\\
	A state $\rho_d$ in $d\otimes d$ dimensional quantum system is represented by a $d^2\times d^2$ density matrix. This $d^2\times d^2$ density matrix may be divided into $d$ submatrices of order $d\times d$. The partial transposition  of the state $\rho_d$ may be calculated by taking the transposition of each of the submatrices. 
	For example, let us consider the density matrix $\rho_2^{AB}$ of a two-qubit composite system $AB$. It is represented by a $4\times 4$ matrix with 4 submatrices of size $2\times 2$. The partial transposition of $\rho_2^{AB}$ can be calculated by taking the transposition of each of the submatrices.
	\begin{eqnarray}
	\rho_2^{AB}=
	\left( \begin{array}{c c|c c}
	a_{11} & a_{12} & a_{13} & a_{14}\\ 
	a_{21} & a_{22} & a_{23} & a_{24}\\
	\hline
	a_{31} & a_{32} & a_{33} & a_{34}\\
	a_{41} & a_{42} & a_{43} & a_{44}
	\end{array}\right) \underrightarrow{T_B}
	\left( \begin{array}{c c|c c}
	a_{11} & a_{21} & a_{13} & a_{23}\\
	a_{12} & a_{22} & a_{14} & a_{24}\\
	\hline
	a_{31} & a_{41} & a_{33} & a_{43}\\
	a_{32} & a_{42} & a_{34} & a_{44}
	\end{array}\right)
	=({\rho_2^{AB}})^{T_B}
	\end{eqnarray} 
	where $T_B$ denote the partial transposition with respect to the subsystem $B$.\\
	\textbf{XXV. Positive Partial Transpose:}
	A bipartite density matrix $\rho$ has a \textbf{positive partial transpose} (or the matrix is \textbf{PPT}) if it's partial transposition has no negative eigenvalues i.e. it is positive semidefinite.\\
	\textbf{XXVI. Negative Partial Transpose:}
	A bipartite density matrix $\rho$ has a \textbf{negative partial transpose} (or the matrix is \textbf{NPT}) if it's partial transposition has atleast one negative eigenvalues.\\
		\textbf{XXVII. Realignment operation on a matrix:} For a matrix $X=(x_{ij})\in \textbf{C}^{m\times n}$, the vector $vec(X)$ is defined as
	\begin{equation}
	vec(X)=(x_{11},...,x_{m1},x_{12},...,x_{m2},...,x_{1n},...,x_{mn})^T
	\end{equation}  
	where $T$ denote the transposition operator.\\
	Let $Y$ be an $m\times m$ block matrix with $n\times n$ subblocks $Y_{i,j}, i,j=1,...,m$. Then the realignment matrix of Y is defined as
	\begin{equation}
	R(Y)=(vec(Y_{1,1})...,vec(Y_{m,1}),...,vec(Y_{1,m}),...,vec(Y_{m,m}))^T
	\end{equation}
	A state in $d\otimes d$ dimensional system may be represented by  $d^2\times d^2$ density matrix and this matrix may be further divided in $d$ submatrices of order $d\times d$. The realigned matrix may be obtained by expressing the element of each submatrix in row.
	For example for a two-qubit density matrix $\rho$, the realigned matrix $\rho^{R}$ may be obtained as
	\begin{eqnarray}
	\rho=
	\left( \begin{array}{c c|c c}
	a_{11} & a_{12} & a_{13} & a_{14}\\ 
	a_{21} & a_{22} & a_{23} & a_{24}\\
	\hline
	a_{31} & a_{32} & a_{33} & a_{34}\\
	a_{41} & a_{42} & a_{43} & a_{44}
	\end{array}\right) \underrightarrow{R}
	\left( \begin{array}{c c c c}
	a_{11} & a_{12} & a_{21} & a_{22}\\
	\hline
	a_{13} & a_{14} & a_{23} & a_{24}\\
	\hline
	a_{31} & a_{32} & a_{41} & a_{42}\\
	\hline
	a_{33} & a_{34} & a_{43} & a_{44}
	\end{array}\right)
	=\rho^{R}
	\end{eqnarray}
	\textbf{XXVIII. Inner product of two operators:}  For two operators $A$ and $B$ in finite dimensional Hilbert space(\textbf{H}), inner product of $A$ and $B$ may be defined as
	\begin{eqnarray}
	\langle A,B\rangle=Tr[A^{\dagger}B]
	\end{eqnarray}
	where $A^{\dagger}$ denotes the complex conjugate of A and the corresponding norm may be defined as
	\begin{eqnarray}
	||A||=\sqrt{\langle A, A\rangle}, ~~A\in \textbf{H}
	\end{eqnarray}
	\textbf{XXIX. Positive and Completely positive map:}
	Let $\gamma:H_1 \rightarrow H_2$ be a linear map. $\gamma$ is said to be a positive map if it maps positive operators of $H_1$ to positive operators of $H_2$, i.e.
	\begin{eqnarray}
	\gamma(A) \geq 0,~\forall A\geq 0
	\end{eqnarray}
	A positive map $\gamma$ is said to be a completely positive if the map defined by,
	\begin{eqnarray}
	\gamma \otimes I_d:H_1 \otimes M_d \rightarrow H_2 \otimes M_d
	\end{eqnarray}
	is also positive for d=2,3,4,... where $I_d$ is the identity matrix of the matrix space $M_d$ of $d\times d$ matrices.	
%	In general,  an operator A $\in$\textbf{A} can be expressed as a
%	matrix with the elements
%	\begin{eqnarray}
%	A_{ij}=\langle e_i|A|e_j\rangle
%	\label{matrixelement}
%	\end{eqnarray}
%	where $e_i$ and $e_j$ are the vectors of an arbitrary basis ${e_i}$ of the Hilbert Space.
%	and same holds for the states, since they are operators.
	\section {Quantum states}
	\noindent Quantum states can be considered as vectors in a Hilbert space (H). The notation used in quantum mechanics for a vector belong to a vector space is given by
	\begin{eqnarray}
	|\psi\rangle=\begin{pmatrix}
	z_1\\
	z_2\\
	.\\
	.\\
	.\\
	z_n
	\end{pmatrix}
	\end{eqnarray}
	where $\psi$ is the label for the vector and the notation $|.\rangle $ denotes the object is a vector. This notation is also known as a ket vector. Here, $z_i\in C, i\in \{1,2,...n\}$ denote the component of the $|\psi\rangle$. Some standard notations related to the description of a quantum mechanical system are given below\\
	(i) $z^{*}$ denote the complex conjugate of $z$.\\
	(ii) $\langle \psi|$ is the dual of $|\psi\rangle$. It is known as bra notation of vector $\psi$.\\
	(iii) $\langle \psi|\phi\rangle$ denote the inner product of two vectors $|\psi\rangle$ and $|\phi\rangle$.\\
	(iv) $|\psi\rangle \otimes |\phi\rangle$ denote the tensor product of two vector $|\psi\rangle$ and $|\phi\rangle$.
	\subsection{Qubit}
	\noindent Quantum bits are used to define quantum states. Qubits, or quantum bits, are the most fundamental unit of quantum information, just as bits are the most fundamental unit in classical information theory. In a two-dimensional Hilbert space, a qubit can be written as a linear combination of two classical bits $|0\rangle$ and $|1\rangle$. Mathematically, a qubit can be expressed as
	\begin{eqnarray}
	|\psi\rangle=\alpha|0\rangle+\beta|1\rangle,~~ |\alpha|^2+|\beta|^2=1
	\label{psi}
	\end{eqnarray}
	where $\alpha$ and $\beta$ are complex numbers. $|\alpha|^2+|\beta|^2=1$ represent the normalization condition. Geometrically, the normalization condition represents all the quantum states  $|\psi\rangle$ that lie on the boundary of the sphere.
	\subsection{Pure and mixed state}
	\noindent A quantum system can be found in any of the following two forms: pure states and mixed states. A pure state is always represented by a vector in a Hilbert space while a mixed state can be expressed as a convex combination of more than one pure state. A mixed state is denoted by an operator $\rho$ which is associated with some ensemble $\{p_i,|\psi_{i}\rangle\}$. Mathematically, a mixed state $\rho$ can be expressed in terms of the ensemble $\{p_i,|\psi_{i}\rangle\}$ as
	\begin{eqnarray}
	\rho=\sum_{i}{p_i|\psi_i\rangle \langle \psi_i|},~~\sum_{i}{p_i}=1,~~0\leq p_i\leq 1,~i=1,2,...
	\end{eqnarray}
	An operator $\rho$ which represents a mixed state is popularly known as a density operator. An operator $\rho$ is said to be the density operator if it satisfies the following properties:
	\begin{eqnarray}
	&&(i)~~Tr[\rho]=1\\&&
	(ii)~~\rho\geq 0~~\text{i.e. $\rho$ has non-negative eigenvalues}
	\end{eqnarray}
	In terms of $Tr[\rho^{2}]$, the pure state and the mixed state may be classified as\\
	(i) $\rho$ is a pure state if and only if $Tr(\rho^{2})=1$.\\
	(ii) $\rho$ is a mixed state if and only if $Tr(\rho^{2})<1$.
	\subsection{Geometrical interpretation of a qubit}
	\noindent Geometrically, a two-level quantum mechanical system can be represented by a Bloch sphere. A two-level quantum mechanical system is described by four parameters in a complex plane. But the normalization condition reduces one parameter and therefore the system finally is described by three parameters which can be represented in a Bloch sphere \cite{bloch,nielsen}. The state lies on the surface of the sphere are pure states of the form
	\begin{eqnarray}
	|\psi\rangle=Cos(\frac{\theta}{2})|0\rangle+e^{i\phi}Sin(\frac{\theta}{2})|1\rangle,~~0\leq \theta \leq \pi,~~0\leq \phi <2\pi
	\end{eqnarray}
	The states lying inside the sphere are mixed states given by
	\begin{eqnarray}
	\rho^{(2)}=\frac{1}{2}(I+\vec{a}.\vec{\sigma})
	\end{eqnarray}
	where $\vec{a}=(a_x,a_y,a_z)\in R^{3}$ is a vector in three dimensional space and $\vec{\sigma}=(\sigma_x,\sigma_y,\sigma_z)$ denotes the Pauli matrices. The center of the sphere represent a maximally mixed state which is described by the identity matrix.
	\begin{figure}
		\centering
		\includegraphics[width=5cm]{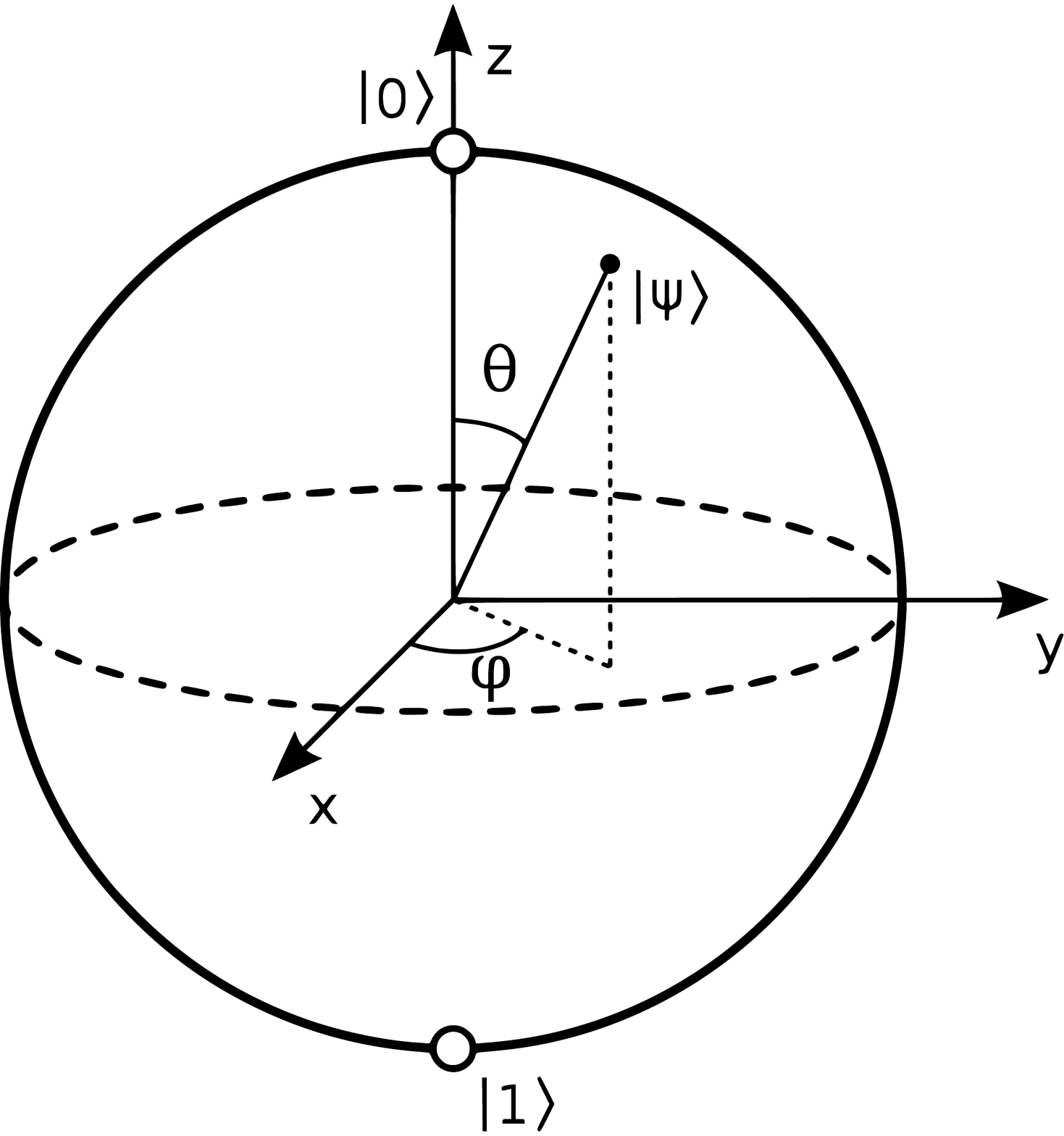}
		\caption{Bloch sphere}
		\label{kk}
	\end{figure}\\
	\subsection{Bloch vectors for $d$-dimensional quantum system}
	\noindent Generalizing the idea of two-dimensional system to $d$-dimensional system, we call the quantum state lying in $d$-dimensional Hilbert space as qudit. Any matrix basis that can be used for Bloch vector decomposition in case of qudits must satisfy the following two properties:\\
	(i) The set of basis elements must contain $I$ and $d-1$ traceless matrices $\{A_i\}$ of order $d\times d$.\\
	(ii) The matrices $\{A_i\}$ must be orthogonal.\\
	The Bloch vector expansion  of the density matrix $\rho^{(d)}$ may be expressed as
	\begin{eqnarray}
	\rho^{(d)}=\frac{1}{d}I+\vec{b}.\vec{\Lambda}
	\label{d dimension}
	\end{eqnarray}
	where $\vec{b}.\vec{\Lambda}$ is a linear combination of $\vec{b} \in R^{d^2-1}$ and the element of basis matrices $\{A_i\}$. The coefficient $b_i$ may be calculated as $b_i=\langle \Lambda_i \rangle=Tr[\rho^{(d)}\Lambda_i]$. The vector $\vec{b}$ is known as the Bloch vector.
	%In the $d\otimes d$ system, there are different bases that can be used to express qudit in $(d^2-1)$ dimensional vector. 
	Unlike the two-qubit case, the map induced is not bijective this means that it is not necessary that every point on the Bloch sphere in $d^2-1$ dimension corresponds to a physical state \cite{kimura}.\\ %However, the pure states lie on the boundary of the Bloch sphere whereas the mixed states lie in the open Bloch ball.\\
	%	The density matrix corresponding to a qudit state  in $d$-dimensional Hilbert space $H^{d}$ may be expressed in terms of the standard basis $\{|k\rangle \}$ where $k=0,1,...,d-1$. 
	%	Thus, the density matrix $\rho$ corresponding to a qudit state can be decomposed using the matrix basis $\{A_i\}$ and \\
	%	\textbf{Note:} A given density matrix can be decomposed into a Bloch vector but it is not necessary that any vector $\rho_{1} $ of the for (\ref{d dimension}) always corresponds to a density matrix, even if $Tr(\rho_{1})=1$ and $Tr(\rho_{1}^{2}) \leq 1$ [since it may not imply $\rho_1 \geq 0$].\\
	The geometric properties of the Bloch sphere in higher dimensions are rather complex as distinct matrix basis induces a different Bloch vector in a Bloch hypersphere, however, every point on the Bloch hypersphere may not correspond to a physical state but all distinct hyper balls are isomorphic as they correspond to the density matrix $\rho$.
	\subsection{Representing a $d$-dimensional quantum state in terms of different basis}
	\noindent Since there are various basis such as the generalized Gell-Mann matrix basis, the polarization operator basis, Weyl operator basis so it is important to know which matrix basis is ideal for a certain goal, like calculating the entanglement degree \cite{bertlmann}. Here we will define the generalized Gell-Mann matrix basis, the polarization operator basis, Weyl operator basis, and then we will compare these three bases for a particular qutrit-qutrit example.\\ \ \\
	\textbf{I. The generalized Gell-Mann matrix (GGM) basis}\\
	The GGM \cite{kimura} are the extensions of Pauli matrices (the basis for a two-qubit system) and the Gell-Mann matrices (the basis for a two-qutrit system). They are the generators of special unitary matrices of order d (SU(d)). They are defined by three different operators as 
	\begin{enumerate}
		\item $\frac{d(d-1)}{2}$ symmetric GGM
		\begin{eqnarray}
		\Lambda_s^{jk}=|j\rangle \langle k|+|k\rangle \langle j|,~~1\leq j< k \leq d
		\end{eqnarray}\item $\frac{d(d-1)}{2}$ antisymmetric GGM
		\begin{eqnarray}
		\Lambda_{a}^{jk}=-i||j\rangle \langle k|+i|k\rangle \langle j|,~~1\leq j< k \leq d
		\end{eqnarray}
		\item $d-1$ diagonal GGM
		\begin{eqnarray}
		\Lambda^{l}=\sqrt{\frac{2}{l(l+1)}}(\sum_{j=1}^{l}|j\rangle \langle j|-l|l+1\rangle \langle l+1|),~~ 1\leq l \leq d-1
		\end{eqnarray}
	\end{enumerate}
	Thus, there are $d^2-1$ hermitian and traceless GGM. Also, they are orthogonal and form a basis for a density matrix of order $d$.\\
	\textbf{Decomposition of standard matrix basis into GGM basis}\\
	The standard matrix basis is formed by using $d\times d$ density matrices such that only one entry has value 1 and all other entries have value 0. These matrices form an orthonormal basis for the Hilbert-Schmidt space. In terms of operators, they can be expressed as
	\begin{eqnarray}
	|j\rangle \langle k|, ~~j,k=1,...,d
	\end{eqnarray} 
	In general, the Bloch vector expansion of the density matrix using the GGM basis is shown below:
	\begin{eqnarray}
	\rho_{GGM}=\frac{1}{d}I+\vec{b}.\vec{\sigma}
	\end{eqnarray}
	where $\vec{b}=(\{b_s^{jk}\},\{b_a^{jk}\},\{b^l\})$ and the components are given by, $b_s^{jk}=Tr[\Lambda_s^{jk}\rho_{GGM}]$, $b_a^{jk}=Tr[\Lambda_a^{jk}\rho_{GGM}]$ and $b^{l}=Tr[\Lambda^{l}\rho_{GGM}]$ for $1\leq j< k\leq d$ and $1\leq l\leq d-1$. All Bloch vectors are contained in a hypersphere of a radius $|\vec{b}|\leq \sqrt{(d-1)/2d}$ \cite{bertlmann}.\\
	The elements of a density matrix in standard matrix basis may be decomposed into GGM basis as
	\begin{eqnarray}
	|j\rangle \langle k|=\begin{cases}\{\frac{1}{2}(\Lambda_s^{jk}+i\Lambda_a^{jk}) & j<k\\
	\frac{1}{2}(\Lambda_s^{kj}-i\Lambda_a^{kj}) & j>k\\
	\sqrt{\frac{j-1}{2j}}\Lambda^{j-1}+\sum_{n=0}^{d-j-1}{\frac{1}{\sqrt{2(j+n)(j+n+1)}}}\Lambda^{j+n}+\frac{1}{d}I & j=k\end{cases}
	\label{GGB}
	\end{eqnarray}\\
	\textbf{II. The polarization operator basis (POB)}\\
	For a $d\times d$ density matrix, POB may be defined as
	\begin{eqnarray}
	T_{LM}=\sqrt{\frac{2L+1}{2s+1}}\sum_{k,l=1}^{d}{C_{sm_l,LM}^{sm_k}{|k\rangle \langle l|}}
	\label{pobtlm}
	\end{eqnarray}
	where $s=\frac{d-1}{2}$, $L=0,1,...,2s$, $M=-L,-L+1,...,L-1,L$, $m_1=s$, $m_2=s-1$,...,$m_d=-s$. The coefficients $C_{sm_l,LM}^{sm_k}$ are Clebsch-Gordan coefficients $C_{j_1m_1,j_2m_2}^{jm}$ of the angular momentum theory and explicit expression are given in \cite{varshalovich}.\\
	The polarization operator is a multiple of the Identity matrix for L=M=0 \cite{varshalovich}. Except for $T_{00}$, all polarization operators are traceless and orthogonal but may or may not be Hermitian. Therefore these $d^2$ polarization operators form an orthonormal matrix basis of $d$ dimensional Hilbert Schmidt space.\\
	Any $d\times d$ density matrix $\rho$ can be decomposed into a Bloch vector using the POB, and it often takes the following form:
	\begin{eqnarray}
	\rho_{POB}=\frac{1}{d}I+\sum_{L=1}^{2s}\sum_{M=-L}^{L}{b_{LM}T_{LM}}=\frac{1}{d}I+\vec{b}.\vec{T}
	\end{eqnarray}
	where $\vec{b}=(b_{1,-1},b_{1,0},b_{1,1},b_{2,-2},b_{2,-1},b_{2,0},...,b_{LM})$ is the Bloch vector and $b_{LM}$ is given by $b_{LM}=Tr[T^{\dagger}_{LM}\rho]$. In general, the POB is not hermitian so, the components $b_{LM}$ are complex. All Bloch vectors are contained in a hypersphere with a radius $|\vec{b}|\leq \sqrt{\frac{d-1}{d}}$.\\ \ \\
	\textbf{Decomposition of standard matrix basis into POB}\\
	The standard matrix basis can be transformed to POB using the following relation
	\begin{eqnarray}
	|i\rangle \langle j|=\sum_{L}\sum_{M}{\sqrt{\frac{2L+1}{2s+1}}C_{sm_j,LM}^{sm_i}T_{LM}}
	\label{pob23}
	\end{eqnarray}\\
	\textbf{III. Weyl operator basis (WOB)}\\
	\noindent For $d$-dimensional Hilbert-Schmidt space, $d^2$ Weyl's operators may be defined by
	\begin{eqnarray}
	W_{nm}=\sum_{k=0}^{d-1}{e^{\frac{2\Pi i}{d}kn}|k\rangle \langle (k+m)mod~d|},~~ n,m=0,1,...,d-1
	\label{weyloperator}
	\end{eqnarray}
	The operators $W_{nm}$ given in (\ref{weyloperator}) are all unitary operators and form an orthonormal basis of the Hilbert spaces \cite{narnhofer,narnhofer1,werner}
	\begin{eqnarray}
	Tr [W_{nm}^{\dagger}W1_{lj}]=d\delta_{nl}\delta_{mj}
	\end{eqnarray}
	The operator $W_{00}=I$. 
	Any $d\times d$ density matrix $\rho$ can be decomposed into a Bloch vector using WOB as
	\begin{eqnarray}
	\rho_{WOB}=\frac{1}{d}I+\sum_{n,m=0}^{d-1}{b_{nm}W_{nm}}=\frac{1}{d}I+\vec{b}.\vec{W},~~ n,m=0,1,...d-1
	\end{eqnarray}
	Assuming $b_{00}=0$, the other components of the Bloch vector $\vec{b}$ are given by $b_{nm}=Tr[W_{nm}\rho_{WOB}]$. In general, Weyl operators are not Hermitian so, $b_{nm}$ are complex numbers and $b_{nm}^{*}$ satisfies $b_{nm}^{*}=e^{\frac{-2\Pi i}{d}nm}b_{-n-m}$. Bloch vectors are contained in a hypersphere with radius $|\vec{b}|\leq \frac{\sqrt{d-1}}{d}$ \cite{bertlmann}.\\ \ \\
	\textbf{Decomposition of standard matrix basis into WOB}\\
	The elements of $d\times d$ density matrix may be expressed in terms of WOB as
	\begin{eqnarray}
	|j\rangle \langle k|=\frac{1}{d}\sum_{i=0}^{d-1}{e^{\frac{-2\Pi i}{d}lj}W_{l(k-j)mod~d}}
	\label{wob}
	\end{eqnarray}
	In particular, for a two-dimensional system, the GGB, WOB, and POB can be expressed in terms of Pauli matrices. The corresponding relationship between the Pauli matrices and the basis GGM, WOB, and POB are given by
	\begin{eqnarray}
	\{W_{00}, W_{01}, W_{10}, W_{11}\}&=&\{I,\sigma_x,i\sigma_y,\sigma_z\},\\
	\{I,\lambda_s^{12},\lambda_a^{12},\lambda^{1}\}&=&\{I,\sigma_x,\sigma_y,\sigma_z\},\\
	\{T_{0,0},T_{1,1},T_{1,0},T_{1,-1}\}&=&\{\frac{1}{\sqrt{2}}I,-\sigma_{+},\frac{1}{\sqrt{2}}\sigma_z,\sigma_{-}\}
	\end{eqnarray}
	where $\sigma_{\pm}=\frac{1}{2}(\sigma_x \pm i\sigma_y)$.
	\subsection{An illustration: A two-qutrit system}
	%	\noindent Consider a bipartite two-qudit density matrix $\rho$ in the Hilbert space $H_A^{d}\otimes H_B^{d}$. Then $\rho$ may be decomposed as
	%	\begin{eqnarray}
	%	\rho^{d\otimes d}=\frac{1}{d}I\otimes I+n_i\Gamma_i\otimes I+m_iI\otimes \Gamma_i+c_{ij}\Gamma_i\otimes \Gamma_j,~~ n_i,m_i,c_{ij}\in C
	%	\end{eqnarray}
	%	where \{$\Gamma_i$\}'s are the basis of the Hilbert space $H^{d}$.\\
	\noindent Let us consider a two-qutrit isotropic state described by the density operator $\rho_{\beta}$. In standard basis, the state $\rho_{\beta}$ may be defined as
	\begin{eqnarray}
	\rho_{\beta}=\beta|\phi_{+}\rangle \langle \phi_{+}|+\frac{1-\beta}{9}I_9, ~~0\leq \beta \leq 1
	\label{betastate}
	\end{eqnarray}
	where $I_9$ denotes the identity matrix of order 9 and the state $|\phi_{+}\rangle$ represents a Bell state in a two-qutrit system and may be expressed as
	\begin{eqnarray}
	|\phi_{+}\rangle=\frac{1}{\sqrt{3}}(|11\rangle +|22\rangle +|33\rangle)
	\label{bellstate}
	\end{eqnarray}
	\textbf{I. Expansion of $\rho_{\beta}$ in terms of GGM basis}\\
	We are now in a position to express a two-qutrit system in terms of GGM basis. For a single-qutrit system, the GGM basis can be expressed as \cite{caves, bertlmann1}\\
	%4	Firstly, we write Gell-Mann matrices in the case of a qutrit system. The eight Gell-Mann matrices for a qutrit system are \\ \ \\
	1. Symmetric GGM\\
	\begin{eqnarray}
	\lambda_s^{12}=
	\begin{pmatrix}
	0 & 1 & 0\\
	1 & 0 & 0\\
	0 & 0 & 0
	\end{pmatrix},
	\lambda_s^{13}=
	\begin{pmatrix}
	0 & 0 & 1\\
	0 & 0 & 0\\
	1 & 0 & 0
	\end{pmatrix},
	\lambda_s^{23}=
	\begin{pmatrix}
	0 & 0 & 0\\
	0 & 0 & 1\\
	0 & 1 & 0
	\end{pmatrix}
	\label{gellmanns}
	\end{eqnarray}
	2. Antisymmetric GGM\\
	\begin{eqnarray}
	\lambda_a^{12}=
	\begin{pmatrix}
	0 & -i & 0\\
	i & 0 & 0\\
	0 & 0 & 0
	\end{pmatrix},
	\lambda_a^{13}=
	\begin{pmatrix}
	0 & 0 & -i\\
	0 & 0 & 0\\
	i & 0 & 0
	\end{pmatrix},
	\lambda_a^{23}=
	\begin{pmatrix}
	0 & 0 & 0\\
	0 & 0 & -i\\
	0 & i & 0
	\end{pmatrix}
	\label{gellmanna}
	\end{eqnarray}
	3. Diagonal GGM\\
	\begin{eqnarray}
	\lambda^{1}=
	\begin{pmatrix}
	1 & 0 & 0\\
	0 & -1 & 0\\
	0 & 0 & 0
	\end{pmatrix},
	\lambda^{2}=\frac{1}{\sqrt{3}}
	\begin{pmatrix}
	1 & 0 & 0\\
	0 & 1 & 0\\
	0 & 0 & -2
	\end{pmatrix}
	\label{gellmannd}
	\end{eqnarray}
	%In case of a two-qutrit system, the components of Bloch vector are $\vec{b}=(b_s^{12},b_s^{13},b_s^{23},b_a^{12},b_a^{13},b_a^{23},b^1,b^2)$ corresponding to GGM (\ref{gellmanns}),(\ref{gellmanna}),(\ref{gellmannd}) and $|\vec{b}|\leq \sqrt{\frac{1}{3}}$.\\
	Let us first write the Bell state $|\phi_{+}\rangle \langle \phi_{+}|$ in the standard basis as
	\begin{eqnarray}
	|\phi_{+}\rangle \langle \phi_{+}|&=&\frac{1}{3}(|11\rangle \langle 11|+|11\rangle \langle 22|+|11\rangle \langle 33|\nonumber\\&+&|22\rangle \langle 11|+|22\rangle \langle 22|+|22\rangle \langle 33|\nonumber\\&+&|33\rangle \langle 11|+|33\rangle \langle 22|+|33\rangle \langle 33|)\nonumber\\&=& A+B
	\label{bellphi}
	\end{eqnarray}
	where $A$ and $B$ may be expressed as
	\begin{eqnarray}
	A&=&\frac{1}{3}(|11\rangle \langle 22|+|11\rangle \langle 33|+|22\rangle \langle 33|)\nonumber\\&+&(|22\rangle \langle 11|+|33\rangle \langle 11|+|33\rangle \langle 22|)\nonumber\\
	B&=& \frac{1}{3}(|11\rangle \langle 11|+|22\rangle \langle 22|+|33\rangle \langle 33|)
	\label{AandB}
	\end{eqnarray} 
	For a qutrit system, the GGM basis is related to the standard basis in the following way
	\begin{eqnarray}
	\Lambda_s^{12}&=&|1\rangle \langle 2|+|2\rangle \langle 1|, \Lambda_s^{13}=|1\rangle \langle 3|+|3\rangle \langle 1|,\Lambda_s^{23}=|2\rangle \langle 3|+|3\rangle \langle 2|\nonumber\\
	\Lambda_a^{12}&=&-i|1\rangle \langle 2|+i|2\rangle \langle 1|, \Lambda_a^{13}=-i|1\rangle \langle 3|+i|3\rangle \langle 1|, \Lambda_a^{23}=-i|2\rangle \langle 3|+i|3\rangle \langle 2|\nonumber\\
	\Lambda^{1}&=&|1\rangle \langle 1|-|2\rangle \langle 2|, \Lambda^{2}=\frac{1}{\sqrt{3}}|1\rangle \langle 1|+|2\rangle \langle 2|-2|3\rangle \langle 3|
	\end{eqnarray}
	%Using (\ref{GGB}), calculating elements of A, we get
	%\begin{eqnarray}
	%|11\rangle \langle 22|=|1\rangle \langle 2|\otimes |1\rangle \langle 2|&=&\frac{1}{4}[(\Lambda_s^{12}+i \Lambda_a^{12})\otimes (\Lambda_s^{12}+i \Lambda_a^{12})]\nonumber\\&=& \frac{1}{4}[\Lambda_s^{12}\otimes \Lambda_s^{12}+i	\Lambda_a^{12}\otimes \Lambda_s^{12}+i	\Lambda_s^{12}\otimes 	\Lambda_a^{12}-	\Lambda_a^{12}\otimes 	\Lambda_a^{12}]\nonumber
	%\end{eqnarray}
	%Simillarly calculating rest of the elements 
	%	\begin{eqnarray}
	%|11\rangle \langle 33|&=&\frac{1}{4}[\Lambda_s^{13}\otimes \Lambda_s^{13}+i	\Lambda_a^{13}\otimes \Lambda_s^{13}+i	\Lambda_s^{13}\otimes 	\Lambda_a^{13}-	\Lambda_a^{13}\otimes 	\Lambda_a^{13}]\nonumber\\
	%|22\rangle \langle 33|&=&\frac{1}{4}[\Lambda_s^{23}\otimes \Lambda_s^{23}+i	\Lambda_a^{23}\otimes \Lambda_s^{23}+i	\Lambda_s^{23}\otimes 	\Lambda_a^{23}-	\Lambda_a^{23}\otimes 	\Lambda_a^{23}]\nonumber\\
	%	|22\rangle \langle 11|&=&\frac{1}{4}[\Lambda_s^{12}\otimes \Lambda_s^{12}-i	\Lambda_a^{12}\otimes \Lambda_s^{12}-i	\Lambda_s^{12}\otimes 	\Lambda_a^{12}-	\Lambda_a^{12}\otimes 	\Lambda_a^{12}]\nonumber\\
	%|33\rangle \langle 11|&=&\frac{1}{4}[\Lambda_s^{13}\otimes \Lambda_s^{13}-i	\Lambda_a^{13}\otimes \Lambda_s^{13}-i	\Lambda_s^{13}\otimes 	\Lambda_a^{13}-	\Lambda_a^{13}\otimes 	\Lambda_a^{13}]\nonumber\\
	%	|33\rangle \langle 22|&=&\frac{1}{4}[\Lambda_s^{23}\otimes \Lambda_s^{23}-i	\Lambda_a^{23}\otimes \Lambda_s^{23}-i	\Lambda_s^{23}\otimes 	\Lambda_a^{23}-	\Lambda_a^{23}\otimes 	\Lambda_a^{23}]
	%	\end{eqnarray}
	The matrices A and B can be re-expressed in terms of GGM basis as 
	\begin{eqnarray}
	A&=&\frac{1}{6}[	\Lambda_s^{12}\otimes 	\Lambda_s^{12}+	\Lambda_s^{13}\otimes 	\Lambda_s^{13}+	\Lambda_s^{23}\otimes 	\Lambda_s^{23}-	\Lambda_a^{12}\otimes 	\Lambda_a^{12}-	\Lambda_a^{13}\otimes 	\Lambda_a^{13}-	\Lambda_a^{23}\otimes 	\Lambda_a^{23}]\nonumber\\
	B&=& \frac{1}{3}[\frac{1}{2}\Lambda^{1}\otimes \Lambda^{1}+\frac{1}{2}\Lambda^{2}\otimes \Lambda^{2}+\frac{1}{3}I\otimes I]
	\end{eqnarray}
	Therefore, the Bell state (\ref{bellphi}), then can be re-expressed in terms of GGM basis as
	\begin{eqnarray}
	|\phi_{+}\rangle \langle \phi_{+}|&=&\frac{1}{6}[\Lambda_s^{12}\otimes 	\Lambda_s^{12}+	\Lambda_s^{13}\otimes 	\Lambda_s^{13}+	\Lambda_s^{23}\otimes 	\Lambda_s^{23}-	\Lambda_a^{12}\otimes 	\Lambda_a^{12}-	\Lambda_a^{13}\otimes 	\Lambda_a^{13}\nonumber\\&-&	\Lambda_a^{23}\otimes 	\Lambda_a^{23}+\Lambda^{1}\otimes \Lambda^{1}+\Lambda^{2}\otimes \Lambda^{2}]+\frac{1}{9}I\otimes I
	\end{eqnarray}
	Thus in terms of GGB, the two-qutrit isotropic state (\ref{betastate}) may be expressed as
	\begin{eqnarray}
	\rho_{\beta}=\frac{\beta}{6}\Lambda+\frac{1}{9}I\otimes I
	\end{eqnarray}
	where $\Lambda=\Lambda_s^{12}\otimes 	\Lambda_s^{12}+	\Lambda_s^{13}\otimes 	\Lambda_s^{13}+	\Lambda_s^{23}\otimes 	\Lambda_s^{23}-	\Lambda_a^{12}\otimes 	\Lambda_a^{12}-	\Lambda_a^{13}\otimes 	\Lambda_a^{13}-	\Lambda_a^{23}\otimes 	\Lambda_a^{23}+\Lambda^{1}\otimes \Lambda^{1}+\Lambda^{2}\otimes \Lambda^{2}$\\
	\textbf{II. Expansion of $\rho_{\beta}$ in terms of POB}\\
	Let us first write the polarizing operators for a three-dimensional system. For a qutrit system, we have $s=1$ and nine polarizing operators are denoted by $T_{L,M}$, $L=0,1,2$ and $M=-L,...,L$ and they are given by %For a qutrit system, $s=1$, $L=0,1,2$, $M=-2,-1,0,1,2$, $m_1=1,m_2=0,m_3=-1$. 
	\begin{eqnarray}
	T_{0,0}&=&\frac{1}{\sqrt{3}}
	\begin{pmatrix}
	1 & 0 & 0\\
	0 & 1 & 0\\
	0 & 0 & 1
	\end{pmatrix},
	T_{1,1}=-\frac{1}{\sqrt{2}}
	\begin{pmatrix}
	0 & 1 & 0\\
	0 & 0 & 1\\
	0 & 0 & 0
	\end{pmatrix},
	T_{1,0}=\frac{1}{\sqrt{2}}
	\begin{pmatrix}
	1 & 0 & 0\\
	0 & 0 & 0\\
	0 & 0 & -1
	\end{pmatrix},\\
	T_{1,-1}&=&\frac{1}{\sqrt{2}}
	\begin{pmatrix}
	0 & 0 & 0\\
	1 & 0 & 0\\
	0 & 1 & 0
	\end{pmatrix},
	T_{2,2}=
	\begin{pmatrix}
	0 & 0 & 1\\
	0 & 0 & 0\\
	0 & 0 & 0
	\end{pmatrix}, 
	T_{2,1}=\frac{1}{\sqrt{2}}
	\begin{pmatrix}
	0 & -1 & 0\\
	0 & 0 & 1\\
	0 & 0 & 0
	\end{pmatrix}\\
	T_{2,0}&=&\frac{1}{\sqrt{6}}
	\begin{pmatrix}
	1 & 0 & 0\\
	0 & -2 & 0\\
	0 & 0 & 1
	\end{pmatrix},
	T_{2,-1}=\frac{1}{\sqrt{2}}
	\begin{pmatrix}
	0 & 0 & 0\\
	1 & 0 & 0\\
	0 & -1 & 0
	\end{pmatrix},
	T_{2,-2}=
	\begin{pmatrix}
	0 & 0 & 0\\
	0 & 0 & 0\\
	1 & 0 & 0
	\end{pmatrix}
	\end{eqnarray} \\
	To express the state $\rho_{\beta}$ defined in (\ref{betastate}) in terms of POB, we first need to express Bell state (\ref{bellstate}) in terms of POB. In the standard basis, Bell state may be written as
	\begin{eqnarray}
	|\phi_{+}\rangle \langle \phi_{+}|&=&\frac{1}{3}[\sum_{i,j=1}^{3}{|i\rangle \langle j}\otimes |i\rangle \langle j|]
	\label{bellstate2}
	\end{eqnarray}
	%Using (\ref{pob}), we first express the elements of a qutrit system in terms of POB.
	Using (\ref{pob23}), the Bell state (\ref{bellstate2}) may be re-expressed as
	\begin{eqnarray}
	|\phi_{+}\rangle \langle \phi_{+}|&=&\frac{1}{3}[\sum_{L}\sum_{M}\sqrt{\frac{2L+1}{3}}C_{sm_j,LM}^{sm_i}T_{LM}\otimes \sum_{L'}\sum_{M}\sqrt{\frac{2L'+1}{3}}C_{sm_j,L'M}^{sm_i}T_{L'M}]
	\label{clebschpob}
	\end{eqnarray}
	Since $M=m_i-m_j$ hold good so $\sum_M$  now reduces to $\sum_{i,j}$. Thus, the above equation (\ref{clebschpob}) reduces to
	\begin{eqnarray}
	|\phi_{+}\rangle \langle \phi_{+}|&=&\frac{1}{3}[\sum_{L,L'=1}^{3}\sqrt{\frac{(2L+1)(2L'+1)}{9}}(\sum_{i,j=1}^{3}C_{sm_j,LM}^{sm_i}C_{sm_j,L'M}^{sm_i})T_{LM}\otimes T_{L'M}]\nonumber\\
	\label{pob1}
	\end{eqnarray}
	The sum rule for the Clebsch-Gordan coefficients \cite{varshalovich} is given by
	\begin{eqnarray}
	\sum_{\alpha,\gamma}{C_{a\alpha,b\beta}^{d\gamma}C_{a\alpha,b'\beta'}^{d\gamma}}=\frac{2d+1}{2b+1}\delta_{bb'}\delta_{\beta\beta'}
	\label{coeffpob}
	\end{eqnarray}
	Using the sum rule for the Clebsch-Gordan coefficients, the equation (\ref{pob1}) can now be purely expressed in terms of POB as
	\begin{eqnarray}
	|\phi_{+}\rangle \langle \phi_{+}| &=&\frac{1}{3}[\sum_{L,L'}^{3}{\sqrt{\frac{(2L+1)(2L'+1)}{9}}}.\frac{3}{2L+1}\delta_{LL'}T_{LM}\otimes T_{L'M}]\nonumber\\
	&=& \frac{1}{3}[\sum_{L,L'=1}^{3}\frac{\sqrt{(2L+1)(2L'+1)}}{2L+1}\delta_{LL'}T_{LM}\otimes T_{L'M}]\nonumber\\&=&
	\frac{1}{3}[\sum_{L=1}^{3}T_{LM}\otimes T_{L'M}]
	\end{eqnarray}
	Using (\ref{pobtlm})
	\begin{eqnarray}
	|\phi_{+}\rangle \langle \phi_{+}|=\frac{1}{9}I\otimes I+\frac{1}{3}(\sum_{L,M\neq 0,0}{T_{LM}\otimes T_{LM}})
	\end{eqnarray}
	Using POB Bloch vector notation, the two-qudit state (\ref{betastate}), may be written as
	\begin{eqnarray}
	\rho_{\beta}=\frac{1}{9}I\otimes I+\frac{\beta}{3}(\sum_{L,M\neq 0,0}{T_{LM}\otimes T_{LM}}) 
	\end{eqnarray}
	\textbf{III. Expansion of $\rho_{\beta}$ in terms of WOB}\\
	To express the state $\rho_{\beta}$ defined in (\ref{betastate}) in terms of WOB, let us first write Weyl's operator for a qutrit system. For a qutrit system $(d=3)$, the Weyl operator may be expressed as
	\begin{eqnarray}
	W_{00}&=&\begin{pmatrix}
	1 & 0 & 0\\
	0 & 1 & 0\\
	0 & 0 & 1
	\end{pmatrix},
	W_{01}=\begin{pmatrix}
	0 & 1 & 0\\
	0 & 0 & 1\\
	1 & 0 & 0
	\end{pmatrix},
	W_{02}=\begin{pmatrix}
	0 & 0 & 1\\
	1 & 0 & 0\\
	0 & 1 & 0
	\end{pmatrix},\\
	W_{10}&=&\begin{pmatrix}
	1 & 0 & 0\\
	0 & e^{\frac{2\Pi i}{3}} & 0\\
	0 & 0 & e^{\frac{-2\Pi i}{3}}
	\end{pmatrix},
	W_{11}=\begin{pmatrix}
	0 & 1 & 0\\
	0 & 0 & e^{\frac{2\Pi i}{3}}\\
	e^{\frac{-2\Pi i}{3}} & 0 & 0
	\end{pmatrix},
	W_{12}=\begin{pmatrix}
	0 & 0 & 1\\
	e^{\frac{2\Pi i}{3}} & 0 & 0\\
	0 & e^{\frac{-2\Pi i}{3}} & 0
	\end{pmatrix},\\
	W_{20}&=&\begin{pmatrix}
	1 & 0 & 0\\
	0 & e^{\frac{-2\Pi i}{3}} & 0\\
	0 & 0 & e^{\frac{-2\Pi i}{3}}
	\end{pmatrix},
	W_{21}=\begin{pmatrix}
	0 & 1 & 0\\
	0 & 0 & e^{\frac{-2\Pi i}{3}}\\
	e^{\frac{2\Pi i}{3}} & 0 & 0
	\end{pmatrix},
	W_{22}=\begin{pmatrix}
	0 & 0 & 1\\
	e^{\frac{-2\Pi i}{3}} & 0 & 0\\
	0 & e^{\frac{2\Pi i}{3}} & 0
	\end{pmatrix}
	\end{eqnarray}
	Let us start with the expression of Bell state (\ref{bellstate}) on standard basis as
	\begin{eqnarray}
	|\phi_{+}\rangle \langle \phi_{+}|=\frac{1}{3}{\sum_{j,k=1}^{3}{|j\rangle \langle k|\otimes |j\rangle \langle k|}}
	\label{bellstatewob}
	\end{eqnarray}
	Equation (\ref{bellstatewob}), in terms of Weyl operators, can be written as
	\begin{eqnarray}
	|\phi_{+}\rangle \langle \phi_{+}|&=&\frac{1}{3}\sum_{j,k=0}^{2}[\frac{1}{3}\sum_{l=0}^{2}{e^{\frac{-2\Pi i}{3}lj}W_{l(k-j)mod 3}}\otimes \frac{1}{3}\sum_{l'=0}^{2}{e^{\frac{-2\Pi i}{3}l'j}W_{l'(k-j)mod 3}}]\nonumber\\
	&=& \frac{1}{27}\sum_{j,k=0}^{2}\sum_{l,l'=0}^{2}{e^{\frac{-2\Pi i}{3}j(l+l')}W_{l(k-j)mod3}\otimes {W_{l'(k-j)mod3}}}
	\end{eqnarray}
	Substituting $m=(k-j)mod 3,~~m=0,1,2$. The above equation may be re-written as
	\begin{eqnarray}
	|\phi_{+}\rangle \langle \phi_{+}|&=&\frac{1}{27}\sum_{m,k=0}^{2}\sum_{l,l'=0}^{2}{e^{\frac{-2\Pi i}{3}(k-m)(l+l')}W_{lm}\otimes {W_{l'm}}}\nonumber\\
	&=&\frac{1}{27}[\sum_{m,k=0}^{2}1.W_{0m}\otimes W_{0m}+\sum_{m,k=0}^{2}\sum_{l,l',l+l'=d}1.W_{lm}\otimes W_{l'm}\nonumber\\&+&\sum_{m}^{2}\sum_{l,l'\neq 0,l+l'=d}^{2}(\sum_k{e^{\frac{-2\Pi i}{3}(k-m)(l+l')}})W_{lm\otimes W_{l'm}}]\nonumber\\
	&=&\frac{1}{9}\sum_m{W_{0m}\otimes W_{0m}}+\frac{1}{9}\sum_m\sum_{l,l',l+l'=d}{W_{lm}\otimes W_{l'm}}\nonumber\\&+&\frac{1}{27}[\sum_m\sum_{l,l',l+l'=d}^{2}(\sum_k{e^{\frac{-2\Pi i}{3}(k-m)(l+l')}})W_{lm}\otimes W_{l'm}]
	\end{eqnarray}
	Using the equivalence
	\begin{eqnarray}
	\sum_{n=0}^{d-1}{e^{\frac{2\Pi i}{d}nx}}=
	\begin{cases}
	d ~~\text{if} ~~x=0\\
	0 ~~\text{if} ~~x\neq 0,
	\end{cases} x\in Z
	\end{eqnarray}	
	We are now in a position to express the Bell state (\ref{bellstate}) in terms of WOB. Using the notation $-l (mod 3)=(3-l)$ and $W_{00}=I$, the Bell state (\ref{bellstate}) in terms of WOB may be expressed as
	%	\begin{eqnarray}
	%	|\phi_{+}\rangle \langle \phi_{+}|&=&\frac{1}{9}\sum_m{W_{0m}\otimes W_{0m}}+\frac{1}{9}\sum_m\sum_{l,l',l+l'=d}{W_{lm}\otimes W_{l'm}}
	%	\end{eqnarray}
	\begin{eqnarray}
	|\phi_{+}\rangle \langle \phi_{+}|&=&\frac{1}{9}I\otimes I+\frac{1}{9}\sum_{l,m=0}^{2}{W_{l,m}\otimes W_{-l,m}},~~ (l,m)\neq (0,0)
	\end{eqnarray}
	The two-qutrit state $\rho_{\beta}$ in terms of WOB may be expressed as
	\begin{eqnarray}
	\rho_{\beta}=\frac{1}{9}I\otimes I+\frac{\beta}{9}\sum_{l,m=0}^{2}{W_{l,m}\otimes W_{-l,m}}
	\end{eqnarray}

	%\section{Linear operators and Matrices}
	%A linear operator between two vector spaces $V_1$ and $V_2$ is defined to be a function A:$V_1$$\rightarrow$ $V_2$ which is linear,
	%\begin{eqnarray}
	%A(\sum_i{a_i|v_i\rangle})=\sum_iA(|v_i\rangle)
	%\end{eqnarray}
	%When we say that a linear operator is defined on a vector space V, we mean that the linear operator is defined from V to V.\\
	%Linear operators may be expressed as matrices and vice-versa. To see this connection, we first need to understand that an m $\times$ n complex matrix A with entries $A_{ij}$ is a linear operator which sends vectors of the vector space $C^{n}$ to the vector space $C^{m}$, under matrix multiplication of A by a vector in $C^{n}$. When we say that the matrix A is a linear operator, we mean,
	%\begin{eqnarray}
	%A(\sum_i{a_i}|v_i\rangle)=\sum_{i}{{a_i}A|v_i\rangle}
	%\end{eqnarray}
	%where the operation is matrix multiplication of A by column vectors. Thus, matrices can be regarded as linear operators.
	%Also, linear operators can be represented as matrices. Suppose $A:V_1 \rightarrow V_2$ be a linear operator between two vector spaces $V_1$ and $V_2$. Let $\{|v_1\rangle, ..., |v_m\rangle\}$ and $\{w_1,....w_n\}$ be a basis for $V_1$ and $V_2$. Then for each j in 1,2,3,...m, there exist complex numbers $A_{1j}$ through $A_{nj}$ such that,
	%\begin{eqnarray}
	%A|v_{j}\rangle =\sum_{i}A_{ij}|w_i\rangle
	%\end{eqnarray}
	%The matrix whose elements whose entries are $A_{ij}$ is said to be the matrix representation of the operator A. The matrix representation of A is completely equivalent to the operator A.
	
	\subsection{Quantum Measurements}
	\noindent By what process does measurement is carried out on a quantum system? This may be discussed in the following way: When a measurement is performed on a quantum system by a measurement apparatus, the quantum state collapses probabilistically into one of a number of eigenstates of the observable. Can this behavior itself be explained by quantum mechanics? These are a few interesting and puzzling issue that has not been settled. It is known as the problem of measurement.\\
	If $O$ denotes the observable and $|s\rangle$ is the corresponding eigenstate, then the measurement process may be described as
	\begin{eqnarray}
	O|s\rangle=\lambda|s\rangle
	\end{eqnarray}
	where $\lambda$ denote the eigenvalue of $O$.\\
	A measurement apparatus is used to calculate the eigenvalue $\lambda$. After the measurement, the uncertainty of the observable $O$ vanishes at least for the instant, when the measurement is performed \cite{greensite}.
	%Quantum measurement is a very complicated problem in quantum mechanics. This is due to the fact that if a quantum system gets interacted with any other system, then the initial state of the quantum state will change.
	One of the most fundamental concepts in quantum physics is the uncertainty principle. The uncertainty principle may be expressed in different ways such as Robertson uncertainty relation \cite{robertson}, Heisenberg uncertainty relation for joint measurements \cite{arthurs1,arthurs2} and Noise-disturbance uncertainty relation \cite{heisenberg1,heisenberg2}. It is widely accepted that any version of the uncertainty principle has a logical connection to the restriction on measuring a system without disturbing it, such as a position measurement often disturbs the momentum.\\
	Let us now explain the idea of measurement when it is performed on a quantum system $|\phi\rangle$. In quantum mechanics, measurements are described by measurement operators $\{M_m\}_{m=1}^n$
	%\begin{eqnarray}
	%\{M_m\}_{m=1}^n
	%\label{measurement operators}
	%\end{eqnarray}
	where $m$ denote the measurement outcome \cite{nielsen}. These operators act on the state space of the system being measured. These measurement operators satisfy the completeness relation
	\begin{eqnarray}
	\sum_{m}{M_m^{\dagger}M_m}=I
	\end{eqnarray}
	where $I$ denotes the identity operator. Consider $|\phi\rangle$ to be the state of the quantum system before measurement then the state defining the system after the measurement is given by,
	\begin{eqnarray}
	|\phi_m\rangle=\frac{M_m|\phi\rangle}{\sqrt{\langle \phi|M_m^{\dagger}M_m|\phi\rangle}}
	\end{eqnarray}
	The probability of getting measurement outcome $m$ is given by
	\begin{eqnarray}
	p(m)=\langle \phi|M_m^{\dagger}M_m|\phi\rangle
	\end{eqnarray}
	%where $|\alpha|^2$ denote the probability of the system collapsing in the state $|0\rangle$ and $|\beta|^2$, denote the probability of the system to be in the state $|1\rangle$.
	\textbf{Projective measurements:} A Projective measurement is described by an observable $M$ that may be decomposed via Spectral decomposition as
	\begin{eqnarray}
	M=\sum_m{mP_m}
	\end{eqnarray}
	where $P_m$ are projectors on the eigenspace of $M$ corresponding to eigenvalue $m$. The projector $P_m$ satisfies the following properties:\\
	(i) $P_m$ is Hermitian i.e. $P_m^{\dagger}=P_m$.\\
	(ii) Idempotent property holds for $P_m$ i.e $P_m^2=P_m$.\\
	(iii) $P_m$ has eigenvalues 0 and 1.\\
	When the projective measurement described by the projectors $P_m$ performed on the state $|\phi\rangle$, the state reduces to
	\begin{eqnarray}
	|\phi_m^{P}\rangle =\frac{P_m|\phi\rangle}{\sqrt{p(m)}}
	\end{eqnarray} 
	where $p(m)$ denotes the probability of getting outcome $m$ and it is given by
	\begin{eqnarray}
	p(m)=\langle \phi|P_m|\phi\rangle
	\end{eqnarray}
	For projective measurement $P$, the average value of the measurement is given by
	\begin{eqnarray}
	E(P)=\sum_m{mp(m)}=\langle \phi|P|\phi\rangle=\langle P\rangle_{\phi}
	\end{eqnarray}
	The standard deviation $(\Delta(P))$ associated with the observations of $P$ may be expressed as
	\begin{eqnarray}
	[\Delta(P)]_{\phi}=\sqrt{\langle (P-\langle P\rangle_{\phi})^2\rangle_{\phi}}=\sqrt{\langle P^2\rangle_{\phi}-\langle P\rangle_{\phi} ^2}
	\end{eqnarray}
	%When the measurement operators are projectors P, such measurements are known as projective measurements. Projectors are the operators which satisfy $P^2=P$. 
	%/;For instance, the measurement operator $M_0=|0\rangle \langle 0|$ is a projection operator [since ($|0\rangle \langle 0|^2=|0\rangle \langle 0||0\rangle \langle 0|=|0\rangle (\langle 0|0\rangle) |0\rangle=|0\rangle \langle 0|$)].\\
	\textbf{von Neumann measurement:} Consider a given quantum system described by the density operator $\rho$ written in the computation basis $\{|n\rangle,~n=0,1,2...\}$. Then the set of projection operators that projects any state onto the basis states may be defined by $P_n=|n\rangle \langle n|$. After the measurement, the state may be expressed as
	\begin{eqnarray}
	\rho^{V}=\frac{P_n\rho P_n}{p(n)}
	\end{eqnarray}
	where $p(n)$ denotes the probability of the system being in state $|n\rangle$ and is defined as
	\begin{eqnarray}
	p(n)=\langle n|\rho|n\rangle=Tr[P_n\rho P_n]=Tr[(P_n)^2\rho]=Tr[P_n(\rho)]
	\end{eqnarray}
	Let us now understand the concept of von Neumann measurement by considering a single qubit state $|\psi\rangle$, which is given by
	\begin{eqnarray}
	|\psi\rangle=\alpha|0\rangle+\beta|1\rangle,~~|\alpha|^2+|\beta|^2=1
	\end{eqnarray}
	$|\alpha|^2+|\beta|^2=1$ denote the normalization condition. Since the state $|\psi\rangle$ is defined on a two-dimensional Hilbert space spanned by $\{|0\rangle, |1\rangle\}$ so, the two measurement operators defined on this state space are given by
	$P_0=|0\rangle \langle 0|$ and $P_1=|1\rangle \langle 1|$. It can be easily shown that the projection operator $P_0$ and $P_1$ satisfy the following properties:\\
	(i) $P_i^2=1, i=0,1$.\\
	(ii) $P_0^{\dagger}P_0+P_1^{\dagger}P_1=P_0+P_1=I$, where $I$ denote the identity operator.\\
	Let us suppose that the state $|\psi\rangle$ is being measured with the measurement operator $P_0$ and $P_1$. When $P_0$ is performed on $|\psi\rangle$, the state after the measurement reduces to $\frac{P_0|\psi\rangle}{|\alpha|}=\frac{\alpha}{|\alpha|}|0\rangle$. This output may be obtained with probability $p(0)$ which is given by
	\begin{eqnarray}
	p(0)=\langle \psi|P_0^{\dagger}P_0|\psi\rangle=\langle \psi|P_0|\psi\rangle=|\alpha|^2
	\end{eqnarray}
	Similarly, when $P_1$ is performed on $|\psi\rangle$, the state after the measurement reduces to $\frac{P_1|\psi\rangle}{|\beta|}=\frac{\beta}{|\beta|}|1\rangle$. This output may be obtained with probability $p(1)$ which is given by
	\begin{eqnarray}
	p(1)=\langle \psi|P_1^{\dagger}P_1|\psi\rangle=\langle \psi|P_1|\psi\rangle=|\beta|^2
	\end{eqnarray}
	%To understand this, suppose there are two-quantum systems, initial and final. The initial system is prepared in some state independent of the final system, and then the systems initial and final are allowed to interact. After the interaction, the initial system becomes correlated with the final system and then von Neumann measurement is performed on the initial system. The measurement of the initial system provides us with information about the final system.\\
	%Consider a quantum system described by vector $|\psi\rangle$ and let $Q$ be the property that is being measured. If $q_n$ is the result obtained after the measurement, then the state of the system immediately after the measurement is $|q_n\rangle$.
	%In terms of projection operators, If we have
	%\begin{eqnarray}
	%P_n=|q_n\rangle \langle q_n|
	%\end{eqnarray}
	%then the state after measurement for which $q_n$ measurement outcome was obtained may be re-expressed as
	%\begin{eqnarray}
	%\frac{P_n}{\sqrt{\langle \psi|P_n|\psi \rangle}}
	%\end{eqnarray}
	%In a Von Neumann measurement, if get the result $E_1$, where $Q$ is some property of the state which is being measured, then the system ends up in the state $|E_1\rangle$.\\
	\textbf{Positive operator-valued measurement (POVM)}: Let $M_m$ be the measurement operators which is acting on a quantum system $|\psi\rangle$. Then, the set of operators $\{E_m=M_m^{\dagger}M_m\}$ is known as POVM. The operators $E_m$ are known as POVM elements. %Here, the operators $E_m$ are positive and satisfy the completeness relation i.e $\sum_m{E_m}=I$ and $p(m)=\langle \psi|E_m|\psi\rangle$. 
	Therefore, we are now in a position to define POVM. POVM may be defined as a set of operators $\{E_m\}$ which satisfies the following two properties:\\
	(i) each operator $E_m~(m=0,1,2,...)$ are positive.\\
	(ii) $\sum_m{E_m}=I$, ~where $I$ is the Identity operator.\\
	%Thus, the operators $\{E_m\}$ are valid measurement operators. 
	In POVM, the probability of getting outcome $m$ is $p(m)=\langle \psi|E_m|\psi\rangle$.
	If the measurement operators and the POVM elements coincide in any measurement then the measurement is a projective measurement.\\
	For instance, suppose Alice send Charlie have one of the two states from the set $\{|\psi_1\rangle, |\psi_2\rangle\}$ where $|\psi_1\rangle=|1\rangle$ and $|\psi_2\rangle=\frac{|0\rangle-|1\rangle}{\sqrt{2}}$. Charlie doesn't know whether he is given $|\psi_1\rangle$ or $|\psi_2\rangle$, so, he performs POVM to determine the received state. To perform POVM, we first need to define the POVM elements. In this case, the POVM elements can be written in the following form
	\begin{eqnarray}
	E_1&=&\frac{\sqrt{2}}{1+\sqrt{2}}|0\rangle \langle 0|\\
	E_2&=&\frac{\sqrt{2}}{1+\sqrt{2}}\frac{(|0\rangle +|1\rangle)(\langle 0|+\langle 1|)}{2}\\
	E_3&=&I-E_1-E_2
	\end{eqnarray}
	where $I$ is $2\times 2$ identity matrix. Then the set $\{E_1,E_2,E_3\}$ form a POVM. Suppose Charlie is given $|\psi_1\rangle$ and he performs the measurement using $\{E_1,E_2,E_3\}$. There is a zero probability that he will get measurement outcome $E_1$. If the result of the measurement outcome comes out to be $E_1$, then Charlie can conclude that he was given $|\psi_2\rangle$. Also, if the measurement outcome comes out to be $E_2$, the state Charlie has must be $|\psi_1\rangle$. If the measurement outcome is $E_3$, then Charlie can infer nothing about the identity of the state but whenever Charlie identifies the state he can never make a mistake in identifying the state.

	%The interior of the Bloch sphere, the open Bloch ball, represents the mixed states of a single qubit, that means mixed states lies within the bloch sphere (in the open bloch ball).	

	\section{EPR Paradox}
	\noindent To explain the EPR paradox, let us discuss the following experiment: Suppose two spin 1/2 particles are ejected in opposite directions each time a button is pressed, and they finally pass through a detector that determines the spin of the particle along a specific axis. The detectors can be rotated which makes it possible to measure the spin along any axis, particularly for x and z axis. After a number of tests, the following observation is made:\\
	\textbf{Observation-1 ($O1$):} To measure spin along the z-axis, both detectors are configured. Half of the time, the particles on the right have spin down and the particles on the left have up spin, and the other half of the time, the particles on the right are up spin and the particles on the left are down spin.\\
	This implies that we may determine the z-component of the spin of the particle on the right by measuring the z-component of the spin of the particle on the left without actually interacting with the particle on the right. Suppose L denotes the particle which moves to the left and R denotes the particles which move to the right. After performing the experiment and knowing that L has spin up, we know that R has spin down. Since the measurement on L couldn't possibly affect R it means that R was spin down even before the measurement on L was performed.\\
	Let us consider the spin state of the particles before entering the detectors which may be denoted by $|\psi_1\rangle$
	\begin{eqnarray}
	|\psi_1\rangle=U_z^{L}D_z^{R}
	\end{eqnarray}
	where $U_z$ and $D_z$ denote up spin and down spin respectively, along the $z$-axis. Here 
	\begin{eqnarray}
	U_z=
	\begin{pmatrix}
	1\\
	0
	\end{pmatrix},
	D_z=
	\begin{pmatrix}
	0\\
	1
	\end{pmatrix}
	\end{eqnarray}
	Similarly, if L has a down spin, then R has an up spin and the spin state of the particles may be written as
	\begin{eqnarray}
	|\psi_2\rangle=D_z^{L}U_z^{R}
	\end{eqnarray}
	From $O1$, half of the time, the particles are in state $|\psi_1\rangle$ and in the other half of the time, the particles are in state $|\psi_2\rangle$. After repeating the experiment many more times we can make another observation that can be stated as follows:\\
	\textbf{Observation-2 ($O2$):} The two particles never have spins that are opposite to one another when both detectors are configured to monitor spin in the x-direction.\\
	The reason is that if the particles were initially in the state $|\psi_1\rangle$, then the probability that measurement would detect both spins up in the x-direction is given by
	\begin{eqnarray}
	|\langle U_x^{L}U_x^{R}|\psi_1\rangle|^2
	\end{eqnarray}
	where $U_x$ and $D_x$ are given by
	\begin{eqnarray}
	U_x=\frac{1}{\sqrt{2}}
	\begin{pmatrix}
	1\\
	1
	\end{pmatrix},
	D_x=\frac{1}{\sqrt{2}}
	\begin{pmatrix}
	1\\
	-1
	\end{pmatrix}
	\end{eqnarray}
	Thus
	\begin{eqnarray}
	|\langle U_x^{L}U_x^{R}|\psi_1\rangle|^2=|\langle U_x^{L}U_x^{R}|U_z^{L}D_z^{R}\rangle|^2=|\langle U_x^{L}|U_z^{L}\rangle \langle U_x^{R}|D_z^{R}\rangle|^2=\frac{1}{4}
	\end{eqnarray}
	If the particles were initially in state $|\psi_2\rangle$, then the probability that both particles have up spin in the x-direction would be
	\begin{eqnarray}
	|\langle U_x^{L}U_x^{R}|\psi_2\rangle|^2=|\langle U_x^{L}U_x^{R}|D_z^{L}U_z^{R}|^2=|\langle U_x^{L}|D_z^{L}\rangle \langle U_x^{R}|U_z^{R}\rangle|^2=\frac{1}{4}
	\end{eqnarray}
	The probability that both the particles with spin up along the x-axis may be defined as the average of the two probabilities that the particles are in state $|\psi_1\rangle$ and the probability that the particle is in the $|\psi_2\rangle$
	\begin{eqnarray}
	\frac{1}{2}|\langle U_x^{L}U_x^{R}|\psi_1\rangle|^2+\frac{1}{2}|\langle U_x^{L}U_x^{R}|\psi_2\rangle|^2=\frac{1}{4}
	\end{eqnarray}
	From $O1$ and $O2$, there is no experiment that satisfies both conclusions.
	EPR pointed out that these outcomes can be obtained for certain quantum mechanical states. Suppose that before reaching the detectors, two particles are in the state
	\begin{eqnarray}
	|\psi\rangle=\frac{1}{\sqrt{2}}(|\psi_1\rangle -|\psi_2\rangle)=\frac{1}{\sqrt{2}}(U_z^{L}D_z^{R}-D_z^{L}U_z^{R})
	\end{eqnarray}
	Then, along the z-axis, the probability of L with up spin and R with down spin may be expressed as
	\begin{eqnarray}
	|\langle U_z^{L}D_z^R|\psi\rangle|^2=\frac{1}{2}|\langle U_z^{L}D_z^R|\psi_1\rangle-\langle U_z^{L}D_z^R|\psi_2\rangle|^2=\frac{1}{2}
	\end{eqnarray}
	Similarly, the probability of L with a down spin and R with an up spin is given by
	\begin{eqnarray}
	|\langle D_z^{L}\alpha_z^R|\psi\rangle|^2=\frac{1}{2}|\langle U_z^{L}D_z^R|\psi_1\rangle-\langle U_z^{L}D_z^R|\psi_2\rangle|^2=\frac{1}{2}
	\end{eqnarray}
	These two probabilities are the same as we got in case $O1$.\\
	The probability of both L and R with spin up along the x-axis is given by
	\begin{eqnarray}
	|\langle U_x^{L}U_x^{R}|\psi \rangle|^{2}=0
	\end{eqnarray}
	which is the same as the result of $O2$.\\
	If $O1$ and $O2$ both are true, then this would mean that before the measurement, R was neither in state $D_z^{R}$ nor in state $U_z^{R}$. So, if we measure that L has spin up along the z-axis, at that very instant we know that R has spin down without actually performing a measurement on R.
	Also, before the measurement, R was neither spin up nor spin down. So, initially, R was not in the spin down direction which means it must have jumped in a down spin as soon as we found that L has an up spin even when L and R are very far away at the time of measurement. But EPR argued that how could a measurement on L influence the state of R even when L and R are spatially separated? This is known as the EPR paradox. Since this paradox is not explained by the theory of quantum mechanics, so, it makes Einstein, Podolsky, and Rosen believe that the quantum theory is not complete. Many experiments have been performed to solve the EPR paradox but the success came in 1964 when J S Bell put forward his theory in terms of an inequality. The inequality derived by Bell is popularly known as Bell's inequality. An experiment has been performed by Alain Aspect et.al.\cite{aspect1, aspect2} to show that the correlation that may exist between linear polarization of pair of photons violates Bell's inequality. This indicates the fact that the correlation exists between the linear polarization of two photons may be non-local in nature.
	%After the development of Bell's inequality the question of completeness of quantum theory shifted to the question that whether or not the quantum theory is compitable with separability and locality.

	% first proposed the concept of quantum entanglement in 1935 \cite{einstein} mechanics, according to Einstein and his colleagues was incomplete because they believed that this special feature of quantum mechanics may not exist in nature. When pairs or groups of particles are created or interact in such a way that the quantum states of each particle cannot be described independently, this is known as quantum entanglement. Entangled particals remains connected in such a way that actions performed on one particle affects the other even when the particles are very far away.
\section{Quantum entanglement}
	% The EPR paradox demonstrates that a measurement can be made on a particle without directly disturbing it by making a measurement on a different entangled particle. After the EPR paradox, Einstein et.al. believed that quantum mechanics is not a complete theory. An important step in the direction to solve EPR problem came in 1964 by Bell. Bell used the idea of EPR and derived theoretical result in form of Bell's inequality which must be satisfied if the assumptions of EPR are true. The experimental realization of Bell's inequality was given by Alain Aspect et.al.\cite{aspect1, aspect2}.\\
	\noindent The term entanglement was first coined by Schrodinger \cite{schrodinger} and at the same time, Einstein, Podolsky, and Rosen recognized this feature as a special feature of quantum mechanics that has no classical analog and called it a "spooky action at a distance" \cite{einstein}. Quantum entanglement has a very complex structure that is very fragile with respect to the environment. In an entangled system, the subsystems are distributed to the regions which are far apart from one another. The amount of entanglement in the system cannot increase on average when a measurement is performed on one subsystem. Let us now understand this complex structure of entanglement that may exist in the bipartite and tripartite systems.
	\subsection{Bipartite and tripartite system}
	\noindent \textbf{I. Bipartite system}\\
	\noindent Let us consider two individual systems $A$ and $B$ consisting of the state vectors residing in the Hilbert spaces $H_A$ and $H_B$ respectively. The composite system of two individual systems can be considered a bipartite system and it is described in the Hilbert space $H_{AB}$. The two individual systems $A$ and $B$ can be referred to as the subsystems of the composite system. A composite Hilbert space $H_{AB}$, may be defined as the tensor product of two subsystems $A$ and $B$, and it is given by
	\begin{eqnarray}
	H_{AB}\equiv H_A^{d_A}\otimes H_B^{d_B}
	\end{eqnarray}
	where $d_A$ and $d_B$ are the dimensions of individual subsystems $A$ and $B$.\\
	If $|\psi_i\rangle_{AB}$ denote the pure state in the Hilbert space $H_{AB}$ then any state in the composite Hilbert space is described by the density operator $\rho_{AB}$
	\begin{eqnarray}
	\rho_{AB}=\sum_{i}p_i|\psi_i\rangle_{AB} \langle \psi_i|,~~0\leq p_i \leq 1,~~\sum_i{p_i}=1
	\end{eqnarray}
	%Bipartite system may be defined as the tensor product of two quantum systems. For a bipartite system, every state is either a separable state or an entangled state. 
	%In a bipartite system, separable states may be defined as,
	%\begin{eqnarray}
	%\rho^{AB}=\sum_{i}{p_i\rho_{i}^{A}\otimes \rho_{i}^{B}}
	%\label{separable}
	%\end{eqnarray}
	%Any bipartite states that are not in the form (\ref{separable}) are called entangled states.\\
	\noindent Let us consider a bipartite system described by a density matrix $\rho_{AB}$. The subsystem $A$ or $B$ may be extracted from the composite system $AB$ by using the mathematical formulation known as the partial trace. The partial trace with respect to subsystem $A$ or $B$ may be defined as
	\begin{eqnarray}
	\rho^{A}&=&Tr_B(\rho_{AB})\nonumber\\
	\rho^{B}&=&Tr_A(\rho_{AB})
	\end{eqnarray}
	Here, the operators $\rho^{A}$ and $\rho^{B}$ are positive semi-definite operators and fulfill the condition $Tr(\rho^{A})=1$ and $Tr(\rho^{B})=1$. Thus, the operators $\rho^{A}$ and $\rho^{B}$ satisfy the condition of a density matrix, and hence, the operators $\rho^{A}$ and $\rho^{B}$ are called reduced density matrices.\\
	\textbf{(a) Entanglement in bipartite system:}\\
	%	In most cases we have mixed states due to environmental noise.
	%The quantum state $\rho$ in a bipartite Hilbert space may be expressed as
	%\begin{eqnarray}
	%\rho=\sum_{i}{p_i|\phi_{i}\langle \phi_{i}|}, ~~\sum_i{p_i}=1~~ \text{and}~~ p_i \geq 0
	%\end{eqnarray}
	%Here, $p_i$ denotes the probability of occurrence of $\rho$ in one of the state $|\phi_{i}\rangle \in \textbf{H}$. Also, the convex combination of two states $\rho_1$ and $\rho_2$ represents a state i.e. if $\rho_1$ and $\rho_2$ represent states of a quantum system, then $\rho=a\rho_1+(1-a)\rho_2$ $(a\in [0,1])$ also represents a quantum system.
	If $\rho^{AB}$ represents the density matrix of a composite system described in the Hilbert space $H_A\otimes H_B$, then it defines a product state if there exist subsystems $\rho^{A}$ and $\rho^{B}$ such that
	\begin{eqnarray}
	\rho^{AB}=\rho^{A}\otimes \rho^{B}
	\end{eqnarray}
	More generally, we can say that the state is separable if it can be written as the convex combination of product states
	\begin{eqnarray}
	\rho^{AB}=\sum_{i}{p_i\rho_{i}^{A}\otimes \rho_{i}^{B}},~~0\leq p_i\leq 1, ~~\sum_{i}{p_i}=1
	\label{separablestate}
	\end{eqnarray}
	where, $p_i$ are the convex weights and $\rho_{i}^{A}\otimes \rho_{i}^{B}$ represent the product states.\\
	If the state $\rho^{AB}$ is not expressed in the above form (\ref{separablestate}), then the state $\rho^{AB}$ is said to be an entangled state. These states cannot be generated via local operation and classical communication (LOCC). For instance, a bipartite two-qubit pure state $|\phi\rangle_{AB}=|0\rangle_A \otimes \frac{|0\rangle_B+|1\rangle_B}{\sqrt{2}}$ represent a pure product state in a two-qubit system.\\
	%\noindent Product states are uncorrelated states whereas the non-product states are correlated states. Separable states have classical correlations between them. This means for the generation of separable states, we need local operation and classical communicatio(LOCC) and entangled states have quantum correlations between them. %It is very important to check whether the state is entangled.
	%A pure state $|\phi^{AB}\rangle$ in the Hilbert space $H\equiv H_A^{d_1}\otimes H_B^{d_2}$ is said to be a pure separable state if it can be written in the form,
	%\begin{eqnarray}
	%|\phi^{AB}\rangle=\sum_i{p_i|\phi_i^A\rangle |\phi_i^B\rangle}
	%\label{pureseparablestate}
	%\end{eqnarray}
	%where $|\phi^{A}\rangle \in$ \textbf{$H_{A}^{d_1}$} and $|\phi^{B}\rangle \in$ \textbf{$H_{B}^{d_2}$}. Otherwise the state is called entangled. For instance, a bipartite two-qubit pure state $|\phi^{AB}\rangle=|0\rangle \otimes \frac{|0\rangle+|1\rangle}{\sqrt{2}}$ is a pure product state in a two-qubit system.
	\textbf{(b) Schmidt Decomposition:} Consider a pure state $|\psi\rangle_{AB}$ of a composite system AB. Then there exist orthonormal states $|i_A\rangle$ and $|i_B\rangle$ for subsystems A and B respectively such that the state $|\psi\rangle_{AB}$ may be written as
	\begin{eqnarray}
	|\psi\rangle_{AB}=\sum_{i}{\lambda_i|i_A\rangle |i_B\rangle},~~\sum_i{\lambda_i}=1
	\end{eqnarray} where $\lambda_i's \geq 0$ are known as Schmidt coefficients.\\
	In particular, when all Schmidt coefficients, except one of the coefficient, are zero, then the state $|\psi\rangle_{AB}$ reduces to a pure product state. When at least two Schmidt coefficients are non-zero, then the pure state $|\psi\rangle_{AB}$ is said to be entangled. In this way, Schmidt's decomposition of any pure state may help us to check whether the given pure state is entangled or separable.\\
	\textbf{(c) Entanglement measures $(\eta(\rho_{AB}))$:} Entanglement measures quantifies the amount of entanglement in a quantum state $\rho_{AB}=\sum_k{p_k({\rho_k})_{AB}}$, where $p_k$ denote the classical probability. $\eta(\rho_{AB})$ can be defined as a non-negative real-valued function that satisfies the following properties, some of which are not satisfied by all entanglement measure \cite{plenio}:\\
	(i) $\eta(\rho_{AB})$ vanishes for the separable state $\rho$.\\
	(ii) A bipartite entanglement measure $\eta(\rho_{AB})$ defines a mapping from the set of density matrices to the set of positive real numbers:
	\begin{eqnarray}
	\rho \rightarrow \eta(\rho)\in R^{+}
	\end{eqnarray}  A normalization
	condition is generally used, for example, for two qudit maximally entangled state of the form
	\begin{equation}
	|\psi\rangle_d=\frac{|0,0\rangle+|1,1\rangle+...|d-1,d-1\rangle}{\sqrt{d}}
	\end{equation}
	The value is given by $\eta(|\psi\rangle_d)=log(d)$.\\
	(iii) $\eta(\rho_{AB})$ does not increase on average under local operation and classical communication (LOCC)
	\begin{eqnarray}
	\eta(\rho)=\sum_i{p_i\eta(\frac{K_i\rho K_i^{\dagger}}{Tr[K_i\rho K_i^{\dagger}]})}
	\end{eqnarray}
	where the $K_i$ are the Kraus operators.\\
	(iv) For pure states $|\psi\rangle \langle \psi|$, $\eta(\rho)$ reduces to entropy of entanglement i.e.
	\begin{eqnarray}
	\eta(|\psi\rangle \langle \psi|)=(S o Tr_B)(|\psi\rangle \langle \psi|)
	\end{eqnarray}
%	(ii) $\eta(\rho_{AB})$ must be invariant under local unitary transformation i.e
%	\begin{eqnarray}
%	\eta(\rho_{AB})=\eta(U_A\otimes U_B\rho_{AB} U_A^{\dagger}\otimes U_B^{\dagger})
%	\end{eqnarray}
%	where $U_A$ and $U_B$ denote the unitary operator for the individual subsystem $A$ and $B$.\\
	
%	(iv) $\eta(\rho_{AB})$ satisfies the convexity property i.e.
%	\begin{eqnarray}
%	\eta(\sum_k{p_k(\rho_k)_{AB}})\leq \sum_k{p_k\eta(\rho_k)_{AB}}
%	\end{eqnarray}
	%(v) $\eta(\rho_{AB})$ satisfies the additivity property i.e. If $\rho_{AB}^{(1)}$ and $\rho_{AB}^{(2)}$ are two quantum states then 
	%\begin{eqnarray}
	%\eta(\rho_{AB}^{(1)}\otimes \rho_{AB}^{(2)})=\eta(\rho_{AB}^{(1)}) +\eta(\rho_{AB}^{(2)})
	%\end{eqnarray}
	%It is not necessary that the entanglement measure satifies all the above properties. For some entanglement measures, the convexity property and the additivity property are not met, or they are difficult to prove.\\	
	\noindent Any function $\eta$ satisfying the first three conditions is known as an entanglement monotone. If $\eta$ satisfies the property (i),(ii), and (iv) and also it does not increase under deterministic LOCC transformations, then $\eta$ is known as an entanglement measure.\\
	\textbf{(i) von Neumann entropy:} Consider a quantum state described by a density operator $\rho_{AB}$, then the entropy of entanglement is a way to quantify the amount of entanglement in a pure state. The entropy of entanglement of $\rho_{AB}$ may be defined as
	\begin{eqnarray}
	E(\rho_{AB})=V(\rho_A)=V(\rho_B)
	\end{eqnarray}
	where $V(\rho)=-Tr[\rho log_2\rho]$ is the von Neumann entropy of $\rho$. Here $\rho_A=Tr_B(\rho_{AB})$ and $\rho_B=Tr_A(\rho_{AB})$ are the reduced matrices obtained by tracing out subsystem $A$ and $B$ respectively.\\
	\textbf{(ii) Entanglement of formation:} The entanglement of formation of a bipartite mixed state $\rho_{AB}$ may be defined as the convex roof of the von Neumann entropy
	\begin{eqnarray}
	\eta_F(\rho_{AB})=\inf_{\{p_k,|\phi_k\rangle\}}\sum_k{p_kV[(\rho_A)_k]}
	\end{eqnarray}
	where $\rho_A$ is the reduced state of $\rho_{AB}$. Physically, it may be understood as the minimum number of singlets needed to create one copy of the state\cite{guhne}.\\
	\textbf{(iii) Concurrence:} For two-qubit pure state $|\psi \rangle_{AB}^{{2\otimes 2}}$, entanglement can be quantified using an entanglement measure known as concurrence \cite{wootters,wootters1,wootters2}, which is defined as
	\begin{eqnarray}
	C(|\psi \rangle_{AB}^{{2\otimes 2}})=\sqrt{2(1-Tr(\rho_{A}^{2}))}
	\end{eqnarray}
	where $\rho_{A}$ is the reduced state of $|\psi \rangle_{AB}^{{2\otimes 2}}$. It may also be generalized to $d\otimes d$ dimensional bipartite pure states as
	\begin{eqnarray}
	C(|\psi\rangle_{AB}^{d_1\otimes d_2})=\sqrt{2\nu_{d_1}\nu_{d_2}[1-Tr(\rho_{A}^{2})]}
	\end{eqnarray}
	where $\rho_{A}$ is the reduced state of $|\psi \rangle_{AB}^{d_1\otimes d_2}$ and $\nu_{d_1}$ and $\nu_{d_2}$ are positive constants.\\
	Concurrence for the two-qubit mixed state described by the density operator $\rho_{AB}$, may be defined as,
	\begin{eqnarray}
	C(\rho_{AB})=max(0,\sqrt{\lambda_{1}}-\sqrt{\lambda_{2}}-\sqrt{\lambda_{3}}-\sqrt{\lambda_{4}})
	\end{eqnarray}
	where	$\lambda_{i}'s$ are the eigenvalues of $\rho_{AB}\widetilde{\rho_{AB}}$ and arranged in descending order. Here, $\widetilde{\rho_{AB}}$=$(\sigma_{y}\otimes\sigma_{y})\rho_{AB}(\sigma_{y}\otimes\sigma_{y})$.\\
	The closed formula for the concurrence of $d\otimes d$ dimensional mixed bipartite system is still not known.\\
	\textbf{(iv) Negativity:} Negativity \cite{vidal} is a popular measure of entanglement used for quantification of entanglement in $d\otimes d$ dimensional bipartite system. For a given density matrix $\rho_{AB}$, it is defined as the sum of negative eigenvalues of the partially transposed matrix $\rho_{AB}^{T_B}$. It may be expressed as \cite{lee}
	\begin{eqnarray}
	N(\rho_{AB})=\frac{||\rho_{AB}^{T_B}||_1-1}{d-1}
	\label{negativity}
	\end{eqnarray}
	where $||.||$ denotes the trace norm.\\ \ \\
	\textbf{II. Tripartite system:}\\
	In the case of a tripartite system, the structure of entanglement is more complex due to the existence of several inequivalent classes of entanglement. For a tripartite system, all states are divided into three inequivalent classes under stochastic local operations and classical communication (SLOCC). These classes are: (i) class of fully separable states (ii) class of biseparable states and (iii) class of genuine entangled states.\\
	\textbf{(i) Class of fully separable states:} Fully separable (FS) three-qubit pure states may be expressed as
	\begin{eqnarray}
	|\phi\rangle_{ABC}^{FS}=|\alpha\rangle_A\otimes|\beta\rangle_B\otimes|\gamma\rangle_C
	\end{eqnarray}
	A fully separable mixed state described by the density operator $\rho^{FS}$ can be expressed as
	\begin{eqnarray}
	\rho^{FS}=\sum_i{p_i|\phi_i^{FS}\rangle \langle \phi_i^{FS}|},~0\leq p_i\leq 1,~ \sum_i{p_i}=1
	\end{eqnarray}
	where each $|\phi_i^{FS}\rangle$ is a fully separable pure state.\\
	\textbf{(ii) Class of biseparable states:}
	Three-qubit pure biseparable states may be expressed as
	\begin{eqnarray}
	|\phi\rangle_{ABC}^{BS}=|a\rangle_A\otimes|d\rangle_{BC}
	\label{onekindbisep}
	\end{eqnarray}
	Here $|d\rangle$ denotes the state which represents an entangled state between the subsystem $B$ and $C$. In the class of biseparable state, two of the three qubits are grouped as one party. Thus, there may exist three possibilities that can be obtained by taking two parties at one time. Hence, we can find three kinds of biseparable states in this class. One kind of biseparable state is given in (\ref{onekindbisep}) and the other two may be defined as
	\begin{eqnarray}
	|\phi\rangle_{ABC}^{BS}&=&|a_1\rangle_B\otimes|d_1\rangle_{AC}\\
	|\phi\rangle_{ABC}^{BS}&=&|a_2\rangle_C\otimes|d_2\rangle_{AB}
	\end{eqnarray}
	A mixed biseparable state can be expressed as a convex combination of different biseparable pure states. Mathematically, biseparable mixed states may be expressed as \cite{guhne}
	\begin{eqnarray}
	\rho^{BS}=\sum_{i}{p_i|\phi_{i}^{BS}\rangle \langle \phi_{i}^{BS}|}
	\end{eqnarray}
	We should note that the state $|\phi_i^{BS}\rangle$ and $|\phi_j^{BS}\rangle$ ($i\neq j$) may not belong to the same partition. For instance, let us consider a three-qubit state described by the density operator $\rho_1$ as
	\begin{eqnarray}
	\rho_1&=&q[|0\rangle_{A} \langle 0| \otimes |\phi^{+}\rangle_{BC} \langle \phi^{+}|]+(1-q)[|1\rangle_{B} \langle 1|\otimes |\phi^{-}\rangle_{AC} \langle \phi^{-}|]
	\label{ex3}
	\end{eqnarray}
	where $0\leq q\leq 1$ and the Bell states $|\phi^{+}\rangle_{BC}$ and $|\phi^{-}\rangle_{AC}$ are given by
	\begin{eqnarray}
	|\phi^{+}\rangle_{BC} &=&\frac{1}{\sqrt{2}}(|00\rangle_{BC}+|11\rangle_{BC}),~~
	|\phi^{-}\rangle_{AC}=\frac{1}{\sqrt{2}}(|00\rangle_{AC}-|11\rangle_{AC})
	\end{eqnarray}
	Identifying $|\phi_1^{BS}\rangle \langle \phi_1^{BS}| =|0\rangle_{A} \langle 0| \otimes |\phi^{+}\rangle_{BC} \langle \phi^{+}|$, a biseparable state in A-BC partition and $|\phi_2^{BS}\rangle \langle \phi_2^{BS}|=|1\rangle_{B} \langle 1|\otimes |\phi^{-}\rangle_{AC} \langle \phi^{-}|$, a biseparable state in another bipartition cut B-AC, we can re-express $\rho_1$ as
	\begin{eqnarray}
	\rho_1=q|\phi_1^{BS}\rangle \langle \phi_1^{BS}|+(1-q)|\phi_2^{BS}\rangle \langle \phi_2^{BS}|
	\end{eqnarray}
	\textbf{(iii) Class of genuinely entangled state:} A three-qubit state is a genuine entangled state if it is neither separable nor biseparable. Therefore, in a genuine three-qubit entangled state, all three qubits must be correlated to each other. For a three-qubit system, there are two SLOCC inequivalent genuine entangled states namely, GHZ states and W states.\\
	Any three-qubit pure state can be written as,
	\begin{eqnarray}
	|\psi\rangle_{ABC}=\lambda_0|000\rangle+\lambda_1e^{i\theta}|100\rangle+\lambda_2|101\rangle+\lambda_3|110\rangle+\lambda_4|111\rangle,\lambda_{i} \geq 0,\sum_{i=0}^{4}{\lambda_i}^{2}=1
	\label{ghz5}
	\end{eqnarray}
	where, $\theta \in [0,\pi]$ and $\{|0\rangle, |1\rangle \}$ denotes the computational basis.\\
	A three-qubit state $|\psi\rangle_{ABC}$ given in (\ref{ghz5}) can be classified as a three-qubit state of GHZ type. A three-qubit W state may be expressed as
	\begin{eqnarray}
	|\psi\rangle_W=\lambda_0|000\rangle+\lambda_1e^{i\theta}|100\rangle+\lambda_2|101\rangle+\lambda_3|110\rangle
	\end{eqnarray}
	%	Pue three-qubit GHZ states may be expressed as
	%	\begin{eqnarray}
	%	|\phi\rangle^{GHZ}=|100\rangle+|010\rangle+|001\rangle
	%	\end{eqnarray}
	%	and GHZ class of states may be expressed as
	%	\begin{eqnarray}
	%	|\phi\rangle^{GHZ}=\frac{1}{\sqrt{2}}[|000\rangle+|111\rangle]
	%	\end{eqnarray}
	To distinguish between GHZ class of states and the W class of states, we need to define a quantity, popularly known as tangle.\\
	\textbf{Tangle:} Tangle \cite{eltschka} is an entanglement measure used to characterize the three-qubit entanglement of the state. %The quantification of entanglement in a multipartite system may be defined using tangle.  %For a three-qubit pure state described by the density operator $\rho_{123}$, a three-tangle may be defined as the residual entanglement in a three-qubit system.
	%	\begin{eqnarray}
	%	\tau=C_{1(23)}^2-C_{12}^2-C_{13}^2
	%	\end{eqnarray}
	Consider a three-qubit pure state defined by $|\phi\rangle_{ABC} \in H_A \otimes H_B \otimes H_C$. In computational basis, the state $|\phi\rangle_{ABC}$ can be expressed as
	\begin{eqnarray}
	|\phi\rangle_{ABC}=a|000\rangle+b|001\rangle+c|010\rangle+d|011\rangle+e|100\rangle+f|101\rangle+g|110\rangle+h|111\rangle
	\end{eqnarray}%
	The state $|\phi\rangle_{ABC}$ is normalised if $|a|^2+|b|^2+|c|^2+|d|^2+|e|^2+|f|^2+|g|^2+|h|^2=1$.\\
	The tangle of a pure state $|\phi\rangle_{ABC}$ may be denoted by $\tau_3(|\phi\rangle_{ABC})$ and defined as
	\begin{eqnarray}
	\tau_3(|\phi\rangle_{ABC})=4|d_1-2d_2+4d_3|
	\end{eqnarray}
	where,
	\begin{eqnarray}
	d_1&=&a^2h^2+b^2g^2+c^2f^2+e^2d^2\\
	\nonumber d_2&=&ahde+ahfc+ahgb+defc+degb+fcgb\\
	%	d_3=\psi_{000}\psi_{110}\psi_{101}\psi_{011}+\psi_{111}\psi_{001}\psi_{010}\psi_{100}
	d_3&=&agfd+hbce
	\end{eqnarray}
	The three-qubit mixed state can be written as a convex combination of pure three-qubit states
	\begin{eqnarray}
	\rho_{ABC}=\sum_{j}{p_j\pi_j}, ~~\pi_j=\frac{|\phi_j\rangle_{ABC} \langle \phi_j|}{\langle \phi_j|\phi_j  \rangle_{ABC}}
	\end{eqnarray}
	The tangle of a three-qubit mixed state may be obtained by the convex roof extension method. Therefore, the tangle for a mixed three-qubit state described by the density operator $\rho_{ABC}$ is given by
	\begin{eqnarray}
	\tau_3(\rho_{ABC})=\displaystyle{\min_{\{p_j,\pi_j\}}}\sum_{j}{P_j\tau_3(\pi_j)}
	\end{eqnarray}
	\textbf{Three-$\pi$}: Three-$\pi$ \cite{cheng} is another measure of entanglement for a three-qubit system. For three-qubit pure states, it may be defined as
	\begin{eqnarray}
	\pi=\frac{\pi_a+\pi_b+\pi_c}{3}
	\end{eqnarray}
	where, $\pi_a$, $\pi_b$ and $\pi_c$ may be defined as
	\begin{eqnarray}
	\pi_a&=& N_{A(BC)}^{2}-N_{AB}^{2}-N_{AC}^2\\
	\pi_b&=& N_{B(AC)}^{2}-N_{BA}^{2}-N_{BC}^2\\
	\pi_c&=& N_{C(AB)}^{2}-N_{CA}^{2}-N_{CB}^2
	\end{eqnarray}
	Here, $N_{ij}=\frac{||\rho^{T_i}||-1}{2},~~i,j=\{A,B,C\}$, where ||.|| represent the trace norm and $N$ represent the negativity of the bipartite system and $N_{i(jk)}=2\sqrt{det([tr_{jk}(\rho_{ijk})])}$.
 
	%These measurement operators acts on the state space of the system being measured. Consider $|\phi\rangle$ to be the state of the quantum system before mesurement then the state defining the system after measurement is given by,
	%\begin{eqnarray}
	%$\frac{M_m|\phi\rangle}{\sqrt{\langle %\phi|M_m^{\dagger}M_m|\phi\rangle}}
	%\end{eqnarray}
	%and the probability of getting result m is given by,
	%\begin{eqnarray}
	%p(m)=\langle \phi|M_m^{\dagger}M_m|\phi\rangle
	%\end{eqnarray}
	%The completeness relation defines the fact that the sum of the probabilities is equal to one,
	%\begin{eqnarray}
	%\sum_m{\langle \phi|M_m^{\dagger}M_m|\phi\rangle}=\sum_{m}{p(m)}=1
	%\end{eqnarray}
	%\section{Multipartite Entanglement}
	%In this section, we will discuss entanglement when two or more subsystems are involved. When two or more subsystems are involved, there are several classes of entangled states. Here, we will mainly concentrate on the separability and entanglement in case of a three-qubit system. Then we will discuss general multipartite system. The main classes of states that may exist in multipartite system such as GHZ class of states, W class of states. They are important classes  because of their entanglement properties and their usefulness in various quantum information processing tasks.
	%\subsection{Entanglement in three-qubit states}
	%\textbf{Pure states:} 
	%\subsection{Families of multipartite entangled states}

\newpage
	%\section{Measurement}
	%
	%If we define $E_m=M_m^{\dagger}M_m$, then $E_m$ is a positive operator. Moreover $\sum_m{E_m}=I$ and $p(m)=\langle \phi|E_m|\phi\rangle$. These operators $E_m$ are known as POVM elements associated with the measurement and the set $\{E_m\}$ is known as a POVM.
	%\newpage
	%The problem of classification of separable state and entangled state is known as separability problem.
	%Using the definition, it is not easy to check whether the given state is separable or entangled since we need to check for every basis. Various criteria have been proposed that can classify whether the given state is entangled or separable.  However, there is no general solution of the separablity problem is available till date.

	\newpage
	\section{Entanglement Detection}
	\noindent Quantum entanglement plays a significant role in the development of quantum information theory. The first revolutionary idea of using entanglement as a resource in quantum teleportation was proposed by Bennett et.al \cite{Bennett1}. Many experimental proposals were put forward to realize quantum teleportation \cite{bouwmeester, boschi, kim, yonezawa, hu}. Afterward, researchers applied the concept of entanglement as a resource in many other quantum information processing tasks such as quantum superdense coding\cite{Bennett2,harrow,barreiro,williams}, remote state preparation\cite{pati,Bennett3,lo,berry}, entanglement swapping\cite{zeilinger,bose,bouwmeester2,su} and quantum cryptography \cite{gisin1,gisin2,gisin3,zhou}. Therefore, entanglement can be considered a vital ingredient in quantum information theory without which the above-mentioned quantum information processing task cannot be performed. Thus, it is necessary to create entangled states to get a better result than classical states in performing quantum information processing tasks. Mere creation of entanglement is not sufficient to fulfill our demand because it is not known whether the generated state is entangled or not.\\
	\noindent Creation and detection of entanglement are thus one of the important problems in quantum information theory. Hence, efficient physically realizable methods for the detection, classification, and quantification of quantum entanglement are of great importance. A lot of progress has been achieved in the creation of entangled states experimentally \cite{tittle,weihs,hagley,sackett,furusawa,howell} but due to the presence of noise in the environment the generated state need not be entangled. Thus, it is important to check whether the state generated in an experiment is entangled or not. The ultimate goal of the detection of entanglement in a given state is to characterize entanglement in a quantitative way and identify the states in which entanglement is maximum. The first step towards this problem is the detection of entanglement in a given state. In the literature, there are many criteria that may help in the detection of entangled states. We can divide these criteria for the detection of entanglement into two categories C1 and C2, which may be given by\\
	C1. Theoretical criteria that may not be realized in an experiment\\
	%\noindent On the mathematical side, the theory of entanglement has revealed very interesting connections with the theory of positive maps \cite{horodecki1, werner}. 
	%Detection of entanglement has a very vast literature. The detection criteria for the 
	C2. Criteria that may be realized in an experiment.
	\subsection{Entanglement detection criterion that may not be implementable in an experiment}
	% ----------------------------------------------------------------
	\textbf{I. The Positive Partial Transposition (PPT) criteria}
	The first necessary criteria for the detection of entanglement was given by Peres \cite{peres}. A bipartite $N\otimes M$ dimensional quantum state described by the density matrix $\rho_{AB}$, may be expressed as,
	\begin{eqnarray}
	\rho_{AB}=\sum_{i,j}^N\sum_{k,l}^M{\rho_{ij,kl}|i\rangle \langle j| \otimes |k\rangle \langle l|}
	\end{eqnarray}
	%Then, partial transposition of the density matrix $\rho$ may be defined with respect to subsystem A and subsystem B. 
	The partial transposition of $\rho_{AB}$ with respect to subsystem A may be expressed as
	\begin{eqnarray}
	\rho_{AB}^{T_A}=\sum_{i,j}^N\sum_{k,l}^M{\rho_{ij,kl}|i\rangle \langle l| \otimes |k\rangle \langle j|}
	\end{eqnarray}
	Similarly, one can define partial transposition of the density matrix $\rho_{AB}$ with respect to subsystem B.\\
	%Any density matrix is said to be a positive partial transpose(PPT), if its partial transposition has no negative eigenvalue i.e. its partial transposition defines a positive semi-definite matrix:
	%	\begin{eqnarray}
	%	\rho^{T_A}\geq 0~~\text{iff}~~\rho^{T_B}\geq 0
	%	\end{eqnarray}
	%If a state is not PPT, then we call it NPT. 
	%PPT Criteria, also known as Peres-Horodecki criteria\cite{} deals with the detection of separable state. It may be stated as,\\
	\textbf{PPT Criteria \cite{peres}:} If a bipartite quantum state described by density operator  $\rho_{AB}$ is separable, then $\rho_{AB}^{T_A}$ (or $\rho_{AB}^{T_B}$) is a positive semidefinite operator.\\
	Later, Horodecki showed that, if $\rho_{AB}$ defines a separable quantum state in $2\otimes 2$ or $2\otimes 3$ dimension, then $\rho_{AB}^{T_A}\geq 0$ and vice-versa. Thus the condition is necessary and sufficient for $2\otimes 2$ or $2\otimes 3$ dimensional quantum system but only necessary condition for $d_1\otimes d_2$  (except $2\otimes 2$ or $2\otimes 3$ or $3\otimes 2$) dimensional system \cite{horodecki1,horodecki2}.\\
	PPT criterion is an important criterion in the sense that this criterion provides (i) necessary and sufficient conditions for $2\otimes 2$ and $2\otimes 3$ dimensional system and (ii) it is based on the spectrum of the partially transposed matrix. This criterion tells us that if the spectrum of the partially transposed matrix has at least one negative eigenvalue, then the given state is entangled.\\
	For instance, consider the two-qubit Werner state 
	\begin{eqnarray}
	\rho_W=\begin{pmatrix}
	\frac{1-F}{4} & 0 & 0 & 0\\
	0 & \frac{1+F}{4} & -\frac{F}{2} & 0\\
	0 & -\frac{F}{2} & \frac{1+F}{4} & 0\\
	0 & 0 & 0 & \frac{1-F}{4}
	\end{pmatrix}, ~~0\leq F\leq 1
	\end{eqnarray}
	Then, the partial transpose $\rho_W^{T_B}$ of the density matrix $\rho_W$ may be calculated as
	\begin{eqnarray}
	\rho_W^{T_B}=
	\begin{pmatrix}
	\frac{1-F}{4} & 0 & 0 & -\frac{F}{2}\\
	0 & \frac{1+F}{4} & 0 & 0\\
	0 & 0 & \frac{1+F}{4} & 0\\
	-\frac{F}{2} & 0 & 0 & \frac{1-F}{4}
	\end{pmatrix}, ~~0\leq F\leq 1
	\end{eqnarray}
	The spectrum of $\rho_W^{T_B}$ is given by $\{\frac{1-3F}{4}, \frac{1+F}{4}, \frac{1+F}{4}, \frac{1+F}{4}\}$. Thus, the minimum eigenvalue $\frac{1-3F}{4}$  of $\rho_W^{T_B}$ is negative for $\frac{1}{3}< F \leq 1$. Hence using PPT criteria, the state $\rho_W$ is entangled for $\frac{1}{3} <F\leq1$.\\
	The PPT criteria define only a necessary condition for a higher dimensional quantum system. to illustrate this, consider a two-qutrit state $\rho_{a,b,c}$ defined by
	\begin{eqnarray}
	\rho_{a,b,c}=
	\begin{pmatrix}
	a & 0 & 0 & 0 & b & 0 & 0 & 0 & b \\
	0 & c & 0 & 0 & 0 & 0 & 0 & 0 & 0 \\
	0 & 0 & a & 0 & 0 & 0 & 0 & 0 & 0 \\
	0 & 0 & 0 & a & 0 & 0 & 0 & 0 & 0 \\
	b & 0 & 0 & 0 & a & 0 & 0 & 0 & 0 \\
	0 & 0 & 0 & 0 & 0 & c & 0 & b & 0 \\
	0 & 0 & 0 & 0 & 0 & 0 & c & 0 & 0 \\
	0 & 0 & 0 & 0 & 0 & b & 0 & a & 0 \\
	b & 0 & 0 & 0 & 0 & 0 & 0 & 0 & a \\
	\end{pmatrix}
	\end{eqnarray}
	Here, $a=\frac{1+\sqrt{5}}{3+9\sqrt{5}}$, $b=\frac{-2}{3+9\sqrt{5}}$ and $c=\frac{-1+\sqrt{5}}{3+9\sqrt{5}}$. The spectrum of $\rho_{a,b,c}^{T_B}$ is given by $\{\frac{3+\sqrt{5}}{3(1+3\sqrt{5})},\\\frac{2\sqrt{5}}{3(1+3\sqrt{5})},\frac{2\sqrt{5}}{3(1+3\sqrt{5})},\frac{1+\sqrt{5}}{3(1+3\sqrt{5})},\frac{1+\sqrt{5}}{3(1+3\sqrt{5})},\frac{-1+\sqrt{5}}{3(1+3\sqrt{5})},\frac{-1+\sqrt{5}}{3(1+3\sqrt{5})},0,0\}$. It is clear that all eigenvalues of $\rho_{a,b,c}^{T_B}$ are positive. But it is known that the state $\rho_{a,b,c}$ is an entangled state \cite{bhattacharya}. This kind of entangled state which is positive under partial transposition operation is known as positive partial transpose entangled states (PPTES) or bound entangled states. The PPTES exist in the higher dimensional system and they are not detected by PPT criteria. For the detection of PPTES, one may use another criterion such as range criteria and realignment criteria.\\
		\textbf{II. Realignment Criteria \cite{chen, rudolph}}
	Let us consider any bipartite state described by the density operator $\rho$,
	\begin{eqnarray}
	\rho=\sum_{ijkl}{a_{ij,kl}|ij\rangle \langle kl|} \in H_A^{d_1} \otimes H_B^{d_2}
	\end{eqnarray} where $H_A^{d_1}$ and $H_B^{d_2}$ denote the Hilbert spaces of dimension $d_1$ and $d_2$ of the subsystems A and B respectively. The realigned matrix $\rho^{R}$ of the density matrix $\rho$ may be expressed as
	\begin{eqnarray}
	\rho^{R}=\sum_{ijkl}{a_{ik,jl}|ik\rangle \langle jl|}
	\end{eqnarray}
	%	\noindent Since, PPT criteria do not detect all entangled states. If PPT criteria fail, various other criteria have been proposed. Here, we will discuss Computable cross-norm criteria(CCNR) or Realignment criteria. 
	The realignment operation can be used to detect entanglement in higher dimensional systems. Especially, it is useful to detect PPTES, which are not detected by PPT criteria. The criteria based on the realignment operation is called realignment criteria.\\
	\textbf{Realignment Criteria:} If the state described by a density operator $\rho$ is separable then $||\rho^{R}||\leq 1$, where $||.||$ denote the trace norm defined as $||H||=Tr(\sqrt{HH^{\dagger}})$.\\
	%In many cases when the PPT fails, entanglement can be checked using the realignment criteria. Also, it is simple to use which makes it a handy tool for the detection of entanglement.\\
	To illustrate the realignment criteria let us consider a two-qutrit state defined in \cite{horodecki4}.
	\begin{eqnarray}
	\rho_{a}=\frac{1}{8a+1}
	\begin{pmatrix}
	a & 0 & 0 & 0 & a & 0 & 0 & 0 & a \\
	0 & a & 0 & 0 & 0 & 0 & 0 & 0 & 0 \\
	0 & 0 & a & 0 & 0 & 0 & 0 & 0 & 0 \\
	0 & 0 & 0 & a & 0 & 0 & 0 & 0 & 0 \\
	a & 0 & 0 & 0 & a & 0 & 0 & 0 & a \\
	0 & 0 & 0 & 0 & 0 & a & 0 & 0 & 0 \\
	0 & 0 & 0 & 0 & 0 & 0 & \frac{1+a}{2} & 0 & \frac{\sqrt{1-a^2}}{2} \\
	0 & 0 & 0 & 0 & 0 & 0 & 0 & a & 0 \\
	a & 0 & 0 & 0 & a & 0 & \frac{\sqrt{1-a^2}}{2} & 0 & \frac{1+a}{2} \\
	\end{pmatrix}, 0\leq a\leq 1
	\end{eqnarray}
	Then, the realigned matrix of the density matrix $\rho_a$ may be calculated as
	\begin{eqnarray}
	\rho_{a}^{R}=\frac{1}{8a+1}
	\begin{pmatrix}
	a & 0 & 0 & 0 & a & 0 & 0 & 0 & a \\
	0 & a & 0 & 0 & 0 & 0 & 0 & 0 & 0 \\
	0 & 0 & a & 0 & 0 & 0 & 0 & 0 & 0 \\
	0 & 0 & 0 & a & 0 & 0 & 0 & 0 & 0 \\
	a & 0 & 0 & 0 & a & 0 & 0 & 0 & a \\
	0 & 0 & 0 & 0 & 0 & a & 0 & 0 & 0 \\
	0 & 0 & 0 & 0 & 0 & 0 & a & 0 & 0 \\
	0 & 0 & 0 & 0 & 0 & 0 & 0 & a & 0 \\
	\frac{1+a}{2} & 0 & \frac{\sqrt{1-a^2}}{2} & 0 & a & 0 & \frac{\sqrt{1-a^2}}{2} & 0 & \frac{1+a}{2} \\
	\end{pmatrix}, 0\leq a\leq 1
	\end{eqnarray}
	It can be seen from the figure given below that $||\rho_{a}^{R}||>1$ for $0< a<1$. Thus the state described by the density matrix $\rho_a$ is entangled for $0<a<1$.
	\begin{figure}[h]
		\centering
		\includegraphics[scale=0.3]{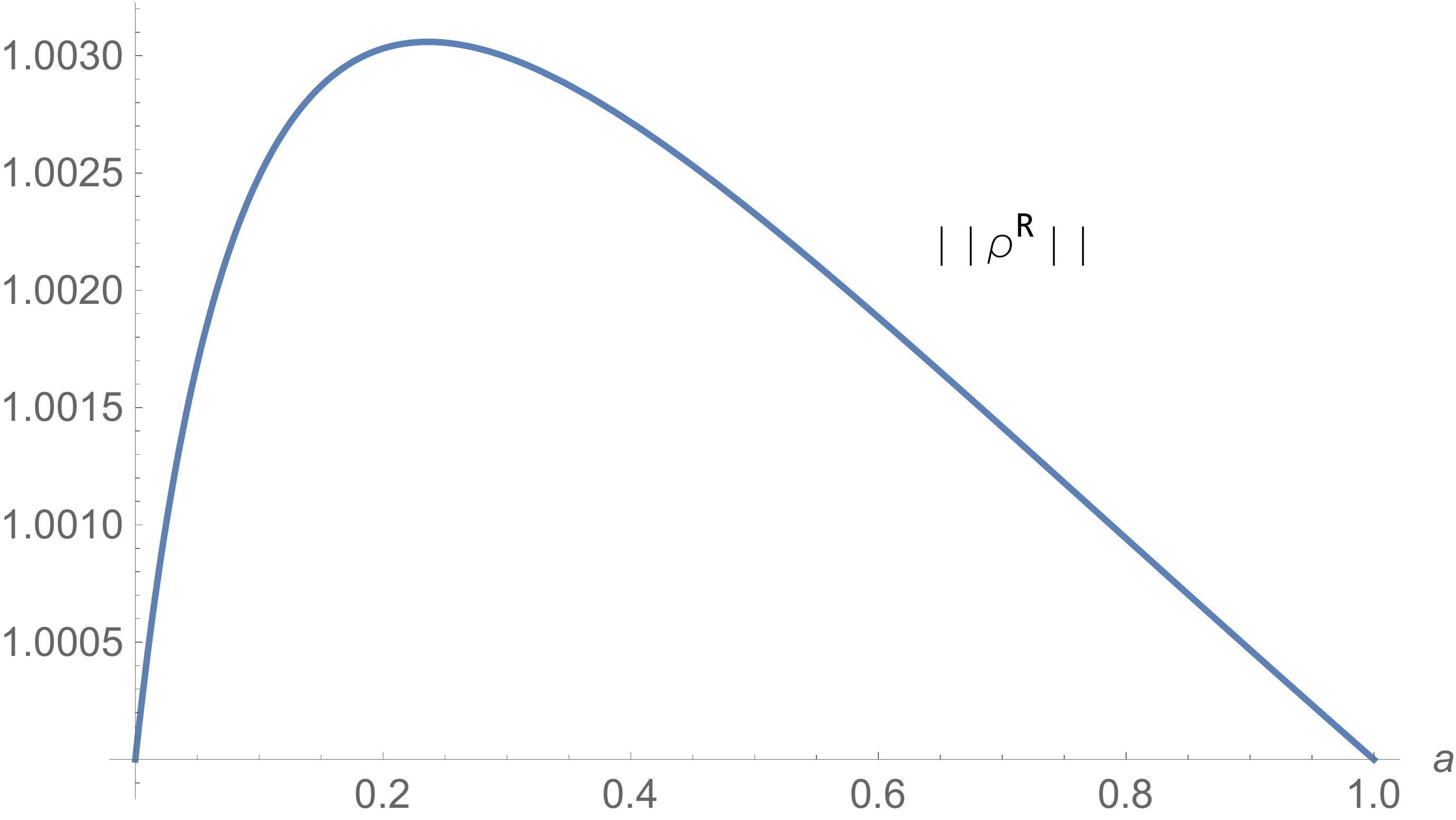}
		%\caption{Here dashed line represents $||(\tilde{\rho})^R||$ and solid line represents 10/28}
	\end{figure}\\
	\textbf{III. Reduction Criteria \cite{horodecki3}}
	Similar to PPT criteria, another entanglement detection criteria can be formulated using a positive but not completely positive map. Like the partial transposition map, there exists another positive but not completely positive map which is known as the reduction map. Reduction map on a single qubit system $X$ may be defined as
	\begin{eqnarray}
	\Lambda^{R}(X)=Tr(X).I-X
	\end{eqnarray}
	\textbf{Reduction criteria:} If a bipartite state described by a density operator $\rho_{AB}$ is separable then $(I_A\otimes \Lambda^{R})(\rho_{AB})=\rho_A\otimes I-\rho_{AB} \geq 0$, where $\Lambda$ denote any positive map.\\
	A reduction map is a so-called decomposable map and may be expressed as
	\begin{eqnarray}
	\Lambda^{R}=P_1+P_2~o~T
	\end{eqnarray}
	where $\{P_1, P_2\}$ are the completely positive maps and $T$ denotes the transposition operator. The reduction criteria is weaker than the PPT criteria because $I_A\otimes\Lambda^{R}$ can never detect entanglement unless $I_A\otimes T$ detects it. It provides a necessary and sufficient condition for separability in the case of a two-qubit quantum system \cite{guhne}.\\
	For illustration, consider a two-qutrit isotropic state defined as
	\begin{eqnarray}
	\rho_{\alpha}=\alpha|\phi_{+}\rangle \langle \phi_{+}|+\frac{1-\alpha}{9}I,~~\alpha\in R,~~0\leq \alpha\leq 1
	\end{eqnarray}
	where $|\phi_{+}\rangle$ is the maximally entangled state and it is given by
	\begin{eqnarray}
	|\phi_{+}\rangle =\frac{1}{\sqrt{3}}(|00\rangle +|11\rangle +|22\rangle)
	\end{eqnarray}
	After applying reduction map on the density matrix $\rho_{\alpha}$, the matrix ${(\rho_{\alpha})}_A\otimes I-\rho_{\alpha}$ becomes
	\begin{eqnarray}
	%	{\rho_{\alpha}}_A\otimes I-\rho_{\alpha}=
	\begin{pmatrix}
	-\frac{2}{9}(-1+\alpha) & 0 & 0 & 0 & -\frac{\alpha}{3} & 0 & 0 & 0 & -\frac{\alpha}{3}\\
	0 & \frac{2+\alpha}{9} & 0 & 0 & 0 & 0 & 0 & 0 & 0\\
	0 & 0 & \frac{2+\alpha}{9} & 0 & 0 & 0 & 0 & 0 & 0\\
	0 & 0 & 0 & \frac{2+\alpha}{9} & 0 & 0 & 0 & 0 & 0\\
	-\frac{\alpha}{3} & 0 & 0 & 0 & -\frac{2}{9}(-1+\alpha) & 0 & 0 & 0 & -\frac{\alpha}{3}\\
	0 & 0 & 0 & 0 & 0 & \frac{2+\alpha}{9} & 0 & 0 & 0\\
	0 & 0 & 0 & 0 & 0 & 0 & \frac{2+\alpha}{9} & 0 & 0\\
	0 & 0 & 0 & 0 & 0 & 0 & 0 & \frac{2+\alpha}{9} & 0\\
	-\frac{\alpha}{3} & 0 & 0 & 0 & -\frac{\alpha}{3} & 0 & 0 & 0 & -\frac{2}{9}(-1+\alpha)
	\end{pmatrix}
	\end{eqnarray}
	The minimum eigenvalue of the resultant matrix ${(\rho_{\alpha}})_A\otimes I-\rho_{\alpha}$ is $-\frac{2}{9}(-1+4\alpha)$ and it is negative for $0.25< \alpha \leq 1$. Thus, using reduction criteria, we can conclude that the state $\rho_{\alpha}$ is entangled for $0.25< \alpha \leq 1$.\\
		\textbf {IV. Range Criteria \cite{bennett4,bruss}}
	Range criteria can be considered one of the earliest criteria for the detection of entangled states. Sometimes, it may detect those higher dimensional entangled states which are not detected by either PPT criteria or realignment criteria. It asserts that if a state described by a density operator $\rho$ is separable, then there exist a set of product vectors $|a_ib_i\rangle$ such that the set $\{|a_ib_i\rangle\}$ spans the range of density matrix $\rho$ and the set $\{|a_i^{*}b_i\rangle \}$ spans the range of $\rho^{T_A}$. This criterion detects many entangled states which are not detected by PPT criteria. But the drawback of this criteria is that it cannot be used when the state is affected by noise because then the density matrix $\rho$ and its partially transposed matrix $\rho^{A}$ will usually have full rank and thus the condition for range criteria is automatically satisfied.
%For any density matrix, using Schmidt decomposition, $\rho$ can be written as,
	%\begin{eqnarray}
	%\rho=\sum_k{\lambda_k{V_k}^{A} \otimes {V_k}^B},~~\lambda_k\geq 0
	%\end{eqnarray}
	%where ${V_k}^{A}$ and ${V_k}^B$ are the orthonormal basis of the Hilbert spaces $H_A$ and $H_B$.
	%\textbf{CCNR Criteria:} If $\rho$ is a separable state, then 
	%\begin{eqnarray}
	%\sum_{k}{\lambda_k} \leq 1
	%\end{eqnarray}
	%Hence, if $\sum_{k}{\lambda_k} > 1$, then the state must be entangled.
	
	\subsection{Experimentally implementable entanglement detection schemes}
	\noindent All the criteria defined above for the detection of entangled states depend on a positive but not completely positive map. It is known that positive but not completely positive maps do not correspond to a physically realizable operator whereas a completely positive map does. Thus, the above-defined entanglement detection criteria may not be realized in an experiment.
	%Also the above methods assume that the density matrix of the state under investigation is known. They all require applying certain operations to the density matrix elements to check whether the state is entangled or not.\\
	We will now discuss some of the important experimentally realizable criteria for the detection of entangled states. The first tool for the detection of entanglement experimentally was given by John Bell in 1964\cite{bell} in the form of Bell's inequality. The objective of this study is to quantitatively express the Einstein-Podolsky-Rosen paradox.\\
	\textbf{I. Bell Inequality \cite{bell,nielsen}}
	\noindent The basic idea behind Bell's inequality is that if measurement is performed by both the parties say, Alice and Bob on their individual subsystems of a bipartite composite system with the assumption that measurement result exists locally before the measurement then it is feasible to obtain bounds on certain quantities composed from the correlation terms of two subsystems of a composite system.\\
	%	\textbf{Assumption of realism:} The assumption that the physical properties $P_a, P_b, P_c,$ and $P_d$ have definite values A, B, C, and D respectively, that exist independent of the observation.\\
	%	\textbf{Assumption of locality:} Alice's measurement does not influence the result of Bob's measurement.\\
	\begin{figure}
		\centering
		\caption{Experimental setup for the Bell inequality}
		\includegraphics[width=16cm]{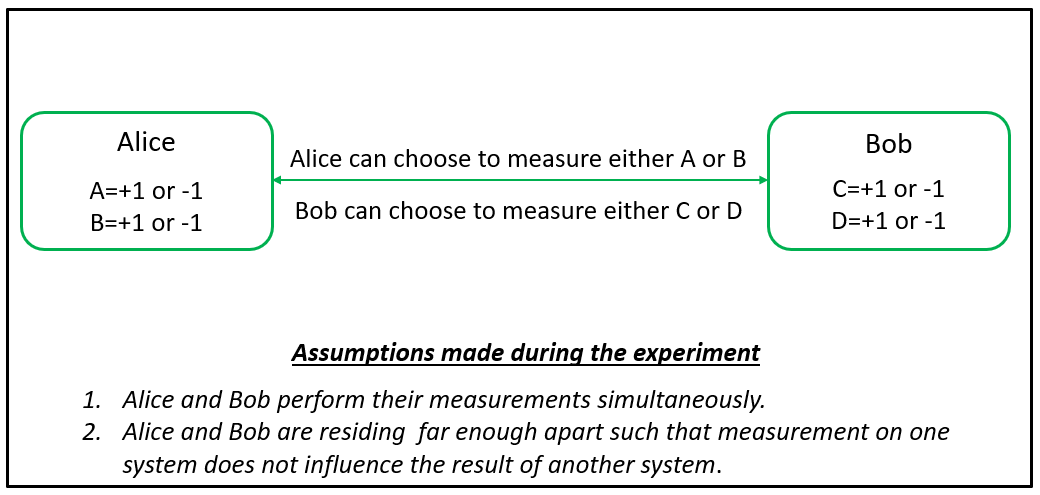}
		%\caption{Experimental setup for the Bell inequality}
		\label{bell's inequality}
	\end{figure}
	\noindent We are now in a position to explain the following experiment described in Figure 1.2. The assumptions made during the experiment are: the assumption of realism and the assumption of locality. Let us assume that Charlie prepares two particles. After finishing the preparation, he sends one particle each to Alice and Bob. As soon as Alice receives her particle, she measures it. Imagine that she has access to two separate measuring apparatuses, and she may select to perform one of the two measurements. Alice chooses the measurement randomly, where the physical properties of the measurements may be described by $P_A$ and $P_B$ respectively. For simplicity, suppose that each of the measurements can have one of the two possible outcomes, +1 or -1. Assume that the attribute $P_A$ of Alice's particle has value A, where A is thought to be an objective feature of Alice's particle that is only disclosed by the measurement, and B stands for the value obtained by measuring the property $P_B$. Similarly, Bob is able to measure one of the two physical properties, $P_C$ or $P_D$, indicating that the property has an objectively existing value C or D, where each takes a value of either +1 or -1. Bob doesn't know in advance which property he will measure, Once he receives the particle, he chooses the measurement randomly. It is assumed that Alice and Bob perform their measurement simultaneously so, Alice's measurement doesn't affect Bob's measurement and vice-versa. We now take the random variables $A, B, C$, and $D$ in such a way that the expression constructed from them is nonfactorizable. Keeping in mind this fact, we may consider the expression $AC-BC+BD+AD$. This expression may be re-expressed as
	\begin{eqnarray}
	AC-BC+BD+AD=(A-B)C+(A+B)D
	\label{belleq1}
	\end{eqnarray} 
	since, $A,B=\pm 1$ thus either $(A-B)C=0$ or $(A+B)D=0$. In either case, we have
	\begin{eqnarray}
	AC-BC+BD+AD=\pm 2
	\end{eqnarray}
	Consider
	\begin{eqnarray}
	E(AC-BC+BD+AD)=\sum_{abcd}{p(a,b,c,d)(ac-bc+bd+ad)}
	\end{eqnarray}
	where $p(a,b,c,d)$ denotes the probability of the system before the measurement, and E(.) denotes the mean of the quantity (.). Thus,
	\begin{eqnarray}
	-\sum_{abcd}{p(a,b,c,d)\times 2}\leq	E(AC-BC+BD+AD)&\leq& \sum_{abcd}{p(a,b,c,d)\times 2}
	\end{eqnarray}
	Since $\sum_{abcd}{p(a,b,c,d)}=1$, so the inequality may be re-written as
	\begin{eqnarray}
	-2\leq	E(AC-BC+BD+AD)\leq 2
	\label{lhs1}
	\end{eqnarray}
	Further, we have
	\begin{eqnarray}
	E(AC-BC+BD+AD)&=&\sum_{abcd}{p(a,b,c,d)ac}-\sum_{abcd}{p(a,b,c,d)bc}\nonumber\\&+&\sum_{abcd}{p(a,b,c,d)bd}+\sum_{abcd}{p(a,b,c,d)ad}\nonumber\\
	&=& E(AC)-E(BC)+E(BD)+E(AD)
	\label{rhs}
	\end{eqnarray}
	Comparing equations (\ref{lhs1}) and (\ref{rhs}), the following inequality is obtained,
	\begin{eqnarray}
	-2\leq E(AC)-E(BC)+E(BD)+E(AD) \leq 2
	\label{bell}
	\end{eqnarray} 
	This inequality is known as Bell's inequality. If Bell's inequality given in (\ref{bell}) is violated by any quantum state, then the state will violate the assumption of local realism. Thus, the state may be considered non-local and thus entangled. All separable states satisfy Bell's inequality but there may also exist entangled states which satisfy Bell's inequality. For higher dimensional system Bell's inequality have been developed in two forms: (i) Clauser-Horne type inequality for two qutrit system and (ii) CHSH type inequality for two arbitrary d-dimensional systems which is now known as Collins-Gisin-Linden-MasserPopoescu (CGLMP) inequalities. The tightness of the
	CGLMP inequality was demonstrated in \cite{kwek}. In a two-qubit system, maximally entangled states violate Bell's inequality maximally whereas, in a higher dimensional system, there may exist non-maximally entangled states that violate Bell's inequality maximally. \cite{kwek}. Acin et.al studied nonlocality in the case of a two-qudit system up to $d=8$ and found that there exists a non-maximally entangled state in which the violation of CGLMP inequalities is more than the maximally entangled state \cite{durt}.\\
	%It was generally felt that maximally entangled states
	%mes would maximally violate the Bell inequality, just as
	%the CHSH inequality has worked for two-qubit. Moreover,
	%the results of Ref. 9 was numerically consistent with Ref.
	%5. However, contrary to prevalent belief, Acín et al. studied
	%the quantum nonlocality of two qutrits as well as two
	%d-dimensional systems up to d=8 and discovered another
	%unexpected result: there existed non-maximally entangled
	%states that lead to greater violation of the CGLMP inequalities
	%compared with maximally entangled states
	%Moreover,
	%it was shown in this study that the maximal violation
	%increases with dimension d, and it reaches 3.1013 for d=8.
	Furthermore, we should note that the expression involved in the derivation of Bell's inequality is not unique and can be modified in such a way that the resulting expression would not be factorizable. Otherwise, the inequality derived from the expression will not take part in the detection of an entangled state.  
	
	\textbf{II. Witness Operator or Entanglement Witnesses \cite{terhal,lewenstein1}}
	\noindent Witness operators are the fundamental tool for the detection of entanglement experimentally. It can be used to detect entangled states in bipartite as well as multipartite quantum systems. %They define a necessary and sufficient condition for entanglement detection in terms of observables. 
	Entanglement witnesses are Hermitian operators and have at least one negative eigenvalue.\\
	An observable $\textbf{W}$ is an entanglement witness if it satisfies the following two conditions:
	\begin{eqnarray}
	&&H_1: Tr(\textbf{W}\rho_s) \geq 0, \text{for all separable states}~\rho_s.\nonumber\\
	&&H_2: Tr(\textbf{W}\rho_e) < 0, \text{for at least one entangled state}~\rho_e.
	\end{eqnarray}
	Thus, if $Tr(\textbf{W}\rho) < 0$, then we say that the state $\rho$ is detected by an entanglement witness \textbf{W}. The first example of witness operator can be considered as $W_B=2I-\textbf{\textit{B}}$ where $\textbf{\textit{B}}$ denote the Bell operator which may be defined as
	\begin{eqnarray}
	\textbf{\textit{B}}=A\otimes C+B\otimes C+B\otimes D-A\otimes D
	\end{eqnarray}
	Indeed it can be shown that $W_B$ represents a witness operator.\\
	Geometrically, entanglement witness $\textbf{W}$ represents a hyperplane that separates at least one entangled state from the set of separable states. This can be seen as the consequence of the Hahn Banach theorem of functional analysis which states that for a convex and closed set S in a finite-dimensional Banach space, there exist a hyperplane that separates $\rho$ from S, when $\rho$ is a point in the Banach space that does not belong to S.\\
	\begin{figure}
		\centering
		\includegraphics[width=8cm]{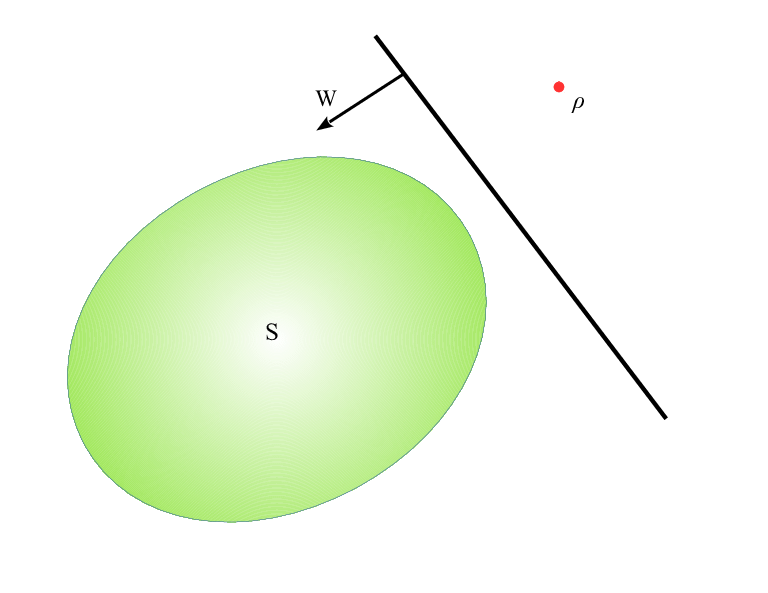}
		\caption{Hahn Banach Theorem}
		%\label{hahn banach}
	\end{figure}
	Let us define three sets $S_0, S_1$, and $S_2$ which are defined by
	\begin{eqnarray}
	S_0=\{\rho:Tr(W\rho)=0\}\nonumber\\
	S_1=\{\rho:Tr(W\rho)>0\}\nonumber\\
	S_2=\{\rho:Tr(W\rho)<0\}
	\end{eqnarray}
	The set $S_0$ contains all those states $\rho$ which are lying on the hyperplane $\textbf{W}$. In this case, the hyperplane \textbf{W} touches the set $S_0$ and thus the hyperplane $\textbf{W}$ can be considered as a tangent plane to the set $S_0$. The set $S_1$ contains all those states $\rho$ which are either separable states or the states which are not detected by the witness $\textbf{W}$. The set $S_2$ denotes the set of entangled states which are detected by $\textbf{W}$.\\
	If $W_1$ and $W_2$ are two entanglement witness then $W_1$ is said to be finer than $W_2$ if $W_1$ detects all entangled states which are detected by $W_2$. An entanglement witness $W_{opt}$ is known as the optimal entanglement witness (OEW) if there exists no entanglement witness which is finer than $W_{opt}$ i.e. if $W_{opt}$ forms a tangent plane to the set of separable states S. Entanglement witnesses may be further classified into two classes as decomposable entanglement witnesses (DEW) and non-decomposable entanglement witnesses (NDEW). Decomposable entanglement witnesses may be expressed as
	\begin{eqnarray}
	W_D=aP+(1-a)Q^{T_B}, ~~a\in[0,1]
	\label{dew}
	\end{eqnarray}
	where P and Q are positive semidefinite operators. The operators which cannot be expressed in the form (\ref{dew}) are non-decomposable entanglement witnesses. DEW cannot detect PPTES. Optimal DEW, W may be expressed as $W=Q^{T_B}$ for some positive semidefinite operator Q which contains no product vectors in its range. Also, for an optimal DEW, $W$, the partial transposition $W^{T_B}$ is not a witness operator. An entanglement witness is non-decomposable if and only if it detects PPTES. Every entanglement witness, however, corresponds to a positive but not completely positive map via the Choi-Jamiolkowski isomorphism\cite{choi1,choi2}. %Let $B(H_1)$ and $B(H_2)$ denote the space of bounded operators on finite-dimensional Hilbert spaces $H_1$ and $H_2$ respectively. Let $d_1$ and $d_2$ be the dimensions of $H_1$ and $H_2$.  For an operator $\Lambda_E\in B(H_1\otimes H_2)$, define $\Lambda_E: B(H_1)\rightarrow B(H_2)$ as 
	\textbf{III. Structural Physical Approximation \cite{horodecki5}}
\noindent Another method for the experimental detection of entanglement is the method of structural physical approximation (SPA). The concept of SPA has been introduced in \cite{horodecki5}. In the SPA method, a positive but not completely positive map is approximated by a completely positive map. The method is useful in the sense that it transforms a non-physical operator to a physical operator and hence makes it experimentally realizable\cite{horodecki6}. A lot of progress has already been done in both theoretical as well as experimental aspects of SPA\cite{korbicz,augusiak,kye1,kye2,hansen,stormer,kye3,kye5}. \\
\noindent The main idea behind SPA is how much proportion of white noise needs to be added to a non-physical operator $\Lambda$, such that the approximated operator $\tilde{\Lambda}$ will be completely positive. Mixing the map $\Lambda$ with the white noise transforms any density matrix into a maximally mixed state, thus resulting map can have no negative values. For any non-physical operator $\Lambda$, the approximated map $\tilde{\Lambda}$ can be written as a convex combination of the operator $\Lambda$ and the depolarizing map $D$ that may be expressed as 
\begin{eqnarray}
\tilde{\Lambda}=(1-p)\Lambda+pD,~~0\leq p\leq 1
\end{eqnarray}
If a $d$-dimensional system is described by a density operator $\rho$ and $\Lambda$ be any positive operator then the approximation of $\Lambda$ denoted by $\tilde{\Lambda}$, which may be evolved as
\begin{eqnarray}
\tilde{\Lambda}(\rho)=(1-p)\Lambda(\rho)+p\frac{I_d}{d},~~0\leq p\leq 1
\label{spatranspose}
\end{eqnarray}
where we have used the fact that $D[\rho]=\frac{I_d}{d}$  and $I_d$ denote the $d$-dimensional identity matrix.\\
In particular, let us consider the positive but not completely positive transposition map $T$, which is defined in $d$-dimensional system. The SPA of $T$ may be expressed as 
\begin{eqnarray}
\tilde{T}=\frac{1}{d+1}T+\frac{d}{d+1}D
\end{eqnarray}
$\widetilde{T}$ is completely positive for $p\geq\frac{d}{d+1}$ \cite{kye5}. In particular, the SPA of the transposition map of a qutrit system has been realized in experiments through prepare and measure strategy \cite{kye5}.\\ %Even for continuous variable system SPA-PT has been constructed \cite{lewenstein1}.\\
\noindent In 2008 \cite{korbicz}, a conjecture has been posed that SPA leads to separable states.
Various examples were given in support of this conjecture\cite{korbicz,augusiak}. For instance, let us take a transposition map acting on one of the subsystems of a composite bipartite system that may lead to entanglement breaking. Further, assume a $d\otimes d$ dimensional maximally entangled state of the form
\begin{eqnarray}
P_{+}=|\phi_{+}\rangle \langle \phi_{+}|,~~|\phi_{+}\rangle=\frac{1}{\sqrt{d}}{\sum_{i=1}^{d}{|ii\rangle}}
\end{eqnarray}
In particular, taking $\Lambda=T$ in (\ref{spatranspose}), it reduces to
\begin{eqnarray}
\widetilde{T}(\rho)=p\frac{I}{d}+(1-p)T({\rho}),~~0<p<1
\end{eqnarray}
where $p$ takes the minimum value for which $\widetilde{T}$ becomes a completely positive map. The corresponding witness operator $\widetilde{E_T}$ acting on the maximally entangled state $P_{+}$,  may be expressed as
\begin{eqnarray}
\widetilde{E_T}(P_{+})=(I\otimes \widetilde{T})(P_{+})=\frac{p}{d^2}I\otimes I +\frac{1-p}{d}[\sum_{i=1}^d{|ii\rangle \langle ii| }+\sum_{i\neq j,i,j=1}^{d}{|ij\rangle \langle ji|}]=\rho_{W}
\label{werner}
\end{eqnarray}
%F(.) denote the flip operator, which may be defined as $F(\psi \otimes \phi)=\phi \otimes \psi$. 
Then, $\widetilde{E_T}$ is positive and thus $\widetilde{T}$ is completely positive when
\begin{eqnarray}
p\geq \frac{d}{d+1}=p_{min}
\end{eqnarray}
%for minimal p such that $\widetilde{E_T}\geq 0$ and 
Using the fact that PPT criteria is necessary and sufficient for the detection of $d\otimes d$-dimensional Werner state described by the density operator $\rho_W$ \cite{korbicz} and from (\ref{werner}), we have
\begin{eqnarray}
(\widetilde{E_T})^{\Gamma}=\frac{p}{d^2}I\otimes I +({1-p})P_{+}
\label{entanglementbreaking}
\end{eqnarray}
where $\Gamma$ denotes the partial transposition. The operator $(\widetilde{E_T})^{\Gamma}$ is non-negative for $0<p<1$ and thus, $\widetilde{E_T}$ represent a separable state. Hence, the SPA of the transposition map is entanglement breaking. Later, this SPA conjecture was disproved by Kye and Ha\cite{kye1,kye2,hansen,stormer}.\\
%Later, In order to disprove SPA conjecture, Ha and Kye\cite{kye1} proposed a decomposable entanglement witness operator whose SPA is entangled and argued that it is optimal. In\cite{augusiak} authors have shown both analytically and numerically that this entanglement witness is not optimal using the method defined in \cite{lewenstein2}. It has been proved that for any positive map ($\gamma$), there exists an entanglement breaking channel ($\phi$) such that SPA of $\gamma$ with the aid of $\phi$ is again an entanglement breaking channel. They have also defined a way for the construction of SPA-PT of a positive map in continuous variable systems. SPA conjucture was disproved by Kye and Ha\cite{kye1,kye2,hansen,stormer}. 
\noindent Now we are in a position to discuss the separability criteria of $d\otimes d$ dimensional system using the structural physical approximation of partial transposition (SPA-PT). 	If we have some prior information about the arbitrary dimensional bipartite system described by the density matrix $\rho$, then the eigenvalues of SPA-PT of $\rho$ can be used to detect whether the state $\rho$ is entangled or not. Using this idea, criteria have been proposed for the detection of negative partial transpose entangled states (NPTES) in the arbitrary dimensional bipartite system. Horodecki et.al. \cite{horodecki5} have studied SPA-PT map extensively to develop the entanglement detection criteria for $d\otimes d$ dimensional system. \\
To understand SPA-PT in $d\otimes d$ dimensional system, let us start our discussion with the partial transposition operation which is a purely mathematical operation and may not be directly implementable in an experiment. 
%Recalling that on the space of matrices of order d $M_d$, a map $\Lambda: M_d\rightarrow M_d$ is positive if $A\geq 0 \implies \Lambda(A)\geq 0$ where $A\geq 0$ means that A has non-negative spectrum. If the induced map $I\otimes \Lambda$ is positive, then the map $\Lambda$ is said to be completely positive where $I$ denotes the Identity operator. 
The partial transposition criteria which is based on the partial transposition operation may be stated as: If a state described by the density operator $\rho^{d\otimes d}$ is separable then
\begin{eqnarray}
(I\otimes T)\rho \geq 0
\end{eqnarray}
where $T$ denotes the transposition operation. Since the map, $I\otimes T$  is not a completely positive map and thus cannot be implemented in a laboratory. This means that the partial transposition operation is a non-physical operation and thus, to make it a physically realizable operator, we approximate it by the method of structural physical approximation. The structural physical approximation of the map $I\otimes T$, where $T$ denotes the transposition operation for $d\otimes d$ dimensional quantum system may be expressed as \cite{bae}
\begin{eqnarray}
\widetilde{(I\otimes T)}=(1-q)I\otimes T+qD\otimes D
\label{spabae}
\end{eqnarray}
where $D(.)=\frac{I_d}{d}$  denote the depolarizing channel. 
When the approximated map $\widetilde{(I\otimes T)}$ acts on the bipartite $d\otimes d$ dimensional density matrix $\rho^{d\otimes d}$, (\ref{spabae}) reduces to
\begin{equation}
\widetilde{(I \otimes T)}(\rho^{d\otimes d})=(1-q_{min})(I\otimes T)(\rho^{d\otimes d})+\frac{q_{min}}{d^2}(I\otimes I)
\label{partialtransposition}
\end{equation}
where $q_{min}$ denotes the minimum value for which  $(\widetilde{I\otimes T})(\rho^{d\otimes d})$  becomes a positive semidefinite matrix. $\widetilde{(I\otimes {T})}(\rho^{d\otimes d})$ becomes a positive semidefinite matrix when
\begin{eqnarray}
q\geq \frac{d^4\lambda}{d^4\lambda+1}\equiv q_{min}
\label{q}
\end{eqnarray}
%Our task is now to find out the minimum value of $q$ for which the map $\widetilde{(I\otimes T)}$ will be completely positive. 
%\begin{eqnarray}
%q=\frac{\lambda d^{4}}{1+\lambda d^4},~~\lambda=-min_{Q>0}Tr[Q(I\otimes T)P_{d}^{+}]
%\end{eqnarray}
where $-\lambda (\lambda>0)$ denote the most negative eigenvalue obtained when the induced map $[(I\otimes I)\otimes (I\otimes T)]$ acts on the maximally entangled state $\frac{1}{d}\sum_{i=1}^{d^2}|i\rangle |i\rangle$. 
Using (\ref{q}) and $\lambda=\frac{1}{d}$, the approximated map (\ref{partialtransposition}) may be re-expressed as
\begin{equation}
[\widetilde{I \otimes T}](\rho^{d\otimes d})=\frac{1}{d^3+1}(I\otimes T)(\rho^{d\otimes d})+\frac{d}{d^3+1}(I\otimes I)
\end{equation}
The physically realizable PPT criteria for $d\otimes d$ dimensional system described by the density operator $\rho$ obtained by applying the SPA-PT, which may be read as: If $\rho$ is separable then \cite{horodecki6}
\begin{equation}
\lambda_{min}[\widetilde{I \otimes T}](\rho^{d\otimes d}) \geq \frac{d}{d^3+1}
\label{spaseparabilitycondition}
\end{equation}
where $\lambda_{min}[\widetilde{I \otimes T}](\rho)$ denote the minimum eigenvalue of SPA-PT of $\rho$.\\
If any $d\otimes d$ dimensional quantum state $\rho$ violate inequality (\ref{spaseparabilitycondition}), then the state $\rho$ is an entangled state.\\
%This means in a bipartite $d\otimes d$ dimensional system, if $\rho$ represent a separable state then $\lambda_{min}([\widetilde{I \otimes T}](\rho))\geq \frac{d}{d^3+1}$.\\ \ \\
%In the partial transposition criteria, we assume that we have prior knowledge about the state but by using SPA this method can be modified so that it is no longer dependent on state estimate and is experimentally viable.\\
%For example, if we consider $T$ to be the transposition map. Then the corresponding  
\noindent Let us now generalize the separability criteria for $d_1\otimes d_2$ dimensional system using the SPA-PT map. To do this, consider a $d_1\otimes d_2$ dimensional system described by a density operator $\rho^{d_1\otimes d_2}$.  Here we assume that $d_1=min\{d_1,d_2\}$. The SPA-PT  for $d_1\otimes d_2$ dimensional system may be defined as
\begin{eqnarray}
\widetilde{(I\otimes T)}=(1-p)I\otimes T+pD_1\otimes D_2
\end{eqnarray}
where 
\begin{eqnarray}
p=\frac{\lambda d_1^{3}d_2}{1+\lambda d_1^{3}d_2},~~\lambda=-\min_{Q>0} Tr[Q(I\otimes T)P_{+}^{d_1}]
\end{eqnarray}
\begin{eqnarray}
P_{+}^{d_1}=|\phi_{+}\rangle \langle \phi_{+}|,~~|\phi_{+}\rangle=\frac{1}{\sqrt{d_1}}{\sum_{i=1}^{d_1}{|ii\rangle}}
\end{eqnarray}
$D_{i}(.)=\frac{I_i}{d_i}$ for i=1,2 denote the depolarizing channel acting on the individual subsystem and the parameter $p$ denote the minimum value for which $\widetilde{(I\otimes T)}$ is completely positive. Thus, the separability criteria for $d_1\otimes d_2$ dimensional system in terms of the minimum eigenvalue of SPA-PT map may be stated as: If $\rho^{d_1\otimes d_2}$ is a separable state in $d_1\otimes d_2$ system, then 
\begin{eqnarray}
\lambda_{min}[\widetilde{(I\otimes T)}(\rho^{d_1\otimes d_2})]\geq \frac{\lambda d_1d_2}{1+\lambda d_1^{3}d_2}, ~~\lambda=-\min_{Q>0} Tr[Q(I\otimes T)P_{+}^{d_1}]
\label{separabilityd1d2}
\end{eqnarray} 
If the inequality (\ref{separabilityd1d2}) is violated then the state $\rho^{d_1\otimes d_2}$ represent an entangled state.\\
From (\ref{separabilityd1d2}), it is clear that to determine whether any arbitrary dimensional quantum state is entangled or not, we need to find the minimum eigenvalue of the SPA-PT of the state under consideration. But it would be a tedious task to determine the matrix elements of the SPA-PT of the higher dimensional quantum system described by the density operator $\rho^{d_1\otimes d_2}$. Thus, we have considered a particular bipartite composite quantum system such as a qutrit-qubit system described by the density matrix $\rho^{3\otimes 2}$ and then calculated the matrix elements of the SPA-PT of $\rho^{3\otimes 2}$ \cite{adhikari3}. Therefore, by constructing the general form of the matrix of SPA-PT of $\rho^{3\otimes 2}$, it would be possible to calculate  $\lambda_{min}[\widetilde{(I\otimes T)}(\rho^{3\otimes 2})]$ at least for the qutrit-qubit system.\\
SPA-PT method not only helps in the detection of entanglement but also has many applications such as to estimate (i) optimal singlet fraction \cite{adhikari1}, (ii) entanglement negativity \cite{adhikari2} of the two-qubit density matrix in an experiment, (iii) concurrence of a two-qubit system using 4 moments \cite{horodecki7}.\\

\section{Classification of entanglement} 
\noindent Another important problem in quantum information theory is the problem of classification of multi-qubit quantum state\cite{horodecki9}, especially when the number of qubits is either equal to three or more than three. In the case of a two-qubit system, the classification of entanglement is not that important as compared to the system where three or more than three qubits are involved because a two-qubit system can be classified as either separable or entangled state\cite{terhal1}. Therefore, when we increase the number of qubits, the complicacy in the structure of the multi-qubit system will also increase, and thus, we may have different classes of entangled states. These classes of entangled states need to be characterized on the basis of the entanglement present in the different subsystems and accordingly, we can use that class of entangled states in appropriate quantum information processing tasks. In particular, three-qubit states may be classified into six inequivalent classes under stochastic local operation and classical communication (SLOCC) as one fully separable state, three biseparable states, and two genuinely entangled states. The fully separable and biseparable states may be expressed in the form as \cite{dur3,guhne1}
\begin{eqnarray}
\rho_{sep}^{ABC}=\sum_{i}p_{i}\rho_{i}^{A}\otimes\rho_{i}^{B}\otimes\rho_{i}^{C},\sum_{i}p_{i}=1
\label{generalseparable}
\end{eqnarray}
\begin{eqnarray}
\rho_{bisep} = p_1 \rho^{A-BC}_{bisep} + p_2 \rho^{B-AC}_{bisep} + p_3 \rho^{C-AB}_{bisep},\sum_{i}p_{i}=1
\label{generalbiseparable}
\end{eqnarray}
where,
\begin{eqnarray}
\rho_{bisep}^{A-BC}=\sum_{i}|a_i\rangle_A \langle a_{i}|\otimes |\phi_{i}\rangle_{BC}\langle \phi_{i}|\nonumber\\
\rho_{bisep}^{B-AC}=\sum_{i}|b_i\rangle_B \langle b_{i}|\otimes |\phi_{i}\rangle_{AC}\langle \phi_{i}|\nonumber\\
\rho_{bisep}^{C-AB}=\sum_{i}|c_i\rangle_C \langle c_{i}|\otimes |\phi_{i}\rangle_{AB}\langle \phi_{i}|\nonumber
\end{eqnarray}
Here, $|a_{i}\rangle $, $|b_{i}\rangle $ and $|c_{i}\rangle $  are (unnormalized) states of systems
A, B and C, respectively and $|\phi_{i}\rangle $ are states of two systems in equations (\ref{generalbiseparable}). If any three-qubit state is neither fully separable nor biseparable state, then the state is said to be a genuine entangled state. There are two classes of genuine three-qubit entangled states namely, W class of states and GHZ class of states.\\
The five-parameter canonical form of three-qubit pure state $|\psi \rangle_{ABC}$ shared between three distant partners $A$, $B$, and $C$ is given by \cite{acin1}
\begin{eqnarray}
|\psi \rangle_{ABC}&=&\lambda_0|000\rangle+\lambda_1e^{i\theta}|100\rangle+\lambda_2|101\rangle+\lambda_3|110\rangle+\lambda_4|111\rangle
\label{canonicalabc}
\end{eqnarray}
with $0\leq \lambda_{i}\leq 1 (i=0,1,2,3,4)$ and $0\leq \theta \leq \pi$. The normalization condition of the state (\ref{canonicalabc}) is given by
\begin{eqnarray}
\lambda_{0}^{2}+\lambda_{1}^{2}+\lambda_{2}^{2}+\lambda_{3}^{2}+\lambda_{4}^{2}=1.
\label{normalization}
\end{eqnarray}
Equation of the type (\ref{canonicalabc}) represents a GHZ state whereas W vectors can be written as
\begin{eqnarray}
|\psi\rangle_W=\lambda_0|000\rangle+\lambda_1e^{i\theta}|100\rangle+\lambda_2|101\rangle+\lambda_3|110\rangle
\label{wabc}
\end{eqnarray}
The tangle for three-qubit state $|\psi \rangle_{ABC}$ is given by
\begin{eqnarray}
\tau_{\psi}=4\lambda_{0}^{2}\lambda_{4}^{2}
\end{eqnarray}
If at least one of the $\lambda_0$ or $\lambda_4$ is zero then the tangle of the given three-qubit state is zero and if none of the $\lambda_0$ or $\lambda_4$ is zero, then the tangle of the state under investigation is non-zero. Thus, if the tangle is non-zero for any three-qubit state, then the state belongs to the class of GHZ states and if the tangle is zero for any three-qubit state then the state can be either separable or biseparable or W state in three qubit system. Thus, the tangle can be used to classify GHZ class of states from the set of separable states, biseparable states and W class of states \cite{dur4}. Now, it is important to further classify separable states, biseparable states, and W states.  In this context, all possible six SLOCC inequivalent three-qubit pure states can be distinguished with the help of observables\cite{datta}. An experiment has been carried out using an NMR quantum information processor, to classify these six SLOCC inequivalent classes\cite{singh}. Classification of genuine three-qubit mixed states has been studied by constructing GHZ and W witness operator in \cite{acin}. Sabin et al.  \cite{sabin} have studied the classification of a three-qubit system based on reduced two-qubit entanglement. Classification of four qubit pure states has been studied in\cite{verstraete,viehmann,zangi}. For multi-qubit pure system, classification has been studied in \cite{miyake,miyake1,li1,chen1}. The complete classification of multipartite systems is still not known due to the complicated nature of multipartite systems.\\
Although there are various methods for the detection and classification of entanglement which exists in the literature, most of them suffer from a serious drawback that they cannot be implemented in an experimental setup. To overcome this defect, we have used the method of witness operator and  SPA-PT method to classify a three-qubit quantum system.

\begin{center}
****************
\end{center}

\chapter{Detection of entanglement via structural physical approximation}\label{ch2}
\vspace{2cm}
\noindent\hrulefill

\noindent\emph{Detection of entanglement is an important problem in quantum information theory and thus various entanglement detection schemes have been proposed to detect entanglement in a mixed bipartite quantum system. In this chapter\;\footnote{ This chapter is based on a published research paper ``Detection of a mixed bipartite entangled state in arbitrary dimension via a structural physical approximation of partial transposition, \emph{Physical Review A} {\bf 100}, 052323 (2019)".}, we have started with the study of SPA-PT map acting on the qutrit-qubit system described by the density operator $\rho^{2\otimes 3}$ to detect entanglement in the given system. Here, it is shown how to apply SPA-PT on $\rho^{2\otimes 3}$ and thus, obtained explicitly the matrix elements of $\widetilde{\rho^{2\otimes 3}}$, where $\widetilde{\rho^{2\otimes 3}}$ denote the SPA-PT of $\rho^{2\otimes 3}$. Using the matrix elements of $\widetilde{\rho^{2\otimes 3}}$, one can find the minimum eigenvalue of the approximated matrix, and using the minimum eigenvalue one can check whether the state under investigation is entangled or not. But the entanglement detection procedure based on the minimum eigenvalue of SPA-PT of the given density matrix will become very tedious when we increase the dimension of the bipartite quantum system. Thus, to check entanglement in an arbitrary dimensional bipartite system, we have proposed three different criteria based on SPA-PT. Also, we have provided a lower and upper bound of concurrence of arbitrary dimensional mixed bipartite NPTES.}

\section{Introduction}\label{section:2.1}
\noindent Entanglement \cite{piani,horodecki9} is a crucial element that enhanced the power of quantum computation \cite{jozsa}. This non-local feature of quantum mechanics plays an important role in the field of quantum information theory. Thus, it is important to check whether the quantum state under investigation is entangled or not.  If in an experiment, one intends to generate an entangled state, then it is not necessary that the generated state at the output is entangled. This happens due to the presence of noise and thus the problem of detection of entanglement will become an important issue.  Several criteria had already been proposed for solving the entanglement detection problem but yet partial result has been achieved \cite{augusiak1,lima,chruscinski1,ganguly,horodecki10,toth1,fei,shen,hou,toth2}.\\
The first entanglement detection criteria i.e. the PPT criteria provides a necessary and sufficient condition for separability in $2\otimes 2$ and $2\otimes 3$ dimensional system. But for $d_{1}\otimes d_{2}$ (except $2\otimes 2$ and $2\otimes 3$) dimensional system, it provides only a necessary condition due to the existence of PPTES. The other weakness of PPT criteria is that it depends on the spectrum of partial transposition of the density matrix. Partial transposition is a positive but not completely positive map and hence, it may not be implemented in a laboratory. This defect can be rectified by using SPA-PT which transforms a positive but not completely positive operator into a completely positive operator and thus, makes the operator experimentally realizable.

%In the SPA-PT method, positive maps can be approximated by completely positive maps and thus transform non-physical partial transposition operations to a physical operation \cite{horodecki3}.\\

\noindent %The main aim of choosing the SPA-PT map method is that it not only detects entanglement but also can be applied to estimate the optimal singlet fraction, and entanglement concurrence of the two-qubit density matrix in an experiment.
H. T. Lim et.al. have studied the SPA-PT of the single qutrit system and shown that it can be implemented in the laboratory with linear optical elements \cite{kye5}. Also, the SPA-PT method has been applied to a two-qubit system to study the entanglement detection problem \cite{adhikari1}. Thus, one may ask that the SPA-PT method works well in a two-qubit system and can be implemented in an experiment but can we use this method to detect entanglement in the higher dimensional system also? To answer this question one first need to determine the density matrix of SPA-PT of any arbitrary dimensional quantum system. Once the density matrix is obtained, one may calculate the minimum eigenvalue of the SPA-PT of the given density matrix and this eigenvalue can be used in the detection of entanglement. But this procedure is very tedious.\\
\noindent In this chapter, we have considered $d_{1}\otimes d_{2}$ dimensional NPTES and derived three different criteria for the detection of entanglement employing the SPA-PT method. It is known that the average fidelity between two quantum states can be realized in an experiment \cite{kwong}. Thus, we have expressed our proposed criteria to detect NPTES, in terms of the average fidelity between two quantum states, and hence the detection criteria can be implemented in an experiment. Among three criteria, two of them are given in terms of the concurrence of the given state so it is essential to find out the concurrence. As far as we know, there does not exist any procedure to calculate the actual value of the concurrence of the given mixed state in arbitrary dimension but in the literature, there exists a lower bound of the concurrence \cite{mintert1}. In the present work also, we have provided a lower and upper bound of the concurrence that can be realized in an experiment. \\ 
\noindent Let us consider a $d_{1}\otimes d_{2}$  dimensional mixed quantum system and perform the partial transposition operation on the second subsystem. Since the partial transposition map $I_A\otimes\textbf{\textsl{T}}$, where $\textbf{\textsl{T}}$ denotes the positive transposition map and $I_A$ represent the identity operator of the individual subsystem $A$, is not a physical map so we consider the approximated map $\widetilde{I_A\otimes\textbf{\textsl{T}}}$, which  is a
completely positive map corresponds to a quantum channel that can be experimentally implementable \cite{bae}. The approximated map can be expressed as \cite{horodecki6}
\begin{eqnarray}
\widetilde{I_A\otimes\textbf{\textsl{T}}}=(1-q^{*})(I_A\otimes\textbf{\textsl{T}})+\frac{q^{*}}{d_{1}d_{2}}I_{A}\otimes
I_{B} \label{spapt}
\end{eqnarray}
$I_B$ represents the identity operator of the individual subsystem $B$ and $q^{*}$ denote the minimum value of $q$ for which $\widetilde{I_A\otimes\textbf{\textsl{T}}}$ become a positive semi-definite operator. Since $\widetilde{I_A\otimes\textbf{\textsl{T}}}$ is a completely positive map so when it operate on a density matrix $\rho^{in}$, it gives another density matrix $\widetilde{\rho}^{out}$ at the output, i.e. $(\widetilde{I_A\otimes\textbf{\textsl{T}}})\rho^{in}=\widetilde{\rho}^{out}$. The minimum eigenvalue of the density matrix $\widetilde{\rho}^{out}$ is important in the sense that there exists a critical value of the minimum eigenvalue $\lambda_{min}(\widetilde{\rho}^{out})$ below which the state described by the density matrix $\rho^{in}$ is entangled. For instance, $2\otimes 2$ dimensional quantum state is entangled when the minimum eigenvalue $\lambda_{min}(\widetilde{\rho}^{out})$ is less than $\frac{2}{9}$. So the analysis of minimum eigenvalue of $\widetilde{\rho}^{out}$ is needed.\\
Using the idea of SPA, a positive map $\Lambda$ can be transformed to a completely positive map $\widetilde{\Lambda}$ and the SPA map $\widetilde{\Lambda}$ may be expressed as
\begin{eqnarray}
\widetilde{\Lambda}=(1-p)\Lambda+pD,~~p\in [0,1]
\end{eqnarray}
where $p$ denotes the minimum value for which $\widetilde{\Lambda}$ represents a completely positive operator. $D[\rho]$ is the depolarizing operator which may be defined as $D[\rho]=\frac{I_d}{d}$, $I_d$ is the $d$-dimensional identity matrix. For linear maps of the form $I\otimes \Lambda$, the SPA may be expressed as 
\begin{eqnarray}
\widetilde{I\otimes \Lambda}=(1-p)I\otimes \Lambda+pD\otimes D,~~p\in[0,1]
\end{eqnarray}
where $p$ denote the minimum value for which $\widetilde{I\otimes \Lambda}$ is positive. Lim et al. have demonstrated the experimental realization of SPA-PT for a two-qubit system using single-photon polarization qubits and linear optical devices. The corresponding decomposition of SPA-PT for a two-qubit system $\rho_{AB}$, may be expressed as \cite{adhikari1,kye3}
\begin{eqnarray}
\widetilde{(I\otimes T)}[\rho_{AB}]=\frac{1}{3}(I\otimes \widetilde{T})[\rho_{AB}]+\frac{2}{3}(\widetilde{\theta}\otimes D)[\rho_{AB}]
\label{spaqubitqubit}
\end{eqnarray}
where $\widetilde{T}$ denote the SPA to the transposition map which may be expressed as
\begin{eqnarray}
\widetilde{T}[\rho_{AB}]=\sum_{k=1}^{4}Tr[M_k\rho_{AB}]|v_k\rangle \langle v_k|
\end{eqnarray}
where $M_k=\frac{|v_k^{*}\rangle \langle v_k^{*}|}{2}$ for $k=1,2,3,4$. The basis $|v_1\rangle$, $|v_2\rangle$, $|v_3\rangle $ and  $|v_4\rangle$ may be expressed as
\begin{eqnarray}
|v_1\rangle = \frac{1}{\sqrt{1+|b_1|^2}}(|0\rangle+b_1|1\rangle)\nonumber\\
|v_2\rangle =\frac{1}{\sqrt{1+|b_2|^2}}(|0\rangle-b_2|1\rangle)\nonumber\\
|v_3\rangle =\frac{1}{\sqrt{1+|b_2|^2}}(|0\rangle+b_2|1\rangle)\nonumber\\
|v_4\rangle =\frac{1}{\sqrt{1+|b_1|^2}}(|0\rangle-b_1|1\rangle)
\end{eqnarray}
where $b_1=\frac{ie^{\frac{2\Pi}{3}{i}}}{i+e^{\frac{-2\Pi}{3}i}}$ and $b_2=\frac{ie^{\frac{2\Pi}{3}{i}}}{i-e^{\frac{-2\Pi}{3}{i}}}$.
In (\ref{spaqubitqubit}), $\widetilde{\theta}$ denote the SPA to the inversion map  where inversion map may be defined as $\theta[\rho_{AB}]=-\rho_{AB}$. The inversion map $\widetilde{\theta}$ may be expressed as 
\begin{eqnarray}
\widetilde{\theta}[\rho_{AB}]=\sum_{k=1}^{4}Tr[M_k\rho_{AB}]\sigma_y|v_k\rangle \langle v_k|\sigma_y
\end{eqnarray}
where $\sigma_y$ is defined in (\ref{paulimatrices})
\section{Preliminary Results}
\textbf{Result-1:}
For any two hermitian (n,n) matrices $X$ and $Y$, we have
\begin{eqnarray}
\lambda_{min}(X)Tr(Y)\leq Tr(XY)\leq \lambda_{max}(X)Tr(Y)
\label{result1}
\end{eqnarray}
\textbf{Proof:} It is known that for any hermitian (n,n) matrices $X$ and $Y$, the following inequality holds \cite{Lasserre}
\begin{eqnarray}
\sum_{i=1}^{n}\lambda_{i}(X)\lambda_{n-i+1}(Y)\leq Tr(XY)\leq \sum_{i=1}^{n}\lambda_{i}(X)\lambda_{i}(Y)
\label{theorem1}
\end{eqnarray}
where $\lambda_{min}=\lambda_{1}\leq \lambda_{2}\leq \lambda_{3}\leq.......\leq \lambda_{n}=\lambda_{max}$.\\
In the LHS of inequality (\ref{theorem1}), if we replace all eigenvalues of $X$ by its minimum eigenvalue, and in the RHS, if we replace all eigenvalues of $X$ by its maximum, then the inequality (\ref{theorem1}) reduces to (\ref{result1}). Hence proved.\\
\textbf{Result-2:} If $W$ represents the witness operator that detects the entangled quantum state described by the density operator $\rho_{AB}$ and
$C(\rho_{AB})$ denote the concurrence of the state $\rho_{AB}$ then the lower bound of concurrence is given by \cite{mintert2}
\begin{eqnarray}
C(\rho_{AB})\geq -Tr[W\rho_{AB}]
\label{result2}
\end{eqnarray}
\textbf{Result-3:} If $\rho_{AB}$ denote the density operator of a bipartite quantum state in any arbitrary dimension then all eigenvalues of $\rho_{AB}^{T_{B}}$ lying within the interval $[\frac{-1}{2},1]$ \cite{rana}.\\
\textbf{Result-4:} If any arbitrary two-qubit density operator $\rho_{AB}$ is given by
\begin{eqnarray}
\rho_{AB}=
\begin{pmatrix}
e_{11} & e_{12} & e_{13} & e_{14} \\
e_{12}^{*} & e_{22} & e_{23} & e_{24} \\
e_{13}^{*} & e_{23}^{*} & e_{33} & e_{34} \\
e_{14}^{*} & e_{24}^{*} & e_{34}^{*} & e_{44}
\end{pmatrix}, \sum_{i=1}^{4}e_{ii}=1
\end{eqnarray}
where $(*)$ denotes the complex conjugate, then the SPA-PT of $\rho_{AB}$ is given by \cite{adhikari1}
\begin{eqnarray}
\widetilde{\rho_{AB}}&=&[\frac{1}{3}(I\otimes\widetilde{T})+\frac{2}{3}(\widetilde{\Theta}\otimes\widetilde{D})]\rho_{AB}\nonumber\\&=&
\begin{pmatrix}
E_{11} & E_{12} & E_{13} & E_{14} \\
E_{12}^{*} & E_{22} & E_{23} & E_{24} \\
E_{13}^{*} & E_{23}^{*} & E_{33} & E_{34} \\
E_{14}^{*} & E_{24}^{*} & E_{34}^{*} & E_{44}
\end{pmatrix}
\label{spa1}
\end{eqnarray}
where
\begin{eqnarray}
&&E_{11}=\frac{1}{9}(2+e_{11}),E_{12}=\frac{1}{9}e_{12}^{*}, E_{13}=\frac{1}{9}e_{13},\nonumber\\&&
E_{14}=\frac{1}{9}e_{23}, E_{22}=\frac{1}{9}(2+e_{22}),E_{23}=\frac{1}{9}e_{14},\nonumber\\&&
E_{24}=\frac{1}{9}e_{24},E_{33}=\frac{1}{9}(2+e_{33}),E_{34}=\frac{1}{9}e_{34}^{*},\nonumber\\&&
E_{44}=\frac{1}{9}(2+e_{44})
\label{spa2a}
\end{eqnarray}
\section{SPA-PT of an arbitrary qutrit-qubit quantum state}
\noindent Using the idea of SPA-PT on a two-qubit system, in this section, we will obtain the SPA-PT of the general density matrix of a qutrit-qubit quantum state.\\
To achieve our goal, let us consider an arbitrary qutrit-qubit quantum state described by the density operator in the
computational basis as
\begin{eqnarray}
\varrho_{AB}=
\begin{pmatrix}
t_{11} & t_{12} & t_{13} & t_{14} & t_{15} & t_{16} \\
t_{12}^{*} & t_{22} & t_{23} & t_{24} & t_{25} & t_{26} \\
t_{13}^{*} & t_{23}^{*} & t_{33} & t_{34} & t_{35} & t_{36} \\
t_{14}^{*} & t_{24}^{*} & t_{34}^{*} & t_{44} & t_{45} & t_{46}\\
t_{15}^{*} & t_{25}^{*} & t_{35}^{*} & t_{45}^{*} & t_{55} & t_{56}\\
t_{16}^{*} & t_{26}^{*} & t_{36}^{*} & t_{46}^{*} & t_{56}^{*} & t_{66}
\end{pmatrix}, \sum_{i=1}^{6}t_{ii}=1
\label{qutrit-qubitstate}
\end{eqnarray}
where $(*)$ denotes the complex conjugate.\\\\
The decomposition of SPA-PT for a qutrit-qubit quantum state
$\varrho_{AB}$ is given by
\begin{eqnarray}
\widetilde{\varrho_{AB}}=[\frac{1}{4}(I\otimes\widetilde{T})+\frac{3}{4}(\widetilde{\Theta}\otimes\widetilde{D})]\rho_{AB}
\label{spa-ptqutrit-qubit}
\end{eqnarray}
The operator $\tilde{T}[.]$ denote the SPA of partial transposition of $(.)$ and it is given by
\begin{eqnarray}
\tilde{T}[.]= \sum_{k=1}^{9}tr[M_{k}(.)]|v_{k}\rangle\langle{v_{k}}|,~~M_{k}=\frac{1}{3}|v_k^{*}\rangle\langle{v_k^{*}|}
\end{eqnarray}
Moreover, the remaining two operators $\widetilde{\Theta}$ and $\widetilde{D}$ denote the SPA of inversion map $\Theta$
and depolarization map $D$ respectively and they can be defined as
\begin{eqnarray}
\widetilde{\Theta}[.] &=& \sum_{i=1}^9 (tr[M_i(.)] \lambda_a |v_k\rangle\langle v_k|\lambda_a)\nonumber\\&=& \lambda_a \Tilde{T}[.] \lambda_a
\end{eqnarray}
\begin{eqnarray}
\widetilde{D}[.] &=& \frac{1}{4}\sum_{i=0,x,y,z}\sigma_{i}[.]\sigma_{i}
\label{spa1}
\end{eqnarray}
where $\sigma_{0}=I$, $\sigma_{i} (i=x,y,z)$ denote the Pauli matrices and
\begin{eqnarray}
&&\lambda_a=a\lambda_{a}^{(01)}+b\lambda_a^{(12)}+c\lambda_a^{(02)},~~a,b,c \in R
\nonumber\\&&\lambda_a^{(01)}=-i|0\rangle\langle1|+i|1\rangle\langle0|
\nonumber\\&&\lambda_a^{(12)}=-i|1\rangle\langle2|+i|2\rangle\langle1|
\nonumber\\&&\lambda_a^{(02)}=-i|0\rangle\langle2|+i|2\rangle\langle0|
\end{eqnarray} and the vectors $|v_i\rangle,~~i\in \{1,2,...,9\}$ may be given by
\begin{eqnarray}
|v_1\rangle&=&\frac{1}{\sqrt{2}}(|0\rangle+w|1\rangle), |v_2\rangle = \frac{1}{\sqrt{2}}(|0\rangle+w^{2}|1\rangle), |v_3\rangle = \frac{1}{\sqrt{2}}(|0\rangle+|1\rangle)\nonumber\\
|v_4\rangle&=&\frac{1}{\sqrt{2}}(|0\rangle+w|2\rangle), |v_5\rangle = \frac{1}{\sqrt{2}}(|0\rangle+w^{2}|2\rangle), |v_6\rangle = \frac{1}{\sqrt{2}}(|0\rangle+|2\rangle)\nonumber\\
|v_7\rangle&=&\frac{1}{\sqrt{2}}(w|0\rangle+|2\rangle), |v_8\rangle = \frac{1}{\sqrt{2}}(w^2|0\rangle+|2\rangle), |v_9\rangle = \frac{1}{\sqrt{2}}(|0\rangle+|2\rangle)
\end{eqnarray}
where $\{1,w,w^2\}$ are cube roots of unity.\\
Let $\widetilde{\varrho}_{AB}$ denote the SPA-PT of $\varrho_{AB}$. The density operator for $\widetilde{\varrho}_{AB}$ is given by
\begin{eqnarray}
\widetilde{\varrho}_{AB}=
\begin{pmatrix}
\widetilde{t}_{11} & \widetilde{t}_{12} & \widetilde{t}_{13} & \widetilde{t}_{14} & \widetilde{t}_{15} & \widetilde{t}_{16} \\
\widetilde{t}_{12}^{*} & \widetilde{t}_{22} & \widetilde{t}_{23} & \widetilde{t}_{24} & \widetilde{t}_{25} & \widetilde{t}_{26} \\
\widetilde{t}_{13}^{*} & \widetilde{t}_{23}^{*} & \widetilde{t}_{33} & \widetilde{t}_{34} & \widetilde{t}_{35} & \widetilde{t}_{36} \\
\widetilde{t}_{14}^{*} & \widetilde{t}_{24}^{*} & \widetilde{t}_{34}^{*} & \widetilde{t}_{44} & \widetilde{t}_{45} & \widetilde{t}_{46}\\
\widetilde{t}_{15}^{*} & \widetilde{t}_{25}^{*} & \widetilde{t}_{35}^{*} & \widetilde{t}_{45}^{*} & \widetilde{t}_{55} & \widetilde{t}_{56}\\
\widetilde{t}_{16}^{*} & \widetilde{t}_{26}^{*} & \widetilde{t}_{36}^{*} & \tilde{t}_{46}^{*} & \widetilde{t}_{56}^{*} & \widetilde{t}_{66}
\end{pmatrix}, \sum_{i=1}^{6}\widetilde{t}_{ii}=1
\label{qutrit-qubit21}
\end{eqnarray}
\begin{eqnarray}
\widetilde{t}_{11}&=&\frac{3}{32}[(a^2+c^2)+a^2(t_{33}+t_{44})+c^2(t_{55}+t_{66})+ac(t_{35}+{t}_{35}^{*}+t_{46}+t_{46}^{*})]\nonumber\\&
+&\frac{1}{4}[\frac{2}{3}t_{11}+\frac{1}{3}t_{22}]
\label{spa123}
\end{eqnarray}
\begin{eqnarray}
\widetilde{t}_{13}&=&\frac{3}{32}[bc(1+t_{55}+t_{66})-a^2(t_{13}+t_{24})-ac(t_{15}+t_{26})+ab(t_{35}^{*}+t_{46}^{*})]
\nonumber\\&+&\frac{1}{4}[\frac{2}{3}t_{13}+\frac{1}{3}t_{24}]
\label{spa2}
\end{eqnarray}
\begin{eqnarray}
\widetilde{t}_{15}&=&\frac{3}{32}[-ab(1+t_{33}+t_{44})-ac(t_{13}+t_{24})-c^2(t_{15}+ t_{26})-bc(t_{35}+t_{46}]\nonumber\\&+&\frac{1}{4}[\frac{2}{3}t_{15}+\frac{1}{3}t_{26}]
\label{spa3}
\end{eqnarray}
\begin{eqnarray}
\widetilde{t}_{22}&=&\frac{3}{32}[(a^2+c^2)+a^2(t_{33}+t_{44})+c^2(t_{55}+t_{66})+ac(t_{35}+{t}_{35}^{*}+t_{46}+t_{46}^{*})]\nonumber\\&+&
\frac{1}{4}[\frac{1}{3}t_{11}+\frac{2}{3}t_{22}]
\label{spa4}
\end{eqnarray}
\begin{eqnarray}
\widetilde{t}_{24}&=&\frac{3}{32}[bc(1+t_{55}+t_{66})-a^2(t_{13}+t_{24})-ac(t_{15}+t_{26})+ab(t_{35}^{*}+t_{46}^{*}]\nonumber\\&+&\frac{1}{4}[\frac{1}{3}t_{13}+\frac{2}{3}t_{24}]
\label{spa5}
\end{eqnarray}
\begin{eqnarray}
\widetilde{t}_{26}&=&\frac{3}{32}[-ab(1+t_{33}+t_{44})-ac(t_{13}+t_{24})-c^2(t_{15}+t_{26})-bc(t_{35}+t_{46})]\nonumber\\&+&\frac{1}{4}[\frac{1}{3}t_{15}
+\frac{2}{3}t_{26}]
\label{spa6}
\end{eqnarray}
\begin{eqnarray}
\widetilde{t}_{33}&=&\frac{3}{32}[a^2+b^2+a^2(t_{11}+t_{22})+b^2(t_{55}+t_{66})-ab(t_{15}+t_{15}^{*})-ab(t_{26}+t_{26}^{*})]\nonumber\\&+&
\frac{1}{4}[\frac{2}{3}t_{33}+\frac{1}{3}t_{44}]
\label{spa7}
\end{eqnarray}
\begin{eqnarray}
\widetilde{t}_{35}&=&\frac{3}{32}[ac(1+t_{11}+t_{22})-bc(t_{15}+t_{26})+ab(t_{13}^{*}+t_{24}^{*})-b^2(t_{35}+t_{46})]\nonumber\\&+&
\frac{1}{4}[\frac{2}{3}t_{35}+\frac{1}{3}t_{46})]
\label{spa8}
\end{eqnarray}
\begin{eqnarray}
\widetilde{t}_{44}&=&\frac{3}{32}[a^2+b^2+a^2(t_{11}+t_{22})+b^2(t_{55}+t_{66})-ab(t_{15}+t_{15}^{*})-ab(t_{26}+t_{26}^{*})]\nonumber\\&+&\frac{1}{4}[\frac{1}{3}t_{33}+\frac{2}{3}t_{44}]
\label{spa9}
\end{eqnarray}
\begin{eqnarray}
\widetilde{t}_{46}&=&\frac{3}{32}[ac(1+t_{11}+t_{22})-bc(t_{15}-t_{26})+ab(t_{13}^{*}+t_{24}^{*})-b^2(t_{35}+t_{46})]\nonumber\\&+&\frac{1}{4}[\frac{1}{3}t_{35}+\frac{2}{3}t_{46}]
\label{spa10}
\end{eqnarray}
\begin{eqnarray}
\widetilde{t}_{55}&=&\frac{3}{32}[(b^2+c^2)+c^2(t_{11}+t_{22})+b^2(t_{33}+t_{44})
+bc(t_{13}+t_{13}^{*}+t_{24}+t_{24}^{*})]\nonumber\\&+&\frac{1}{4}[\frac{2}{3}t_{55}+\frac{1}{3}t_{66}]
\label{spa11}
\end{eqnarray}
\begin{eqnarray}
\widetilde{t}_{66}&=&\frac{3}{32}[(b^2+c^2)+c^2(t_{11}+t_{22})+b^2(t_{33}+t_{44})
+bc(t_{13}+t_{13}^{*}+t_{24}+t_{24}^{*})]\nonumber\\&+&\frac{1}{4}[\frac{1}{3}t_{55}+\frac{2}{3}t_{66}]
\label{spa12}
\end{eqnarray}
\begin{eqnarray}
\widetilde{t}_{12}&=&\frac{1}{12}t_{12}^{*},~~~\widetilde{t}_{14}=\frac{1}{12}t_{23},~~~
\widetilde{t}_{16}=\frac{1}{12}t_{25},~~~\widetilde{t}_{23}=\frac{1}{12}t_{14},~~~ \widetilde{t}_{25}=\frac{1}{12}t_{16},\nonumber\\
\widetilde{t}_{34}&=&\frac{1}{12}t_{34}^{*},~~~\widetilde{t}_{36}=\frac{1}{12}t_{45},~~~ \widetilde{t}_{45}=\frac{1}{12}t_{36},~~~
\widetilde{t}_{56}=\frac{1}{12}t_{56}^{*}
\label{spa13}
\end{eqnarray}
The value of the parameters $a$,$b$,$c$ can be chosen in such a way that $Tr(\widetilde{\varrho}_{AB})=1$.
\section{Criteria for the detection of bipartite Negative Partial Transpose Entangled states in arbitrary dimension}
\noindent In this section, we derive the criteria for the detection of bipartite NPTES in arbitrary dimensions employing the method of SPA-PT. In the first part, we will derive the lower and upper bound of the minimum eigenvalue of SPA-PT of the bipartite mixed quantum state in arbitrary dimension, and then using this lower and upper bound in the second part, we will derive the criterion for detection of NPTES in arbitrary dimensional bipartite systems.
\subsection{Lower and Upper bound of the minimum eigenvalue of SPA-PT of a bipartite mixed quantum state in arbitrary dimension}
Let us consider a bipartite mixed quantum state in arbitrary dimension described by the density operator $\rho_{AB}$. If $\widetilde{\rho}_{AB}$ denote the SPA-PT of $\rho_{AB}$ and $Q$ be any positive semi-definite operator such that $Tr(Q)=1$ then the quantity $Tr[(\widetilde{\rho}_{AB}+Q^{T_B})\rho_{AB}]$, where $T_{B}$ denote the partial transposition with respect to the subsystem B, may be expressed as
\begin{eqnarray}
Tr[(\widetilde{\rho}_{AB}+Q^{T_B})\rho_{AB}]&=&Tr[\widetilde{\rho}_{AB}\rho_{AB}+Q^{T_B}\rho_{AB}]\nonumber\\&=&Tr[\widetilde{\rho}_{AB}\rho_{AB}]+Tr[Q^{T_B}\rho_{AB}]
\nonumber\\&=&Tr[\widetilde{\rho}_{AB}\rho_{AB}]+Tr[Q\rho_{AB}^{T_B}]
\label{quantity}
\end{eqnarray}
Taking $X=\widetilde{\rho}_{AB}$ and $Y=\rho_{AB}$ in the LHS part of (\ref{result1}), we get
\begin{eqnarray}
\lambda_{min}(\widetilde{\rho}_{AB})\leq Tr(\widetilde{\rho}_{AB}\rho_{AB})
\label{ineqtr}
\end{eqnarray}
Similarly, applying Result-1 with $X=\rho_{AB}^{T_B}$ and $Y=Q$ in (\ref{result1}), we obtain
\begin{eqnarray}
\lambda_{min}(\rho_{AB}^{T_B})\leq Tr[Q\rho_{AB}^{T_B}]=Tr[Q^{T_{B}}\rho_{AB}]
\label{ineq2}
\end{eqnarray}
Adding (\ref{ineqtr}) and (\ref{ineq2}), we get
\begin{eqnarray}
&&\lambda_{min}(\widetilde{\rho}_{AB})+\lambda_{min}(\rho_{AB}^{T_B})\leq Tr[\widetilde{\rho}_{AB}\rho_{AB}]+Tr[Q^{T_{B}}\rho_{AB}]\nonumber\\&&
\Rightarrow \lambda_{min}(\rho_{AB}^{T_B})\leq G
\label{ineq3}
\end{eqnarray}
where $G=Tr[\widetilde{\rho}_{AB}\rho_{AB}]+Tr[Q^{T_{B}}\rho_{AB}]-\lambda_{min}(\widetilde{\rho}_{AB})$.\\
Therefore, the above inequality gives the upper bound on the minimum eigenvalue of $\rho_{AB}^{T_B}$.\\
A bipartite density operator $\rho_{AB}$ in any arbitrary dimension, represent a NPTES if and only if $\lambda_{min}(\rho_{AB}^{T_B})$ is negative. Further, if we assume that NPTES described by the density operator $\rho_{AB}$
detected by the witness operator $W=Q^{T_{B}}$ then we have
\begin{eqnarray}
&&\lambda_{min}(\widetilde{\rho}_{AB})\geq Tr[\widetilde{\rho}_{AB}\rho_{AB}]+Tr[W\rho_{AB}]
\label{lb}
\end{eqnarray}
Again, inequality (\ref{ineq3}) can be re-expressed as
\begin{eqnarray}
\lambda_{min}(\widetilde{\rho}_{AB}) &\leq& -\lambda_{min}(\rho_{AB}^{T_B})+Tr[\widetilde{\rho}_{AB}\rho_{AB}]\nonumber\\&+&Tr[Q^{T_{B}}\rho_{AB}]
\label{ineq4}
\end{eqnarray}
Using Result-3 in (\ref{ineq4}) and $W=Q^{T_{B}}$, the inequality (\ref{ineq4}) reduces to
\begin{eqnarray}
\lambda_{min}(\widetilde{\rho}_{AB})\leq \frac{1}{2}+Tr[\widetilde{\rho}_{AB}\rho_{AB}]+Tr[W\rho_{AB}]
\label{ub}
\end{eqnarray}
Combining inequalities (\ref{lb}) and (\ref{ub}), we get the lower bound (L) and upper bound (U) of the minimum eigenvalue of the SPA-PT of $\rho_{AB}$ and they are given by
\begin{eqnarray}
L \leq \lambda_{min}(\widetilde{\rho}_{AB}) \leq U
\label{ub1}
\end{eqnarray}
where
\begin{eqnarray}
L= Tr[\widetilde{\rho}_{AB}\rho_{AB}]+Tr[W\rho_{AB}]
\label{lu}
\end{eqnarray}
\begin{eqnarray}
U= \frac{1}{2}+Tr[\widetilde{\rho}_{AB}\rho_{AB}]+Tr[W\rho_{AB}]
\label{ub100}
\end{eqnarray}
We note that the lower bound given by $L$ can be negative also but since the minimum eigenvalue
$\lambda_{min}(\widetilde{\rho}_{AB})$ of the positive semi-definite operator $\widetilde{\rho}_{AB}$ is always positive
so the inequality (\ref{ub1}) can be re-expressed as
\begin{eqnarray}
max\{L,0\} \leq \lambda_{min}(\widetilde{\rho}_{AB}) \leq U
\label{bound}
\end{eqnarray}
where $L$ and $U$ are given by (\ref{lu}) and (\ref{ub100}) respectively.
\subsection{Criteria for the detection of $d_{1}\otimes d_{2}$ dimensional bipartite NPTES}
The entanglement of a bipartite mixed quantum state $\rho_{AB}$ in $d_{1}\otimes d_{2}$ dimension can be detected by computing the value of $L$. To see this,
let us consider the second term of $L$, which is given by $Tr(W\rho_{AB})=Tr(Q^{T_{B}}\rho_{AB})$. Since the witness operator has been constructed
by taking the partial transpose of a positive semi-definite operator so it is not physically realizable and so we would like to approximate
the witness operator $W$ in such a way that it would become a completely positive operator. Therefore, if $\widetilde{W}$ is the approximation of the witness operator $W$ then it can be expressed as
\begin{eqnarray}
\widetilde{W}=pW+\frac{1-p}{d_{1}d_{2}}I, 0\leq p\leq1,
\label{approxw}
\end{eqnarray}
The value of the parameter $p$ should be chosen in such a way that $\widetilde{W}$ becomes a positive semi-definite operator. Further, we note
that $Tr(\widetilde{W})=1$. Thus, the operator $\widetilde{W}$ represents a quantum state.\\
\textbf{Criterion-1:} A bipartite $d_{1}\otimes d_{2}$ dimensional quantum state described by the density operator $\rho_{AB}$ represent a NPTES iff
\begin{eqnarray}
Tr(\widetilde{W}\rho_{AB})<\frac{1-p}{d_{1}d_{2}}=R
\label{criteria1}
\end{eqnarray}
\textbf{Proof:} Using (\ref{approxw}), the relation between $Tr(W\rho)$ and $Tr(\widetilde{W}\rho)$ can be established as
\begin{eqnarray}
Tr(W\rho_{AB})=\frac{1}{p}[Tr(\widetilde{W}\rho_{AB})-\frac{1-p}{d_{1}d_{2}}]
\label{expectation1}
\end{eqnarray}
Since $W$ denote the witness operator that detect the NPTES $\rho_{AB}$ so $Tr(W\rho_{AB})<0$
and hence proved the required criterion.\\
Since $\widetilde{W}$ have all the properties of a quantum state so $Tr(\widetilde{W}\rho_{AB})$ can be considered same as
the average fidelity between two mixed quantum state $\widetilde{W}$ and $\rho_{AB}$ and therefore, it is given by \cite{kwong}
\begin{eqnarray}
Tr(\widetilde{W}\rho_{AB})=F_{avg}(\widetilde{W},\rho_{AB})
\label{averagefidelity}
\end{eqnarray}
It is also known that the average fidelity $F_{avg}(\widetilde{W},\rho_{AB})$
can be estimated experimentally by Hong-Ou-Mandel interferometry and thus equation (\ref{criteria1}) can be re-expressed
as
\begin{eqnarray}
F_{avg}(\widetilde{W},\rho_{AB})<\frac{1-p}{d_{1}d_{2}}
\label{criteria1mod}
\end{eqnarray}
\textbf{Criterion-2:} The bipartite state $\rho_{AB}$ of any arbitrary dimension is NPTES iff
\begin{eqnarray}
\lambda_{min}(\widetilde{\rho}_{AB})\geq F_{avg}(\rho_{AB},\widetilde{\rho}_{AB})-C(\rho_{AB})
\label{entcrit2}
\end{eqnarray}
where $C(\rho_{AB})$ denote the concurrence of the density operator $\rho_{AB}$. The lower and upper bound of $C(\rho_{AB})$ is given by
\begin{eqnarray}
\frac{1-p}{pd_{1}d_{2}}-\frac{F_{avg}(\widetilde{W},\rho_{AB})}{p}\leq C(\rho_{AB})\leq F_{avg}(\rho_{AB},\widetilde{\rho}_{AB})
\label{concbounded}
\end{eqnarray}
\textbf{Proof:} Let us first recall the lower bound $L$ of the minimum eigenvalue $\lambda_{min}(\widetilde{\rho}_{AB})$ and using Result-2 in (\ref{lu}), we get
\begin{eqnarray}
L \geq Tr[\widetilde{\rho}_{AB}\rho_{AB}]-C(\rho_{AB})
\label{lbconc}
\end{eqnarray}
The RHS of the inequality (\ref{lbconc}) may take a non-negative or negative value. Consider the case
when $Tr[\widetilde{\rho}_{AB}\rho_{AB}]-C(\rho_{AB})$ is non-negative i.e.
\begin{eqnarray}
C(\rho_{AB})\leq Tr[\widetilde{\rho}_{AB}\rho_{AB}]=F_{avg}(\rho_{AB},\widetilde{\rho}_{AB})
\label{conc100}
\end{eqnarray}
Then using (\ref{lbconc}), we can re-write (\ref{lb}) as
\begin{eqnarray}
\lambda_{min}(\widetilde{\rho}_{AB})\geq F_{avg}(\rho_{AB},\widetilde{\rho}_{AB})-C(\rho_{AB})
\label{entcrit200}
\end{eqnarray}
Moreover, using Result-2, (\ref{expectation1}) and (\ref{averagefidelity}), we get
\begin{eqnarray}
C(\rho_{AB})\geq \frac{1}{p}[\frac{1-p}{d_{1}d_{2}}-F_{avg}(\widetilde{W},\rho_{AB})]
\label{concurrence1}
\end{eqnarray}
Combining (\ref{conc100}) and (\ref{concurrence1}), we get
\begin{eqnarray}
\frac{1-p}{pd_{1}d_{2}}-\frac{F_{avg}(\widetilde{W},\rho_{AB})}{p}\leq C(\rho_{AB})\leq F_{avg}(\rho_{AB},\widetilde{\rho_{AB}})
\label{concbounded}
\end{eqnarray}
Hence proved.\\
Since the average fidelity $F_{avg}(\widetilde{W},\rho_{AB})$ and $F_{avg}(\rho_{AB},\widetilde{\rho}_{AB})$
can be estimated experimentally so the lower bound and upper bound of $C(\rho_{AB})$ can be estimated experimentally.\\
\textbf{Criterion-3:} The bipartite state $\rho_{AB}$ of any arbitrary dimension is NPTES iff
\begin{eqnarray}
U^{ent}<\frac{1}{2}
\label{entcrit3}
\end{eqnarray}
where $U^{ent}=\frac{1}{2}+F_{avg}(\rho_{AB},\widetilde{\rho_{AB}})-C(\rho_{AB})$ and $C(\rho_{AB})$ is given by
\begin{eqnarray}
C(\rho_{AB})> F_{avg}(\rho_{AB},\widetilde{\rho_{AB}})
\label{concbounded100}
\end{eqnarray}
\textbf{Proof:} Let us recall the upper bound $U$ of the minimum eigenvalue $\lambda_{min}(\widetilde{\rho}_{AB})$ and using Result-2 in (\ref{ub100}), we get
\begin{eqnarray}
U \geq \frac{1}{2}+ Tr[\widetilde{\rho}_{AB}\rho_{AB}]-C(\rho_{AB})=U^{ent}
\label{ubconc}
\end{eqnarray}
If the quantity $Tr[\widetilde{\rho}_{AB}\rho_{AB}]-C(\rho_{AB})$ is non-negative then $U \geq \frac{1}{2}$. Also if the state is separable i.e. if $C(\rho_{AB})=0$ then the value of $U$ again comes out to be greater than $\frac{1}{2}$. So it would be difficult to detect
the entangled state when $Tr[\widetilde{\rho}_{AB}\rho_{AB}]-C(\rho_{AB})\geq0$. If we now consider the case when $Tr[\widetilde{\rho}_{AB}\rho_{AB}]-C(\rho_{AB})<0$, which indeed may be the case, then we can obtain
$U^{ent}<\frac{1}{2}$. Thus, we can infer when $U^{ent}<\frac{1}{2}$ for $C(\rho_{AB})>Tr[\widetilde{\rho}_{AB}\rho_{AB}]=F_{avg}(\rho_{AB},\widetilde{\rho_{AB}})$, the state is NPTES. Hence the criterion.
\section{Illustrations}
In this section, we will verify our entanglement detection criteria given in the previous section by taking the example of a class of qubit-qubit system and qutrit-qubit system.
\subsection{Qubit-Qubit system}
Let us consider a large class of two-qubit systems described by the density operator ${\rho^{(1)}_{AB}}$ given as \cite{connor}
\begin{eqnarray}
\rho^{(1)}_{AB}=
\begin{pmatrix}
a & 0 & 0 & 0 \\
0 & b & f & 0 \\
0 & f^{*} & b & 0 \\
0 & 0 & 0 & a
\end{pmatrix}, a+b=\frac{1}{2}
\label{twoqubitexample1}
\end{eqnarray}
where $(*)$ denotes the complex conjugate.\\
The two-qubit density matrix of the form (\ref{twoqubitexample1}) has been studied by many authors \cite{bruss2,verstraete1,munro} as this form of density matrix has high entanglement. In particular, Ishizaka and Hiroshima \cite{ishizaka} have studied such density matrix and maximized the entanglement for a fixed set of eigenvalues with one of the eigenvalues being zero.\\
The density operator $\rho_{AB}^{(1)}$ is positive semi-definite only when
\begin{eqnarray}
|f| \leq b
\label{psd}
\end{eqnarray}
The partial transpose of $\rho^{(1)}_{AB}$ is given by
\begin{eqnarray}
(\rho^{(1)}_{AB})^{T_B}=
\begin{pmatrix}
a & 0 & 0 & f \\
0 & b & 0 & 0 \\
0 & 0 & b & 0 \\
f^{*} & 0 & 0 & a
\end{pmatrix}
\end{eqnarray}
$(\rho^{(1)}_{AB})^{T_B}$ has negative eigenvalue if
\begin{eqnarray}
|f|>a
\label{nev}
\end{eqnarray}
Therefore, the state $\rho^{(1)}_{AB}$ is an entangled state iff
\begin{eqnarray}
a<|f|\leq b~~~~ \textrm{and}~~~~ a+b=\frac{1}{2}
\label{entstate}
\end{eqnarray}
The concurrence of the state $\rho^{(1)}_{AB}$ is given by
\begin{eqnarray}
C(\rho^{(1)}_{AB})=|f|-a
\label{entstate}
\end{eqnarray}
Let us now assume that if $\rho^{(1)}_{AB}$ is an entangled state then it is detected by the witness operator $W^{(1)}$. The witness operator $W^{(1)}$ can be expressed as
\begin{eqnarray}
W^{(1)}=|\psi\rangle^{T_{B}}\langle\psi|,~|\psi\rangle=\frac{1}{\sqrt{1+|k|^{2}}}(k|00\rangle+|11\rangle)
\label{witness1}
\end{eqnarray}
where $k=\frac{-f}{|f|}$.\\
Since it is not possible to implement partial transposition operation experimentally so we use Result-4 to obtain the SPA-PT of $\rho^{(1)}_{AB}$ and $W^{(1)}$. SPA-PT of $\rho^{(1)}_{AB}$ and $W^{(1)}$ are given in the following form
\begin{eqnarray}
\widetilde{\rho}^{(1)}_{AB}=
\begin{pmatrix}
\frac{2+a}{9} & 0 & 0 & \frac{f}{9}\\
0 & \frac{2+b}{9} & 0 & 0\\
0 & 0 & \frac{2+b}{9} & 0\\
\frac{f^{*}}{9} & 0 & 0 & \frac{2+a}{9}
\end{pmatrix}
\end{eqnarray}
\begin{eqnarray}
\widetilde{W}^{(1)}=
\begin{pmatrix}
\frac{1}{3} & 0 & 0 & 0\\
0 & \frac{1}{6} & \frac{k}{6} & 0\\
0 & \frac{k^{*}}{6} & \frac{1}{6} & 0\\
0 & 0 & 0 & \frac{1}{3}
\end{pmatrix}
\end{eqnarray}
Also, the average fidelities between two pair of mixed quantum state $(\rho^{(1)}_{AB}, \widetilde{\rho}^{(1)}_{AB})$ and $(\widetilde{W}^{(1)},\rho^{(1)}_{AB})$ respectively, are given by
\begin{eqnarray}
F_{avg}(\rho_{AB},\widetilde{\rho}^{(1)}_{AB})= \frac{2a}{9}(2+a)+\frac{2b}{9}(2+b)
\label{avfidstate}
\end{eqnarray}
\begin{eqnarray}
F_{avg}(\widetilde{W}^{(1)},\rho^{(1)}_{AB})=\frac{2a+b-|f|}{3}
\label{avstatewitness}
\end{eqnarray}
Now we are in a position to discuss criterion-1, criterion-2, and criterion-3 for a large class of qubit-qubit systems described by the
density operator $\rho^{(1)}_{AB}$.\\
Criterion-1 for the density operator $\rho^{(1)}_{AB}$ takes the form as
\begin{eqnarray}
&&F_{avg}(\widetilde{W}^{(1)},\rho^{(1)}_{AB})<\frac{1}{6}
\label{avstatewitness1}
\end{eqnarray}
The satisfaction of the above criterion is given in Table 2.1.
\begin{table}
	\begin{center}
		\caption{Table varifying (\ref{avstatewitness1}) for different values of the state parameters (a, b, f)}
		\begin{tabular}{|c|c|c|c|}\hline
			State parameter & $F_{avg}(\widetilde{W}^{(1)},\rho^{(1)}_{AB})$ & Criterion-1 & Nature of $\rho^{(1)}_{AB})$ \\ (a, b, f) & & & \\  \hline
			(0.05, 0.45, 0.4+0.1i)  & 0.04589 & satisfied & Entangled\\\hline
			(0.1, 0.4, 0.25+0.25i)  & 0.08214 & satisfied & Entangled\\\hline
			(0.15, 0.35, 0.24+0.2i) & 0.11253 & satisfied & Entangled\\\hline
			(0.2, 0.3, 0.27+0.13i) & 0.13344 & satisfied & Entangled\\\hline
		\end{tabular}
	\end{center}
\end{table}
Next, let us illustrate criterion-2 for the density operator $\rho^{(1)}_{AB}$. Criterion-2 can be re-written for $\rho^{(1)}_{AB}$ as
\begin{eqnarray}
\lambda_{min}(\widetilde{\rho}^{(1)}_{AB})\geq F_{avg}(\rho^{(1)}_{AB},\widetilde{\rho}^{(1)}_{AB})-C(\rho^{(1)}_{AB})
\label{entcrit2ex}
\end{eqnarray}
where $C(\rho_{AB})$ is given by
\begin{eqnarray}
\frac{1}{2}-3F_{avg}(\widetilde{W}^{(1)},\rho^{(1)}_{AB})\leq C(\rho^{(1)}_{AB})\leq F_{avg}(\rho^{(1)}_{AB},\widetilde{\rho}^{(1)}_{AB})
\label{concboundedex}
\end{eqnarray}
We can now verify (\ref{entcrit2ex}) and (\ref{concboundedex}) by taking different values of the parameters of the state $\rho^{(1)}_{AB}$. The newly derived lower and upper bound of the concurrence $C(\rho^{(1)}_{AB})$ and the criterion-2 can be verified by Table 2.2 and Table 2.3 respectively.
\begin{table}
	\begin{center}
		\caption{Table varifying (\ref{concboundedex}) for different values of the state parameters (a, b, f)}
		\begin{tabular}{|c|c|c|c|}\hline
			State parameter & $F_{avg}(\widetilde{W}^{(1)},\rho^{(1)}_{AB})$ & $F_{avg}(\widetilde{\rho}^{(1)}_{AB},\rho^{(1)}_{AB})$ & C($\rho^{(1)}_{AB})$) \\ (a, b, f) & & & \\  \hline
			(0.05, 0.45, 0.2+0.2i)  & 0.08905 & 0.26777 & 0.23284\\\hline
			(0.1, 0.4, 0.25+0.25i)  & 0.08215 & 0.26 & 0.25355\\\hline
			(0.15, 0.35, 0.24+0.2i) & 0.11253 & 0.25444 & 0.16241\\\hline
			(0.2, 0.3, 0.27+0.13i) & 0.13344 & 0.25111 & 0.09966\\\hline
		\end{tabular}
	\end{center}
\end{table}
\begin{table}
	\begin{center}
		\caption{Table varifying (\ref{entcrit2ex}) for differnt values of the state parameters (a, b, f)}
		\begin{tabular}{|c|c|c|c|}\hline
			State parameter & $\lambda_{min}(\widetilde{\rho}^{(1)}_{AB})$ & Criterion-2 & Nature of $\rho^{(1)}_{AB}$   \\ (a, b, f) & & & \\  \hline
			(0.05, 0.45, 0.2+0.2i)  & 0.19635 & satisfied & Entangled\\\hline
			(0.1, 0.4, 0.25+0.25i)  & 0.19405 & satisfied & Entangled\\\hline
			(0.15, 0.35, 0.24+0.2i) & 0.20417 & satisfied & Entangled\\\hline
			(0.2, 0.3, 0.27+0.13i) & 0.21114 & satisfied & Entangled\\\hline
		\end{tabular}
	\end{center}
\end{table}
\subsection{Qutrit-Qubit system}
It is known that certain qutrit-qubit entangled state shows the property of time-invariant entanglement under collective dephasing. Further,
it has been observed that all dimension of Hilbert space studied so far does not exhibit simultaneously two important properties such as
time-invariant entanglement and freezing dynamics. The qutrit-qubit system given in the example below is important in the sense that it exhibits time-invariant entanglement, as well as freezing dynamics of entanglement under collective dephasing \cite{ali,karpat}.\\
Let us now consider an example of a qutrit-qubit quantum state described by the density operator $\rho^{(2)}_{AB}$,
\begin{eqnarray}
\rho^{(2)}_{AB}=
\begin{pmatrix}
0 & 0 & 0 & 0 & 0 & 0\\
0 & \frac{\alpha}{2} & 0 & 0 & \frac{\alpha}{2} & 0 \\
0 & 0 & \frac{1-\alpha}{2} & 0 & 0 & \frac{1-\alpha}{2} \\
0 & 0 & 0 & 0 & 0 & 0\\
0 & \frac{\alpha}{2} & 0 & 0 & \frac{\alpha}{2} & 0 \\
0 & 0 & \frac{1-\alpha}{2} & 0 & 0 & \frac{1-\alpha}{2} \\
\end{pmatrix}, 0 \leq \alpha \leq 1
\end{eqnarray}
It has been found that the state $\rho^{(2)}_{AB}$ is entangled for $0\leq \alpha \leq 1$ \cite{ali}. This result can be verified by
the witness operator method. The witness operator that detects the quantum state $\rho^{(2)}_{AB}$ as an entangled state is given by
\begin{eqnarray}
W^{(2)}=|\chi\rangle^{T_{B}}\langle\chi|,~|\chi\rangle=\frac{1}{\sqrt{1+|\kappa|^{2}}}(-\kappa|11\rangle+|20\rangle)
\label{witness2}
\end{eqnarray}
where $\kappa=\frac{\alpha+\sqrt{4-8\alpha+5\alpha^{2}}}{2(1-\alpha)}$.\\
The operator $W^{(2)}$ has been constructed based on the idea of partial transposition but since it is not practically
implementable operation so we use our criteria for the detection of entanglement, which can be implemented in the laboratory.\\
To implement our first criteria, we will use the SPA-PT of $W^{(2)}$, which is given by
\begin{eqnarray}
\widetilde{W}^{(2)}=
\begin{pmatrix}
\frac{1}{8} & 0 & 0 & 0 & 0 & 0\\
0 & \frac{1}{8} & 0 & 0 & 0 & 0\\
0 & 0 & \frac{1}{8} & 0 & 0 & -r\kappa\\
0 & 0 & 0 & \frac{1}{8}+r|\kappa|^2 & 0 & 0\\
0 & 0 & 0 & 0 & \frac{1}{8}+r & 0\\
0 & 0 & -r\kappa^{*} & 0 & 0 & \frac{1}{8}
\end{pmatrix}
\end{eqnarray}
where $r=\frac{1}{4(1+|\kappa|^2)}$.\\
Therefore, criterion-1 for the density operator $\rho^{(2)}_{AB}$ reduces to
\begin{eqnarray}
&&F_{avg}(\widetilde{W}^{(2)},\rho^{(2)}_{AB})-\frac{1}{8}<0
\label{avstatewitness2}
\end{eqnarray}
where $F_{avg}(\widetilde{W}^{(2)},\rho^{(2)}_{AB})$ is given by
\begin{eqnarray}
F_{avg}(\widetilde{W}^{(2)},\rho^{(2)}_{AB})=\frac{1-4r\alpha}{8}
\label{avgwtildequtritqubit}
\end{eqnarray}
%\begin{figure}[h]
%\centering
%\includegraphics[scale=0.5]{Figure-1.eps}
%\caption{Plot of D versus $\alpha$ where $D=F_{avg}(\tilde{W}^{(2)},\rho^{(2)}_{AB})-\frac{1}{8}$.}
%\end{figure}
Since the parameters $\alpha$ and $r$ lying in $[0,1]$ so $F_{avg}(\widetilde{W}^{(2)},\rho^{(2)}_{AB})$ is always
less than zero in this range of parameters. Hence the state $\rho^{(2)}_{AB}$ is entangled for $0\leq \alpha \leq 1$.\\
%Figure-1 tells us that the above inequality is satisfied for $0\leq \alpha \leq 1$.
Criterion 2 not only detects the entanglement but also provides the lower and upper bound of the measure of entanglement characterized by concurrence. Therefore, our next task is to calculate the lower and upper bound of the concurrence of the state described by the density operator $\rho^{(2)}_{AB}$. To achieve our goal, we use (\ref{qutrit-qubit21},\ref{spa123}-\ref{spa13}) to obtain the SPA-PT of $\rho^{(2)}_{AB}$, which can be expressed as
\begin{eqnarray}
\widetilde{\rho}^{(2)}_{AB}=
\begin{pmatrix}
\widetilde{t}_{11} & 0 & \widetilde{t}_{13} & 0 & \widetilde{t}_{15} & \widetilde{t}_{16}\\
0 & \widetilde{t}_{22} & 0 & \widetilde{t}_{24} & 0 & \widetilde{t}_{26} \\
\widetilde{t}_{13}^{*} & 0 & \widetilde{t}_{33} & 0 & \widetilde{t}_{35} & 0\\
0 & \widetilde{t}_{24}^{*} & 0 & \widetilde{t}_{44} & \widetilde{t}_{45} & \widetilde{t}_{46}\\
\widetilde{t}_{15}^{*} & 0 & \widetilde{t}_{35}^{*} & \widetilde{t}_{45}^{*} & \widetilde{t}_{55} & 0 \\
\widetilde{t}_{16}^{*} & \widetilde{t}_{26}^{*} & 0 & \widetilde{t}_{46}^{*} & 0 & \widetilde{t}_{66} \\
\end{pmatrix}, 0 \leq \alpha \leq 1
\end{eqnarray}
where
\begin{eqnarray}
&&\widetilde{t}_{11}=\frac{54}{384}+\frac{7\alpha}{384},~~\widetilde{t}_{13}=\frac{9}{128},~~\widetilde{t}_{15}=-\frac{9}{128}+\frac{3\alpha}{128},\nonumber\\&&\widetilde{t}_{16}=\frac{\alpha}{24},\widetilde{t}_{22}=\frac{9}{64}+\frac{23\alpha}{384},~~\widetilde{t}_{24}=\frac{9}{128},\nonumber\\&&\widetilde{t}_{26}=-\frac{9}{128}+\frac{3\alpha}{128},\widetilde{t}_{33}=\frac{77}{384}-\frac{23\alpha}{384},\widetilde{t}_{35}=\frac{3}{64}+\frac{3\alpha}{128},\nonumber\\&&\widetilde{t}_{44}=\frac{61}{384}-\frac{7\alpha}{384},\widetilde{t}_{45}=\frac{1-\alpha}{24},\widetilde{t}_{46}=\frac{3}{64}+\frac{3\alpha}{128}, \nonumber\\&&\widetilde{t}_{55}=\frac{61}{384}+\frac{\alpha}{24},\widetilde{t}_{66}=\frac{77}{384}-\frac{\alpha}{24}
\end{eqnarray}
The lower and upper bound of the concurrence $C(\rho^{(2)}_{AB})$ is given by
\begin{eqnarray}
\frac{1}{2}-4F_{avg}(\widetilde{W}^{(2)},\rho^{(2)}_{AB})\leq C(\rho^{(2)}_{AB}) \leq F_{avg}(\rho^{(2)}_{AB},\widetilde{\rho}^{(2)}_{AB})
\label{concqutritqubit}
\end{eqnarray}
where $F_{avg}(\rho^{(2)}_{AB},\widetilde{\rho}^{(2)}_{AB})$ is given by
\begin{eqnarray}
F_{avg}(\rho^{(2)}_{AB},\widetilde{\rho}^{(2)}_{AB})= \frac{78\alpha^{2}-78\alpha+154}{768}
\label{avgrhotildequtritqubit}
\end{eqnarray}
The lower and upper bound of the concurrence $C(\rho^{(2)}_{AB})$ is shown in Figure 2.1.
\begin{figure}[h]
	\centering
	\includegraphics[scale=0.6]{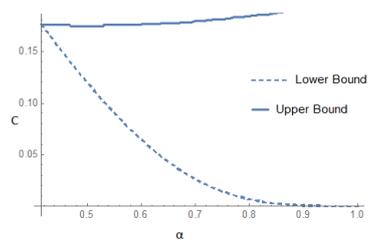}
	\caption{Plot of concurrence (C) versus $\alpha$. The dotted line shows the upper bound and the solid line
		represent the lower bound of the concurrence of the qutrit-qubit state described by the density
		matrix $\rho^{(2)}_{AB}$ }
\end{figure}
\section{Conclusion}
\noindent In this chapter, we have discussed the method of SPA-PT on a two-qubit system. Using the idea for SPA-PT on a two-qubit system, we perform a SPA map on the qutrit-qubit system and then explicitly calculated the elements of the SPA-PT of the qutrit-qubit system. We have obtained the criteria based on SPA-PT for the detection of NPTES in arbitrary dimensional Hilbert space. The first criterion detects NPTES if and only if the average fidelity of two quantum states described by the density matrix $\rho_{AB}$ and SPA-PT of the witness operator $\widetilde{W}$ is less than a quantity $R$. The quantity $R$ depends on (i) the dimension of the composite Hilbert space and (ii) the parameter that makes the SPA-PT of the witness operator a positive semi-definite. The second criterion tells us that the given state is NPTES if and only if the minimum eigenvalue of the SPA-PT of the given state is greater or equal to the difference between the concurrence of the given state and the average fidelity between the given state and its SPA-PT. Since it would not be possible to find out the exact value of the concurrence of $d_1\otimes d_2$ dimensional mixed bipartite state so we have derived the lower and upper bound of the concurrence of the given mixed state which may be realized in an experiment also. The third criterion also dealt with the detection of entanglement. Moreover, using the elements of a qutrit qubit system one can find the minimum eigenvalue of the SPA-PT in the qutrit-qubit system. Then, by analyzing the minimum eigenvalue, one can check whether the given qutrit-qubit composite system is entangled or not.

\begin{center}
	****************
\end{center}

\chapter{Classification of multipartite states through witness operator}\label{ch3}
\vspace{2cm}
\noindent\hrulefill

\noindent \emph{GHZ class of states can be considered an important class of three-qubit pure states in the sense that, it can serve as an efficient quantum channel for quantum communication purposes. In this chapter\;\footnote{Chapter 3 is based on a research paper entitled ``Classification witness operator for the classification of different subclasses of three-qubit GHZ class, \emph{Quantum Information processing} {\bf 20}, 316 (2021)''}, we have considered this important class of three-qubit pure states, divide it into different subclasses according to their utility in quantum information processing task and then studied the classification of different subclasses of GHZ class of pure states. We have started with the definition of different subclasses of the canonical form of a pure three-qubit GHZ state and then constructed a classification witness operator to classify these subclasses. The defined subclasses may be denoted as $S_{1}$, $S_{2}$, $S_{3}$, $S_{4}$. The motivation for the classification of the pure GHZ class of states into four subclasses is that the GHZ state belonging to $S_{1}$ may be more efficient than the three-qubit GHZ state belonging to $S_{2}$ or $S_{3}$ or $S_{4}$ and vice-versa with respect to some quantum information processing task. We have constructed different witness operators that can classify the subclasses $S_{i}, i=2,3,4$ from $S_{1}$ and further, shown that the constructed witness operator can be decomposed into Pauli matrices. The decomposition of the witness operator into Pauli matrices may make it possible to realize the witness operators in an experiment.}

\section{Introduction}
Entanglement is a purely quantum mechanical phenomenon that plays a vital role in the advancement of quantum information theory. The two basic problems of quantum information theory are: (i) detection of n-qubit entangled states and (ii) classification of n-qubit entangled states. The complexity of the system will increase as the number of qubits in the system increases and thus, the difficulty level of the above-mentioned two basic problems also increases. Hence, we restrict ourselves here, to study the classification of three-qubit entangled states. Three-qubit states may be classified as one separable, three biseparable, and two genuine entangled states \cite{dur4}. These six classes of three-qubit pure state can be considered inequivalent classes under SLOCC. The two SLOCC inequivalent genuine entangled classes are GHZ class and W class. In the literature, it has been shown that there exist observables that can be used to distinguish the above-mentioned six inequivalent classes of three-qubit pure states \cite{datta}. The experiment using an NMR quantum information processor has been carried out to classify six inequivalent classes under SLOCC \cite{singh}. Monogamy score can also be used to classify pure tripartite systems \cite{bera}.\\
The first result for the classification of mixed three-qubit states was given by  Acin et. al \cite{acin}. They have classified mixed three-qubit states through the method of constructing the witness operator. Sabin et. al. \cite{sabin} have studied the classification of pure as well as mixed three-qubit entanglement based on reduced two-qubit entanglement.\\
The classification of the system containing more than three qubits has also been studied in the literature. The classification of different classes of four qubit pure states has been studied in \cite{verstraete,viehmann,zangi}. The number of different classes of an n-qubit system increases when we increase the number of qubits and the discrimination of these different classes of the n-qubit system has been studied in \cite{miyake,chen1,li1,miyake1}.\\
In this chapter, we are focusing on the classification of different subclasses of GHZ class. To define different subclasses of GHZ class, let us consider the five-parameter canonical form of three-qubit pure state $|\psi \rangle_{ABC}$ shared between three distant partners $A$, $B$ and $C$. The three-qubit pure state $|\psi \rangle_{ABC}$ is given by \cite{acin1}
\begin{eqnarray}
|\psi \rangle_{ABC}&=&\lambda_0|000\rangle+\lambda_1e^{i\theta}|100\rangle+\lambda_2|101\rangle+\lambda_3|110\rangle+\lambda_4|111\rangle
\label{canonical}
\end{eqnarray}
where $\lambda_i \in R$ and $0\leq \lambda_{i}\leq 1 (i=0,1,2,3,4)$; $\theta$ ($0\leq \theta \leq \pi$) denote the phase parameter.\\
The normalization condition of the state (\ref{canonical}) is given by
\begin{eqnarray}
\lambda_{0}^{2}+\lambda_{1}^{2}+\lambda_{2}^{2}+\lambda_{3}^{2}+\lambda_{4}^{2}=1.
\label{normalization}
\end{eqnarray}
%The three-tangle $\tau_{\psi}$ for a pure three-qubit state $|\psi \rangle_{ABC}$ can be defined as \cite{coffman}
%\begin{eqnarray}
%\tau_{\psi}= C_{A(BC)}^{2}-C_{AB}^{2}-C_{AC}^{2}
%\label{tangledef}
%\end{eqnarray}
%where $C_{AB}$, $C_{AC}$ represent the partial concurrences between the pairs $(A,B)$, $(A,C)$ and $C_{A(BC)}$ denote the entanglement of qubit $A$ with the joint state of qubits $B$ and $C$. $C_{A(BC)}$ may be interpreted as residual entanglement\cite{coffman}, which is not captured by two-qubit entanglement. $C_{AB}$, $C_{AC}$ and $C_{A(BC)}$ may be expressed as       
%\begin{eqnarray}
%C_{AB}&=&C[tr_C(\rho_{ABC})]\nonumber\\
%C_{AC}&=&C[tr_B(\rho_{ABC})]\nonumber\\
%C_{A(BC)}&=&C(\rho_{ABC})=2\sqrt{det([tr_{BC}(\rho_{ABC})])}
%\end{eqnarray}
%where $\rho_{ABC}=|\psi\rangle_{ABC}\langle \psi|$.\\
The three-tangle $\tau_{\psi}$ for a pure three-qubit state $|\psi \rangle_{ABC}$ may be defined as \cite{datta,torun}
%For a pure three-qubit state $|\psi \rangle_{ABC}$, The tangle $\tau_{\psi}$ can also be calculated as\cite{datta}
\begin{eqnarray}
\tau_{\psi}= 4\lambda_{0}^{2}\lambda_{4}^{2}
\label{tanglecal}
\end{eqnarray}
The tangle $\tau_{\psi}\neq 0$ for GHZ class and $\tau_{\psi}= 0$ for W class of states. To define the subclasses of GHZ class, we make the following assumption: (i) the state parameters $\lambda_{0}$ and $\lambda_{4}$ are not equal to zero, and (ii) the phase factor $\theta=0$.\\
It may be noted that similar calculations can be performed by taking $\theta \neq 0$ also. We are now in a position to divide the three-qubit pure GHZ class of states given in (\ref{canonical}) into four subclasses as:\\
\begin{eqnarray}
&&\underline{\textbf{Subclass-I}:}\nonumber\\&&
S_{1}=\{|\psi_{S}\rangle\}, \textrm{where}\nonumber\\&&
|\psi_{S}\rangle = \lambda_0|000\rangle+\lambda_4|111\rangle
\label{class1}
\end{eqnarray}
\begin{eqnarray}
&&\underline{\textbf{Subclass-II}:}\nonumber\\&&
S_{2}=\{|\psi_{\lambda_{1}}\rangle,|\psi_{\lambda_{2}}\rangle,|\psi_{\lambda_{3}}\rangle\},\textrm{where}\nonumber\\&&
|\psi_{\lambda_{1}}\rangle=\lambda_0|000\rangle+\lambda_1|100\rangle+\lambda_4|111\rangle,\nonumber\\&&
|\psi_{\lambda_{2}}\rangle=\lambda_0|000\rangle+\lambda_2|101\rangle +\lambda_4|111\rangle,\nonumber\\&&
|\psi_{\lambda_{3}}\rangle=\lambda_0|000\rangle+\lambda_3|110\rangle +\lambda_4|111\rangle\}
\label{class2}
\end{eqnarray}
\begin{eqnarray}
&&\underline{\textbf{Subclass-III}:}\nonumber\\&&
S_{3}=\{|\psi_{\lambda_{1},\lambda_{2}}\rangle,|\psi_{\lambda_{1},\lambda_{3}}\rangle,|\psi_{\lambda_{2},\lambda_{3}}\rangle\},\textrm{where}\nonumber\\&&
|\psi_{\lambda_{1},\lambda_{2}}\rangle=\lambda_0|000\rangle+\lambda_1|100\rangle+\lambda_2|101\rangle+\lambda_4|111\rangle,\nonumber\\&& |\psi_{\lambda_{1},\lambda_{3}}\rangle=\lambda_0|000\rangle+\lambda_1|100\rangle+\lambda_3|110\rangle+\lambda_4|111\rangle,\nonumber\\&&
|\psi_{\lambda_{2},\lambda_{3}}\rangle=\lambda_0|000\rangle+\lambda_2|101\rangle+\lambda_3|110\rangle+\lambda_4|111\rangle
\label{class3}
\end{eqnarray}
\begin{eqnarray}
&&\underline{\textbf{Subclass-IV $(S_{4})$}:}\nonumber\\&&
S_{4}=\{|\psi_{\lambda_1,\lambda_2,\lambda_3}\rangle\}, \textrm{where}\nonumber\\&&
|\psi_{\lambda_1,\lambda_2,\lambda_3}\rangle= \lambda_0|000\rangle+\lambda_1|100\rangle+\lambda_2|101\rangle+\lambda_3|110\rangle+\nonumber\\&&\lambda_4|111\rangle
\label{class4}
\end{eqnarray}
Different subclasses of GHZ class of states are distributed in four different sets $S_{1}$, $S_{2}$, $S_{3}$, $S_{4}$. Classification of these subclasses can be diagrammatically shown in Figure 3.1. In Figure 3.1, the outermost circle represents GHZ states belonging to subclass IV, the second outermost circle represents the GHZ states belonging to subclass III, the third outermost circle represents the GHZ states belonging to subclass II and the innermost circle represents the standard GHZ class of states belonging to subclass-I. We should note here that these subclasses are not inequivalent under SLOCC. To transform a state from one subclass to another, we need to perform local quantum operations that depend on the state which is to be transformed. So it is necessary to know the state or at least the subclass in which the state belongs.\\
\begin{figure}[h]
	\centering
	\includegraphics[scale=0.3]{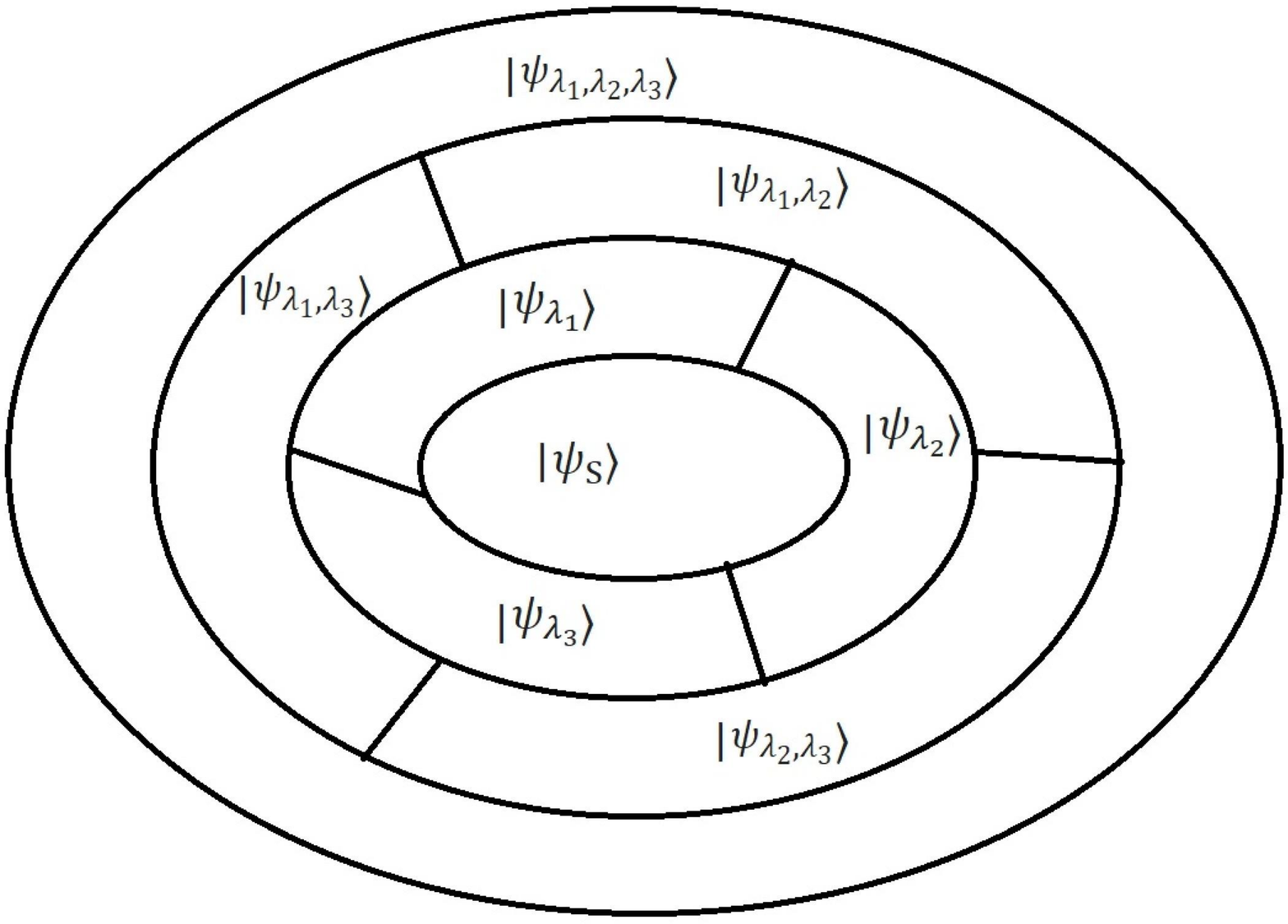}
	\caption{Classification of different subclasses of GHZ class of states described by the four sets $S_{1},S_{2}, S_{3}, S_{4}$}
\end{figure}
The motivation of this chapter is divided into three parts: \textbf{M1},\textbf{M2} and \textbf{M3}.\\
%The motivation for the work is as follows: Firstly, 
\textbf{M1: Comparing the teleportation fidelities}\\ Let us consider the teleportation scheme introduced by Lee et.al. \cite{soojoonlee}. According to this teleportation scheme, a single-qubit measurement has been performed either on the qubit $A$ or qubit $B$ or qubit $C$ of the pure three-qubit state. After the measurement, the pure three-qubit state reduces to a two-qubit state at the output. Then the resulting two-qubit state can be used as a resource state for quantum teleportation. The efficiency of the resource state is provided by teleportation fidelity.
In particular, if the single-qubit measurement is performed on either qubit $A$ or qubit $B$ or qubit $C$ of the state $|\psi_{S}\rangle \in S_{1}$ then the corresponding maximal teleportation fidelities are given by \cite{soojoonlee}
\begin{eqnarray}
F_{A}^{(|\psi_{S}\rangle)}=F_{B}^{(|\psi_{S}\rangle)}=F_{C}^{(|\psi_{S}\rangle)}=\frac{2(1+\lambda_0\lambda_4)}{3}
\label{ghz1}
\end{eqnarray}
where $F_{i}^{(|\psi_{S}\rangle)}$, $i=A,B,C$ denote the maximal teleportation fidelity when measurement is performed on qubit $i$.
In a similar fashion, if the single-qubit measurement is performed on the state $|\psi_{\lambda_{1}^{'}}\rangle \in S_{2}$, then the corresponding maximal teleportation fidelities are given by \cite{soojoonlee}
\begin{eqnarray}
F_A^{(|\psi_{\lambda_{1}^{'}}\rangle )}&=&\frac{2(1+\lambda_4^{'}\sqrt{(\lambda_{0}^{'})^{2}+(\lambda_{1}^{'})^{2})}}{3}\nonumber\\
F_B^{(|\psi_{\lambda_{1}^{'}}\rangle )}&=&F_C^{(|\psi_{\lambda_{1}^{'}}\rangle )}=\frac{2(1+\lambda_{0}^{'}\lambda_{4}^{'})}{3}
\label{ghz2}
\end{eqnarray}
Again, if the single-qubit measurement is performed on the state $|\psi_{\lambda_{1}^{''}}\rangle \in S_{3}$, then the corresponding maximal teleportation fidelities are 
\begin{eqnarray}
F_A^{(|\psi_{\lambda_{1,2}^{''}}\rangle )}&=&\frac{2(1+\lambda_4^{''}\sqrt{(\lambda_{0}^{''})^{2}+(\lambda_{1}^{''})^{2})}}{3}\nonumber\\
F_B^{(|\psi_{\lambda_{1,2}^{''}}\rangle )}&=&\frac{2(1+\lambda_0^{''}\sqrt{(\lambda_{2}^{''})^{2}+(\lambda_{4}^{''})^{2})}}{3}\nonumber\\
F_C^{(|\psi_{\lambda_{1,2}^{''}}\rangle )}&=&\frac{2(1+\lambda_{0}^{''}\lambda_{4}^{''})}{3}
\label{ghz3}
\end{eqnarray}
and if the single-qubit measurement is performed on the state $|\psi_{\lambda_{1}^{'''}}\rangle \in S_{4}$, then the corresponding maximal teleportation fidelities are 
\begin{eqnarray}
F_A^{(|\psi_{\lambda_{1,2,3}^{'''}}\rangle )}&=&\frac{2(1+\sqrt{y})}{3}\nonumber\\
F_B^{(|\psi_{\lambda_{1,2,3}^{'''}}\rangle )}&=&\frac{2(1+\lambda_0^{'''}\sqrt{(\lambda_{2}^{'''})^{2}+(\lambda_{4}^{'''})^{2})}}{3}\nonumber\\
F_C^{(|\psi_{\lambda_{1,2,3}^{'''}}\rangle )}&=&\frac{2(1+\lambda_{0}^{'''}\sqrt{(\lambda_{3}^{'''})^{2}+(\lambda_{4}^{'''})^{2})}}{3}
\label{ghz4}
\end{eqnarray} 
where
\begin{eqnarray}
y=(\lambda_{0}^{'''})^{2}(\lambda_{4}^{'''})^{2}+(\lambda_{1}^{'''})^{2}(\lambda_{4}^{'''})^{2}+(\lambda_{2}^{'''})^{2}(\lambda_{3}^{'''})^{2}-4\lambda_{1}^{'''}\lambda_{2}^{'''}\lambda_{3}^{'''}\lambda_{4}^{'''}
\end{eqnarray}
It can be easily seen that there exist state parameters $\lambda_{0}^{'''},\lambda_{1}^{'''},\lambda_{2}^{'''}, \lambda_{3}^{'''}, \lambda_{4}^{'''}, \lambda_{0}^{''},\lambda_{1}^{''},\lambda_{2}^{''}, \lambda_{4}^{''}, \lambda_{0}^{'},\lambda_{1}^{'}, \lambda_{4}^{'}, \lambda_{0}$ and $\lambda_{4}$ such that the inequalities
\begin{eqnarray}
F_A^{(|\psi_{\lambda_{1,2}^{''}}\rangle )}&\geq& F_{A}^{(|\psi_{\lambda_{1}^{'}}\rangle)}, F_A^{(|\psi_{\lambda_{1,2,3}^{'''}}\rangle )}\geq F_{A}^{(|\psi_{\lambda_{1,2}^{''}}\rangle)}\nonumber\\ F_A^{(|\psi_{\lambda_{1,2,3}^{'''}}\rangle )}&\geq& F_{A}^{(|\psi_{\lambda_{1}^{'}}\rangle)},
F_A^{(|\psi_{\lambda_{1}^{'}}\rangle )}\geq F_{A}^{(|\psi_{S}\rangle)} \nonumber\\ F_A^{(|\psi_{\lambda_{1,2}^{''}}\rangle )}&\geq& F_{A}^{(|\psi_{S}\rangle)}, F_A^{(|\psi_{\lambda_{1,2,3}^{'''}}\rangle )}\geq F_{A}^{(|\psi_{S}\rangle)}
\label{compfid}
\end{eqnarray}
holds.\\
In this way, we can compare the teleportation fidelities of the GHZ states belonging to different subclasses.
%It can be easily seen that there exist state parameters $\lambda_{0}^{'},\lambda_{1}^{'}$ and $\lambda_{4}^{'}$ such that the inequality
%\begin{eqnarray}
%F_A^{(|\psi_{\lambda_{1}^{'}}\rangle )}\geq F_{A}^{(|\psi_{S}\rangle)}
%\label{compfid}
%\end{eqnarray}
%holds.\\
We can conclude from (\ref{compfid}) that the pure three-qubit state $|\psi_{\lambda_{1}^{'}}\rangle \in S_{2}$ is more efficient than $|\psi_{S}\rangle \in S_{1}$ in the teleportation scheme \cite{soojoonlee}. In the same way, we can say that the states belonging to subclass $S_3$ are more efficient than the states belonging to $S_2$ or $S_1$. Also, it can be observed that the states belonging to any of the defined subclasses are GHZ states. Thus it is necessary to discriminate the pure three-qubit states belonging to different subclasses of GHZ class.\\
\textbf{M2: Comparing the concurrence of the reduced two-qubit state}\\% We can compare the entanglement and the tangle in these subclasses.\\
We can compare the entanglement between the reduced two-qubit mixed states obtained after tracing out either subsystem A or subsystem B or subsystem C in the following way:
If we have GHZ state belonging to subclass $S_1$, then after tracing out one qubit, the concurrence of the resulting two-qubit system will become zero, that is, $C_{AB}=C_{AC}=C_{BC}=0$. Thus, after tracing out one subsystem, the remaining two-qubit state will become a separable state. Now, if we consider GHZ state belonging to subclass $S_2$, then we have exactly one of the concurrences either $C_{AB}$ or $C_{AC}$ or $C_{BC}$ of the two-qubit mixed reduced system is non-zero. Thus, if we require any two-qubit entangled state in some quantum information processing protocol, then we can obtain it by tracing out one qubit from a three-qubit GHZ state belonging to subclass $S_2$. For example, if we need any two-qubit shared entangled state between Alice and Bob, then we can use three-qubit GHZ state$(|\psi_{ABC}\rangle=\lambda_0|000\rangle+\lambda_3|110\rangle +\lambda_4|111\rangle)$, lying in subclass $S_2$.It is possible, since, the concurrence of the reduced state $\rho_{AB}=Tr_C(|\psi\rangle_{ABC}\langle \psi|)$ is not equal to zero. But this type of situation will not arise in the case of a three-qubit GHZ state belonging to subclass $S_1$. Not only the subclass $S_2$, but we can use other subclasses such as $S_3$ and $S_4$ to get the entangled mixed two-qubit state.\\
\textbf{M3: Comparing the three-tangle of the three-qubit state}\\
We can see changes in the tangle in these subclasses as follows:\\
For a GHZ state belonging to $S_1$, we have only two parameters $\lambda_0$ and $\lambda_4$. But for the GHZ state belonging to $S_2$ have parameters $\lambda_0^{'}$, $\lambda_1^{'}$ and $\lambda_4^{'}$. Due to the normalization condition, the values of parameters get distributed. Thus, using normalization condition we get, $\lambda_0\lambda_4 > \lambda_0^{'}\lambda_4^{'}$. From (\ref{tanglecal}), we can say that the tangle of the three-qubit GHZ state belonging to $S_1$ will be more than the tangle of the GHZ state belonging to $S_2$. Again, if we compare the tangle of the three-qubit GHZ state belonging to subclass $S_3$ and subclass $S_2$, then we will find that the tangle of the GHZ state belonging to subclass $S_2$ will be more than the tangle of the GHZ state belonging to $S_3$. In this way, we can conclude that
\begin{eqnarray}
\tau^{(2)} \geq \tau^{(3)} \geq \tau^{(4)} \geq \tau^{(5)}
\end{eqnarray}
where, $\tau^{(2)}$ is the tangle of the GHZ state belonging to subclass $S_1$, $\tau^{(3)}$ is the tangle of the GHZ state belonging to subclass $S_2$, $\tau^{(4)}$ is the tangle of the GHZ state belonging to subclass $S_3$ and $\tau^{(5)}$ is the tangle of the GHZ state belonging to subclass $S_4$.\\
Therefore, \textbf{M1}, \textbf{M2} and \textbf{M3} provide us sufficient motivation to classify different subclasses of GHZ class of states.
%This paper is organized as follows: In Sec. II, we have revisited the correlation tensor for the canonical form of the three-qubit pure state which will be needed in the later section. In Sec. III, we have constructed a witness operator that can detect different subclasses of the three-qubit pure GHZ class of states. In Sec. IV, we have verified our results with some examples. We conclude in Sec. V.
\section{Derivation of the inequality required for the construction of classification witness operator}
\noindent In this section, we will construct the Hermitian matrices from the component of the correlation tensor and then use its minimum and maximum eigenvalues to derive the required inequality for the construction of the classification witness operator.\\
To start with, let us consider any arbitrary three-qubit state described by the density operator $\rho$. The three-qubit state $\rho$ may be expressed as
\begin{eqnarray}
\rho&=&\frac{1}{8}[I\otimes I\otimes I+\vec{l}.\vec{\sigma}\otimes I\otimes I+I\otimes \vec{m}.\vec{\sigma} \otimes I+I\otimes I\otimes \vec{n}{\vec{\sigma}}+ \vec{u}{\vec{\sigma}}\otimes \vec{v}{\vec{\sigma}}\otimes I+\vec{u}{\vec{\sigma}}\otimes I\otimes \vec{w}{\vec{\sigma}}\nonumber\\&+&I\otimes \vec{v}{\vec{\sigma}}\otimes \vec{w}{\vec{\sigma}}+\sum_{i,j,k=x,y,z}{t_{ijk}\sigma_i\otimes \sigma_j \otimes \sigma_k}]
\end{eqnarray} 
where 
\begin{eqnarray}
l_i&=& \text{Tr}(\rho(\sigma_i \otimes I \otimes I)),\nonumber\\ m_i&=&\text{Tr}(\rho(I \otimes \sigma_i\otimes I))\nonumber\\
n_i&=&\text{Tr}(\rho(I \otimes I\otimes \sigma_i)),\nonumber\\
u_iv_i&=&\text{Tr}(\rho(\sigma_i \otimes \sigma_i \otimes I))\nonumber\\
u_iw_i&=&\text{Tr}(\rho(\sigma_i \otimes I\otimes \sigma_i)),\nonumber\\
v_iw_i&=&\text{Tr}(\rho(I \otimes \sigma_i \otimes \sigma_i)), (i=x,y,z)
\end{eqnarray}
The correlation coefficient of the state $\rho$ can be obtained as
\begin{eqnarray}
t_{ijk}=\text{Tr}(\rho(\sigma_i \otimes \sigma_j \otimes \sigma_k)), (i,j,k=x,y,z)
\end{eqnarray}
Then the correlation tensor $\vec{T}$ can be defined as $\vec{T}=(T_x,T_y,T_z)$, where
\begin{eqnarray}
T_x=
\begin{pmatrix}
t_{xxx} & t_{xyx} & t_{xzx}\\
t_{xxy} & t_{xyy} & t_{xzy}\\
t_{xxz} & t_{xyz} & t_{xzz}\\
\end{pmatrix}
\end{eqnarray}
and
\begin{eqnarray}
T_y=
\begin{pmatrix}
t_{yxx} & t_{yyx} & t_{yzx}\\
t_{yxy} & t_{yyy} & t_{yzy}\\
t_{yxz} & t_{yyz} & t_{yzz}\\
\end{pmatrix}
\end{eqnarray}
and
\begin{eqnarray}
T_z=
\begin{pmatrix}
t_{zxx} & t_{zyx} & t_{zzx}\\
t_{zxy} & t_{zyy} & t_{zzy}\\
t_{zxz} & t_{zyz} & t_{zzz}\\
\end{pmatrix}
\end{eqnarray}
\subsection{Correlation tensor for the canonical form of three-qubit pure state}
Let us consider a three-qubit pure state described by the density operator $\rho_{\psi}$
\begin{eqnarray}
\rho_{\psi}=|\psi \rangle_{ABC} \langle \psi|
\end{eqnarray}
where $|\psi \rangle_{ABC}$ is given by (\ref{canonical}).\\
The components $T_{x},T_{y}$ and $T_{z}$ of the correlation tensor $\vec{T}$ for the state $\rho_{\psi}$ is given by
\begin{eqnarray}
T_x=
\begin{pmatrix}
2\lambda_0\lambda_4 & 0 & 2\lambda_0\lambda_2\\
0 & -2\lambda_0\lambda_4 & 0\\
2\lambda_0\lambda_3 & 0 & 2\lambda_0\lambda_1 cos\theta\\
\end{pmatrix}
\label{tx}
\end{eqnarray}
\begin{eqnarray}
T_y=
\begin{pmatrix}
0 & -2\lambda_0\lambda_4 & 0\\
-2\lambda_0\lambda_4 & 0 & -2\lambda_0\lambda_2\\
0 & -2\lambda_0\lambda_3 & 2\lambda_0\lambda_1 sin\theta\\
\end{pmatrix}
\label{ty}
\end{eqnarray}
\begin{eqnarray}
T_z=
\begin{pmatrix}
t_{zxx} & t_{zyx} & t_{zzx}\\
t_{zxy} & t_{zyy} & t_{zzy}\\
t_{zxz} & t_{zyz} & t_{zzz}
\end{pmatrix}
\label{tz}
\end{eqnarray}
where $t_{zxx}=-2(\lambda_2\lambda_3+\lambda_1\lambda_4 cos\theta)$,$t_{zyx}=2\lambda_1\lambda_4 sin\theta$, $t_{zzx}=2(\lambda_3\lambda_4-\lambda_1\lambda_2 cos\theta)$,$t_{zxy}=2\lambda_1\lambda_4 sin\theta$, $t_{zyy}=2(\lambda_1\lambda_4 cos\theta-\lambda_2\lambda_3)$,$t_{zzy}=2\lambda_1\lambda_2 sin\theta$, $t_{zxz}=2(\lambda_2\lambda_4-\lambda_1\lambda_3 cos\theta)$, $t_{zyz}=2\lambda_1\lambda_3 sin\theta$, $t_{zzz}=\lambda_0^2-\lambda_1^2+\lambda_2^2+\lambda_3^2-\lambda_4^2$.\\
The Hermitian matrices can be constructed from $T_{x}$  and $T_{y}$ as
\begin{eqnarray}
T_x^{T}T_{x}=
\begin{pmatrix}
a_{x} & 0 & b_{x}\\
0 & c_{x} & 0\\
b_{x} & 0 & d_{x}\\
\end{pmatrix}
\label{hermitian2}
\end{eqnarray}
where $a_{x}=4\lambda_0^{2}(\lambda_4^{2}+\lambda_3^{2})$, $b_{x}=4\lambda_0^{2}(\lambda_2\lambda_4+\lambda_1\lambda_3 cos\theta)$,
$c_{x}=4\lambda_0^{2}\lambda_4^{2}$, $d_{x}=4\lambda_0^{2}(\lambda_2^{2}+\lambda_1^{2} cos^{2}\theta)$.\\
and
\begin{eqnarray}
T_y^{T}T_{y}=
\begin{pmatrix}
a_{y} & 0 & b_{y}\\
0 & c_{y} & d_{y}\\
b_{y} & d_{y} & e_{y}\\
\end{pmatrix}
\label{hermitian3}
\end{eqnarray}
where $a_{y}=4\lambda_0^{2}\lambda_4^{2}$, $b_{y}=4\lambda_0^{2}\lambda_2\lambda_4$, $c_{y}=4\lambda_0^{2}(\lambda_4^{2}+\lambda_3^{2})$,
$d_{y}=-4\lambda_0^{2}\lambda_1\lambda_3 sin\theta$, $e_{y}=4\lambda_0^{2}(\lambda_2^{2}+\lambda_1^{2} sin^{2}\theta)$.\\
The superscript $T$ refers to the simple matrix transposition operation.
\subsection{Inequality for the construction of classification witness operator}
Let us recall the canonical form of a three-qubit state $|\psi\rangle_{ABC}$ given in (\ref{canonical}). The invariants with respect to the state $|\psi\rangle_{ABC}$ under local unitary transformations are given by \cite{adhikari,torun}
\begin{eqnarray}
&&\lambda_{0}\lambda_{4}=\frac{\sqrt{\tau_{\psi}}}{2}\nonumber\\&&
\lambda_{0}\lambda_{2}=\frac{C_{AC}}{2}\nonumber\\&&
\lambda_{0}\lambda_{3}=\frac{C_{AB}}{2}\nonumber\\&&
|\lambda_{2}\lambda_{3}-e^{i\varphi}\lambda_{1}\lambda_{4}|=\frac{C_{BC}}{2}
\label{invariant}
\end{eqnarray}
Here $\tau_{\psi}$ denote the three-tangle of the state $|\psi\rangle_{ABC}$ whereas $C_{AB}$, $C_{AC}$ and $C_{BC}$ represent the partial concurrences between the pairs $(A,B)$, $(A,C)$ and $(B,C)$ respectively.\\
Furthermore, the other invariants of three-qubit states under local unitary transformations have been studied in \cite{sudbery} and the invariants are given by
\begin{eqnarray}
&&I_1=\langle \psi|\psi\rangle_{ABC} \nonumber\\&&
I_2=tr(\rho_C^{2})=2(\lambda_1\lambda_2+\lambda_3\lambda_4)^2\nonumber\\&&
I_3=tr(\rho_B^{2})=2(\lambda_1\lambda_3+\lambda_2\lambda_4)^2\nonumber\\&&
I_4=tr(\rho_A^{2})=2\lambda_0^2\lambda_1^2\nonumber\\&&
I_5=\frac{1}{4}\tau_{\psi}^{2}=4\lambda_0^{4}\lambda_4^{4}
\end{eqnarray}
where $\rho_{A}=Tr_{BC}(|\psi\rangle_{ABC}\langle \psi|)$, $\rho_{B}=Tr_{AC}(|\psi\rangle_{ABC}\langle \psi|)$, $\rho_{C}=Tr_{AB}(|\psi\rangle_{ABC}\langle \psi|)$ denote reduced density matrices of a single qubit.\\
Further, recalling the Hermitian matrices $T_{x}^{T}T_{x}$ and $T_{y}^{T}T_{y}$ from (\ref{hermitian2}) and (\ref{hermitian3}), we calculate the traces of the Hermitian matrices as
\begin{eqnarray}
Tr({T_x}^{T}T_x)&=&8{\lambda_0}^2{\lambda_4}^2+4{\lambda_0}^2{\lambda_3}^2+4{\lambda_0}^2{\lambda_2}^2
+4{\lambda_0}^2{\lambda_1}^2cos^{2}\theta
\label{Tr1}
\end{eqnarray}
and
\begin{eqnarray}
Tr({T_y}^{T}T_y)&=&8{\lambda_0}^2{\lambda_4}^2+4{\lambda_0}^2{\lambda_3}^2+4{\lambda_0}^2{\lambda_2}^2
+4{\lambda_0}^2{\lambda_1}^2sin^{2}\theta
\label{Tr2}
\end{eqnarray}
Adding (\ref{Tr1}) and (\ref{Tr2}), we get
\begin{eqnarray}
Tr[({T_x}^{T}T_x)+({T_y}^{T}T_y)]&=&16{\lambda_0}^2{\lambda_4}^2+8{\lambda_0}^2{\lambda_3}^2+8{\lambda_0}^2{\lambda_2}^2+4{\lambda_0}^2{\lambda_1}^2
\label{Tracecond1}
\end{eqnarray}
The expression for $Tr[({T_x}^{T}T_x)+({T_y}^{T}T_y)]$ can be re-expressed in terms of invariants such as three-tangle and partial concurrences as
\begin{eqnarray}
Tr[({T_x}^{T}T_x)+({T_y}^{T}T_y)]&=&4\tau_{\psi}+2{C_{AB}}^2+2{C_{AC}}^2+4{\lambda_0}^2{\lambda_1}^2
\label{Tracecond2}
\end{eqnarray}
In terms of expectation of the operators, the expression (\ref{Tracecond2}) can further be written as
\begin{eqnarray}
Tr[({T_x}^{T}T_x)+({T_y}^{T}T_y)]&=&(\langle O_1\rangle_{\psi_{ABC}})^2+\frac{1}{2}(\langle O_3\rangle_{\psi_{ABC}})^2\nonumber\\&+&\frac{1}{2}(\langle O_2 \rangle_{\psi_{ABC}})^2+4{\lambda_0}^2{\lambda_1}^2
\label{Tracecond3}
\end{eqnarray}
\begin{eqnarray}
Tr[({T_x}^{T}T_x)+({T_y}^{T}T_y)]&\leq& [\langle O_1\rangle_{\psi_{ABC}}+\frac{1}{\sqrt{2}}\langle O_3\rangle_{\psi_{ABC}}\nonumber\\&+&\frac{1}{\sqrt{2}}\langle O_2 \rangle_{\psi_{ABC}}+2{\lambda_0}{\lambda_1}]^2
\label{upperbound}
\end{eqnarray}

where
\begin{eqnarray}
O_1&=&2(\sigma_x \otimes \sigma_x \otimes \sigma_x)\nonumber\\
O_2&=&2(\sigma_x \otimes \sigma_z \otimes \sigma_x)\nonumber\\
O_3&=&2(\sigma_x \otimes \sigma_x \otimes \sigma_z)
\end{eqnarray}
The expection values of the operator $O_1$, $O_2$, $O_3$ may be written in terms of invariants as\cite{datta},
\begin{eqnarray}
\langle O_1\rangle =4\lambda_0\lambda_4=2\sqrt{\tau_{\psi}}\nonumber\\
\langle O_2\rangle =4\lambda_0\lambda_2=\frac{C_{AC}}{2}\nonumber\\
\langle O_3\rangle =4\lambda_0\lambda_3=\frac{C_{AB}}{2}
\end{eqnarray}
Using (\ref{upperbound}), the upper bound (U) and the lower bound (L) of $Tr[({T_x}^{T}T_x)+({T_y}^{T}T_y)]$ is given by
\begin{eqnarray}
L \leq Tr[({T_x}^{T}T_x)+({T_y}^{T}T_y)]\leq U
\label{bound}
\end{eqnarray}
where $L=\mu_{max}({T_x}^{T}T_x)+\mu_{min}({T_y}^{T}T_y)$ and $U=(4\lambda_0\lambda_4+2\sqrt{2}\lambda_0\lambda_3+2\sqrt{2}\lambda_0\lambda_2+2\lambda_0\lambda_1)^2$. The lower bound $L$ can be obtained using Weyl's result \cite{horn}. $\mu_{max}({T_x}^{T}T_x)$ and $\mu_{min}({T_y}^{T}T_y)$ denote the maximum and minimum eigenvalue of $T_x^{T}T_x$ and $T_y^{T}T_y$ respectively.\\
Thus, equation (\ref{bound}) can be re-written as
\begin{eqnarray}
&&[\mu_{max}({T_x}^{T}T_x)+\mu_{min}({T_y}^{T}T_y)]^{\frac{1}{2}} \leq \nonumber\\&& 4\lambda_0\lambda_4+2\sqrt{2}\lambda_0\lambda_3+2\sqrt{2}\lambda_0\lambda_2+2\lambda_0\lambda_1
\label{inequality}
\end{eqnarray}
%The equation (\ref{inequality}) can be simplified as
%\begin{eqnarray}
%&&-[\mu_{max}({T_x}^{T}T_x)+\mu_{min}({T_y}^{T}T_y)]^{\frac{1}{2}} \leq \nonumber\\&& [\langle O_1\rangle_{\psi}+\frac{1}{\sqrt{2}}\langle O_2\rangle_{\psi}+\frac{1}{\sqrt{2}}\langle O_3\rangle_{\psi}+2\lambda_0\lambda_1] \nonumber\\&\leq& [\mu_{max}({T_x}^{T}T_x)+\mu_{min}({T_y}^{T}T_y)]^{\frac{1}{2}}
%\label{inequality1}
%\end{eqnarray}
If $0\leq \mu_{max}({T_x}^{T}T_x)+\mu_{min}({T_y}^{T}T_y)\leq 1$ then the inequality (\ref{inequality}) reduces to
\begin{eqnarray}
&&\mu_{max}({T_x}^{T}T_x)+\mu_{min}({T_y}^{T}T_y) \leq \nonumber\\&& 4\lambda_0\lambda_4+2\sqrt{2}\lambda_0\lambda_3+2\sqrt{2}\lambda_0\lambda_2+2\lambda_0\lambda_1
\label{inequality1}
\end{eqnarray}

The derived inequality (\ref{inequality1}) will be useful in constructing the Hermitian operators for the classification of states that lies within the subclasses of GHZ class.
\section{Construction of classification witness operator}
Let $|\chi\rangle\in S_{1}$  and $|\omega\rangle$ be any state belong to either $S_{2}$ or $S_{3}$ or $S_{4}$. Then,  the Hermitian operator $H$ is said to be the classification witness operator if
\begin{eqnarray}
&& (a) Tr(H|\chi\rangle\langle \chi|) \geq 0, \forall~~ |\chi\rangle \in S_{1} \nonumber\\&&
(b) Tr(H|\omega\rangle\langle \omega|) < 0, \textrm{for at least one}~ |\omega\rangle \in S_{i},~~
(i=2,3,4)
\label{cwo1}
\end{eqnarray}
If the above condition holds then the classification witness operator $H$ classifies the states between (i) subclass-I and subclass-II (ii) subclass-I and subclass-III (iii) subclass-I and subclass-IV. 
%We will now discuss the procedure of constructing the different classification witness operators that can classify the states residing in (i) subclass-I and subclass-II (ii) subclass-I and subclass-III (iii) subclass-I and subclass-IV.

\subsection{Classification witness operator for the classification of states contained in subclass-I and subclass-II}
We are now in a position to construct the classification witness operator that can classify the states residing in subclass-I and subclass-II.\\
\textbf{(I) Classification of states confined in subclass-II with state parameters $\lambda_0$, $\lambda_1$ and $\lambda_4$ and subclass-I:}\\
The GHZ class of state within subclass-II with state parameters $\lambda_0$, $\lambda_1$, and $\lambda_4$ is given by
\begin{eqnarray}
|\psi_{\lambda_1}\rangle=\lambda_0|000\rangle+\lambda_1|100\rangle+\lambda_4|111\rangle
\label{psi1}
\end{eqnarray}
with the normalization condition $\lambda_0^{2}+\lambda_1^{2}+\lambda_4^{2}=1$.\\
In particular, for $\lambda_1=0$, the state $|\psi_{\lambda_1}\rangle$ reduces to $|\psi_{\lambda_1=0}\rangle \in S_{1}$
where
\begin{eqnarray}
|\psi_{\lambda_1=0}\rangle=\lambda_0|000\rangle+\lambda_4|111\rangle, \lambda_0^{2}+\lambda_4^{2}=1
\label{psi0}
\end{eqnarray}
The Hermitian matrices $T_{x}^{T}T_{x}$ and $T_{y}^{T}T_{y}$ for the state $\rho_{\lambda_1}=|\psi_{\lambda_1 }\rangle \langle \psi_{\lambda_1}|$ is given by
\begin{eqnarray}
T_{x}^{T}T_{x}=
\begin{pmatrix}
4\lambda_0^{2}\lambda_4^{2} & 0 & 0\\
0 & 4\lambda_0^{2}\lambda_4^{2} & 0\\
0 & 0 & 4\lambda_0^{2}\lambda_1^{2}\\
\end{pmatrix}
\end{eqnarray}
\begin{eqnarray}
T_{y}^{T}T_{y}=
\begin{pmatrix}
4\lambda_0^{2}\lambda_4^{2} & 0 & 0\\
0 & 4\lambda_0^{2}\lambda_4^{2} & 0\\
0 & 0 & 0\\
\end{pmatrix}
\end{eqnarray}
The expression for $Tr[({T_x}^{T}T_x)+({T_y}^{T}T_y)]$ is given by
\begin{eqnarray}
Tr[({T_x}^{T}T_x)+({T_y}^{T}T_y)]&=&16{\lambda_0}^2{\lambda_4}^2+4{\lambda_0}^2{\lambda_1}^2
\label{Tracecond21}
\end{eqnarray}
The maximum eigenvalue of $T_{x}^{T}T_{x}$ is given by
\begin{eqnarray}
\mu_{max}({T_x}^{T}T_x)=max\{4\lambda_0^{2}\lambda_1^{2},4\lambda_0^{2}\lambda_4^{2}\}
\label{maxeigenval1}
\end{eqnarray}
The minimum eigenvalue of $T_{y}^{T}T_{y}$ is given by
\begin{eqnarray}
\mu_{min}({T_y}^{T}T_y)=0
\label{mineigenval1}
\end{eqnarray}
It can be easily observed that in this case $0\leq \mu_{max}({T_x}^{T}T_x)+\mu_{min}({T_y}^{T}T_y)\leq 1$ holds.\\
Since $\mu_{max}({T_x}^{T}T_x)$ depends on the value of the two parameters $\lambda_1$ and $\lambda_4$ so we will investigate two cases independently.\\
\underline{\textbf{Case-I: $\lambda_{4}>\lambda_{1}$}}\\
If $\lambda_{4}>\lambda_{1}$ then $\mu_{max}({T_x}^{T}T_x)=4\lambda_0^{2}\lambda_4^{2}$.\\
The inequality (\ref{inequality1}) then can be re-expressed in terms of the expectation value of the operators $O_{1}$ as
\begin{eqnarray}
-2\lambda_0\lambda_1 \leq \langle O_1\rangle_{\psi_{\lambda_1}}-\frac{ \langle O_1\rangle_{\psi_{\lambda_1}}^{2}}{4}
\label{inequality11}
\end{eqnarray}
If $\lambda_1=0$ then the R.H.S of the inequality (\ref{inequality11}) is always positive. Further, it can be observed that since $0\leq \langle O_1\rangle_{\psi_{\lambda_1}}\leq 1$ so the R.H.S of the inequality (\ref{inequality11}) still positive even for $\lambda_1 \neq 0$. Thus the R.H.S of the inequality is positive for every state belonging to $S_{2}$. Hence, for $\lambda_{4}>\lambda_{1}$, it is not possible to make a distinction between the class of states $|\psi_{\lambda_{1}}\rangle \in S_{2} $ and $|\psi_{\lambda_{1}=0}\rangle \in S_{1}$ using the inequality (\ref{inequality11}).\\
\underline{\textbf{Case-II: $\lambda_{4}<\lambda_{1}$}}\\
If $\lambda_{4}<\lambda_{1}$ then $\mu_{max}({T_x}^{T}T_x)=4\lambda_0^{2}\lambda_1^{2}$.\\
The inequality (\ref{inequality1}) can be re-written as
\begin{eqnarray}
-2\lambda_0\lambda_1 \leq \langle O_1\rangle_{\psi_{\lambda_1}}-4\lambda_0^{2}\lambda_1^{2}
\label{inequality12}
\end{eqnarray}
We can now define a Hermitian operator $H_{1}$ as
\begin{eqnarray}
H_{1}=O_1-\frac{1}{4}\langle O_4\rangle_{\psi_{\lambda_1}}^{2}I
\label{h1}
\end{eqnarray}
where,
\begin{eqnarray}
O_4=2(\sigma_{x} \otimes I \otimes I)
\end{eqnarray}
The expectation value of the operator $O_4$, in terms of invariants may be written as,
\begin{eqnarray}
\langle O_4\rangle=4\lambda_0\lambda_1=4\sqrt{\frac{I_4}{\sqrt{2}}}
\end{eqnarray}
Therefore, the inequality (\ref{inequality12}) can be re-formulated as
\begin{eqnarray}
-2\lambda_0\lambda_1 \leq \langle H_{1}\rangle_{\psi_{\lambda_1}}
\label{inequality13}
\end{eqnarray}
If $\lambda_1=0$ then $\langle H_{1}\rangle_{\psi_{\lambda_1}}\geq 0$ for all states $|\psi_{\lambda_1=0}\rangle \in S_{1}$.\\
For $\lambda_1\neq0$, we can calculate $\langle H_{1}\rangle_{\psi_{\lambda_1}}
\label{inequality13}=Tr(H_{1}\rho_{\lambda_1})$ which is given by
\begin{eqnarray}
Tr(H_{1}\rho_{\lambda_1})&=&4\lambda_0(\lambda_4-\lambda_0{\lambda_1}^2)
\label{tr12}
\end{eqnarray}
It can be easily shown that there exist state parameters $\lambda_0,\lambda_1,\lambda_4$ for which $\lambda_4-\lambda_0{\lambda_1}^2<0$ and thus $Tr(H_{1}\rho_{\lambda_1})<0$. For instance, if we take $\lambda_0=0.4$, $\lambda_1=0.911043$ and $\lambda_4=0.1$, Then $Tr(H_1\rho_{\lambda_1})=-0.3712$, which is negative.\\
Thus the Hermitian operator $H_{1}$ discriminate the class $|\psi_{\lambda_{1}}\rangle \in S_{2}$ from $|\psi_{\lambda_1=0}\rangle \in S_{1}$.\\ \ \\
\textbf{(II) Classification of states confined in subclass-II with state parameters $\lambda_0$, $\lambda_i (i=2,3)$ and $\lambda_4$ and subclass-I:}\\
The GHZ class of state within subclass-II with state parameters ($\lambda_0$, $\lambda_2$, $\lambda_4$) and ($\lambda_0$, $\lambda_3$, $\lambda_4$) are given by
\begin{eqnarray}
|\psi_{\lambda_2}\rangle=\lambda_0|000\rangle+\lambda_2|101\rangle+\lambda_4|111\rangle
\label{psi2}
\end{eqnarray}
with $\lambda_0^{2}+\lambda_2^{2}+\lambda_4^{2}=1$ and
\begin{eqnarray}
|\psi_{\lambda_3}\rangle=\lambda_0|000\rangle+\lambda_3|110\rangle+\lambda_4|111\rangle
\label{psi3}
\end{eqnarray}
with $\lambda_0^{2}+\lambda_3^{2}+\lambda_4^{2}=1$.\\
The Hermitian matrices $T_{x}^{T}T_{x}$ and $T_{y}^{T}T_{y}$ for the state $\rho_{\lambda_2}=|\psi_{\lambda_2 }\rangle \langle \psi_{\lambda_2}|$ may be given by
\begin{eqnarray}
T_{x}^{T}T_{x}= T_{y}^{T}T_{y}=
\begin{pmatrix}
4\lambda_0^{2}\lambda_4^{2} & 0 & 4\lambda_0^{2}{\lambda_2}{\lambda_4}\\
0 & 4\lambda_0^{2}\lambda_4^{2} & 0\\
4{\lambda_0}^2{\lambda_2}{\lambda_4} & 0 & 4\lambda_0^{2}\lambda_2^{2}\\
\end{pmatrix}
\end{eqnarray}
The Hermitian matrices $T_{x}^{T}T_{x}$ and $T_{y}^{T}T_{y}$ for the state $\rho_{\lambda_3}=|\psi_{\lambda_3 }\rangle \langle \psi_{\lambda_3}|$ is given by
\begin{eqnarray}
T_{x}^{T}T_{x}=
\begin{pmatrix}
4\lambda_0^{2}(\lambda_3^{2}+\lambda_4^{2}) & 0 & 0\\
0 & 4\lambda_0^{2}\lambda_4^{2} & 0\\
0 & 0 & 0\\
\end{pmatrix}
\end{eqnarray}
\begin{eqnarray}
T_{y}^{T}T_{y}=
\begin{pmatrix}
4\lambda_0^{2}\lambda_4^{2} & 0 & 0\\
0 & 4\lambda_0^{2}(\lambda_3^{2}+\lambda_4^{2}) & 0\\
0 & 0 & 0\\
\end{pmatrix}
\end{eqnarray}
The maximum eigenvalue of $T_{x}^{T}T_{x}$ for the state $|\psi_{\lambda_i}\rangle (i=2,3)$  is given by
\begin{eqnarray}
\mu_{max}({T_x}^{T}T_x)=4\lambda_0^{2}(\lambda_i^{2}+\lambda_4^{2}), i=2,3
\label{maxeigenval1}
\end{eqnarray}
The minimum eigenvalue of $T_{y}^{T}T_{y}$ for the state $|\psi_{\lambda_i}\rangle (i=2,3)$ is given by
\begin{eqnarray}
\mu_{min}({T_y}^{T}T_y)=0
\label{mineigenval1}
\end{eqnarray}
%\begin{eqnarray}
%T_{y}^{T}T_{y}=
%\begin{pmatrix}
%4\lambda_0^{2}\lambda_4^{2} & 0 & 4\lambda_0^{2}{\lambda_2}{\lambda_4}\\
%0 & 4\lambda_0^{2}\lambda_4^{2} & 0\\
%4{\lambda_0}^2{\lambda_2}{\lambda_4} & 0 & 4{\lambda_0}^2{\lambda_2}^2\\
%\end{pmatrix}
%\end{eqnarray}e
For the state either described by the density operator $\rho_{\lambda_2}=|\psi_{\lambda_2}\rangle \langle \psi_{\lambda_2}|$ or $\rho_{\lambda_3}=|\psi_{\lambda_3 }\rangle \langle \psi_{\lambda_3}|$, the expression of $Tr[({T_x}^{T}T_x)+({T_y}^{T}T_y)]$ is given by
\begin{eqnarray}
Tr[({T_x}^{T}T_x)+({T_y}^{T}T_y)]&=&16\lambda_0^{2}\lambda_4^{2}+8\lambda_0^{2}\lambda_i^{2},\\&& (i=2,3)\nonumber
\label{Tracecond21}
\end{eqnarray}
%It can be easily observed that in this case $0\leq \mu_{max}({T_x}^{T}T_x)+\mu_{min}({T_y}^{T}T_y)\leq 1$ holds.\\

The inequality (\ref{inequality1}) then can be re-expressed in terms of the expectation value of the operators $O_{1}$ and $O_{4}$ as
\begin{eqnarray}
-2\sqrt{2}\lambda_0\lambda_i &\leq& \langle O_1\rangle_{\psi_{\lambda_i}}-\frac{1}{4}\langle O_1\rangle_{\psi_{\lambda_i}}^{2}-\frac{1}{4}\langle O_i\rangle_{\psi_{\lambda_i}}^{2}, (i=2,3)
\label{inequality21}
\end{eqnarray}
We can now define classification witness operators $H_{i}, (i=2,3)$ as
\begin{eqnarray}
H_{i}=O_1-\frac{1}{4}[\langle O_i\rangle_{\psi_{\lambda_i}}^{2}+\langle O_1\rangle_{\psi_{\lambda_i}}^{2}]I
\label{h2}
\end{eqnarray}
Therefore, the inequality (\ref{inequality21}) can be re-formulated as
\begin{eqnarray}
-2\sqrt{2}\lambda_0\lambda_i \leq \langle H_{i}\rangle_{\psi_{\lambda_i}}, i=2,3
\label{inequality22}
\end{eqnarray}
If $\lambda_i=0, (i=2,3) $ then $\langle H_{i}\rangle_{\psi_{\lambda_i}}\geq 0$ for all states $|\psi_{\lambda_i=0}\rangle (i=2,3) \in S_{1}$.\\
For $\lambda_i\neq0, (i=2,3)$, we can calculate $Tr(H_{i}\rho_{\lambda_i}) (i=2,3)$ which is given by
\begin{eqnarray}
Tr(H_{i}\rho_{\lambda_i})&=&4\lambda_0\lambda_4(1-\lambda_0\lambda_4)-4{\lambda_0}^2{\lambda_i}^2,i=2,3
\label{tr12}
\end{eqnarray}
It can be easily shown that there exist state parameters $\lambda_0,\lambda_i (i=2,3),\lambda_4$ for which $Tr(H_{i}\rho_{\lambda_i})<0$. For instance, if we  take $\lambda_0=0.4$,$\lambda_i$=0.894427$ (i=2,3)$ and $\lambda_4=0.2$, we get $Tr[H_i\rho_{\lambda_i}]=-0.2176$.
Therefore, the classification witness operator $H_{i} (i=2,3)$ classify the class of states $\psi_{\lambda_i} (i=2,3) \in S_{2}$ given in (\ref{psi2}) from the class $|\psi_{\lambda_i=0}\rangle (i=2,3) \in S_{1}$.
\subsection{Classification witness operator for the classification of states contained in subclass-I and subclass-III}
In this subsection, we will construct a classification witness operator to discriminate subclass-I from subclasses of GHZ class spanned by four basis states.\\ \ \\
\textbf{(I) Classification of states confined in subclass-III with state parameters $\lambda_0$, $\lambda_2$, $\lambda_3$ and $\lambda_4$ and subclass-I:}\\
The GHZ class of state within subclass-III with state parameters $\lambda_0$, $\lambda_2$, $\lambda_3$, and $\lambda_4$ is given by
\begin{eqnarray}
|\psi_{\lambda_{2},\lambda_{3}}\rangle=\lambda_0|000\rangle+\lambda_2|101\rangle+\lambda_3|110\rangle+\lambda_4|111\rangle
\label{psi23}
\end{eqnarray}
with $\lambda_0^{2}+\lambda_2^{2}+\lambda_3^{2}+\lambda_4^{2}=1$.\\
The Hermitian matrices $T_{x}^{T}T_{x}$ and $T_{y}^{T}T_{y}$ for the state $\rho_{\lambda_{2},\lambda_{3}}=|\psi_{\lambda_{2},\lambda_{3}}\rangle \langle \psi_{\lambda_{2},\lambda_{3}}|$ is given by,
\begin{eqnarray}
T_{x}^{T}T_{x}=
\begin{pmatrix}
4\lambda_0^{2}(\lambda_3^{2}+\lambda_4^{2}) & 0 & 4\lambda_0^{2}\lambda_2\lambda_4\\
0 & 4\lambda_0^{2}\lambda_4^{2} & 0\\
4\lambda_0^{2}\lambda_2\lambda_4 & 0 & 4\lambda_0^{2}\lambda_2^{2}\\
\end{pmatrix}
\end{eqnarray}
\begin{eqnarray}
T_{y}^{T}T_{y}=
\begin{pmatrix}
4\lambda_0^{2}\lambda_4^{2} & 0 & 4\lambda_0^{2}\lambda_2\lambda_4\\
0 & 4\lambda_0^{2}(\lambda_3^{2}+\lambda_4^{2}) & 0\\
4\lambda_0^{2}\lambda_2\lambda_4 & 0 & 4\lambda_0^{2}\lambda_2^{2}\\
\end{pmatrix}
\end{eqnarray}
The maximum eigenvalue of $T_{x}^{T}T_{x}$ is given by
\begin{eqnarray}
\mu_{max}({T_x}^{T}T_x)&=&max\{u,v_{1}\}
\label{maxeigenval23}
\end{eqnarray}
where $u=4\lambda_0^{2}\lambda_4^{2}$ and $v_{1}=2\lambda_0^{2}(1-\lambda_0^{2}+\sqrt{-4\lambda_2^2\lambda_3^{2}+(1-\lambda_0^{2})^2})$.\\
The minimum eigenvalue of $T_{y}^{T}T_{y}$ is given by
\begin{eqnarray}
\mu_{min}({T_y}^{T}T_y)=0
\label{mineigenval1}
\end{eqnarray}

The expression for $Tr[({T_x}^{T}T_x)+({T_y}^{T}T_y)]$ is given by
\begin{eqnarray}
Tr[({T_x}^{T}T_x)+({T_y}^{T}T_y)]&=&16\lambda_0^{2}\lambda_4^{2}+8\lambda_0^{2}\lambda_2^{2}+8\lambda_0^{2}\lambda_{3}^{2}
\label{Tracecond121}
\end{eqnarray}
%It can be easily observed that in this case $0\leq \mu_{max}({T_x}^{T}T_x)+\mu_{min}({T_y}^{T}T_y)\leq 1$ holds.\\
\textbf{Case-I:} If $\mu_{max}({T_x}^{T}T_x)=u=4\lambda_0^{2}\lambda_4^{2}$ and $\mu_{min}({T_y}^{T}T_y)=0$.
The inequality (\ref{inequality1}) then can be re-expressed in terms of the expectation value of the operators $O_{1}$ as
\begin{eqnarray}
-2\sqrt{2}\lambda_0(\lambda_2+\lambda_3) &\leq& \langle O_1\rangle_{\psi_{\lambda_2,\lambda_3}}-\frac{\langle O_1\rangle_{\psi_{\lambda_2,\lambda_3}}^{2}}{4}
\label{inequality231}
\end{eqnarray}
If $-2\sqrt{2}\lambda_0(\lambda_2+\lambda_3)=0$, then RHS of inequality (\ref{inequality231}) is always positive for every state $|\psi \rangle$. Thus, in this case, it is not possible to discriminate between the class of states $|\psi_{\lambda_{2}=0,\lambda_{3}=0}\rangle \in S_{1}$ and the class of  states $|\psi_{\lambda_{2},\lambda_{3}}\rangle \in S_{3}$.\\
\textbf{Case-II:}
If $\mu_{max}({T_x}^{T}T_x)= v_{1}$ and $\mu_{min}({T_y}^{T}T_y)=0$
then the inequality (\ref{inequality1}) reduces to
\begin{eqnarray}
-2\sqrt{2}\lambda_0(\lambda_2+\lambda_3) &\leq& \langle O_1\rangle_{\psi_{_{\lambda_2,\lambda_3}}}-P_1
\label{inequality232}
\end{eqnarray}
where,
\begin{eqnarray}
P_1&=&2\lambda_0^{2}(1-\lambda_0^2+\sqrt{-4\lambda_2^2\lambda_3^{2}+(1-\lambda_0^2)^2})\nonumber\\&=&2\langle O_5\rangle_{\psi_{\lambda_2,\lambda_3}}(1-\langle O_5\rangle_{\psi_{\lambda_2,\lambda_3}}\nonumber\\&+&\sqrt{-\frac{1}{4}\langle O_6\rangle_{\psi_{\lambda_2,\lambda_3}}^{2}+(1-\langle O_5\rangle_{\psi_{\lambda_2,\lambda_3}})^{2})}
\end{eqnarray}
where,
\begin{eqnarray}
O_5&=&\frac{1}{8}(I+\sigma_{z} \otimes I+\sigma_{z} \otimes I+\sigma_{z})\nonumber\\
O_6&=& 2(I \otimes \sigma_{y} \otimes \sigma _{y})
\end{eqnarray}
The expectation values of the operators $O_5$ and $O_6$, in terms of invariants may be written as,
\begin{eqnarray}
\langle O_5\rangle&=&\lambda_0^{2}=\frac{C_{AC}C_{AB}}{2C_{BC}}-\frac{1}{C_{BC}}\sqrt{\frac{2I_{4}I_{5}}{\tau_{\psi}}}\nonumber\\
\langle O_6\rangle&=&4(\lambda_2\lambda_3-\lambda_1\lambda_4)=2C_{BC}
\end{eqnarray}
We can now define a Hermitian operator $H_{5}$ as
\begin{eqnarray}
H_{4}&=&O_1-P_1I
\label{h4}
\end{eqnarray}
Therefore, the inequality (\ref{inequality232}) can be re-formulated as
\begin{eqnarray}
-2\sqrt{2}\lambda_0(\lambda_2+\lambda_3) \leq \langle H_{4}\rangle_{\psi_{\lambda_2,\lambda_3}}
\label{inequality233}
\end{eqnarray}
If $\lambda_2+\lambda_3=0$ then $\langle H_{4}\rangle_{\psi_{\lambda_2,\lambda_3}} \geq 0$ for all states $|\psi_{\lambda_{2}=0,\lambda_{3}=0}\rangle$.\\
For $-2\sqrt{2}\lambda_0(\lambda_2+\lambda_3)\neq0$, we can calculate $Tr(H_{4}\rho_{\lambda_2,\lambda_{3}})$, which is given by
\begin{eqnarray}
Tr(H_{4}\rho_{\lambda_{2},\lambda_{3}})&=&4\lambda_0\lambda_4-2{\lambda_0}^2(1-{\lambda_0}^2+\sqrt{T_1})
\label{tr23}
\end{eqnarray}
where,
\begin{eqnarray}
T_1={\lambda_2}^4-2{\lambda_2}^2{\lambda_3}^2+2{\lambda_2}^2{\lambda_4}^2+({\lambda_3}^2+{\lambda_4}^2)^2
\end{eqnarray}
It can be easily shown that there exist state parameters $(\lambda_0,\lambda_2, \lambda_3, \lambda_4)$ for which $Tr(H_{4}\rho_{\lambda_{2},\lambda_{3}})<0$. For instance, if we take $\lambda_0=0.35$, $\lambda_2=0.3$, $\lambda_3=0.864581$ and $\lambda_4=0.2$, we get $Tr[H_4\rho_{\lambda_{2},\lambda_{3}}]=-0.108386$. Therefore, the classification operator $H_{4}$ classify the class of states $\rho_{\lambda_{2},\lambda_{3}}\in S_{3}$ and the class of states $\rho_{\lambda_{2}=0,\lambda_{3}=0}\in S_{1}$.\\ \ \\
\textbf{(II) Classification of states confined in subclass-III with state parameters $\lambda_0$, $\lambda_1$, $\lambda_i (i=2,3)$ and $\lambda_4$ and subclass-I:}\\
The GHZ class of state within subclass-III with state parameters ($\lambda_0$, $\lambda_1$, $\lambda_2$, $\lambda_4$) and ($\lambda_0$, $\lambda_1$, $\lambda_3$, $\lambda_4$) are given by
\begin{eqnarray}
|\psi_{\lambda_{1},\lambda_{2}}\rangle=\lambda_0|000\rangle+\lambda_1|100\rangle+\lambda_2|101\rangle+\lambda_4|111\rangle
\label{psi102}
\end{eqnarray}
with $\lambda_0^{2}+\lambda_1^{2}+\lambda_2^{2}+\lambda_4^{2}=1$.
\begin{eqnarray}
|\psi_{\lambda_{1},\lambda_{3}}\rangle=\lambda_0|000\rangle+\lambda_1|100\rangle+\lambda_3|110\rangle+\lambda_4|111\rangle
\label{psi13}
\end{eqnarray}
with $\lambda_0^{2}+\lambda_1^{2}+\lambda_3^{2}+\lambda_4^{2}=1$.\\
The Hermitian matrices $T_{x}^{T}T_{x}$ and $T_{y}^{T}T_{y}$ for the state $\rho_{\lambda_{1},\lambda_{2}}=|\psi_{\lambda_{1},\lambda_{2}}\rangle \langle \psi_{\lambda_{1},\lambda_{2}}|$ is given by

\begin{eqnarray}
T_{x}^{T}T_{x}=
\begin{pmatrix}
4\lambda_0^{2}\lambda_4^{2} & 0 & 4\lambda_0^{2}\lambda_2\lambda_4\\
0 & 4\lambda_0^{2}\lambda_4^{2} & 0\\
4\lambda_0^{2}\lambda_2\lambda_4 & 0 & 4\lambda_0^{2}(\lambda_1^{2}+\lambda_2^{2})\\
\end{pmatrix}
\end{eqnarray}
\begin{eqnarray}
T_{y}^{T}T_{y}=
\begin{pmatrix}
4\lambda_0^{2}\lambda_4^{2} & 0 & 4\lambda_0^{2}\lambda_2\lambda_4\\
0 & 4\lambda_0^{2}\lambda_4^{2} & 0\\
4\lambda_0^{2}\lambda_2\lambda_4 & 0 & 4\lambda_0^{2}\lambda_2^{2}\\
\end{pmatrix}
\end{eqnarray}
The Hermitian matrices $T_{x}^{T}T_{x}$ and $T_{y}^{T}T_{y}$ for the state $\rho_{\lambda_{1,3}}=|\psi_{\lambda_{1,3}}\rangle \langle \psi_{\lambda_{1,3}}|$ is given by
\begin{eqnarray}
T_{x}^{T}T_{x}=
\begin{pmatrix}
4\lambda_0^{2}(\lambda_3^{2}+\lambda_4^{2}) & 0 & 4\lambda_0^{2}\lambda_1\lambda_3\\
0 & 4\lambda_0^{2}\lambda_4^{2} & 0\\
4\lambda_0^{2}\lambda_1\lambda_3 & 0 & 4\lambda_0^{2}\lambda_1^{2}\\
\end{pmatrix}
\end{eqnarray}
\begin{eqnarray}
T_{y}^{T}T_{y}=
\begin{pmatrix}
4\lambda_0^{2}\lambda_4^{2} & 0 & 0\\
0 & 4\lambda_0^{2}(\lambda_3^{2}+\lambda_4^{2}) & 0\\
0 & 0 & 0\\
\end{pmatrix}
\end{eqnarray}
The maximum eigenvalue of $T_{x}^{T}T_{x}$ is given by
\begin{eqnarray}
\mu_{max}({T_x}^{T}T_x)&=&max\{u,v_{i}\}, i=2,3
\label{maxeigenval23}
\end{eqnarray}
where $v_{i}=2\lambda_0^{2}(\lambda_1^{2}+\lambda_{i}^{2}+\lambda_4^{2}+\sqrt{(\lambda_1^{2}+\lambda_{i}^{2}+\lambda_4^{2})^2-4\lambda_1^2\lambda_4^{2}})$,~i=2,3.\\
The minimum eigenvalue of $T_{y}^{T}T_{y}$ is given by
\begin{eqnarray}
\mu_{min}({T_y}^{T}T_y)=0
\label{mineigenval1}
\end{eqnarray}
The expression for $Tr[({T_x}^{T}T_x)+({T_y}^{T}T_y)]$ is given by
\begin{eqnarray}
Tr[({T_x}^{T}T_x)+({T_y}^{T}T_y)]&=&16\lambda_0^{2}\lambda_4^{2}+8\lambda_0^{2}\lambda_i^{2}+
4\lambda_0^{2}\lambda_1^{2},~~(i=2,3)
\label{Tracecond131}
\end{eqnarray}
\textbf{Case-I:} If $\mu_{max}({T_x}^{T}T_x)=4\lambda_0^{2}\lambda_4^{2}$ and $\mu_{min}({T_y}^{T}T_y)=0$.
The inequality (\ref{inequality1}) then can be re-expressed in terms of the expectation value of the operators $O_{1}$ as
\begin{eqnarray}
-2\lambda_0(\sqrt{2}\lambda_i+\lambda_1) &\leq& \langle O_1\rangle_{\psi_{\lambda_1,\lambda_i}}-\frac{(\langle O_1\rangle_{\psi_{\lambda_1,\lambda_i}})^{2}}{4},~i=2,3
\label{inequality1211}
\end{eqnarray}
If $\sqrt{2}\lambda_i+\lambda_1=0 (i=2,3)$, then RHS of inequality (\ref{inequality1211}) is always positive. Thus, the R.H.S of the inequality is positive for every state $|\psi\rangle$.  But since $0\leq \langle O_1\rangle_{\psi_{\lambda_1,\lambda_i}}\leq 1$ so the R.H.S of the inequality (\ref{inequality1211}) still positive even for $\sqrt{2}\lambda_i+\lambda_1 \neq 0, (i=2,3)$. Thus it is not possible to differentiate between the class of states $|\psi_{\lambda_{1}=0,\lambda_{i}=0}\rangle \in S_{1} (i=2,3)$ and $|\psi_{\lambda_{1},\lambda_{i}}\rangle \in S_{3} (i=2,3)$, using the inequality (\ref{inequality1211}) for this case.\\
\textbf{Case-II:}
If $\mu_{max}({T_x}^{T}T_x)=2\lambda_0^{2}(\lambda_1^{2}+\lambda_{i}^{2}+\lambda_4^{2}+\sqrt{(\lambda_1^{2}+\lambda_{i}^{2}+\lambda_4^{2})^2-4\lambda_1^2\lambda_4^{2}})$,~i=2,3 and\\ $\mu_{min}({T_y}^{T}T_y)=0$.\\
Then the inequality (\ref{inequality1}) can be re-written as
\begin{eqnarray}
-2\lambda_0(\sqrt{2}\lambda_{i}+\lambda_1) &\leq& \langle O_1\rangle_{\psi_{\lambda_1,\lambda_i}}-P_{i},i=2,3
\label{inequality1221}
\end{eqnarray}
where, for i=2,3
\begin{eqnarray}
P_i&=&2\lambda_0^{2}(\lambda_1^{2}+\lambda_{i}^{2}+\lambda_4^{2}+\sqrt{(\lambda_1^{2}+\lambda_{i}^{2}+\lambda_4^{2})^2-4\lambda_1^2\lambda_4^{2}})
\end{eqnarray}
$P_{i}(i=2,3)$ can also be re-expressed in terms of the expectation values of the operators $O_1$, $O_4$ and $O_{i}(i=2,3)$ as
\begin{eqnarray}
P_i&=&q+\sqrt{q^2-\frac{1}{16}\langle O_1\rangle_{\psi_{\lambda_1,\lambda_i}}^2\langle O_4\rangle_{\psi_{\lambda_1,\lambda_i}}^2}
\end{eqnarray}
where $q=\frac{1}{8}[\langle O_1\rangle_{\psi_{\lambda_1,\lambda_i}}^{2}+\langle O_i\rangle_{\psi_{\lambda_1,\lambda_i}}^{2}+\langle O_4\rangle_{\psi_{\lambda_1,\lambda_i}}^{2}]$
We can now define an Hermitian operator $H_{k}, k=5,6$ as
\begin{eqnarray}
H_{k}&=&O_1-P_iI, ~k=5,6, i=2,3, i=k
\label{h5}
\end{eqnarray}
Therefore, the inequality (\ref{inequality1221}) can be re-formulated as
\begin{eqnarray}
-2\lambda_0
(\sqrt{2}\lambda_i+\lambda_1) \leq \langle H_{k}\rangle_{\psi_{\lambda_1,\lambda_i}},~~i=2,3, k=5,6
\label{inequality123}
\end{eqnarray}
If $\sqrt{2}\lambda_{i}+\lambda_1=0,i=2,3$ then $\langle H_{k}\rangle_{\psi_{\lambda_1,\lambda_i}} \geq 0, k=5,6$ for all states $|\psi_{\lambda_{1}=0,\lambda_{i}=0}\rangle \in S_{1} (i=2,3)$.\\
For $\sqrt{2}\lambda_{i}+\lambda_1 \neq0, i=2,3$, we can calculate $Tr(H_{k}\rho_{\lambda_{1},\lambda_{i}}),(i=2,3, k=5,6)$, which is given by
\begin{eqnarray}
Tr(H_{k}\rho_{\lambda_{1},\lambda_{i}})&=&4\lambda_0\lambda_4-2{\lambda_0}^2(\lambda_1^2+\lambda_{i}^2+\lambda_4^{2}+\sqrt{T_{i}}), i=2,3, k=5,6
\label{tr12}
\end{eqnarray}
where
\begin{eqnarray}
T_{i}={\lambda_1}^4+2{\lambda_1}^2({\lambda_{i}}^2-{\lambda_4}^2)+({\lambda_{i}}^2+{\lambda_4}^2)^2,i=2,3
\end{eqnarray}
It can be easily shown that there exist state parameters $(\lambda_0,\lambda_1, \lambda_{i}, \lambda_4), (i=2,3)$ for which $Tr(H_{k}\rho_{\lambda_1,\lambda_{i}})<0$ for k=5,6. For instance, if we take $\lambda_0=0.5$, $\lambda_1=0.83666$, $\lambda_i=0.2,(i=2,3)$ and $\lambda_4=0.1$, we get $Tr[H_k\rho_{\lambda_{1},\lambda_{i}}]=-0.540548, (k=5,6)$. Thus, the Hermitian operator $H_{k}$, k=\{5,6\} serves as a classification witness operator and classify the class of states described by the density operator $\rho_{\lambda_{1},\lambda_{i}},(i=2,3) \in S_{3}$ and the class of states $\rho_{\lambda_{1}=0,\lambda_{i}=0},(i=2,3) \in S_{1}$.

%It can be easily observed that in this case $0\leq \mu_{max}({T_x}^{T}T_x)+\mu_{min}({T_y}^{T}T_y)\leq 1$ holds.\\

\subsection{Classification of states confined in subclass-IV with state parameters $(\lambda_0$, $\lambda_1$,$\lambda_2$ $\lambda_3$, $\lambda_4)$ and subclass-I}
The GHZ class of state within subclass-IV with state parameters ($\lambda_0$, $\lambda_1$, $\lambda_2$, $\lambda_3$, $\lambda_4$) is given by
\begin{eqnarray}
|\psi_{\lambda_{1},\lambda_{2},\lambda_{3}}\rangle=\lambda_0|000\rangle+\lambda_1|100\rangle+\lambda_2|101\rangle+\lambda_3|110\rangle + \lambda_4|111\rangle
\label{psi123}
\end{eqnarray}
with $\lambda_0^{2}+\lambda_1^{2}+\lambda_2^{2}+\lambda_3^{2}+\lambda_4^{2}=1$.\\
The Hermitian matrices $T_{x}^{T}T_{x}$ and $T_{y}^{T}T_{y}$ for the state $\rho_{\lambda_{1},\lambda_{2},\lambda_{3}}=|\psi_{\lambda_{1},\lambda_{2},\lambda_{3}}\rangle \langle \psi_{\lambda_{1},\lambda_{2},\lambda_{3}}|$ is given by

\begin{eqnarray}
T_{x}^{T}T_{x}=
\begin{pmatrix}
4\lambda_0^{2}(\lambda_3^{2}+\lambda_4^{2}) & 0 & 4\lambda_0^{2}(\lambda_1\lambda_3+\lambda_2\lambda_4)\\
0 & 4\lambda_0^{2}\lambda_4^{2} & 0\\
4\lambda_0^{2}(\lambda_1\lambda_3+\lambda_2\lambda_4) & 0 & 4\lambda_0^{2}(\lambda_1^{2}+\lambda_2^{2})\\
\end{pmatrix}
\end{eqnarray}
\begin{eqnarray}
T_{y}^{T}T_{y}=
\begin{pmatrix}
4\lambda_0^{2}\lambda_4^{2} & 0 & 4\lambda_0^{2}\lambda_2\lambda_4\\
0 & 4\lambda_0^{2}(\lambda_3^{2}+\lambda_4^{2}) & 0\\
4\lambda_0^{2}\lambda_2\lambda_4 & 0 & 4\lambda_0^{2}\lambda_2^{2}\\
\end{pmatrix}
\end{eqnarray}
The maximum eigenvalue of $T_{x}^{T}T_{x}$ is given by
\begin{eqnarray}
\mu_{max}({T_x}^{T}T_x)&=&max\{u,v_{4}\}
\label{maxeigenval23}
\end{eqnarray}
where $v_{4}=2{\lambda_0}^2(k+\sqrt{-4(\lambda_2\lambda_3-\lambda_1\lambda_4)^2+k^{2}})$, $k=1-\lambda_0^{2}$.\\
The minimum eigenvalue of $T_{y}^{T}T_{y}$ is given by
\begin{eqnarray}
\mu_{min}({T_y}^{T}T_y)=0
\label{mineigenval123}
\end{eqnarray}

The expression for $Tr[({T_x}^{T}T_x)+({T_y}^{T}T_y)]$ is given by
\begin{eqnarray}
Tr[({T_x}^{T}T_x)+({T_y}^{T}T_y)]&=&16\lambda_0^{2}\lambda_4^{2}+8\lambda_0^{2}\lambda_3^{2}+8\lambda_0^{2}\lambda_2^{2}
+4\lambda_0^{2}\lambda_1^{2}
\label{Tracecond1231}
\end{eqnarray}
%It can be easily observed that in this case $0\leq \mu_{max}({T_x}^{T}T_x)+\mu_{min}({T_y}^{T}T_y)\leq 1$ holds.\\
\textbf{Case-I:} If $\mu_{max}({T_x}^{T}T_x)=u=4\lambda_0^{2}\lambda_4^{2}$ and $\mu_{min}({T_y}^{T}T_y)=0$.
The inequality (\ref{inequality1}) then can be re-expressed in terms of the expectation value of the operators $O_{1}$ as
\begin{eqnarray}
-2\lambda_0(\sqrt{2}\lambda_2+\sqrt{2}\lambda_3+\lambda_1) &\leq& \langle O_1\rangle_{\psi_{\lambda_{1},\lambda_{2},\lambda_{3}}}-
\frac{\langle O_1\rangle_{\psi_{\lambda_{1},\lambda_{2},\lambda_{3}}}^{2}}{4}
\label{inequality1231}
\end{eqnarray}
If $\sqrt{2}\lambda_2+\sqrt{2}\lambda_3+\lambda_1=0$, then RHS of inquality (\ref{inequality1231}) is always positive irrespective of the values of the state parameter $(\lambda_1,\lambda_2,\lambda_3)$ . Thus it is not possible to differentiate between the class of states $|\psi_{\lambda_{1},\lambda_{2},\lambda_{3}}\rangle \in S_{4}$ and the class of states $|\psi_{\lambda_{1}=0,\lambda_{2}=0,\lambda_{3}=0}\rangle \in S_{1}$.\\
\textbf{Case-II:}
$\mu_{max}({T_x}^{T}T_x)=2{\lambda_0}^2(1-\lambda_0^{2}+\sqrt{-4(\lambda_2\lambda_3-\lambda_1\lambda_4)^2+(1-\lambda_0^{2})^{2}})$ and\\ $\mu_{min}({T_y}^{T}T_y)=0$.
Then the inequality (\ref{inequality1}) can be re-written as
\begin{eqnarray}
-2\lambda_0(\sqrt{2}\lambda_2+\sqrt{2}\lambda_3+\lambda_1) &\leq& \langle O_1\rangle_{\psi_{\lambda_{1},\lambda_{2},\lambda_{3}}}-P_4
\label{inequality1232}
\end{eqnarray}
where,
\begin{eqnarray}
P_4&=&2{\lambda_0}^2(1-\lambda_0^2+\sqrt{-4(\lambda_2\lambda_3-\lambda_1\lambda_4)^2+(1-\lambda_0^{2})^{2}})
\nonumber\\&=& 2 \langle O_{5}\rangle_{\psi_{\lambda_{1},\lambda_{2},\lambda_{3}}}(1-\langle O_{5}\rangle_{\psi_{\lambda_{1},\lambda_{2},\lambda_{3}}}+\sqrt{-\frac{\langle O_{6}\rangle_{\psi_{\lambda_{1},\lambda_{2},\lambda_{3}}}^{2}}{4}+(1-\langle O_{5}\rangle_{\psi_{\lambda_{1},\lambda_{2},\lambda_{3}}})^{2}})
\end{eqnarray}
We can now define a Hermitian operator $H_{7}$ as
\begin{eqnarray}
H_{7}&=&O_1-P_4I
\label{h7}
\end{eqnarray}
Therefore, the inequality (\ref{inequality1232}) can be re-formulated as
\begin{eqnarray}
-2\lambda_0(\sqrt{2}\lambda_2+\sqrt{2}\lambda_3+\lambda_1) \leq \langle H_{7}\rangle_{\psi_{\lambda_{1},\lambda_{2},\lambda_{3}}}
\label{inequality1233}
\end{eqnarray}
If $-2\lambda_0(\sqrt{2}\lambda_2+\sqrt{2}\lambda_3+\lambda_1)=0$ then $\langle H_{7}\rangle_{\psi_{\lambda_{1},\lambda_{2},\lambda_{3}}}\geq 0$ for all states $|\psi_{\lambda_{1}=0,\lambda_{2}=0,\lambda_{3}=0}\rangle \in S_{1}$.\\
For $-2\lambda_0(\sqrt{2}\lambda_2+\sqrt{2}\lambda_3+\lambda_1)\neq0$, we can calculate $Tr(H_{7}\rho_{\lambda_{1},\lambda_{2},\lambda_{3}})$ which is given by
\begin{eqnarray}
Tr(H_{7}\rho_{\lambda_{1},\lambda_{2},\lambda_{3}})&=&4\lambda_0\lambda_4-2{\lambda_0}^2({\lambda_1}^2+{\lambda_2}^2+{\lambda_3}^2+{\lambda_4}^2
\nonumber\\&+&\sqrt{T_4})
\label{tr13}
\end{eqnarray}
where,
\begin{eqnarray}
T_4&=&{\lambda_1}^4+{\lambda_2}^4+{\lambda_3}^4+{\lambda_4}^4+8\lambda_1\lambda_2\lambda_3\lambda_4-2{\lambda_2}^2{\lambda_3}^2+
\nonumber\\&&2{\lambda_2}^2{\lambda_4}^2+2{\lambda_1}^2{\lambda_2}^2+2{\lambda_1}^2{\lambda_3}^2-2{\lambda_1}^2{\lambda_4}^2
\nonumber\\&&+2{\lambda_3}^2{\lambda_4}^2
\end{eqnarray}
It can be easily shown that there exist state parameters $(\lambda_0,\lambda_1, \lambda_2, \lambda_3, \lambda_4)$ for which $Tr(H_{7}\rho_{\lambda_{1},\lambda_{2},\lambda_{3}})<0$. For instance, if we take $\lambda_0=0.6$, $\lambda_1=0.785812$, $\lambda_2=0.1$, $\lambda_3=0.05$ and $\lambda_4=0.1$, we get $Tr[H_7\rho_{\lambda_{123}}]=-0.303798$. Thus, the classification witness operator $H_7$ classify the
class of states described by the density operator $\rho_{\lambda_{1},\lambda_{2},\lambda_{3}} \in S_{4}$ and the class of states described by $\rho_{\lambda_{1}=0,\lambda_{2}=0,\lambda_{3}=0} \in S_{1}$

\section{Examples}
In this section, we have provided a few examples of three-qubit states for which we construct classification witness operators.\\
\textbf{Example-1:} The three-qubit maximal slice state is given by \cite{ghoses},
\begin{eqnarray}
|MS\rangle=\frac{1}{\sqrt{2}}(|000\rangle+c|110\rangle+d|111\rangle),~~ c^2+d^2=1
\label{MS}
\end{eqnarray}
Let us consider the classification witness operators $H_{1}$, $H_{2}$ and $H_{3}$. The classification witness operator $H_{3}$ for (\ref{MS}) is now reduces to
\begin{eqnarray}
W_{MS}=O_1-I
\end{eqnarray}
The expectation value of $W_{MS}$ with respect to the state $|MS\rangle$ can be evaluated as
\begin{eqnarray}
Tr(W_{MS}\rho_{MS})=(2d-1)
\end{eqnarray}
Therefore, we can verify that $Tr(W_{MS}|MS\rangle\langle MS|)<0$ for the state parameter $d<\frac{1}{2}$. Further, it is easy to verify that the expectation value of the witness operator $H_{3}$ is positive for all states belonging to subclass-I. Since the given state is detected by the classification witness operator $H_{3}$ so state (\ref{MS}) belongs to subclass-II. To investigate the form of the given state lying within subclass II, we need to further classify it from the other classes of states belonging to subclass II. We can check that in the same range of the state parameter $d$ i.e. for $d<\frac{1}{2}$, the value of $Tr(H_1\rho_{MS})$ and $Tr(H_2\rho_{MS})$ are non-negative. Thus, we can say that the classification witness operators $H_3$ discriminate the maximal slice state from subclass-I, and also it detects the state in the form (\ref{psi3}).\\
\textbf{Example-2:} Let us consider another three-qubit state defined as
\begin{eqnarray}
|\phi\rangle=\sqrt{p}|G\rangle-\sqrt{1-p}|K\rangle
\label{ex2state}
\end{eqnarray}
where,
\begin{eqnarray}
|G\rangle&=&a|000\rangle+b|111\rangle,~~a^{2}+b^{2}=1\nonumber\\
|K\rangle&=&c|110\rangle+d|101\rangle,~~c^{2}+d^{2}=1
\end{eqnarray}
Now our task is to construct a classification witness operator that may distinguish it from the state belonging to subclass-I and also detect the form of the given state that belongs to a particular class within subclass-III. To accomplish our task, let us consider classification witness operators $H_{4}$, $H_{5}$ and $H_{6}$ given in (\ref{h4}) and (\ref{h5}). We find that the expectation value of the witness operator $H_{4}$ is positive for all states belonging to subclass-I but it gives a negative value for some states belonging to subclass-III. Hence the state (\ref{ex2state}) belongs to subclass-III. Moreover, we have investigated this classification problem within subclass-III by constructing a table below. It shows that the expectation value of classification witness operator $H_{4}$ is negative for some range of the state parameter $p$ while the expectation value of other classification witness operators $H_{5}$ and $H_{6}$ gives positive values for the same range of the state parameters. This means that the given state (\ref{ex2state}) belongs to subclass-III and it takes the form (\ref{psi23}).
In Table 3.1, we have found the range of p where the witness operator $H_4$ detects the GHZ state given in the example whereas $H_5$ and $H_6$ do not detect the given GHZ state.
\begin{table}
	\begin{center}
		\caption{Range of the parameter p for which the classification witness operator $H_4$ classifies the given GHZ state within the subclass-III}
		\begin{tabular}{|c|c|c|c|c|}\hline
			State parameter & p & $Tr[H_5{\rho}]$ & $Tr[H_4{\rho}]$ & $Tr[H_6{\rho}]$ \\ \hline (a, c) & & & \\  \hline
			(0.8, 0.3)  & (.291,.3) & $>0$ & $<0$ & $>0$ \\\hline
			(0.9, 0.4)  & (.548,.57) & $>0$ & $<0$ & $>0$ \\\hline
			(0.91, 0.8)  & (.4,.51) & $>0$ & $<0$ & $>0$ \\\hline
			(0.85, 0.35)  & (.43,.45) & $>0$ & $<0$ & $>0$ \\\hline
			(0.88, 0.8)  & (.25,.385) & $>0$ & $<0$ & $>0$ \\\hline
			(0.78, 0.3)  & (.208,.22) & $>0$ & $<0$ & $>0$ \\\hline
			(0.95, 0.4)  & (.69,.7) & $>0$ & $<0$ & $>0$ \\\hline
			(0.83, 0.45)  & (.26,.31) & $>0$ & $<0$ & $>0$ \\\hline
		\end{tabular}
	\end{center}
\end{table}

%Here,
%\begin{eqnarray}
%\lambda_{max}(T_x^{\dagger}T_x)&=&2a^2p[c^2+d^2+b^2p-c^2p-d^2p+\sqrt{F}]
%\end{eqnarray}
%where,
%\begin{eqnarray}
%F=-4c^2d^2(-1+p)^2+(c^2(-1+p)+d^2(-1+p)-b^2p)^2
%\end{eqnarray}
%Then,
%\begin{eqnarray}
%Tr(W\rho_{\phi})&=&4abp+\sqrt{2T^{*}}(-1+p)(c^2+d^2)\nonumber\\&-&\sqrt{2T^{*}}p(a^2+b^2)
%\end{eqnarray}
%where,
%\begin{eqnarray}
%T^{*}=a^2p(c^2+d^2+b^2p-c^2p-d^2p+\sqrt{F}
%\end{eqnarray}
%Then,
%\begin{eqnarray}
%Tr(W\rho_{\phi})<0~~\text{when}~~b<\frac{a}{2}
%\end{eqnarray}

%Thus, the witness operator $W_{\phi}$ detects the given GHZ state when$b<\frac{a}{2}$
%\begin{figure}
%\centering
%\includegraphics[scale=1]{123}
%\caption{Comparison between lower bound of tangle}
%\label{Figure1}
%\end{figure}

\section{Classification witness operator for the classification of states contained in subclass-II and subclass-III}
The GHZ class of state within subclass-III with state parameters ($\lambda_0$, $\lambda_1$, $\lambda_2$, $\lambda_4$) is given by
\begin{eqnarray}
|\psi_{\lambda_{1},\lambda_{2}}\rangle=\lambda_0|000\rangle+\lambda_1|100\rangle+\lambda_2|101\rangle+\lambda_4|111\rangle
\label{psi12}
\end{eqnarray}
with $\lambda_0^{2}+\lambda_1^{2}+\lambda_2^{2}+\lambda_4^{2}=1$.\\
The Hermitian matrices $T_{x}^{T}T_{x}$ and $T_{y}^{T}T_{y}$ for the state $\rho_{\lambda_{1},\lambda_{2}}=|\psi_{\lambda_{1},\lambda_{2}}\rangle \langle \psi_{\lambda_{1},\lambda_{2}}|$ is given by

\begin{eqnarray}
T_{x}^{T}T_{x}=
\begin{pmatrix}
4\lambda_0^{2}\lambda_4^{2} & 0 & 4\lambda_0^{2}\lambda_2\lambda_4\\
0 & 4\lambda_0^{2}\lambda_4^{2} & 0\\
4\lambda_0^{2}\lambda_2\lambda_4 & 0 & 4\lambda_0^{2}(\lambda_1^{2}+\lambda_2^{2})\\
\end{pmatrix}
\end{eqnarray}
\begin{eqnarray}
T_{y}^{T}T_{y}=
\begin{pmatrix}
4\lambda_0^{2}\lambda_4^{2} & 0 & 4\lambda_0^{2}\lambda_2\lambda_4\\
0 & 4\lambda_0^{2}\lambda_4^{2} & 0\\
4\lambda_0^{2}\lambda_2\lambda_4 & 0 & 4\lambda_0^{2}\lambda_2^{2}\\
\end{pmatrix}
\end{eqnarray}
The Hermitian matrices $T_{x}^{T}T_{x}$ and $T_{y}^{T}T_{y}$ for the state $\rho_{\lambda_{1,3}}=|\psi_{\lambda_{1,3}}\rangle \langle \psi_{\lambda_{1,3}}|$ is given by
\begin{eqnarray}
T_{x}^{T}T_{x}=
\begin{pmatrix}
4\lambda_0^{2}(\lambda_3^{2}+\lambda_4^{2}) & 0 & 4\lambda_0^{2}\lambda_1\lambda_3\\
0 & 4\lambda_0^{2}\lambda_4^{2} & 0\\
4\lambda_0^{2}\lambda_1\lambda_3 & 0 & 4\lambda_0^{2}\lambda_1^{2}\\
\end{pmatrix}
\end{eqnarray}
\begin{eqnarray}
T_{y}^{T}T_{y}=
\begin{pmatrix}
4\lambda_0^{2}\lambda_4^{2} & 0 & 0\\
0 & 4\lambda_0^{2}(\lambda_3^{2}+\lambda_4^{2}) & 0\\
0 & 0 & 0\\
\end{pmatrix}
\end{eqnarray}
The maximum eigenvalue of $T_{x}^{T}T_{x}$ is given by
\begin{eqnarray}
\mu_{max}({T_x}^{T}T_x)&=&max\{u,v_{i}\}, i=2,3
\label{maxeigenval23}
\end{eqnarray}
where $v_{i}=2\lambda_0^{2}(\lambda_1^{2}+\lambda_{i}^{2}+\lambda_4^{2}+\sqrt{(\lambda_1^{2}+\lambda_{i}^{2}+\lambda_4^{2})^2-4\lambda_1^2\lambda_4^{2}})$,~i=2,3.\\
The minimum eigenvalue of $T_{y}^{T}T_{y}$ is given by
\begin{eqnarray}
\mu_{min}({T_y}^{T}T_y)=0
\label{mineigenval1}
\end{eqnarray}
The expression for $Tr[({T_x}^{T}T_x)+({T_y}^{T}T_y)]$ is given by
\begin{eqnarray}
Tr[({T_x}^{T}T_x)+({T_y}^{T}T_y)]&=&16\lambda_0^{2}\lambda_4^{2}+8\lambda_0^{2}\lambda_2^{2}+
4\lambda_0^{2}\lambda_1^{2}
\label{Tracecond131}
\end{eqnarray}

\textbf{Case-I:} If $\mu_{max}({T_x}^{T}T_x)=4\lambda_0^{2}\lambda_4^{2}$ and $\mu_{min}({T_y}^{T}T_y)=0$.
The inequality (\ref{inequality1}) then can be re-expressed in terms of the expectation value of the operators $O_{1}$ as
\begin{eqnarray}
-2\lambda_0(\sqrt{2}\lambda_2+\lambda_1) &\leq& \langle O_1\rangle_{\psi_{\lambda_1,\lambda_2}}-\frac{\langle O_1\rangle_{\psi_{\lambda_1,\lambda_2}}^{2}}{4}
\end{eqnarray}
If $\lambda_2=0$, then above inequality becomes,
\begin{eqnarray}
-2\lambda_0\lambda_1 \leq \langle O_1\rangle_{\psi_{\lambda_1,\lambda_2}}-\frac{\langle O_1\rangle_{\psi_{\lambda_1,\lambda_2}}^{2}}{4}
\label{inequality121}
\end{eqnarray}
The R.H.S of the inequality is positive for every state $|\psi\rangle$. Thus it is not possible to differentiate between the class of states $|\psi_{\lambda_{1},\lambda_{2}=0}\rangle \in S_{2}$ and $|\psi_{\lambda_{1},\lambda_{2}}\rangle \in S_{3}$, using the inequality (\ref{inequality121}) for this case.\\
\textbf{Case-II:}
If $\mu_{max}({T_x}^{T}T_x)=2\lambda_0^{2}(\lambda_1^{2}+\lambda_{i}^{2}+\lambda_4^{2}+\sqrt{(\lambda_1^{2}+\lambda_{i}^{2}+\lambda_4^{2})^2-4\lambda_1^2\lambda_4^{2}})$ and\\ $\mu_{min}({T_y}^{T}T_y)=0$, then the inequality (\ref{inequality1}) can be re-written as
\begin{eqnarray}
-2\lambda_0(\sqrt{2}\lambda_{2}+\lambda_1) &\leq& \langle O_1\rangle_{\psi_{\lambda_1,\lambda_i}}-P_{2}
\label{inequality122}
\end{eqnarray}
We can now define a Hermitian operator $H_{8}$, as
\begin{eqnarray}
H_{8}&=& O_1-P_2I+\frac{O_4}{2}
\label{h7}
\end{eqnarray}
Therefore, the inequality (\ref{inequality122}) can be re-formulated as
\begin{eqnarray}
-2\sqrt{2}\lambda_0\lambda_2
\leq \langle H_{8}\rangle_{\psi_{\lambda_1,\lambda_2}}
\label{inequality123}
\end{eqnarray}
If $\lambda_2=0$ then $\langle H_{8}\rangle_{\psi_{\lambda_1,\lambda_2}} \geq 0$, for all states $|\psi_{\lambda_{1}=0,\lambda_{2}=0} \rangle \in S_{1}$ and $|\psi_{\lambda_{1},\lambda_{2}=0}\rangle \in S_{2}$.\\
For $\lambda_2 \neq 0$, then there exist state parameters $(\lambda_0,\lambda_1, \lambda_{2}, \lambda_4)$ for which $Tr(H_{8}\rho_{\lambda_1,\lambda_{2}})<0$. For instance, if we take $\lambda_0=0.01$, $\lambda_1=0.948631$, $\lambda_2=0.3$ and $\lambda_4=0.1$, we get $Tr[H_8\rho_{\lambda_{1},\lambda_{2}}]=-0.129027$. Thus, the Hermitian operator $H_{8}$ serves as a classification witness operator and classify GHZ class of states described by the density operator $\rho_{\lambda_{1},\lambda_{2}} \in S_{3}$ and the GHZ of states $\rho_{\lambda_{1}=0,\lambda_{2}=0} \in S_{1}$ or $\rho_{\lambda_{1},\lambda_{i}=0},(i=2,3) \in S_{2}$.\\
Simillarly, we can construct witness operator that can classify GHZ states belonging to subclass-III and subclass-IV.

\section{Conclusion}
To summarize, we have defined systematically different subclasses of the pure three-qubit GHZ class. The subclass-I denoted by $S_{1}$ contain the states of the form $\lambda_{0}|000\rangle + \lambda_{1}|111\rangle$. In particular, if $\lambda_{0}=\lambda_{1}=\frac{1}{\sqrt{2}}$ then the three-qubit state reduces to standard GHZ state and it is known that this state is very useful in various quantum information processing task. In this chapter, it has been shown that there exist states either belong to subclass-II (denoted by $S_{2}$) or subclass-III (denoted by $S_{3}$) or subclass-IV (denoted by $S_{4}$), that may be more useful in some teleportation scheme in comparison to the states belong to $S_{1}$. Also, we found that the tangle of the GHZ states belonging to subclass $S_1$ is more as compared to GHZ states belonging to subclass $S_2$. This observation gives the motivation to discriminate the states belonging to $S_{i},i=2,3,4$ from the family of states belonging to $S_{1}$. We have prescribed the method for the construction of the witness operator to study the classification of the states belonging to $S_{i},i=2,3,4$.

\begin{center}
	****************
\end{center}

\chapter{Classification of Three-qubit States using $l_1$-norm of Coherence}\label{ch4}
\vspace{1cm}
\noindent\hrulefill

\noindent\emph{In this chapter\;\footnote{This chapter is based on the research paper entitled ``Detection and Classification of Three-qubit States Using $l_{1}$-norm of Coherence''.}, we have studied the problem of classification of three-qubit states using $l_1$-norm of coherence. In the literature, the problem of classification of three-qubit entangled states has been studied through three-qubit entanglement measures such as tangle. But since tangle is non-zero for only GHZ class of states and zero for the set of separable states, biseparable states, and W class of states, so, it cannot be used to classify the states belonging to separable, biseparable, or W class of states. To overcome this issue, we have used $l_1$-norm of coherence for the detection and classification of three-qubit entangled states. In this chapter, we have derived coherence based inequalities that may be satisfied by biseparable states. The inequality satisfied by the biseparable states is denoted by $I_1$. The violation of $I_1$ indicates that the state under investigation is either separable or genuine entangled. Furthermore, we have derived an expression $E_1$ which is only satisfied by the separable states. If this expression is not satisfied, then the state may be either biseparable or a genuine entangled state. Thus if there exists a three-qubit state which violates the inequality $I_1$ and also does not satisfy $E_1$, then the state under probe is a genuine entangled state.}

\noindent\hrulefill
\newpage

\section{Introduction}

\noindent Quantum entanglement and coherence are two fundamental features that arise from the superposition principle of quantum mechanics. One notable difference between these two features is that coherence may exist in a single system while more than one system or more than one degree of freedom is required for entanglement. Secondly, the non-zero off-diagonal elements in the density matrix signify the presence of coherence in the quantum system while it may not ensure the existence of entanglement in the given composite quantum system. Both quantum entanglement and quantum coherence can be used as a resource \cite{horodecki4,streltsov}. M. Hillery has shown that coherence may act as a useful resource in the Deutsch-Jozsa algorithm \cite{hillery1}. On the other hand, quantum entanglement has many application in quantum information processing tasks, notably, quantum teleportation \cite{Bennett1}, quantum superdense coding \cite{Bennett2}, quantum remote state preparation \cite{pati}, quantum cryptography \cite{gisin3} etc. Extensive research has already been carried out to understand the non-classical feature of bipartite system \cite{peres,horodecki3,wootters,vidal,plenio,horodecki4,zhang2}. Eventually, when we increase the number of parties in the system, the complications in the shared system between the parties also increase. Therefore, it is indispensable to understand the entanglement properties of the shared multipartite system.\\
In this chapter, we will consider the problem of classification of entanglement and we have studied this problem specifically in the case of a tripartite system. We will derive coherence-based inequalities to classify a three-qubit entangled system.\\
Let us start our discussion on the tripartite system by assuming that the Hilbert spaces $H_{A}$, $H_{B}$ and $H_{C}$, which describe each subsystem $A$, $B$, $C$ of a tripartite system, is spanned by the computational basis states $|0\rangle$ and $|1\rangle$ respectively.\\
Any three-qubit state may be classified as fully separable states, biseparable states, or genuinely entangled states \cite{dur4}. If the three-qubit state is shared by three distant parties Alice, Bob and Charlie then the shared state may be either in the form of a fully separable or biseparable state or a genuine entangled state. The fully separable and biseparable state may be expressed in the form as \cite{dur3,toth2}
\begin{eqnarray}
\rho_{sep}^{ABC}=\sum_{i}p_{i}\rho_{i}^{A}\otimes\rho_{i}^{B}\otimes\rho_{i}^{C},\sum_{i}p_{i}=1
\label{generalseparable}
\end{eqnarray}
\begin{eqnarray}
\rho_{bisep} = p_1 \rho^{A-BC}_{bisep} + p_2 \rho^{B-AC}_{bisep} + p_3 \rho^{C-AB}_{bisep},~~ 0\leq p_i\leq 1,~~\sum_{i}p_{i}=1
\label{generalbiseparable}
\end{eqnarray}
where,
\begin{eqnarray}
\rho_{bisep}^{A-BC}=\sum_{i}|a_i\rangle_A \langle a_{i}|\otimes |\phi_{i}\rangle_{BC}\langle \phi_{i}|\nonumber\\
\rho_{bisep}^{B-AC}=\sum_{i}|b_i\rangle_B \langle b_{i}|\otimes |\phi_{i}\rangle_{AC}\langle \phi_{i}|\nonumber\\
\rho_{bisep}^{C-AB}=\sum_{i}|c_i\rangle_C \langle c_{i}|\otimes |\phi_{i}\rangle_{AB}\langle \phi_{i}|\nonumber
\end{eqnarray}
Here, $|a_{i}\rangle $, $|b_{i}\rangle $ and $|c_{i}\rangle $  are (unnormalized) states of systems
A, B, and C, respectively, and $|\phi_{i}\rangle $ represent the states of two systems jointly. The density operators $\rho_{i}^{x},~x=A,B,C$ lying on the Hilbert spaces of dimension 2.\\ %whereas the density operators $\rho_{i}^{yz},~y,z=A,B,C$ lying on the composite Hilbert space of two qubits. If any three-qubit state does not fall under the above forms given by (\ref{generalseparable}) and (\ref{generalbiseparable}) then the state is a genuine three-qubit entangled state.\\
It has been observed that the three-qubit biseparable states are important and useful in many contexts. The importance of biseparable states comes from the fact that they may be used as unextendible biseparable bases (UBB) which are proved to be useful to construct genuinely entangled subspaces \cite{agrawal}. Instead of a genuine tripartite entangled state, it has been shown that a biseparable state is enough to use in the controlled quantum teleportation protocol as a resource state \cite{barasinski}. Barasinski et.al. \cite{artur2}  have analyzed the fidelity of the controlled quantum teleportation via mixed biseparable state and have concluded that a statistical mixture of biseparable states can be suitable for the perfect controlled quantum teleportation. Further, it has been shown that there exists a special class of biseparable state i.e. a non-maximally entangled mixed biseparable X state, which can be useful as a resource state for the attainment of high fidelity in controlled quantum teleportation \cite{paulson}. Recently, it is shown that the biseparable states can also be used in obtaining the non-zero conference key \cite{carrara}.\\
Since the three-qubit biseparable state has potential applications in quantum information theory, it is crucial to classify the three-qubit biseparable state. There are a lot of earlier works in this line of research but we mention here a few of them. Eggeling and Werner \cite{eggeling} provided the necessary and sufficient criteria in terms of the projection parameters to detect the biseparable state. But this result is true for only one bipartition cut $A-BC$. In \cite{zhaofei}, it has been shown that the witness operator may be constructed to distinguish fully separable state, biseparable state, and genuine entangled states for a multipartite system. An n-partite inequality is presented in \cite{bancal}, whose violation by a state implies that the state under investigation is not a biseparable state. A non-linear entanglement witness operator has been constructed to identify all three types of three-qubit states \cite{zhao12}. Using entanglement measures, biseparability in mixed three-qubit systems has also been analyzed in detail in  \cite{lohmayer,salay}.
The necessary and sufficient condition for the detection of permutationally invariant three-qubit biseparable states has been studied in \cite{novo}. The multipartite biseparable entangled states under any bipartite partitions can be detected using linear contraction methods \cite{shen1}. A set of Bell inequalities was introduced in \cite{das}, which can distinguish separable, biseparable, and genuinely entangled pure three-qubit states. The method of construction of biseparable state is given in \cite{han} and the classification of three-qubit pure states has been studied in \cite{datta}. Recently, it has been shown that a particular class of three-qubit GHZ and W class of states can be discriminated using coherence-based inequality \cite{kairon}.\\
%\section{Detection of three-qubit biseparable states using $l_1$-norm of coherence}
\noindent In this chapter, we will derive the coherence-based inequality which can be used to detect the given three-qubit state as a biseparable and separable state. To accomplish this task, we first find out the formula for the $l_{1}-$ norm of coherence of the tensor product of two quantum states, which are subsystems of an n-partite system. Then, we use the derived formula to establish the required inequality for the classification of biseparable and separable states.
\section{$l_1$-norm of coherence of the tensor product of bipartite states}
\noindent The coherence can be measured by different measures such as distance measure, relative entropy of coherence, and $l_{p}(p=1,2)$-norms. Any function C which maps the set of all states to the non-negative real
numbers and satisfies the following properties is a valid measure of coherence \cite{baumgratz}:\\
($C_1$) It vanishes for incoherent states i.e.
\begin{eqnarray}
C(\rho)=0,~ \text{whenever}~~\rho~\text{is an incoherent state}
\end{eqnarray}\\
($C_2$) Any valid coherence measure C, should not increase under incoherent operations.\\
($C_3$) Monotonicity under incoherent completely positive and trace-preserving maps i.e
\begin{eqnarray}
C(\rho) \geq C(\phi_{ICPTP}(\rho))
\end{eqnarray}
for all incoherent completely positive and trace-preserving maps $\phi_{ICPTP}$.\\
($C_4$)  Monotonicity under selective measurements on average i.e.
\begin{eqnarray}
C(\rho) \geq \sum_n{p_nC(\rho_n)}~~ \forall~~ \{K_n\}
\end{eqnarray}
where $\rho_n=\frac{K_n\rho{K_n}^{\dagger}}{p_n}$ represent the state corresponding to the outcome $n$ and the probability of occurrence of the state $\rho_n$ is given by $p_n=Tr[K_n\rho{K_n}^{\dagger}]$. $\{K_n\}$ are the Kraus operators such that
$\sum_n{{K_n}^{\dagger}K_n=}{I}$ and ${K_n}\textbf{I}{K_n}^{\dagger}\subset \textbf{I}$, $I$ denote the identity operator and $\textbf{I}$ represent the set of incoherent states.\\
($C_{5}$) A valid coherence monotone should satisfy the convexity property i.e.
\begin{eqnarray}
\sum_n{p_n C(\rho_n)}\geq C(\sum_n{p_n\rho_n}),~0\leq p_n\leq 1, \sum_n{p_n}=1
\end{eqnarray}
for any set of states ${\rho_n}$. Here $\rho_n=\frac{K_n\rho{K_n}^{\dagger}}{p_n}$ and $p_n=Tr[K_n\rho{K_n}^{\dagger}]$. $\{K_n\}$ are the Kraus operators.\\
In this chapter, we will use the $l_1$ norm of coherence which is defined as
\begin{eqnarray}
C_{l_1}(\rho)=\sum_{i,j,i\neq j}|\rho_{ij}|
\end{eqnarray}
where $\rho_{ij}$ denotes the complex numbers corresponding to the ij-th entry of the density matrix $\rho$. $l_1$-norm of coherence of a state depends on the choice of basis in which the given state is expressed and thus from now on, we are considering only computational basis to describe the density matrix of the given state. $l_1$-norm of coherence satisfies all the properties ($C_1$-$C_5$) and thus, it can be considered as a valid coherence monotone. Therefore, it may serve as a useful measure of coherence \cite{baumgratz}.\\
Now, we are in a position to state the result on the $l_1$-norm of coherence of the tensor product of two quantum states which are the subsystems of an n-partite system.\\
\textbf{Result-1:} If the density operators $\rho_{A_1,A_2,...,A_M}$ and $\rho_{A_{M+1},A_{M+2},...,A_N}$ denote the subsystem of a n-partite system, then the $l_1$-norm of coherence of the tensor product of $\rho_{A_1,A_2,...,A_M}$ and $\rho_{A_{M+1},A_{M+2},...,A_N}$ is given by
\begin{eqnarray}
C_{l_1}(\rho_{A_1,A_2,...,A_M}\otimes \rho_{A_{M+1},A_{M+2},...,A_N})&=&C_{l_1}(\rho_{A_1,A_2,...,A_M})+C_{l_1}(\rho_{A_{M+1},A_{M+2},...,A_N})\nonumber\\&+&C_{l_1}(\rho_{A_1,A_2,...,A_M}).C_{l_1}(\rho_{A_{M+1},A_{M+2},...,A_N})
\label{result-1}
\end{eqnarray}
\textbf{Proof:}	Consider two quantum states described by the density matrices $\rho_{A_1,A_2,...,A_M}$ and $\rho_{A_{M+1},A_{M+2},...,A_N}$. The matrix representation of the density operators  $\rho_{A_1,A_2,...,A_M}$ and $\rho_{A_{M+1},A_{M+2},...A_N}$ are given by
\begin{eqnarray}
\rho_{A_1,A_2,...,A_M}=
\begin{pmatrix}
a_{1,1} & a_{1,2} & . & . & . & a_{1,m}\\
a_{2,1} & a_{2,2} & . & . & . & a_{2,m}\\
. & . & . & . & . & .\\
. & . & . & . & . & .\\
. & . & . & . & . & .\\
a_{m,1} & a_{m,2} & . & . & . & a_{m,m}
\end{pmatrix}
\end{eqnarray}
where, ~~$a_{1,1} +a_{2,2}+.+.+.+a_{m,m}=1$,
\begin{eqnarray}
\rho_{A_{M+1},A_{M+2},...A_N}=
\begin{pmatrix}
a_{m+1,m+1} & a_{m+1,m+2} & . & . & . & a_{m+1,n}\\
a_{m+2,m+1} & a_{m+2,m+2} & . & . & . & a_{m+2,n}\\
. & . & . & . & . & .\\
. & . & . & . & . & .\\
. & . & . & . & . & .\\
a_{n,m+1} & a_{n,m+2} & . & . & . & a_{n,n}
\end{pmatrix}
\end{eqnarray}
where, ~~$a_{m+1,m+1}+a_{m+2,m+2}+...+a_{n,n}=1$ and $a_{j,i}$ denotes the complex conjugate of $a_{i,j}$.\\
Then, the $l_1$ norm of coherence of the density matrices $\rho_{A_1,A_2,...,A_M}$ and $\rho_{A_{M+1},A_{M+2},...,A_N}$ are given by
\begin{eqnarray}
C_{l_1}(A_1,A_2,...,A_M)&=&\sum_{{i,j=1},{i\neq j}}^{m}{|a_{i,j}|}
\label{norm1111}
\end{eqnarray}
and
\begin{eqnarray}
C_{l_1}(\rho_{A_{M+1},A_{M+2},...,A_N})&=&\sum_{{i,j=m+1,i\neq j}}^{n}|a_{i,j}|
\label{norm1122}
\end{eqnarray}
Then,
\begin{eqnarray}
C_{l_1}(\rho_{A_1,A_2,...,A_M}\otimes \rho_{A_{M+1},A_{M+2},...,A_N})&=&|a_{1,1}|(\sum_{{i,j=m+1},{i\neq  j}}^{n}{|a_{i,j}|})+|a_{1,2}|(\sum_{i,j=m+1}^{n}{|a_{i,j}|})\nonumber\\&+&...+|a_{1,m}|(\sum_{i,j=m+1}^{n}{|a_{i,j}|})+|a_{2,1}|(\sum_{i,j=m+1}^{n}|a_{i,j}|)\nonumber\\&+&|a_{2,2}|(\sum_{{i,j=m+1},{i\neq j}}^{n}|a_{i,j}|)+...+|a_{2,n}|(\sum_{i,j=m+1}^{n}|a_{i,j}|)\nonumber\\&+&...+|a_{m,1}|(\sum_{i,j=m+1}^{n}{|a_{i,j}|})+...+|a_{m,m}|(\sum_{{i,j=m+1},{i\neq j}}^{n}{|a_{i,j}|})\nonumber\\
\end{eqnarray}
Simplifying the above equation, we get
\begin{eqnarray}
C_{l_1}(\rho_{A_1,A_2,...,A_M}\otimes \rho_{A_{M+1},A_{M+2},...,A_N})&=&\sum_{{i,j=m+1},{i\neq j}}^{n}{|a_{i,j}|}[|a_{1,1}|+|a_{2,2}|+...+|a_{m,m}|]\nonumber\\&+&\sum_{i,j=m+1}^{n}{|a_{i,j}|}[|a_{1,2}|+|a_{1,3}|+...+|a_{1,m}|+|a_{2,1}|\nonumber\\&+&|a_{2,3}|+...+|a_{2,n}|+...+|a_{m,1}|+...+|a_{m,m-1}|]\nonumber\\
\end{eqnarray}

Using normalization condition of $\rho_{A_1,A_2,...,A_M}$ and $\rho_{A_{M+1},A_{M+2},...,A_N}$, we get
\begin{eqnarray}
C_{l_1}(\rho_{A_1,A_2,...,A_M}\otimes \rho_{A_{M+1},A_{M+2},...,A_N})&=&\sum_{{i,j=m+1},{i\neq j}}^{n}{|a_{i,j}|}+[\sum_{i,j=m+1}^{n}{|a_{i,j}|}].[\sum_{{i,j=m+1},{i\neq j}}^{n}{|a_{i,j}|}]\nonumber\\&=&\sum_{{i,j=m+1},{i\neq j}}^{n}{|a_{i,j}|}+[1+\sum_{{i,j=m+1},{i\neq j}}^{n}{|a_{i,j}|}]\times\nonumber\\&&[\sum_{{i,j=m+1},{i\neq j}}^{n}{|a_{i,j}|}]\nonumber\\
\end{eqnarray}
From equations (\ref{norm1111}) and (\ref{norm1122}), we get,
\begin{eqnarray}
&&C_{l_1}(\rho_{A_1,A_2,...,A_M}\otimes \rho_{A_{M+1},A_{M+2},...,A_N})=C_{l_1}(\rho_{A_1,A_2,...,A_M})\nonumber\\&+&[1+C_{l_1}(\rho_{A_{M+1},A_{M+2},...,A_N})].[C_{l_1}(\rho_{A_1,A_2,...,A_M})]\nonumber
\end{eqnarray}
Thus, the $l_1$-norm of coherence of the tensor product of an m-qubit and an (n-m) qubit, which are a part of an n-partite quantum system, is given by
\begin{eqnarray}
&&C_{l_1}(\rho_{A_1,A_2,...,A_M}\otimes \rho_{A_{M+1},A_{M+2},...,A_N})\nonumber\\&=&C_{l_1}(\rho_{A_1,A_2,...,A_M})+C_{l_1}(\rho_{A_{M+1},A_{M+2},...,A_N})\nonumber\\&+&C_{l_1}(\rho_{A_1,A_2,...,A_M}).C_{l_1}(\rho_{A_{M+1},A_{M+2},...,A_N})
\end{eqnarray}
\textbf{Corollary-1:} For any two single qubit quantum states described by the density operators $\rho_1$ and $\rho_2$, the $l_1$ norm of coherence of the tensor product of $\rho_1$ and $\rho_2$ is given by
\begin{eqnarray}
C_{l_1}(\rho_1 \otimes \rho_2)=C_{l_1}(\rho_1)+C_{l_1}(\rho_2)+C_{l_1}(\rho_1).C_{l_1}(\rho_2)
\label{result-1}
\end{eqnarray}
\textbf{Proof:} Any two single-qubit quantum states described by the density matrices $\rho_1$ and $\rho_2$ are given by
\begin{eqnarray}
\rho_i=
\begin{pmatrix}
a_i & b_i\\
b_i^{*} & d_i
\end{pmatrix}, a_i+d_i=1,~~i=1,2
\end{eqnarray}
The $l_1$ norm of coherence of the density matrices $\rho_1$ and $\rho_2$ are given by
\begin{eqnarray}
C_{l_1}(\rho_i)=2|b_{i}|,~~i=1,2
\label{norm1}
\end{eqnarray}
The tensor product of $\rho_1$ and $\rho_2$ may be defined as
\begin{eqnarray}
\rho_1 \otimes \rho_2&=&
\begin{pmatrix}
a_1 & b_1\\
b_1^{*} & d_1
\end{pmatrix}
\otimes
\begin{pmatrix}
a_2 & b_2\\
b_2^{*} & d_2
\end{pmatrix}\nonumber\\&=&\begin{pmatrix}
a_{1}a_{2} & a_{1}b_{2} & b_{1}a_{2} & b_{1}b_{2}\\
a_{1}b_{2}^{*} & a_{1}d_{2} & b_{1}b_{2}^{*} & b_{1}d_{2}\\
b_{1}^{*}a_{2} & b_{1}^{*}b_{2} & d_{1}a_{2} & d_{1}b_{2}\\
b_{1}^{*}b_{2}^{*} & b_{1}^{*}d_{2} & d_{1}b_{2}^{*} & d_{1}d_{2}
\end{pmatrix}
\end{eqnarray}
The $l_1$-norm of coherence of the tensor product $\rho_1 \otimes \rho_2$ is given by
\begin{eqnarray}
C_{l_1}(\rho_1 \otimes \rho_2)&=& a_{1}|b_{2}|+|b_{1}|a_{2}+|b_{1}||b_{2}|+a_{1}|b_{2}|\nonumber\\&+&|b_{1}||b_{2}|+
|b_{1}|d_{2}+|b_{1}|a_{2}+|b_{1}||b_{2}|\nonumber\\&+&d_{1}|b_{2}|+|b_{1}||b_{2}|+|b_{1}|d_{2}+d_{1}|b_{2}|
\nonumber\\&=&2(a_{1}+d_{1})|b_2|
+2(a_{2}+d_{2})|b_1|+4|b_{1}|.|b_2|\nonumber\\&=&2|b_{1}|+2|b_{2}|+4|b_{1}|.|b_{2}|
\label{rho1trho2}
\end{eqnarray}
Using equation (\ref{norm1}), the equation (\ref{rho1trho2}) reduces to
\begin{eqnarray}
C_{l_1}(\rho_1 \otimes \rho_2)=C_{l_1}(\rho_1)+C_{l_1}(\rho_2)+C_{l_1}(\rho_1).C_{l_1}(\rho_2)
\end{eqnarray}
Hence proved.
%The general proof of $l_1$ norm of coherence of the tensor product of a single qubit and an (n-1) qubit quantum system is given in the -1.\\
\\
\textbf{Corollary-2:} If the  three-qubit biseparable system described either by the density operator $\rho_{A-BC}\equiv \rho_{A}\otimes \rho_{BC}$ or $\rho_{B-AC}\equiv \rho_{B}\otimes \rho_{AC}$ or $\rho_{C-AB}\equiv \rho_{C}\otimes \rho_{AB}$, then the $l_{1}$-norm of coherence for the density operator $\rho_{A-BC}$, $\rho_{B-AC}$ and $\rho_{C-AB}$ are given by
\begin{eqnarray}
C_{l_1}(\rho_{A-BC})=C_{l_1}(\rho_A)+C_{l_1}(\rho_{BC})+C_{l_1}(\rho_A).C_{l_1}(\rho_{BC})
\label{coherence1}
\end{eqnarray}
\begin{eqnarray}
C_{l_1}(\rho_{B-AC})=C_{l_1}(\rho_{B})+C_{l_1}(\rho_{AC})+C_{l_1}(\rho_{B}).C_{l_1}(\rho_{AC})
\label{coherence2}
\end{eqnarray}
\begin{eqnarray}
C_{l_1}(\rho_{C-AB})=C_{l_1}(\rho_{C})+C_{l_1}(\rho_{AB})+C_{l_1}(\rho_{C}).C_{l_1}(\rho_{AB})
\label{coherence3}
\end{eqnarray}
\begin{enumerate}
	\item If the equality given by (\ref{coherence1}) is violated by any three-qubit state, then the state under investigation is not a biseparable state of the form $\rho_{A}\otimes \rho_{BC}$.
	\item If the equality given by (\ref{coherence2}) is violated by any three-qubit state, then the state under investigation is not a biseparable state of the form $\rho_{B} \otimes \rho_{AC}$.
	\item If any three-qubit state does not satisfy the equality given by (\ref{coherence3}), then the given state is not a biseparable state of the form $\rho_{C}\otimes \rho_{AB}$.
\end{enumerate}
\textbf{Corollary-3:} If the three-qubit system represent the separable system described by the density operator $\rho_{A-B-C}\equiv \rho_{A}\otimes \rho_{B}\otimes \rho_{C}$, then the $l_{1}$-norm of coherence for the density operator $\rho_{A-B-C}$ is given by
\begin{eqnarray}
C_{l_1}(\rho_{A-B-C})&=&C_{l_1}(\rho_A)+C_{l_1}(\rho_{B})+C_{l_1}(\rho_C)\nonumber\\&+&C_{l_1}(\rho_A).C_{l_1}(\rho_{B})+
C_{l_1}(\rho_A).C_{l_1}(\rho_{C})\nonumber\\&+&C_{l_1}(\rho_B).C_{l_1}(\rho_{C})\nonumber\\&+&C_{l_1}(\rho_A).C_{l_1}(\rho_{B}).C_{l_1}(\rho_{C})
\label{coherence4}
\end{eqnarray}
If the equality given by (\ref{coherence4}) is violated by any three-qubit state, then the state under probe is not a separable state.
\subsection{Coherence-based inequality for the detection of a particular form of three-qubit biseparable states}
%In the previous section, we derived the lower bound of $l_{1}$ norm of coherence of the biseparable state, and it is observed that the derived bound is dependent on the coherence of the single-qubit subsystem and the concurrence of the two-qubit entangled subsystem. Since the classification of different biseparable systems is not possible with the lower bound of $l_{1}$ norm of coherence so we need another criterion to detect biseparable systems.\\
In this subsection, we deduce  coherence based inequality for the detection of three-qubit biseparable state of the form $\rho^{i-jk} (i\neq j\neq k;i,j,k=A,B,C)$. To verify this inequality, we need only the information on the density matrix elements of the given three-qubit system under investigation.\\
\textbf{Result-2:} If the three-qubit state described by the density operator $\rho^{A-BC}=\sum_{i}{p_i}{\rho_{A}^{i}\otimes \rho_{BC}^{i}}$ is biseparable such that the $l_1$-norm of coherence of at least one of the reduced system is non zero, then the $l_{1}$ norm of coherence of the biseparable state satisfies
\begin{eqnarray}
C_{l_1}(\rho^{A-BC})\leq \sum_{i}{p_i}(\frac{X_{i}^2}{4}+X_i)
\label{result2}
\end{eqnarray}
where 
\begin{eqnarray}
X_{i}=C_{l_1}({\rho_A^{i}})+C_{l_1}({\rho_{BC}^{i}}),~~ i=1,2,3,...
\end{eqnarray}

\textbf{Proof:} Let us consider a biseparable state for $A-BC$ partition. The biseparable state in this partition is given by
\begin{eqnarray}
\rho^{A-BC}=\sum_{i}{p_i}{\rho_A^{i}\otimes \rho_{BC}^{i}}
\end{eqnarray}
Then, $l_1$ norm of coherence of $\rho^{A-BC}$ is given by
\begin{eqnarray}
C_{l_1}({\rho^{A-BC}})&=&C_{l_1}[{\sum_{i}{p_i}{\rho_A^{i}\otimes \rho_{BC}^{i}}}]\nonumber\\
&\leq&\sum_{i}{p_i}{C_{l_1}[{{\rho_A^{i}\otimes \rho_{BC}^{i}}}]}\nonumber\\
&=&\sum_{i}{p_i}(C_{l_1}({\rho_A^{i}})+C_{l_1}({\rho_{BC}^{i}})+C_{l_1}({\rho_A^{i}}).C_{l_1}({\rho_{BC}^{i}}))
\label{eql1}
\end{eqnarray}
We have used the convexity property of the $l_1$-norm of coherence in the second step. The last step follows from Corollary-1.\\
Now, Arithmetic mean (AM) and Geometric mean (GM) of $C_{l_1}({\rho_A^{i}})$ and $C_{l_1}({\rho_{BC}^{i}})$ are given by
\begin{eqnarray}
\frac{C_{l_1}({\rho_A^{i}})+C_{l_1}({\rho_{BC}^{i}})}{2}~~ \text{and}~~ \sqrt{C_{l_1}({\rho_A^{i}}).C_{l_1}({\rho_{BC}^{i}})}
\label{am-gm}
\end{eqnarray}
Using AM-GM inequality\cite{horn} on $C_{l_1}({\rho_A^{i}})$ and $C_{l_1}({\rho_{BC}^{i}})$, we get
\begin{eqnarray}
\frac{(C_{l_1}({\rho_A^{i}})+C_{l_1}({\rho_{BC}^{i}}))^2}{4}\geq C_{l_1}({\rho_A^{i}}).C_{l_1}({\rho_{BC}^{i}})
\end{eqnarray}
From (\ref{eql1}), we get
\begin{eqnarray}
C_{l_1}(\rho^{A-BC})&\leq& \sum_{i}{p_i}(C_{l_1}({\rho_A^{i}})+C_{l_1}({\rho_{BC}^{i}})+\frac{(C_{l_1}({\rho_A^{i}})+C_{l_1}({\rho_{BC}^{i}}))^2}{4})\nonumber\\&=&\sum_{i}{p_i}(X_i+\frac{X_{i}^2}{4})
\label{result2proof}
\end{eqnarray}
where $X_{i}=C_{l_1}({\rho_A^{i}})+C_{l_1}({\rho_{BC}^{i}})$ for i=1,2,3,...\\
Hence proved.
%\textbf{Corollary-3:} If any three-qubit state violates the inequality (\ref{result2}) then the given three-qubit state may be either a separable state or genuine entangled state but it cannot be a biseparable state.\\
%\textbf{Corollary-4:} If any three-qubit state does not satisfy the equality condition (\ref{coherence4}) and violates the inequality (\ref{result2}) then the three-qubit state under investigation is a genuine entangled state.\\
%\textbf{Corollary-5:} If any three-qubit state does not satisfy the equality condition (\ref{coherence4}) and simultaneously violate the inequalities (\ref{ineq2}), (\ref{ineq25}), (\ref{ineq26}) then the given three-qubit state is a genuine entangled state.
\subsection{Example}
\noindent \textbf{Example-1:} Let us consider a biseparable state described by the density operator $\rho=|\psi\rangle_{ABC} \langle \psi|$, where $|\psi\rangle_{ABC}$ is given by
\begin{eqnarray}
|\psi\rangle_{ABC}=\lambda_{0}|101\rangle+\lambda_{1}|110\rangle+\lambda_{2}|111\rangle
\label{ex11}
\end{eqnarray}
where $\lambda_{0},\lambda_{1},\lambda_{2}\in R^{+}$ and $\lambda_{0}^{2}+\lambda_{1}^{2}+\lambda_{2}^{2}=1$.
Let us assume that $\lambda_{0}\geq\lambda_{1}\geq\lambda_{2}$.\\
The value of $C_{l_1}(\rho_A)$ and $C_{l_1}(\rho_{BC})$ is given by
\begin{eqnarray}
&&C_{l_1}(\rho_A)=0,~~
C_{l_1}(\rho_{BC})=2(\lambda_0\lambda_1+\lambda_1\lambda_2+\lambda_0\lambda_2)
\label{coh1}
\end{eqnarray}
For the state $|\psi\rangle_{ABC}$ given in $(\ref{ex11})$, we can calculate $C_{l_1}(\rho^{A-BC})$ as
\begin{eqnarray}
C_{l_1}(\rho^{A-BC})=2(\lambda_0\lambda_1+\lambda_1\lambda_2+\lambda_0\lambda_2)
\label{exp1}
\end{eqnarray}
Using (\ref{coh1}) and (\ref{exp1}), it can be shown that the inequality (\ref{result2proof}) is satisfied.\\
\textbf{Example-2:} Consider a $|W\rangle_{ABC}$ state of the form
\begin{eqnarray}
|W\rangle_{ABC}=\frac{1}{\sqrt{3}}(|100\rangle_{ABC}+|010\rangle_{ABC}+|001\rangle_{ABC})
\label{w1}
\end{eqnarray}
The $l_{1}$-norm of coherence of the $|W\rangle_{ABC}$ state is given by $C_{l_{1}}(|W\rangle_{ABC})=2$.\\
The reduced single-qubit state may be expressed as
\begin{eqnarray}
\rho^{W}_{i}=Tr_{jk}(|W\rangle_{ABC})=\frac{1}{3}(2|0\rangle_{A}\langle0|+|1\rangle_{A}\langle1|),~~i\neq j\neq k,i,j,k\in \{A,B,C\}
\label{w2}
\end{eqnarray}
Since the single-qubit density operators $\rho^{W}_{A}$, $\rho^{W}_{B}$ and $\rho^{W}_{C}$ does not contain any off-diagonal elements so the $l_{1}$-norm of coherence for these single qubit states is given by $C_{l_{1}}(\rho^{W}_{A})=C_{l_{1}}(\rho^{W}_{B})=C_{l_{1}}(\rho^{W}_{C})=0$. Therefore, it can be easily shown that the equality condition given in (\ref{coherence4}) is not maintained for the state (\ref{w1}). Further, it can be shown that the set of equality conditions given by (\ref{coherence1}), (\ref{coherence2}) and (\ref{coherence3}) are not satisfied by the state (\ref{w1}). Thus, the state (\ref{w1}) is neither a fully separable state of the form $\rho_A^{W}\otimes \rho_B^{W} \otimes \rho_C^{W}$ nor a biseparable state of the form $\rho^{W}_{A}\otimes\rho^{W}_{BC}$ or $\rho^{W}_{B}\otimes\rho^{W}_{CA}$ or $\rho^{W}_{C}\otimes\rho^{W}_{AB}$.
%Hence we can conclude that the state (\ref{w1}) is a genuine entangled state.
\section{Detection of general three-qubit biseparable states and separable states}
\noindent A mixed state is said to be fully separable if it can be written as the convex combination of fully separable pure states.
A mixed state is said to be biseparable if it is not fully separable and it can be written as a convex combination of biseparable pure states.
Let us recall the three-qubit mixed biseparable state given in (\ref{generalbiseparable}) and re-write it as
\begin{eqnarray}
\sigma_{bisep} = p_1 \sigma^{A-BC}_{bisep} + p_2 \sigma^{B-AC}_{bisep} + p_3 \sigma^{C-AB}_{bisep},~~0\leq p_i\leq 1,~~\sum_{i}p_{i}=1
\label{bisep11}
\end{eqnarray}
\subsection{Coherence-based inequality for the classification of general three-qubit biseparable states}
\noindent Classification of three-qubit mixed states has been studied by constructing the witness operator \cite{acin}. In this section, we will study the classification of three-qubit mixed biseparable states using coherence-based inequality.\\
\textbf{Result-3:} If the three-qubit mixed state described by the density operator in (\ref{bisep11}) is biseparable then it satisfies the inequality
\begin{eqnarray}
1+C_{l_1}(p_1\sigma_1^{A-BC}+p_2\sigma_2^{B-CA}+p_3\sigma_3^{C-AB}) \leq \frac{1}{4}\sum_{i=1}^{3}p_i(X_i+2)^2
\label{result3}
\end{eqnarray}
where,
\begin{eqnarray}
X_1=C_{l_1}(\sigma_1^{A})+C_{l_1}(\sigma_1^{BC})\nonumber\\
X_2=C_{l_1}(\sigma_2^{B})+C_{l_1}(\sigma_2^{CA})\nonumber\\
X_3=C_{l_1}(\sigma_3^{C})+C_{l_1}(\sigma_3^{AB})
\label{x1x2x3}
\end{eqnarray}

\textbf{Proof:} Let us consider a mixed three-qubit biseparable state whose density matrix is given by $p_1\sigma_1^{A-BC}+p_2\sigma_2^{B-CA}+p_3\sigma_3^{C-AB}$. Using $l_1$-norm of coherence of $p_1\sigma_1^{A-BC}+p_2\sigma_2^{B-CA}+p_3\sigma_3^{C-AB}$ and the covexity property of $l_1$-norm of coherence, we have,
\begin{eqnarray}
C_{l_1}(p_1\sigma_1^{A-BC}+p_2\sigma_2^{B-CA}+p_3\sigma_3^{C-AB})
&\leq& p_1C_{l_1}(\sigma_1^{A-BC})+p_2C_{l_1}(\sigma_2^{B-CA})\nonumber\\&+&p_3C_{l_1}(\sigma_3^{C-AB})
\end{eqnarray}
Using the relation (\ref{coherence1}) or other related coherence relation like (\ref{coherence2}) or (\ref{coherence3}), we get
\begin{eqnarray}
C_{l_1}(p_1\sigma_1^{A-BC}+p_2\sigma_2^{B-CA}+p_3\sigma_3^{C-AB}) &\leq& p_1[C_{l_1}(\sigma_1^{A}+C_{l_1}(\sigma_1^{BC})+C_{l_1}(\sigma_1^{A}).C_{l_1}(\sigma_1^{BC})]\nonumber\\&+&p_2[C_{l_1}(\sigma_2^{B}+C_{l_1}(\sigma_2^{CA})+C_{l_1}(\sigma_2^{B}).C_{l_1}(\sigma_2^{CA})]\nonumber\\&+&p_3[C_{l_1}(\sigma_3^{C}+C_{l_1}(\sigma_3^{AB})+C_{l_1}(\sigma_3^{C}).C_{l_1}(\sigma_3^{AB})]\nonumber\\
\label{ineq43}
\end{eqnarray}
Recalling expressions of AM and GM of $C_{l_1}(\sigma_{i}^{A})$ and $C_{l_1}(\sigma_{i}^{BC})$ from (\ref{am-gm}) and applying $AM\geq GM$ on $C_{l_1}(\sigma_{i}^{A})$ and $C_{l_1}(\sigma_{i}^{BC})$, inequality (\ref{ineq43}) reduces to,
\begin{eqnarray}
C_{l_1}(p_1\sigma_1^{A-BC}+p_2\sigma_2^{B-CA}+p_3\sigma_3^{C-AB}) &\leq& p_1[C_{l_1}(\sigma_1^{A})+C_{l_1}(\sigma_1^{BC})+\frac{1}{4}(C_{l_1}(\sigma_1^{A})+C_{l_1}(\sigma_1^{BC}))^2]\nonumber\\&+&p_2[C_{l_1}(\sigma_2^{B})+C_{l_1}(\sigma_2^{CA})+\frac{1}{4}(C_{l_1}(\sigma_2^{B})+C_{l_1}(\sigma_2^{CA}))^2]\nonumber\\&+&p_3[C_{l_1}(\sigma_3^{C})+C_{l_1}(\sigma_3^{AB})+\frac{1}{4}(C_{l_1}(\sigma_3^{C})+C_{l_1}(\sigma_3^{AB}))^2]\nonumber\\
\end{eqnarray}
Thus, we have
\begin{eqnarray}
C_{l_1}(p_1\sigma_1^{A-BC}+p_2\sigma_2^{B-CA}+p_3\sigma_3^{C-AB}) \leq p_1(X_1+\frac{X_1^{2}}{4})+p_2(X_2+\frac{X_2^{2}}{4})+p_3(X_3+\frac{X_3^{2}}{4})
\end{eqnarray}
Therefore,
\begin{eqnarray}
C_{l_1}(p_1\sigma_1^{A-BC}+p_2\sigma_2^{B-CA}+p_3\sigma_3^{C-AB}) &\leq&p_1(X_1+\frac{X_1^{2}}{4})+p_2(X_2+\frac{X_2^{2}}{4})+p_3(X_3+\frac{X_3^{2}}{4})\nonumber\\&=&\frac{1}{4}[p_1(X_1+2)^2+p_2(X_2+2)^2+p_3(X_3+2)^2]-1\nonumber\\
\label{ineq8}
\end{eqnarray}
where $X_{1}, X_2$ and $X_3$ are given by (\ref{x1x2x3}).\\
Simplifying (\ref{ineq8}), we get
\begin{eqnarray}
1+C_{l_1}(p_1\sigma_1^{A-BC}+p_2\sigma_2^{B-CA}+p_3\sigma_3^{C-AB}) \leq \frac{1}{4}\sum_{i=1}^{3}p_i(X_i+2)^2
\end{eqnarray}
which is the required result. Hence proved.\\
\\
\textbf{Corollary-4:} If any three-qubit mixed state violates the inequality (\ref{result3}), then the given state is not a biseparable state.
\subsection{Coherence-based inequality for the classification of three-qubit mixed separable states}
\noindent In this subsection, we will discuss how to classify the given state as a separable state which belongs to the set of mixed three-qubit states using the $l_1$-norm of coherence of a single qubit. Let us consider a mixed separable state, which can be expressed as
\begin{eqnarray}
\sigma^{A-B-C}=\sum_{i}p_{i}\sigma_{i}^{A}\otimes\sigma_{i}^{B}\otimes\sigma_{i}^{C}
\label{sepmix100}
\end{eqnarray}
The $l_{1}$-norm of the coherence of the state (\ref{sepmix100}) is given by
\begin{eqnarray}
C_{l_1}(\sigma^{A-B-C})&=& C_{l_{1}}(\sum_{i}p_{i}\sigma_{i}^{A}\otimes\sigma_{i}^{B}\otimes\sigma_{i}^{C})\nonumber\\
&\leq& \sum_{i}p_{i}C_{l_{1}}(\sigma_{i}^{A}\otimes\sigma_{i}^{B}\otimes\sigma_{i}^{C})\nonumber\\&=& \sum_{i}p_{i}[\sum_{x=A,B,C}C_{l_{1}}(\sigma_{i}^{x})+\sum_{x\neq y,x,y=A,B,C}C_{l_{1}}(\sigma_{i}^{x})C_{l_{1}}(\sigma_{i}^{y})\nonumber\\&+&C_{l_{1}}(\sigma_{i}^{A})C_{l_{1}}(\sigma_{i}^{B})C_{l_{1}}(\sigma_{i}^{C})]
\label{separabilityc}
\end{eqnarray}
The inequality in the second step follows from the convexity property of $l_{1}$-norm of coherence.\\
We are now in a position to state the condition for separability for a mixed three-qubit system. The statement for separability is as follows:\\
\textbf{Result-4:} If the three-qubit mixed state described by the density operator $\sigma^{A-B-C}=\sum_{i}p_{i}\sigma_{i}^{A}\otimes\sigma_{i}^{B}\otimes\sigma_{i}^{C}$ is separable then it satisfies the inequality
\begin{eqnarray}
C_{l_1}(\sigma^{A-B-C})&\leq& \sum_{i}p_{i}[\sum_{x=A,B,C}C_{l_{1}}(\sigma_{i}^{x})+\sum_{x\neq y,x,y=A,B,C}C_{l_{1}}(\sigma_{i}^{x})C_{l_{1}}(\sigma_{i}^{y})\nonumber\\&+&C_{l_{1}}(\sigma_{i}^{A})C_{l_{1}}(\sigma_{i}^{B})C_{l_{1}}(\sigma_{i}^{C})]
\label{sepmix2000}
\end{eqnarray}
\textbf{Corollary-5:} If any three-qubit mixed state violates the inequality (\ref{sepmix2000}) then the given state is not a separable state.
\subsection{Illustrations}
\textbf{Example-1:} Consider a mixed three-qubit biseparable state described by the density matrix $\rho_1$, which is given by
\begin{eqnarray}
\rho_1&=&q|0\rangle^{A} \langle 0|) \otimes |\phi^{+}\rangle^{BC} \langle \phi^{+}|+(1-q)|1\rangle^{B} \langle 1|\otimes |\phi^{-}\rangle^{AC} \langle \phi^{-}|
\label{exphi}
\end{eqnarray}
where $0\leq q\leq 1$ and the Bell states $|\phi^{+}\rangle^{BC}$ and $|\phi^{-}\rangle^{AC}$ are given by
\begin{eqnarray}
|\phi^{+}\rangle^{BC} &=&\frac{1}{\sqrt{2}}(|00\rangle^{BC}+|11\rangle^{BC})\nonumber\\
|\phi^{-}\rangle^{AC}&=&\frac{1}{\sqrt{2}}(|00\rangle^{AC}-|11\rangle^{AC})
\end{eqnarray}
Comparing equation (\ref{exphi}) with general mixed three-qubit biseparable state, we have $p_1=q$, $p_2=1-q$ and $p_3=0$. $l_1$-norm of coherence for the given state, defined in (\ref{exphi}), is given by
\begin{eqnarray}
C_{l_1}(\rho_{1})=1
\end{eqnarray}
For the given state $\rho_1$, $C_{l_1}(\rho_1^{A})=C_{l_1}(\rho_1^{B})=C_{l_1}(\rho_1^{C})=0$, $C_{l_1}(\rho_1^{BC})=1$, $C_{l_1}(\rho_1^{AC})=1$ and $C_{l_1}(\rho_1^{AB})=0$.
The quantity $X_{i}'s$, i=1,2,3 can be calculated as
\begin{eqnarray}
X_{1}&=&C_{l_1}(\rho_{1}^{A})+C_{l_1}(\rho_{1}^{BC})=1\nonumber\\
X_{2}&=&C_{l_1}(\rho_{1}^{B})+C_{l_1}(\rho_{1}^{AC})=1\nonumber\\
X_{3}&=&C_{l_1}(\rho_{1}^{C})+C_{l_1}(\rho_{1}^{AB})=0
\label{X1ex1}
\end{eqnarray}
Substituting values of $p_i's$, $X_i's$ and $l_1$-norm of coherence of $\rho_1$, we can see that equation (\ref{result3})
is satisfied for the state (\ref{exphi}).\\
\textbf{Example-2:} Let us consider the mixed three-qubit state described by the density operator $\varrho^{ABC}$ as
\begin{eqnarray}
\varrho^{ABC}&=&q|GHZ\rangle \langle GHZ|)+(1-q)|W\rangle \langle W|
\label{mixedex2}
\end{eqnarray}
where $0\leq q\leq 1$ and the three-qubit states $|GHZ\rangle$ and $|W\rangle$ are given by
\begin{eqnarray}
|GHZ\rangle&=&\frac{1}{\sqrt{2}}(|000\rangle+|111\rangle)\\
|W\rangle&=&\frac{1}{\sqrt{3}}(|001\rangle+|010\rangle+|100\rangle
\end{eqnarray}
Compairing equation (\ref{mixedex2}) with general three-qubit mixed biseparable state, we get $p_1=q$, $p_2=1-q$ and $p_3=0$.\\
The $l_1$-norm of coherence of the state (\ref{mixedex2}) is given by
\begin{eqnarray}
C_{l_1}(\varrho^{ABC})=3
\label{Mex2}
\end{eqnarray}
For the given state $\varrho^{ABC}$, $C_{l_1}(\varrho^{A})= C_{l_1}(\varrho^{B})=C_{l_1}(\varrho^{C})=0$ and $ C_{l_1}(\varrho^{BC})=C_{l_1}(\varrho^{AC})=C_{l_1}(\varrho^{AB})=\frac{2}{3}$. Therefore, the quantities $X_i's$, i=1,2,3 can be written as,
\begin{eqnarray}
X_{1}=X_{2}=X_{3}=C_{l_1}(\rho_{1}^{i})+C_{l_1}(\rho_{1}^{jk})=\frac{2}{3},~~ i\neq j\neq k,~~i,j,k\in \{A,B,C\}
\end{eqnarray}
Substituting values of $p_i's$, $X_i's$ and $C_{l_1}(\varrho^{ABC})$ in equation (\ref{result3}), we can observe that the inequality given in Result-3 is violated for any q. Thus, the given state $\varrho^{ABC}$ is not a biseparable state. Moreover, a simple calculation also shows that the inequality (\ref{separabilityc}) is violated for any $q$. So we can infer that the given state (\ref{mixedex2}) is not a separable state. Thus, we find that the given state $\varrho^{ABC}$ is neither a biseparable state nor a separable state. Hence, we may conclude that the given state is a genuine mixed three-qubit entangled state.
\section{An idea to generalize the results to a four-qubit system}
\noindent The results we have obtained in this chapter can be generalized to the multipartite system also. For instance, if we consider a four-qubit system, Result-3 for mixed biseparable states may be re-stated as:\\
If the four-qubit mixed state described by the density operator
\begin{eqnarray}
\rho=p_1\sigma_1^{A-BCD}+p_2\sigma_2^{B-CAD}+p_3\sigma_3^{C-ABC}+p_4\sigma_4^{D-ABC}
\label{bisep4}
\end{eqnarray} is biseparable then it satisfies the inequality
\begin{eqnarray}
1+C_{l_1}(p_1\sigma_1^{A-BCD}+p_2\sigma_2^{B-CAD}+p_3\sigma_3^{C-ABD}+p_4\sigma_4^{D-ABC}) \leq \frac{1}{4}\sum_{i=1}^{4}p_i(X_i+2)^2
\label{result31}
\end{eqnarray}
where,
\begin{eqnarray}
X_1=C_{l_1}(\sigma_1^{A})+C_{l_1}(\sigma_1^{BCD})\nonumber\\
X_2=C_{l_1}(\sigma_2^{B})+C_{l_1}(\sigma_2^{CAD})\nonumber\\
X_3=C_{l_1}(\sigma_3^{C})+C_{l_1}(\sigma_3^{ABD})\nonumber\\
X_4=C_{l_1}(\sigma_4^{D})+C_{l_1}(\sigma_4^{ABC})
\label{x1x2x3}
\end{eqnarray}
To verify this result, let us consider a mixed biseparable state in a four-qubit system described by the density matrix $\rho_{ABCD}$, which is given by
\begin{eqnarray}
\rho_{ABCD}&=&\frac{1}{2}[|0\rangle^{A} \langle 0|) \otimes |\phi^{+}\rangle^{BCD} \langle \phi^{+}|]+\frac{1}{2}[|1\rangle^{B} \langle 1|\otimes |\phi^{-}\rangle^{ACD} \langle \phi^{-}|]
\label{ex}
\end{eqnarray}
where the states $|\phi^{+}\rangle^{BCD}$ and $|\phi^{-}\rangle^{ACD}$ are given by
\begin{eqnarray}
|\phi^{+}\rangle^{BCD} &=&\frac{1}{\sqrt{2}}(|100\rangle^{BCD}+|010\rangle^{BCD})\nonumber\\
|\phi^{-}\rangle^{ACD}&=&\frac{1}{\sqrt{2}}(|100\rangle^{ACD}-|010\rangle^{ACD})
\end{eqnarray}
Comparing equation (\ref{ex}) with general mixed four qubit biseparable state (\ref{bisep4}), we have $p_1=\frac{1}{2}$, $p_2=\frac{1}{2}$, $p_3=0$ and $p_4=0$. $l_1$-norm of coherence for the given state, defined in (\ref{ex}), is given by
\begin{eqnarray}
C_{l_1}(\rho_{ABCD})=1
\end{eqnarray}
For the given state $\rho_1$, $C_{l_1}(\rho_{ABCD}^{A})=C_{l_1}(\rho_{ABCD}^{B})=C_{l_1}(\rho_{ABCD}^{C})=C_{l_1}(\rho_{ABCD}^{D})=0$, $C_{l_1}(\rho_{ABCD}^{BCD})=1$, $C_{l_1}(\rho_{ABCD}^{ACD})=1$, $C_{l_1}(\rho_{ABCD}^{ABD})=1$ and $C_{l_1}(\rho_{ABCD}^{ABC})=0$.
The quantity $X_{i}'s$, i=1,2,3,4 can be calculated as
\begin{eqnarray}
X_{1}&=&C_{l_1}(\rho_{ABCD}^{A})+C_{l_1}(\rho_{ABCD}^{BCD})=1\nonumber\\
X_{2}&=&C_{l_1}(\rho_{ABCD}^{B})+C_{l_1}(\rho_{ABCD}^{ACD})=1\nonumber\\
X_{3}&=&C_{l_1}(\rho_{ABCD}^{C})+C_{l_1}(\rho_{ABCD}^{ABD})=0\nonumber\\
X_{4}&=&C_{l_1}(\rho_{ABCD}^{D})+C_{l_1}(\rho_{ABCD}^{ABC})=0
\label{X1ex}
\end{eqnarray}
Substituting the values of $p_i's$, $X_i's$ and $l_1$-norm of coherence of $\rho_{ABCD}$ in (\ref{result31}), it can be easily seen that (\ref{result31})
is satisfied for the state defined in (\ref{ex}).\\
Moreover, Result-4 for four-qubit mixed separable states may be re-stated as:\\
If the four-qubit mixed state described by the density operator $\sigma^{A-B-C-D}=\sum_{i}p_{i}\sigma_{i}^{A}\otimes\sigma_{i}^{B}\otimes\sigma_{i}^{C}\otimes\sigma_{i}^{D}$ is separable then it satisfies the inequality
\begin{eqnarray}
C_{l_1}(\sigma^{A-B-C-D})&\leq& \sum_{i}p_{i}[\sum_{x=A,B,C,D}C_{l_{1}}(\sigma_{i}^{x})+\sum_{x\neq y,x,y=A,B,C,D}C_{l_{1}}(\sigma_{i}^{x})C_{l_{1}}(\sigma_{i}^{y})\nonumber\\&+&\sum_{x\neq y\neq z,x,y,z=A,B,C,D}C_{l_{1}}(\sigma_{i}^{x})C_{l_{1}}(\sigma_{i}^{y})C_{l_{1}}(\sigma_{i}^{z})\nonumber\\&+&C_{l_{1}}(\sigma_{i}^{A})C_{l_{1}}(\sigma_{i}^{B})C_{l_{1}}(\sigma_{i}^{C})C_{l_{1}}(\sigma_{i}^{D})]
\label{sepmix200}
\end{eqnarray}
To verify the above result for a four-qubit mixed separable state, let us consider a separable state in a four-qubit system, described by the density operator $\rho_2$,
\begin{eqnarray}
\rho_2&=&\frac{1}{4}|0000\rangle^{ABCD}\langle 0000|+\frac{1}{4}|0011\rangle^{ABCD} \langle 0011|\nonumber\\&+&\frac{1}{4}|1000\rangle^{ABCD}\langle 1000|+\frac{1}{4}|1111\rangle^{ABCD} \langle 1111|
\label{ex24}
\end{eqnarray}
Then, for the state defined in (\ref{ex24}), $C_{l_1}(\rho_2)=0$ and $C_{l_1}(\rho_2^{A})=C_{l_1}(\rho_2^{B})=C_{l_3}(\rho_2^{C})=C_{l_1}(\rho_2^{D})=0$, thus we can say that the equation (\ref{sepmix200}) is verified for the state $\rho_2$.
\section{An idea to generalize the results to a three-qutrit system}
\noindent The results, we have obtained in this chapter can be generalized to higher dimensional quantum systems. For example, let us consider a three-qutrit state described by the density operator $\rho=|\psi\rangle \langle \psi|$, where $|\psi\rangle$ is given by,
\begin{eqnarray}
|\psi\rangle=|0\rangle_A \otimes \frac{1}{\sqrt{3}}(|12\rangle_{BC} +|01\rangle_{BC}+|20\rangle_{BC})
\label{ex3}
\end{eqnarray}
Comparing equation (\ref{ex3}) with general mixed three-qutrit biseparable state, we have $p_1=1$, $p_2=0$ and $p_3=0$. $l_1$-norm of coherence for the given state, defined in (\ref{ex3}), is given by
\begin{eqnarray}
C_{l_1}(\rho)=2
\end{eqnarray}
For the given state $\rho$, $C_{l_1}(\rho^{A})=C_{l_1}(\rho^{B})=C_{l_1}(\rho^{C})=0$, $C_{l_1}(\rho^{BC})=2$, $C_{l_1}(\rho^{AC})=0$ and $C_{l_1}(\rho^{AB})=0$.
The quantity $X_{i}'s$, i=1,2,3 can be calculated as
\begin{eqnarray}
X_{1}&=&C_{l_1}(\rho^{A})+C_{l_1}(\rho^{BC})=2\nonumber\\
X_{2}&=&C_{l_1}(\rho^{B})+C_{l_1}(\rho^{AC})=0\nonumber\\
X_{3}&=&C_{l_1}(\rho^{C})+C_{l_1}(\rho^{AB})=0
\label{X1ex1}
\end{eqnarray}
Substituting values of $p_i's$, $X_i's$ and $l_1$-norm of coherence of $\rho$, we can see that equation (\ref{result3}) is satisfied for the state (\ref{ex3}). Thus, the given states is a biseparable states. Hence, the result obtained in this chapter may be generalized to higher dimensional systems.
\section{Conclusion}
\noindent In this chapter, we have used $l_{1}$-norm of coherence for the classification of three-qubit entangled states. Coherence is a basic phenomenon that arises from the superposition principle of quantum mechanics. It can be measured by different measures such as distance measure, relative entropy of coherence, and $l_{p}$ norms. $l_1$-norm is a valid coherence monotone, serves as a useful measure of coherence \cite{baumgratz} and also, we have shown that it may be useful in classifying the multi-qubit system. We have calculated the $l_{1}$-norm of coherence for the tensor product $m-qubit \otimes (n-m)-qubit$. In order to obtain criteria for the classification of three-qubit states, we have obtained inequalities for biseparable states and separable states in terms of $l_{1}$-norm of coherence. Further, we have shown that if the obtained inequality is violated by any three-qubit state then the state under investigation is neither a biseparable state nor a separable state. Since, for a three-qubit system, we have only three categories of state and if we find that the given state is neither a separable state nor a biseparable state then we can conclude that the state is a genuinely entangled state. The results obtained in this chapter are supported by examples. At the end of this chapter, we have provided an idea to generalize the obtained results to higher dimensional and multi-qubit systems.

%Simillarly, for a three-qubit system, we have,
%\begin{eqnarray}
%||\rho_{X-YZ}||_1 \leq C_{l_1}(\rho_{X-YZ}),~~X,Y,Z=A,B,C
%\end{eqnarray}
%inequality is followed.\\
%In this way, if have a k-qubit state defined by $\rho_{A_1,A_2,...,A_k}$ and a (n-k) qubit state defined by $\rho_{A_{k+1},A_{k+2},...,A_N})$, then $l_1$ norm of coherence of $\rho=\rho_{A_1,A_2,...,A_k} \otimes \rho_{A_{k+1},A_{k+2},...,A_N}$ may be defined as
%\begin{eqnarray}
%&&C_{l_1}(\rho_{A_1,A_2,...,A_k}\otimes \rho_{A_{k+1},A_{k+2},...,A_N})\nonumber\\&=&C_{l_1}(\rho_{A_1,A_2,...,A_k})+C_{l_1}(\rho_{A_{k+1},A_{k+2},...,A_N})\nonumber\\&+&C_{l_1}(\rho_{A_1,A_2,...,A_k}).C_{l_1}(\rho_{A_{k+1},A_{k+2},...,A_N})

\chapter{Classification of Three-qubit States using SPA-PT}\label{ch5}
\vspace{1cm}
\noindent\hrulefill

\noindent\emph{In this chapter\;\footnote{This chapter is based on the research papers entitled ``Structural physical approximation of partial transposition makes possible to distinguish SLOCC inequivalent classes of three-qubit system, \emph{European Physical Journal D} {\bf76} 73 (2022)''}, we have studied the problem of classification of three-qubit states using a structural physical approximation of partial transposition (SPA-PT). We have exploited the concept of SPA-PT so that the classification of a three-qubit entangled state may be realized in an experiment. To study the classification problem of the three-qubit system, we have constructed a SPA-PT map for the three-qubit quantum system and then the matrix elements of the density matrix describing the SPA-PT of a three-qubit system have been calculated. We have proposed criteria for the classification of all possible SLOCC inequivalent classes of pure as well as mixed three-qubit states through the SPA-PT map.}

%	\\To start our study, we have applied partial transposition operation on one of the qubit of the three-qubit system and then studied the entanglement properties of the three-qubit system, which is under investigation.
%	\\
%	\noindent In the second part of this chapter, we have used $l_{1}$-norm of coherence for the detection of three-qubit entangled states. Coherence is a basic phenomenon that arises from the superposition principle of quantum mechanics. It can be measured by different measures such as distance measure, relative entropy of coherence, and $l_{p}$ norms. $l_1$ norm is a valid coherence monotone and serves as a useful measure of coherence \cite{baumgratz}. In this chapter, we have calculated the $l_{1}$ norm of coherence of the tensor product of two quantum systems. In order to obtain criteria for the detection and classification of three-qubit states, we have obtained inequalities for biseparable states and separable states in terms of $l_{1}$-norm of coherence. Further, we have shown that if the obtained inequality is violated by any three-qubit state then the state under investigation is neither a biseparable state nor a separable state. Since, for a three-qubit system, we have only three categories of state and if we find that the given state is neither a separable state nor a biseparable state then we can conclude that the state is a genuinely entangled state. At the end of this chapter, we have provided an idea to generalize the obtained results to higher dimensional and multi-qubit systems.}

\noindent\hrulefill
\newpage

\section{Introduction}
\noindent Quantum entanglement\cite{einstein} is a physical phenomenon in which the state of each particle in the group cannot be described independently of the state of the others, even when the particles are separated by a great distance. The entangled quantum system in $d_{1}\otimes d_{2}$ dimensional Hilbert space may be useful in various quantum information processing tasks such as quantum teleportation\cite{Bennett1}, remote state preparation\cite{pati}, entanglement swapping\cite{pan}, secret sharing\cite{hillery} and quantum repeater\cite{li}. We require entangled states to perform quantum information processing tasks in an efficient way, but the process of generation of entangled states is not an easy task. Even if we generate a quantum state in an experiment, we can ask two more questions: (i) whether the generated multiparticle state is an entangled state or not. (ii) If we found that the multiparticle state is an entangled state, then what type of entangled state it is? For instance, in the case of a three-qubit system, if we know that the three-qubit entangled state is generated at the output, then it is necessary to know whether the generated three-qubit entangled state belongs to a biseparable state or genuine entangled state. Further, if we know that the generated three-qubit entangled state is a genuine entangled state, then it is important to classify it further as two SLOCC inequivalent classes i.e GHZ class and W class. The classification of three-qubit genuine entanglement into these subclasses is important from the quantum communication point of view.  Now, we can answer the above questions, if we are able to proceed a little bit further toward the problem of "classification of multi-qubit entangled states". The classification problem starts with a three-qubit system. A three-qubit pure system can be classified as one fully separable state, three biseparable states, and two genuinely entangled states. Genuine entangled states have entanglement in all the subsystems whereas biseparable states have entanglement in two of the three subsystems. The classification of genuine entangled states and biseparable states is equally important as they have their own merits. One of the possible merits of genuine entangled states and biseparable states is the following: genuine entangled states have potential applications in quantum communication \cite{barasinski} whereas biseparable states are useful in obtaining the non-zero conference key \cite{carrara}.\\
In chapter 4, we have defined a way to classify a three-qubit system using $l_1$-norm of coherence but that method may not be realized physically. To overcome this issue, in this chapter, we have used the method of SPA-PT for the classification of a three-qubit system. %In the first part of this chapter, we will use partial transposition operation to investigate the problem of detection of a three-qubit entangled system. Then we will study the SPA-PT map for the three-qubit system to classify its different SLOCC inequivalent classes.

\section{Studying the Effect of Partial Transposition Operation on one of the qubit of a Three-Qubit System}
\noindent Here, we will study the effect of partial transposition operation on any one of the qubit of a three-qubit system shared between Alice, Bob, and Charlie. Let us assume that any three-qubit state is described by the density operator $\rho_{ABC}$. If the entries of the three-qubit state $\rho_{ABC}$ are represented by the $2\times 2$ block matrices then it is given by
\begin{eqnarray}
\rho_{ABC}=
\begin{pmatrix}
A     & B     & C     & D \\
B^{*} & E     & F     & G \\
C^{*} & F^{*} & H     & I \\
D^{*} & G^{*} & I^{*} & J
\end{pmatrix}
\end{eqnarray}
where $A$,$B$,$C$,$D$,$E$,$F$,$G$,$H$,$I$,$J$ denote the $2\times 2$ block matrices.\\
When the partial transposition operation acts on the first qubit $A$ of the state $\rho_{ABC}$, the state transformed as
\begin{eqnarray}
\rho_{ABC}\rightarrow \rho_{ABC}^{T_{A}}\equiv [T\otimes I\otimes I](\rho_{ABC})
\label{PTA}
\end{eqnarray}
The partial transposition with respect to the second and third qubit respectively reduces the state $\rho_{ABC}$ to
\begin{eqnarray}
\rho_{ABC}\rightarrow \rho_{ABC}^{T_{B}}\equiv [I\otimes T\otimes I](\rho_{ABC})
\label{PTB}
\end{eqnarray}
\begin{eqnarray}
\rho_{ABC}\rightarrow \rho_{ABC}^{T_{C}}\equiv [I\otimes I\otimes T](\rho_{ABC})
\label{PTC}
\end{eqnarray}
The partial transposed states $\rho_{ABC}^{T_{A}}$, $\rho_{ABC}^{T_{B}}$, $\rho_{ABC}^{T_{C}}$ can be expressed in terms of block matrices as
\begin{eqnarray}
\rho_{ABC}^{T_{A}}=
\begin{pmatrix}
A & B & C^{*} & F^{*}\\
B^{*} & E & D^{*} & G^{*}\\
C & D & H & I\\
F & G & I^{*} & J
\end{pmatrix}
\label{PTA1}
\end{eqnarray}
\begin{eqnarray}
\rho_{ABC}^{T_{B}}=
\begin{pmatrix}
A & B^{*} & C & F\\
B & E & D & G\\
C^{*} & D^{*} & H & I^{*}\\
F^{*} & G^{*} & I & J
\end{pmatrix}
\label{PTB1}
\end{eqnarray}
\begin{eqnarray}
\rho_{ABC}^{T_{C}}=
\begin{pmatrix}
A^{*} & B^{*} & C^{*} & D^{*}\\
B & E^{*} & F^{*} & G^{*}\\
C & F & H^{*} & I^{*}\\
D & G & I & J^{*}
\end{pmatrix}
\label{PTC1}
\end{eqnarray}
It is well known that the partial transposition criterion is necessary and sufficient for $2 \otimes 2$ and $2 \otimes 3$ system while it is only necessary condition for the system $m \otimes n,~~(m\geq2,n\geq 3,~~\textrm{If}~~ m=2~~ \textrm{then}~~ n\neq 3)$ and for the multipartite system also.\\
We now consider the simplest tripartite system i.e. $2 \otimes 2 \otimes 2$ quantum system to classify its different SLOCC inequivalent classes through partial transposition operation on any one of the single qubit of the given three-qubit system.\\
\textbf{(i)} Let us choose a particular form of an arbitrary state lying in the $GHZ$-class, which is given by
\begin{eqnarray}
|GHZ\rangle_{ABC}=\alpha |000\rangle_{ABC}+ \beta|111\rangle_{ABC},~~|\alpha|^{2}+|\beta|^{2}=1
\label{ghz}
\end{eqnarray}
The density operator $\rho_{GHZ}=|GHZ\rangle_{ABC}\langle GHZ|$ can be expressed as
\begin{eqnarray}
\rho^{GHZ}=
\begin{pmatrix}
A_{1} & B_{1} & C_{1} & D_{1}\\
B_{1}^{*} & E_{1} & F_{1} & G_{1}\\
C_{1}^{*} & F_{1}^{*} & H_{1} & I_{1}\\
D_{1}^{*} & G_{1}^{*} & I_{1}^{*} & J_{1}
\end{pmatrix}
\end{eqnarray}
where $A_{1}=\begin{pmatrix}
|\alpha|^{2} & 0\\
0 & 0
\end{pmatrix}, D_{1}=\begin{pmatrix}
0 & \alpha\beta^{*}\\
0 & 0
\end{pmatrix},J_{1}=\begin{pmatrix}
0 & 0 \\
0 & |\beta|^{2}
\end{pmatrix}$ and all other $2 \times 2$ block matrices are null matrices.\\
If we apply partial transposition operation on the qubit $A$ of the state described by the density operator $(\rho^{GHZ}_{ABC})^{T_A}$ then the partially transposed state $(\rho^{GHZ}_{ABC})^{T_A}$ at the output can be obtained by the prescription given in (\ref{PTA1}). The eigenvalues of $(\rho^{GHZ}_{ABC})^{T_A}$ are given by $\{0,0,0,0,|\alpha|^{2},|\beta|^{2},|\alpha||\beta|,-|\alpha||\beta|\}$. Thus, $(\rho^{GHZ}_{ABC})^{T_A}$ has one negative eigenvalue. The minimum eigenvalue of $(\rho^{GHZ}_{ABC})^{T_A}$ is given by
\begin{eqnarray}
\lambda_{min}(\alpha,\beta)=-|\alpha||\beta|
\label{mineigenvalue}
\end{eqnarray}
Since the minimum eigenvalue of $(\rho^{GHZ}_{ABC})^{T_A}$ is negative so the state $\rho_{ABC}^{GHZ}$ under investigation is an entangled state for all non-zero values of the state parameter $\alpha$ and $\beta$. The most negative eigenvalue is important for more than one reason, which will be clear in the later stage. The most negative eigenvalue can be obtained for $|\alpha|=\frac{1}{\sqrt{2}}$ and $|\beta|=\frac{1}{\sqrt{2}}$. Therefore, the minimum most eigenvalue of $(\rho^{GHZ}_{ABC})^{T_A}$ is given by
\begin{eqnarray}
\lambda_{min}[(\rho^{GHZ}_{ABC})^{T_A}]=-\frac{1}{2}
\label{minmostAeigenvalue}
\end{eqnarray}
For the same value of $\alpha$ and $\beta$ i.e. for $|\alpha|=\frac{1}{\sqrt{2}}$ and $|\beta|=\frac{1}{\sqrt{2}}$, we can obtain the maximum value of tangle $\tau$ which is given by $\tau=1$.\\
Proceeding in a similar way, we can obtain the minimum most eigenvalue of $\rho^{T_{B}}_{GHZ}$ and $\rho^{T_{C}}_{GHZ}$ and they are given by
\begin{eqnarray}
\lambda_{min}[{(\rho^{GHZ}_{ABC})^{T_B}}]=\lambda_{min}[(\rho^{GHZ}_{ABC})^{T_C}]=-\frac{1}{2}
\label{minmostBCeigenvalue}
\end{eqnarray}
Since the minimum most eigenvalues of the partially transposed state with respect to the qubits $A$, $B$, and $C$ are the same so we denote it by $\lambda_{min}[{\rho_{ABC}^{GHZ}}]$. Thus, we have $\lambda_{min}[(\rho^{GHZ}_{ABC})^{T_A}]=\lambda_{min}[(\rho^{GHZ}_{ABC})^{T_B}]=\lambda_{min}[(\rho^{GHZ}_{ABC})^{T_C}]\equiv\lambda_{min}[{\rho_{ABC}^{GHZ}}]$.\\
\textbf{(ii)} Next, we will choose a particular form of an arbitrary state belonging to the $W$-class, which is given by
\begin{eqnarray}
|W\rangle_{ABC}&=&\lambda_{0} |001\rangle_{ABC}+ \lambda_{1} |010\rangle_{ABC}+ \lambda_{2} |100\rangle_{ABC}
\label{wstate}
\end{eqnarray}
where the state parameters $\lambda_{i} (i=0,1,2)$ are real numbers satisfying $\lambda_{0}^{2}+\lambda_{1}^{2}+\lambda_{2}^{2}=1$. We may note here that one may choose a general form of $W$ class of states for detailed analysis but may face difficulty in finding the analytical form of eigenvalues. Thus, we restrict ourselves to studying one of the particular forms defined in (\ref{wstate}).
The density operator $\rho_{ABC}^{W}$ can be expressed as
\begin{eqnarray}
\rho_{ABC}^{W}=
\begin{pmatrix}
A & B & C & D\\
B^{*} & E & F & G\\
C^{*} & F^{*} & H & I\\
D^{*} & G^{*} & I^{*} & J\\
\end{pmatrix}
\end{eqnarray}
where,
\begin{eqnarray}
A=
\begin{pmatrix}
0 & 0\\
0 & \lambda_0^{2}\\
\end{pmatrix},
B=
\begin{pmatrix}
0 & 0\\
\lambda_0\lambda_1 & 0\\
\end{pmatrix},
C=
\begin{pmatrix}
0 & 0\\
\lambda_0\lambda_2 & 0\\
\end{pmatrix},\nonumber\\
E=
\begin{pmatrix}
\lambda_1^{2} & 0\\
0 & 0\\
\end{pmatrix},
F=
\begin{pmatrix}
\lambda_1\lambda_2 & 0\\
0 & 0\\
\end{pmatrix},
H=
\begin{pmatrix}
\lambda_2^{2} & 0\\
0 & 0\\
\end{pmatrix},
\end{eqnarray}
D, G, I, and J are zero matrices.\\
The eigenvalues of $(\rho^{W}_{ABC})^{T_A}$ can be calculated as $\{0,0,0,0,\lambda_{0}^{2}+\lambda_{1}^{2},\lambda_{2}^{2},\lambda_{2}\sqrt{\lambda_{0}^{2}+\lambda_{1}^{2}},\\
-\lambda_{2}\sqrt{\lambda_{0}^{2}+\lambda_{1}^{2}}\}$. Thus, $(\rho^{W}_{ABC})^{T_A}$ has one negative eigenvalue irrespective of the sign of the real parameters $\lambda_{0}$, $\lambda_{1}$, $\lambda_{2}$. The minimum eigenvalue of $(\rho^{W}_{ABC})^{T_A}$ is given by
\begin{eqnarray}
\lambda_{min}(\lambda_{0},\lambda_{1},\lambda_{2})&=&-|\lambda_{2}|\sqrt{\lambda_{0}^{2}+\lambda_{1}^{2}}\nonumber\\&=&
-|\lambda_{2}|\sqrt{1-\lambda_{2}^{2}}
\label{mineigenvalueW}
\end{eqnarray}
In this case also, we find that the minimum eigenvalue of $(\rho^{W}_{ABC})^{T_A}$ is negative so the state described by the density operator $\rho_{ABC}^{W}$ is an entangled state for all non-zero values of the state parameter $\lambda_{i} (i=0,1,2)$.\\
The most negative eigenvalue can be obtained for $|\lambda_{2}|=\frac{7}{10}$ and for any value of $\lambda_{0}$ and $\lambda_{1}$ satisfying $\lambda_{0}^{2}+\lambda_{1}^{2}=0.51$. Thus, the minimum most eigenvalue of $(\rho^{W}_{ABC})^{T_A}$ is given by
\begin{eqnarray}
\lambda[(\rho^{W}_{ABC})^{T_A}]=-0.4999
\label{minmostAeigenvaluew}
\end{eqnarray}
Proceeding in a similar way, the eigenvalues of $(\rho_{ABC}^{W})^{T_B}$ and $(\rho_{ABC}^{W})^{T_C}$ can be calculated as $\{0,0,0,0,\lambda_1^{2}, \lambda_0^{2}+\lambda_2^{2}, \lambda_1\sqrt{\lambda_0^{2}+\lambda_2^{2}}, -\lambda_1\sqrt{\lambda_0^{2}+\lambda_2^{2}}
\}$ and $\{0,0,0,0,\lambda_0^{2}, \lambda_1^{2}+\lambda_2^{2},\\ \lambda_0\sqrt{\lambda_1^{2}+\lambda_2^{2}}, -\lambda_0\sqrt{\lambda_1^{2}+\lambda_2^{2}}\}$ respectively. Following the same procedure, we can obtain the minimum most eigenvalue of $(\rho_{ABC}^{W})^{T_B}$ and $(\rho_{ABC}^{W})^{T_C}$ respectively as
\begin{eqnarray}
\lambda_{min}[(\rho_{ABC}^{W})^{T_B}]=\lambda_{min}[(\rho_{ABC}^{W})^{T_C}]=-0.4999
\label{minmostBCeigenvaluew}
\end{eqnarray}
Since the minimum most eigenvalue of the partial transposed state with respect to the qubits $A$, $B$ and $C$ are same so we can denote it by $\lambda_{min}[{\rho_{ABC}^{W}}]$. Thus, we have $\lambda_{min}^{\rho^{T_{A}}_{W}}=\lambda_{min}^{\rho^{T_{B}}_{W}}=\lambda_{min}^{\rho^{T_{C}}_{W}}=\lambda_{min}^{\rho_{W}}$.

\section{Structural physical approximation of partial transposition (SPA-PT) of a single qubit in a three-qubit system}
%Since the partial transposition map $T\otimes I\otimes I$ is not completely positive so we can approximate the partial transposition map in such a way that the resulting map will be completely positive.
\noindent Let us consider a map that may be defined as the convex combination of the depolarizing map and the partial transposition map. We can mix the depolarizing map with the partial transposition map with respect to any one of the qubits in such a way that the resulting map is completely positive. Let us now start our discussion on SPA-PT of the three-qubit when the partial transposition operation has been performed with respect to the qubit $A$. Therefore, the newly constructed SPA-PT map, when partial transposition is taken with respect to the qubit $A$ is denoted by $\widetilde{[T\otimes I\otimes I]}$.\\
After applying the SPA-PT map on the qubit $A$ of the three-qubit state $\rho_{ABC}$, the state transformed as
\begin{eqnarray}
\widetilde{[T\otimes I\otimes I]}\rho_{ABC}\equiv \widetilde{\rho_{ABC}^{T_{A}}}&=&\frac{p_{A}}{8}(I\otimes I\otimes I)+(1-p_{A})[T\otimes I\otimes I](\rho_{ABC})
\label{SPA-PT1}
\end{eqnarray}
where $0\leq p_{A} \leq 1$.\\
In a similar way, SPA-PT with respect to the qubit $B$ and $C$ respectively transformed the state $\rho_{ABC}$ as
\begin{eqnarray}
\widetilde{[I\otimes T\otimes I]}\rho_{ABC}\equiv \widetilde{\rho_{ABC}^{T_{B}}}&=&\frac{p_{B}}{8}(I\otimes I\otimes I)+(1-p_{B})[I\otimes T\otimes I](\rho_{ABC})
\label{SPA-PT2}
\end{eqnarray}
\begin{eqnarray}
\widetilde{[I\otimes I\otimes T]}\rho_{ABC}\equiv \widetilde{\rho_{ABC}^{T_{C}}}&=&\frac{p_{C}}{8}(I\otimes I\otimes I)+(1-p_{C})[I\otimes I\otimes T](\rho_{ABC})
\label{SPA-PT3}
\end{eqnarray}
where $0\leq p_{B},p_{C} \leq 1$.\\
\subsection{When Structural Physical Approximation Map will be Completely Positive?}
\noindent In this section, we derive the condition for which the SPA-PT map is completely positive. To start deducing the condition, we first consider the approximation of partial transposition operation with respect to the qubit $A$. In a similar fashion, one can deduce the same condition by approximating the partial transposition with respect to the other two qubits $B$ and $C$ respectively. \\
We say that the SPA-PT map with respect to the qubit $A$ is positive if $\lambda_{min}(\widetilde{\rho_{ABC}^{T_{A}}})\geq 0$ holds. Therefore, using (\ref{SPA-PT1}), we can write the expression of minimum eigenvalue of the operator $\widetilde{\rho_{ABC}^{T_{A}}}$ as
\begin{eqnarray}
\lambda_{min}(\widetilde{\rho_{ABC}^{T_{A}}})&=&\lambda_{min}[\frac{p_{A}}{8}(I\otimes I\otimes I)+(1-p_{A})(T\otimes I\otimes I)\rho_{ABC}]
\label{mineigvalA}
\end{eqnarray}
Further, the R.H.S of (\ref{mineigvalA}) can be reduced using Weyl's inequality as
\begin{eqnarray}
\lambda_{min}(\widetilde{\rho_{ABC}^{T_{A}}}) &\geq& \frac{p_{A}}{8}\lambda_{min}(I\otimes I\otimes I)+(1-p_{A})\lambda_{min}[(T\otimes I\otimes I)\rho_{ABC}]
\label{ineq1}
\end{eqnarray}
If $\lambda_{min}[(T\otimes I\otimes I)\rho_{ABC}]\equiv\lambda_{min}(\rho_{ABC}^{T_{A}})\geq 0$, then the above inequality (\ref{ineq1}) reduces to
\begin{eqnarray}
\lambda_{min}(\widetilde{\rho_{ABC}^{T_{A}}}) &\geq& \frac{p_{A}}{8}
\label{SPA2}
\end{eqnarray}
Now, our task is to find out the minimum value of $p_{A}$ for which the operator
$\widetilde{(T\otimes I\otimes I)}$ will be completely positive. Since the partial transposition operator is not a completely positive operator so the induced map $[(I \otimes I \otimes I) \otimes (T \otimes I \otimes I)]$ generates at least one negative eigenvalue. The most negative eigenvalue generated when the induced map $[(I \otimes I \otimes I) \otimes (T \otimes I \otimes I)]$ is applying on the state $[(I \otimes I \otimes I) \otimes |GHZ\rangle_{ABC}]$, where $|GHZ\rangle_{ABC}=\frac{1}{\sqrt{2}}(|000\rangle+|111\rangle)$. Thus, if we suitably choose the minimum value of $p_{A}$ for which the positive eigenvalues of the maximally mixed three-qubit state generated by the depolarizing map dominate over the minimum most negative eigenvalue generated by the induced map then we can make the operator $\widetilde{(T\otimes I\otimes I)}$ completely positive. Therefore, the map $\widetilde{(T\otimes I\otimes I)}$ is completely positive and hence physically implementable when
\begin{eqnarray}
p_{A}\geq \frac{4}{5}
\label{cond1}
\end{eqnarray}
In a similar way, it can be shown that if we take the partial transposition with respect to system $B$ and $C$ then the SPA-PT map will be completely positive when
\begin{eqnarray}
p_{B}\geq \frac{4}{5}
\label{cond2}
\end{eqnarray}
\begin{eqnarray}
p_{C}\geq \frac{4}{5}
\label{cond3}
\end{eqnarray}
\subsection{Determination of the matrix elements of the density matrix after SPA-PT operation}
\noindent In this section, we study how the entries of the approximated partial transposed density matrix denoted by $\widetilde{\varrho_{ABC}^{T_A}}$ are related with the entries of the original matrix described by the density matrix $\varrho_{ABC}$. If we have an arbitrary three-qubit state described by the density operator $\varrho_{ABC}$ then after the application of SPA-PT operation with respect to qubit A on it, the density matrix has been changed and changes to $\widetilde{\rho_{ABC}^{T_A}}$. As a consequence, the elements of the matrix $\widetilde{\rho_{ABC}^{T_A}}$ can be expressed in terms of the matrix elements of $\varrho_{ABC}$. Thus, the determination of the matrix elements of the density matrix $\widetilde{\rho_{ABC}^{T_A}}$ is important because the entanglement properties of $\varrho_{ABC}$ can be studied using the matrix elements of $\widetilde{\rho_{ABC}^{T_A}}$.
To start with, we consider an arbitrary three-qubit quantum state described by the density matrix $\varrho_{ABC}$, which is given by
\begin{eqnarray}
\varrho_{ABC}=
\begin{pmatrix}
t_{11} & t_{12} & t_{13} & t_{14} & t_{15} & t_{16} & t_{17} & t_{18} \\
t_{12}^{*} & t_{22} & t_{23} & t_{24} & t_{25} & t_{26} & t_{27} & t_{28} \\
t_{13}^{*} & t_{23}^{*} & t_{33} & t_{34} & t_{35} & t_{36} & t_{37} & t_{38} \\
t_{14}^{*} & t_{24}^{*} & t_{34}^{*} & t_{44} & t_{45} & t_{46} & t_{47} & t_{48}\\
t_{15}^{*} & t_{25}^{*} & t_{35}^{*} & t_{45}^{*} & t_{55} & t_{56} & t_{57} & t_{58}\\
t_{16}^{*} & t_{26}^{*} & t_{36}^{*} & t_{46}^{*} & t_{56}^{*} & t_{66} & t_{67} & t_{68}\\
t_{17}^{*} & t_{27}^{*} & t_{37}^{*} & t_{47}^{*} & t_{57}^{*} & t_{67}^{*} & t_{77} & t_{78}\\
t_{18}^{*} & t_{28}^{*} & t_{38}^{*} & t_{48}^{*} & t_{58}^{*} & t_{68}^{*} & t_{78}^{*} & t_{88}
\end{pmatrix}, \sum_{i=1}^{8}t_{ii}=1
\label{threequbitstate}
\end{eqnarray}
where $(*)$ denotes the complex conjugate.\\\\
The SPA-PT with respect to qubit $A$ of a three-qubit quantum state $\varrho_{ABC}$ is given by
\begin{eqnarray}
\widetilde{\rho_{ABC}^{T_{A}}}=[\frac{1}{10}(I\otimes I\otimes I)+\frac{1}{5}(T\otimes I\otimes I)\rho_{ABC}]
\label{spaptthreequbit1}
\end{eqnarray}
where $T$ denotes the transposition operator acting on qubit $A$.\\
The matrix representation of $\widetilde{\rho_{ABC}^{T_{A}}}$ is given by
\begin{eqnarray}
\widetilde{\rho_{ABC}^{T_{A}}}=
\begin{pmatrix}
\tilde{t}_{11} & \tilde{t}_{12} & \tilde{t}_{13} & \tilde{t}_{14} & \tilde{t}_{15} & \tilde{t}_{16} & \tilde{t}_{17} & \tilde{t}_{18}\\
\tilde{t}_{12}^{*} & \tilde{t}_{22} & \tilde{t}_{23} & \tilde{t}_{24} & \tilde{t}_{25} & \tilde{t}_{26}  & \tilde{t}_{27} & \tilde{t}_{28} \\
\tilde{t}_{13}^{*} & \tilde{t}_{23}^{*} & \tilde{t}_{33} & \tilde{t}_{34} & \tilde{t}_{35} & \tilde{t}_{36}  & \tilde{t}_{37} & \tilde{t}_{38} \\
\tilde{t}_{14}^{*} & \tilde{t}_{24}^{*} & \tilde{t}_{34}^{*} & \tilde{t}_{44} & \tilde{t}_{45} & \tilde{t}_{46}  & \tilde{t}_{47} & \tilde{t}_{48}\\
\tilde{t}_{15}^{*} & \tilde{t}_{25}^{*} & \tilde{t}_{35}^{*} & \tilde{t}_{45}^{*} & \tilde{t}_{55} & \tilde{t}_{56}  & \tilde{t}_{57} & \tilde{t}_{58}\\
\tilde{t}_{16}^{*} & \tilde{t}_{26}^{*} & \tilde{t}_{36}^{*} & \tilde{t}_{46}^{*} & \tilde{t}_{56}^{*} & \tilde{t}_{66}  & \tilde{t}_{67} & \tilde{t}_{68}\\
\tilde{t}_{17}^{*} & \tilde{t}_{27}^{*} & \tilde{t}_{37}^{*} & \tilde{t}_{47}^{*} & \tilde{t}_{57}^{*} & \tilde{t}_{67}^{*}  & \tilde{t}_{77} & \tilde{t}_{78}\\
\tilde{t}_{18}^{*} & \tilde{t}_{28}^{*} & \tilde{t}_{38}^{*} & \tilde{t}_{48}^{*} & \tilde{t}_{58}^{*} & \tilde{t}_{68}^{*}  & \tilde{t}_{78}^{*} & \tilde{t}_{88}\\
\end{pmatrix}, \sum_{i=1}^{8}\tilde{t}_{ii}=1
\label{qutrit-qubit2}
\end{eqnarray}
%For three qubit system, we have d=2 and $p=\frac{4}{5}$. From (\ref{spaptthreequbit}), SPA-PT of a three qubit quantum system can be defined as
%\begin{eqnarray}
%\widetilde{\varrho_{123}}=[\frac{1}{10}(I\otimes I\otimes I)+\frac{1}{5}(T\otimes I\otimes I)\rho_{123}]
%\label{spaptthreequbit1}
%\end{eqnarray}
where the entries of the density matrix $\widetilde{\rho^{T_{A}}}$ are given by,
\begin{eqnarray}
&&\tilde{t}_{11}=\frac{1}{10}+\frac{t_{11}}{5}, \tilde{t}_{12}=\frac{t_{12}}{5}, \tilde{t}_{13}=\frac{t_{13}}{5}, \tilde{t}_{14}=\frac{t_{14}}{5}, \tilde{t}_{15}=\frac{t_{15}^{*}}{5}, \tilde{t}_{16}=\frac{t_{25}^{*}}{5}, \tilde{t}_{17}=\frac{t_{35}^{*}}{5}, \tilde{t}_{18}=\frac{t_{45}^{*}}{5}\nonumber\\&& \tilde{t}_{22}=\frac{1}{10}+\frac{t_{22}}{5}, \tilde{t}_{23}=\frac{t_{23}}{5}, \tilde{t}_{24}=\frac{t_{24}}{5}, \tilde{t}_{25}=\frac{t_{16}^{*}}{5}, \tilde{t}_{26}=\frac{t_{26}^{*}}{5}, \tilde{t}_{27}=\frac{t_{36}^{*}}{5}, \tilde{t}_{28}=\frac{t_{46}^{*}}{5}, \tilde{t}_{33}=\frac{1}{10}+\frac{t_{33}}{5}\nonumber\\&& \tilde{t}_{34}=\frac{t_{34}}{5}, \tilde{t}_{35}=\frac{t_{17}^{*}}{5}, \tilde{t}_{36}=\frac{t_{27}^{*}}{5}, \tilde{t}_{37}=\frac{t_{37}^{*}}{5}, \tilde{t}_{38}=\frac{t_{47}^{*}}{5}, \tilde{t}_{44}=\frac{1}{10}+\frac{t_{44}}{5}, \tilde{t}_{45}=\frac{t_{18}^{*}}{5}, \tilde{t}_{46}=\frac{t_{28}^{*}}{5}\nonumber\\&& \tilde{t}_{47}=\frac{t_{38}^{*}}{5}, \tilde{t}_{48}=\frac{t_{48}^{*}}{5}, \tilde{t}_{55}=\frac{1}{10}+\frac{t_{55}}{5},  \tilde{t}_{56}=\frac{t_{56}}{5}, \tilde{t}_{57}=\frac{t_{57}}{5}, \tilde{t}_{58}=\frac{t_{58}}{5}, \tilde{t}_{66}=\frac{1}{10}+\frac{t_{66}}{5},  \tilde{t}_{67}=\frac{t_{67}}{5}\nonumber\\&& \tilde{t}_{68}=\frac{t_{68}}{5}, \tilde{t}_{77}=\frac{1}{10}+\frac{t_{77}}{5},  \tilde{t}_{78}=\frac{t_{78}}{5}, \tilde{t}_{88}=\frac{1}{10}+\frac{t_{88}}{5}
\label{spa1}
\end{eqnarray}
Following the same procedure, one can determine the matrix elements of the density matrix resulting from the application of completely positive maps $\widetilde{[I\otimes T\otimes I]}\rho_{ABC}$ and $\widetilde{[I\otimes I\otimes T]}\rho_{ABC}$ respectively.\\
In the next section, we will show that the minimum eigenvalue of $\widetilde{\rho_{ABC}^{T_{A}}}$, $\widetilde{\rho_{ABC}^{T_{B}}}$ and $\widetilde{\rho_{ABC}^{T_{C}}}$ is the entity that may detect whether the given three-qubit state $\varrho_{ABC}$ possess the property of entanglement or not, so, it is very essential to extract the information about the entries of the matrix $\widetilde{\rho_{ABC}^{T_{A}}}$, $\widetilde{\rho_{ABC}^{T_{B}}}$ and $\widetilde{\rho_{ABC}^{T_{C}}}$. Thus the matrix elements given by (\ref{spa1}) play a vital role in detecting the entanglement of a three-qubit system when the SPA-PT operation is performed with respect to the system $A$.

\section{Necessary condition for the separability (either in the form of a full separability or biseparability) of a three-qubit state}
\noindent In this section, we will derive the necessary condition for the full separability and biseparability of a three-qubit state. Thus, if any three-qubit state violates the necessary condition then we can infer that the given three-qubit state is a genuine entangled state.\\
To move forward in this direction, we consider any three-qubit state shared between three distant parties $Alice (A)$, $Bob (B)$, and $Charlie (C)$ and ask whether the shared state is entangled or not.
%If it comes out to be entangled then one may ask: what type of entanglement (biseparable or genuine) is contained in the shared state?
To detect the entanglement in a three-qubit system, one may follow the partial transposition criterion and thus apply partial transposition operation on any one of the qubits of the given three-qubit system. To overcome the difficulty of the real implementation of the partial transposition map in an experiment, we approximate the partial transposition operation by the method of SPA. We have already shown in the previous section that the SPA-PT map can serve as a completely positive map and thus can be implemented in a real experimental setup. Now we are in a position to give the statement of a necessary condition of the separability and biseparability of a three-qubit state.\\
\textbf{Theorem-1:} If the state described by the density operator $\rho_{ABC}$ denoting either a separable state of the form $A-B-C$ or a biseparable state of the form $A-BC$ then the following inequality is satisfied
\begin{eqnarray}
\lambda_{min}(\widetilde{\rho_{ABC}^{T_{A}}}) &\geq& \frac{1}{10}
\label{finalcond1}
\end{eqnarray}
\textbf{Proof:} The required inequality (\ref{finalcond1}) follows from (\ref{SPA2}) and (\ref{cond1}).\\
\textbf{Theorem-2:} If the state described by the density operator $\rho_{ABC}$ denoting either a separable state of the form $A-B-C$ or a biseparable state of the form $B-AC$ then the following inequality is satisfied
\begin{eqnarray}
\lambda_{min}(\widetilde{\rho_{ABC}^{T_{B}}}) &\geq& \frac{1}{10}
\label{finalcond2}
\end{eqnarray}
\textbf{Theorem-3:} If the state described by the density operator $\rho_{ABC}$ denoting either a separable state of the form $A-B-C$ or a biseparable state of the form $C-AB$ then the following inequality is satisfied
\begin{eqnarray}
\lambda_{min}(\widetilde{\rho_{ABC}^{T_{C}}}) &\geq& \frac{1}{10}
\label{finalcond3}
\end{eqnarray}
Let us now provide a few results that may help in classifying the given three-qubit state $\rho_{ABC}$ as either a separable state, a biseparable state, or a genuine entangled state.
To do this task, we assume that $\lambda_{min}(\widetilde{\rho_{ABC}})=max\{\lambda_{min}(\widetilde{\rho_{ABC}^{T_{A}}}), \lambda_{min}(\widetilde{\rho_{ABC}^{T_{B}}}), \lambda_{min}(\widetilde{\rho_{ABC}^{T_{C}}})\}$.\\
\textbf{Result-1:} If $\lambda_{min}(\widetilde{\rho_{ABC}}) <\frac{1}{10}$, then $\rho_{ABC}$ is a genuine entangled state.\\
\textbf{Result-2:} If $\lambda_{min}(\widetilde{\rho_{ABC}^{T_{A}}})\geq\frac{1}{10}$ and either $\lambda_{min}(\widetilde{\rho_{ABC}^{T_{B}}})<\frac{1}{10}$ or  $\lambda_{min}(\widetilde{\rho_{ABC}^{T_{C}}})<\frac{1}{10}$ or both $\lambda_{min}(\widetilde{\rho_{ABC}^{T_{B}}}),\lambda_{min}(\widetilde{\rho_{ABC}^{T_{C}}})<\frac{1}{10}$ holds, then $\rho_{ABC}$ is biseparable in $A-BC$ cut.\\
\textbf{Result-3:} If $\lambda_{min}(\widetilde{\rho_{ABC}^{T_{B}}})\geq\frac{1}{10}$ and either $\lambda_{min}(\widetilde{\rho_{ABC}^{T_{A}}})<\frac{1}{10}$ or  $\lambda_{min}(\widetilde{\rho_{ABC}^{T_{C}}})<\frac{1}{10}$ or both $\lambda_{min}(\widetilde{\rho_{ABC}^{T_{A}}}),\lambda_{min}(\widetilde{\rho_{ABC}^{T_{C}}})<\frac{1}{10}$ holds, then $\rho_{ABC}$ is biseparable in $B-AC$ cut.\\
\textbf{Result-4:} If $\lambda_{min}(\widetilde{\rho_{ABC}^{T_{C}}})\geq\frac{1}{10}$ and either $\lambda_{min}(\widetilde{\rho_{ABC}^{T_{A}}})<\frac{1}{10}$ or  $\lambda_{min}(\widetilde{\rho_{ABC}^{T_{B}}})<\frac{1}{10}$ or both $\lambda_{min}(\widetilde{\rho_{ABC}^{T_{A}}}),\lambda_{min}(\widetilde{\rho_{ABC}^{T_{B}}})<\frac{1}{10}$ holds, then $\rho_{ABC}$ is biseparable in $C-AB$ cut.\\
\textbf{Result-5:} If $\lambda_{min}(\widetilde{\rho_{ABC}^{T_{A}}})\geq\frac{1}{10}$, $\lambda_{min}(\widetilde{\rho_{ABC}^{T_{B}}})\geq \frac{1}{10}$ and  $\lambda_{min}(\widetilde{\rho_{ABC}^{T_{C}}}) \geq \frac{1}{10}$ holds, then $\rho_{ABC}$ is a fully separable state.\\

\section{A Few Examples}
\noindent In this section, we discuss a few examples of three-qubit genuine entangled states and three-qubit biseparable states that can be detected by the results given in the previous section.
\subsection{Genuine Entangled States}
\textbf{Example-1:} Let us consider the state $|\psi_{G_{1}}\rangle$ described by the density operator $\rho_{G_{1}}=|\psi_{G_{1}}\rangle\langle \psi_{G_{1}}|$, where $|\psi_{G_{1}}\rangle =\alpha|000\rangle +\beta|111\rangle,~~ |\alpha|^2+|\beta|^2=1$. We now proceed to calculate the minimum eigenvalue of $\widetilde{\rho_{G_{1}}^{T_{A}}}$, $\widetilde{\rho_{G_{1}}^{T_{B}}}$ and $\widetilde{\rho_{G_{1}}^{T_{C}}}$.\\
The eigenvalues are given by
\begin{eqnarray}
\lambda_{min}(\widetilde{\rho_{G_{1}}^{T_{A}}})=\lambda_{min}(\widetilde{\rho_{G_{1}}^{T_{B}}})=\lambda_{min}(\widetilde{\rho_{G_{1}}^{T_{C}}})&\equiv&
\lambda_{min}(\widetilde{\rho_{G_{1}}})\nonumber\\&=&\frac{1-2\alpha\beta}{10}
\label{ex1}
\end{eqnarray}
Thus, we can easily find that $\lambda_{min}(\widetilde{\rho_{G_{1}}})<\frac{1}{10}$ when $\alpha\beta>0$. Thus it satisfies Result-1 and hence, we can say that the state $|\psi_{G_{1}}\rangle$ represent genuine entangled state when $\alpha\beta>0$.\\

\noindent \textbf{Example-2:}  Let us consider a pure three-qubit state which is given by
\begin{eqnarray}
|\psi_{G_{2}}\rangle =\frac{1}{\sqrt{5}}[|000\rangle+|100\rangle+|101\rangle+|110\rangle+|111\rangle]
\label{ex2}
\end{eqnarray}
For the given state $|\psi_{G_{2}}\rangle$, we have
\begin{eqnarray}
&&\lambda_{min}(\widetilde{|\psi_{G_{2}}\rangle^{T_{A}}\langle\psi_{G_{2}}|})=0.030718,
\lambda_{min}(\widetilde{|\psi_{G_{2}}\rangle^{T_{B}}\langle\psi_{G_{2}}|})=\nonumber\\&&
\lambda_{min}(\widetilde{|\psi_{G_{2}}\rangle^{T_{C}}\langle\psi_{G_{2}}|})=0.0434315
\label{ex21}
\end{eqnarray}
Therefore, we find that $\lambda_{min}(\widetilde{|\psi_{G_{2}}\rangle\langle\psi_{G_{2}}|})=max\{0.030718,0.0434315\}=0.0434315$. Thus, $\lambda_{min}(\widetilde{|\psi_{G_{2}}\rangle\langle\psi_{G_{2}}|})<\frac{1}{10}$. Hence, the given state $|\psi_{G_{2}}\rangle\langle\psi_{G_{2}}|$ is a genuine entangled state.\\\\
\textbf{Example-3:}  Let us take another state defined by $\rho_{G_{3}}=|\psi_{G_{3}}\rangle\langle \psi_{G_{3}}|$, where $|\psi_{G_{3}}\rangle =\lambda_0|000\rangle+\lambda_1|100\rangle +\lambda_2|111\rangle,0 \leq \lambda_i \leq 1, \sum_{i=0}^{2}\lambda_i^{2}=1$.
For the given state $\rho_{G_{3}}$, we can easily verify that for all values of $\lambda_0$, $\lambda_1$ and $\lambda_2$ lying between 0 and 1, we have, $\lambda_{min}(\widetilde{\rho_{G_{3}}})<\frac{1}{10}$. Further, we have calculated the values of $\lambda_{min}(\widetilde{\rho_{G_{3}}})$ by taking some values of $\lambda_0$, $\lambda_1$ and $\lambda_2$ and those values are tabulated in the Table 5.1 for the verification of our result. Thus from Result-1, the given state $\rho_{G_{3}}$ is a genuine three-qubit entangled state.
\begin{table}
	\begin{center}
		\caption{Table varifying Result-1 for $\rho_{G_3}$ for differnt values of the state parameters $(\lambda_0, \lambda_1, \lambda_2)$}
		\begin{tabular}{|c|c|c|c|}\hline
			State parameter & Minimum  & Minimum  & $\lambda_{min}(\widetilde{\rho_{G_{3}}})$\\ $(\lambda_0, \lambda_1, \lambda_2)$ & eigenvalue of &  eigenvalue of   & \\ & SPA-PT state w.r.t  & SPA-PT state w.r.t  & \\ & qubit $A$ and $C$ &  qubit $B$   & \\ & $\lambda_{min}(\widetilde{\rho_{G_{3}}^{T_{A}}})$,$\lambda_{min}(\widetilde{\rho_{G_{3}}^{T_{C}}})$ & $\lambda_{min}(\widetilde{\rho_{G_{3}}^{T_{B}}})$& \\   \hline
			(0.7, 0.1, 0.707107)  & 0.00101 & $1.295\times 10^{-18}$ & 0.00101\\\hline
			(0.3,0.4,0.866)  & 0.048 & 0.0134 & 0.048\\\hline
			(0.7,0.3, 0.648) & 0.0093 & 0.0013 & 0.0093\\\hline
			(0.1, 0.2, 0.9747) & 0.0805 & 0.056 & 0.0805\\\hline
			(0.2, 0.4, 0.8944) & 0.0642 & 0.02 &  0.0642\\\hline
		\end{tabular}
	\end{center}
\end{table}\\ \ \\
\textbf{Example-4}  Consider the state defined by $\rho_{GHZ,W}=q|GHZ\rangle \langle GHZ|+(1-q)|W\rangle \langle W|, ~~0 \leq q \leq 1$, where $|GHZ\rangle=\frac{1}{\sqrt{2}}[|000\rangle+|111\rangle]$ and $|W\rangle =\frac{1}{\sqrt{3}}[|001\rangle+|010\rangle+|100\rangle]$. For the given state described by the density operator $\rho_{GHZ, W}$, the minimum eigenvalues are given by
\begin{eqnarray}
\lambda_{min}(\widetilde{\rho_{GHZ,W}^{T_{A}}})&=&\lambda_{min}(\widetilde{\rho_{GHZ,W}^{T_{B}}})=\lambda_{min}(\widetilde{\rho_{GHZ,W}^{T_{C}}})
\nonumber\\&=&min\{Q_{1},Q_{2}\}
\end{eqnarray}
where $Q_{1}=\frac{1}{30}(4-q-\sqrt{1-2q+10q^2})$, and $Q_{2}=\frac{1}{60}(6+3q-\sqrt{32-64q+41q^2})$. It can be easily seen that $min\{Q_{1},Q_{2}\}<\frac{1}{10}$ and thus, we have $\lambda_{min}(\widetilde{\rho_{GHZ,W}})<\frac{1}{10}$. Hence from Result-1, We can say that the given state $\rho_{GHZ,W}$ is a genuine entangled state for all $q \in [0,1]$.
\subsection{Biseparable states}
\textbf{Example-1:}  Consider the state defined by the density matrix $\rho_{B_{1}}=q|0\rangle\langle 0| \otimes |\phi^{+}\rangle \langle \phi^{+}|+(1-q)|1\rangle \langle 1|\otimes |\phi^{-}\rangle \langle \phi^{-}|, ~~0 \leq q \leq 1$, where $|\phi^{\pm}\rangle=\frac{1}{\sqrt{2}}[|00\rangle \pm |11\rangle]$. For the given state $\rho_{B_{1}}$, the minimum eigenvalues of the partial transposed state are given by $\lambda_{min}(\widetilde{\rho_{B_{1}}^{T_{A}}})=\frac{1}{10},\lambda_{min}(\widetilde{\rho_{B_{1}}^{T_{B}}})=\lambda_{min}(\widetilde{\rho_{B_{1}}^{T_{C}}})=\min\{\frac{q}{10}, \frac{1-q}{10}\}$. When the state parameter $q$ satisfies the inequality $0 \leq q \leq 1$, we observe that the minimum eigenvalue satisfy
\begin{eqnarray}
\lambda_{min}(\widetilde{\rho_{B_{1}}^{T_{A}}})=\frac{1}{10},~~\lambda_{min}(\widetilde{\rho_{B_{1}}^{T_{B}}})=
\lambda_{min}(\widetilde{\rho_{B_{1}}^{T_{C}}})<\frac{1}{10}
\end{eqnarray}
Therefore, we can infer from Result-2, that the given state $\rho_{B_{1}}$ is biseparable in $A-BC$ cut.\\
\textbf{Example-2}  Let us take a pure state, which is defined by $\rho_{B_{2}}=|\psi\rangle_{B_{2}}\langle \psi|$, where $|\psi\rangle_{B_{2}} =\lambda_0|001\rangle+\lambda_1|101\rangle +\lambda_2|111\rangle,~~\sum_{i=0}^{2}\lambda_{i}^{2}=1,~~0 \leq \lambda_{i} \leq 1, (i=0,1,2)$.
We find that for the given state described by the density operator $\rho_{B_{2}}$ that for all values of $\lambda_0, \lambda_1,\lambda_2 \in [0,1]$, we have, $\lambda_{min}(\widetilde{\rho_{B_{2}}^{T_{A}}})<\frac{1}{10}$, $\lambda_{min}(\widetilde{\rho_{B_{2}}^{T_{B}}})<\frac{1}{10}$ and $\lambda_{min}(\widetilde{\rho_{B_{2}}^{T_{C}}})\geq \frac{1}{10}$ . We have constructed Table 5.2 to clarify our result. Thus using Result-4, we can conclude that the given state $\rho_{B_{2}}$ is biseparable in $AB-C$ cut.\\
\begin{table}
	\begin{center}
		\caption{Table varifying Result-4 for $\rho_{B_2}$ different values of the state parameters $(\lambda_0, \lambda_1, \lambda_2)$}
		\begin{tabular}{|c|c|c|c|}\hline
			State parameter & Minimum  & Minimum  & $\lambda_{min}(\widetilde{\rho_{B_{2}}})$\\ $(\lambda_0, \lambda_1, \lambda_2)$ & eigenvalue of &  eigenvalue of   & \\ & SPA-PT state w.r.t  & SPA-PT state w.r.t  & \\ & qubit $A$ and $B$ &  qubit $C$   & \\ & $\lambda_{min}(\widetilde{\rho_{B_{2}}^{T_{A}}})$,$\lambda_{min}(\widetilde{\rho_{B_{2}}^{T_{B}}})$ & $\lambda_{min}(\widetilde{\rho_{B_{2}}^{T_{C}}})$& \\ \hline
			(0.1, 0.4, 0.911)  & 0.0818 & 0.1 & 0.1\\\hline
			(0.2, 0.4, 0.8944)  & 0.0642 & 0.1 & 0.1\\\hline
			(0.6, 0.1, 0.7937) & 0.00475 & 0.1 & 0.1\\\hline
			(0.5, 0.4, 0.7681) & 0.0232 & 0.1 & 0.1\\\hline
		\end{tabular}
	\end{center}
\end{table}
\subsection{Separable States}
\textbf{Example-1:} Let us consider the state also known as Kay state \cite{akay,kourbolagh}, which is defined by
\begin{eqnarray}
\rho_{S_{1}}=\frac{1}{8+8a}
\begin{pmatrix}
4+a & 0 & 0 & 0 & 0 & 0 & 0 & 2 \\
0 & a & 0 & 0 & 0 & 0 & 2 & 0 \\
0 & 0 & a & 0 & 0 & -2 & 0 & 0 \\
0 & 0 & 0 & a & 2 & 0 & 0 & 0\\
0 & 0 & 0 & 2 & a & 0 & 0 & 0\\
0 & 0 & -2 & 0 & 0 & a & 0 & 0\\
0 & 2 & 0 & 0 & 0 & 0 & a & 0\\
2 & 0 & 0 & 0 & 0 & 0 & 0 & 4+a
\end{pmatrix}, ~~a\geq 2
\end{eqnarray}
The state $\rho_{S_{1}}$ is a fully separable for $a\geq 4$ and is a PPT entangled state in all possible partitions for $2 \leq a <2\sqrt{2}$ \cite{akay}.\\
For the given state $\rho_{S_{1}}$, we find that $\lambda_{min}(\widetilde{\rho_{S_{1}}^{T_{A}}})=\lambda_{min}(\widetilde{\rho_{S_{1}}^{T_{B}}})=\lambda_{min}(\widetilde{\rho_{S_{1}}^{T_{C}}})=
\frac{2+5a}{40(1+a)}$. Thus, $\lambda_{min}(\widetilde{\rho_{S_{1}}})=\frac{2+5a}{40(1+a)}$. Hence from Result-5, we conclude that the given state $\rho_{S_{1}}$ is fully separable for $a\geq 4$.\\\\
%It has been found that for $a\geq 4$, $\frac{2+5a}{40(1+a)}>\frac{1}{10}$.
\textbf{Example-2:}  Consider another state defined by $\rho_{S_{2}}=(1-\alpha)|GHZ\rangle \langle GHZ|+\frac{\alpha}{8}I_8, ~~0 \leq \alpha \leq 1$, where $|GHZ\rangle=\frac{1}{\sqrt{2}}[|000\rangle+|111\rangle]$. For the given state $\rho_{S_{2}}$, the minimum eigenvalues of $\rho_{S_{2}}^{T_{A}}$, $\rho_{S_{2}}^{T_{B}}$ and $\rho_{S_{2}}^{T_{C}}$ are given by $\lambda_{min}(\widetilde{\rho_{S_{2}}^{T_{A}}})=\lambda_{min}(\widetilde{\rho_{S_{2}}^{T_{B}}})=\lambda_{min}(\widetilde{\rho_{S_{2}}^{T_{C}}})=
\frac{\alpha+4}{40}$. Thus,  $\lambda_{min}(\widetilde{\rho_{S_{2}}})=\frac{\alpha+4}{40}$. For $0.8 <\alpha  \leq 1$, we can check that $\lambda_{min}(\widetilde{\rho_{S_{2}}})>0.1$. From Result-5, we can say that the given state $\rho_{S_{2}}$ is fully separable for the state parameter $\alpha$ satisfying $0.8 <\alpha  \leq 1$.\\\\
\textbf{Example-3:}  Let us consider the state described by the density operator $\rho_{S_{3}}=q|\psi\rangle \langle \psi|+(1-q)|111\rangle \langle 111|, ~~0 \leq q \leq 1$, where $|\psi\rangle=\frac{1}{\sqrt{2}}[|001\rangle+|101\rangle]$. The minimum eigenvalues of the partial transposed states are given by $\lambda_{min}(\widetilde{\rho_{S_{3}}^{T_{A}}})=\lambda_{min}(\widetilde{\rho_{S_{3}}^{T_{B}}})=\lambda_{min}(\widetilde{\rho_{S_{3}}^{T_{C}}})=
\frac{1}{10}$. Thus, $\lambda_{min}(\widetilde{\rho_{S_{3}}})= \frac{1}{10}$. Therefore, Result-5 tells us that $\rho_{S_{3}}$ is a fully separable state.
\subsection{Genuine/Biseparable/Separable}
\textbf{Example-1}  Let us consider the state defined by $\rho_{1}=q|000\rangle \langle 000|+(1-q)|GHZ\rangle \langle GHZ|, ~~0 \leq q \leq 1$, where $|GHZ\rangle=\frac{1}{\sqrt{2}}[|000\rangle+|111\rangle]$. For the given state $\rho_{1}$, we find that $\lambda_{min}(\widetilde{\rho_{1}^{T_{A}}})=\lambda_{min}(\widetilde{\rho_{1}^{T_{B}}})=\lambda_{min}(\widetilde{\rho_{1}^{T_{C}}})=\frac{q}{10}$. Thus,
\begin{eqnarray}
\lambda_{min}(\widetilde{\rho_{1}})&=&\frac{q}{10},~ 0 \leq q <1 \nonumber\\
&=&\frac{1}{10},~ q=1
\end{eqnarray}
Hence, $\rho_{1}$ is a genuine entangled state for $0\leq q<1$ and fully separable state for $q=1$ .\\
\textbf{Example-2:} Let us consider a mixed state, which is a convex combination of $GHZ$, $W$ and $\widetilde{W}$ state, and it is defined as\cite{jung}
\begin{eqnarray}
\rho_{2}&=&q_1 |GHZ\rangle \langle GHZ|+q_2|W\rangle \langle W|\nonumber\\&+&(1-q_1-q_2)|\widetilde{W} \rangle \langle \widetilde{W}|, 0\leq q_{1},q_{2}\leq 1
\end{eqnarray}
where,
\begin{eqnarray}
|GHZ\rangle&=&\frac{1}{\sqrt{2}}[|000\rangle +|111\rangle]\nonumber\\
|W\rangle&=&\frac{1}{\sqrt{3}}[|001\rangle+|010\rangle +|100\rangle ]\nonumber\\
|\widetilde{W}\rangle&=&\frac{1}{\sqrt{3}}[|110\rangle +|101\rangle +|011\rangle]
\end{eqnarray}
The minimum eigenvalue of SPA-PT of the state $\rho_{2}$ is given by $\lambda_{min}(\widetilde{\rho_{2}^{T_{A}}})=\lambda_{min}(\widetilde{\rho_{2}^{T_{B}}})=\lambda_{min}(\widetilde{\rho_{2}^{T_{C}}})
=\frac{1}{30}(4-q_1-\sqrt{1-2q_1+10q_1^2-4q_2+4q_1q_2+4q_2^{2}})$. If the state parameters $q_{1}$ lying in the range $0.25 \leq q_1 \leq 1$ and $q_2=\frac{1-q_1}{n}$, where $n$ denote a positive integer, then within this range of parameters, the minimum eigenvalues of $\widetilde{\rho_{2}^{T_{A}}}$, $\widetilde{\rho_{2}^{T_{B}}}$ and $\widetilde{\rho_{2}^{T_{C}}}$ satisfying the inequality given by $\lambda_{min}(\widetilde{\rho_{2}^{T_{A}}})=\lambda_{min}(\widetilde{\rho_{2}^{T_{B}}})=\lambda_{min}(\widetilde{\rho_{2}^{T_{C}}})< \frac{1}{10}$. Thus, applying Result-1, we find that $\rho_{2}$ is a genuine entangled state.\\
Now, our task is to classify two inequivalent classes of genuine entangled states by considering different cases.\\
\textbf{Case-I:} If the state parameters $q_{1}$ lying in the range $0.25 \leq q_1 \leq 0.6269$ and $q_2=\frac{1-q_1}{n}$ then it has been shown that the three tangle of $\rho_{2}$ is zero \cite{jung}. Therefore, we find a sub-region in which not only the three-tangle of $\rho_{2}$ vanishes but also the state $\rho_{2}$ is genuinely entangled. Hence, we can conclude that the state described by the density operator $\rho_{2}$ represent a $W$ class of state when the state parameters $q_{1}$ and $q_{2}$ satisfying $0.25 \leq q_1 \leq 0.629,~~q_2=\frac{1-q_1}{n}$.\\
\textbf{Case-II:} If the state parameters $q_{1}$ lying in the range $0.6269 \leq q_1 \leq 1$ and $q_2=\frac{1-q_1}{n}$ then the three tangle of genuine entangled state $\rho_{2}$ is non-zero. Therefore, the state $\rho_{2}$ represent a $GHZ$ class of state when the state parameters $q_{1}$ and $q_{2}$ satisfying $0.6269 < q_1 \leq 1,~~q_2=\frac{1-q_1}{n}$.

\section{Conclusion}
\noindent To conclude, we have used the SPA-PT map to classify a three-qubit system as six SLOCC inequivalent classes. We have started our study by investigating the effect of partial transposition operation on one qubit of a three-qubit system. We have provided a matrix representation of three-qubit partially transposed states, in terms of $2\times 2$ block matrices, when partial transposition operation is performed with respect to the first qubit or the second qubit, or the third qubit. Then, we studied the application of the SPA-PT map on a three-qubit system and explicitly calculated the matrix elements of the matrix corresponding to the SPA-PT of a three-qubit state. Later, we proposed different criteria for the classification of all possible SLOCC inequivalent classes of pure as well as mixed three-qubit states. Since our classification criterion is based on the method of the SPA-PT map, so, we can realize it in an experiment. Thus, using our experimental-friendly criterion, one can classify all possible SLOCC inequivalent classes in a three-qubit system.\\

\begin{center}
	****************
\end{center}

\chapter{Detection and quantification of entanglement}\label{ch5}
\vspace{1cm}
\noindent\hrulefill

\noindent \emph{In this chapter,\;\footnote{This chapter is based on a research paper ``Structured Negativity: A physically realizable measure of entanglement based on structural physical approximation, \emph{Annals of Physics} {\bf 446} 169113 (2022)''} we have studied the problem of detection and quantification of entangled state $\rho$ in $d\otimes d$ dimensional bipartite quantum system by defining a physically realizable quantity, which we named as structured negativity $(N_S(\rho))$. The defined quantity $(N_S(\rho))$ depends on the minimum eigenvalue of the SPA-PT of the given density matrix $\rho$ and the dimension of the system $d$. Later, we proved that the introduced quantity $(N_S(\rho))$ satisfies the properties of a valid entanglement monotone. Thereafter, we established a relationship between negativity and structured negativity. We conjecture that the negativity and structured negativity of $\rho$ coincide when the number of negative eigenvalues of the partially transposed matrix $\rho^{T_B}$ is $\frac{d(d-1)}{2}$}.\\
\noindent\hrulefill

\newpage
%Quantum entanglement \cite{horodecki11} has been considered the most important non-classical feature of quantum information theory. We may realize its usefulness when we think of quantum networks that are based on the non-local feature of the entanglement that may provide the basis of some speculated applications such as large-scale quantum computation and distributed quantum computing \cite{leent}.
\noindent Quantification of entanglement is one of the crucial tasks in quantum information theory. The importance of this problem can be understood if we consider a simple instance in which we study the relationship between the amount of entanglement present in the shared arbitrary dimensional bipartite resource state and the fidelity of teleportation \cite{sazim}. The quantification problem has already been studied for a two-qubit system, bipartite higher dimensional, and multi-qubit system but still there exist a few problems in the higher dimensional bipartite mixed system that need to be addressed. There exist various entanglement measures such as concurrence \cite{wootters1,wootters2,wootters}, negativity\cite{vidal}, relative entropy of entanglement \cite{plenio2}, the geometric measure of entanglement\cite{wei} that can quantify the amount of entanglement in a two-qubit as well as higher dimensional bipartite pure and mixed state. Now, the question arises whether the entanglement measures already existed in the literature can quantify the amount of entanglement for any arbitrary dimensional bipartite system practically.\\
In the case of a two-qubit state, entanglement of formation \cite{wootters1} can be measured without prior state reconstruction \cite{horodecki7}. Also, it has been shown that a single observable is not sufficient to determine the entanglement of a given unknown pure two-qubit state \cite{sancho}, nevertheless, the amount of entanglement in a pure two-qubit state can be determined experimentally with a minimum of two copies of the state \cite{walborn}. The situation will become more complex when we will consider the problem of quantification of entanglement for a higher dimensional bipartite system. For higher dimensional bipartite pure state, there exist some measures of entanglement such as generalized concurrence \cite{rungta2}, negativity\cite{lee}, the geometric measure of entanglement that may quantify the amount of entanglement in the given pure state but on the contrary, we have a handful of measures of entanglement which work for the higher dimensional bipartite mixed state. This is due to the fact that till today we don't have any closed formula for the concurrence of higher dimensional bipartite mixed state. Secondly, an easily computable measure of entanglement, namely, negativity may be used to quantify the amount of entanglement in higher dimensional bipartite pure as well as mixed states but the problem with this measure is that it depends on the negative eigenvalues of the non-physical partial transposition operation. Thus, negativity does not correspond to a completely positive map, and hence, difficult to implement it in the laboratory. Recently, the generalized geometric measure of entanglement has been defined for multipartite mixed states \cite{sen} but it is not yet clear whether it can be a realizable quantity in an experiment or not.\\
Another way of quantification of entanglement is by using witness operators \cite{brandao, guhne} that can be employed for any arbitrary $d_1 \otimes d_2$ dimensional system. Although the witness operators are physically realizable, in general, it is not easy to construct a witness operator for the detection of an entangled state. Thus, we should look for another way of measuring entanglement in a given $d_1\otimes d_2$ dimensional system.\\
Horodecki \cite{horodecki7} has proposed a protocol based on the SPA-PT map, which directly measures the concurrence of $2\otimes 2$  system using four moments only but to calculate the expectation of these four moments, the method needs at most 20 copies of the state. This method is efficient in comparison to quantum state tomography with respect to the estimation of state parameters but on the other hand, this method will show its inefficiency if we compare the number of copies required in the above-mentioned methods. This motivates us to define a new measure of entanglement using the structural physical approximation of partial transposition.\\
In this chapter, we define a physically realizable measure of entanglement which is based on the minimum eigenvalue of the structural physical approximation of partial transposition operation. 
\section{Measures of entanglement: Concurrence and Negativity}
\noindent \textbf{Concurrence:} A very popular measure for the quantification of bipartite quantum correlations is the concurrence\cite{wootters1,wootters2,wootters}. For any two-qubit pure state $|\psi \rangle_{AB}^{{2\otimes 2}}$, it is defined as,
\begin{eqnarray}
C(|\psi \rangle_{AB}^{{2\otimes 2}})=\sqrt{2(1-Tr(\rho_{A}^{2}))}
\end{eqnarray}
where $\rho_{A}$ is the reduced state of $|\psi \rangle_{AB}^{{2\otimes 2}}$.\\
Concurrence for the two-qubit mixed state described by the density operator $\rho$, may be defined as,
\begin{eqnarray}
C(\rho)=max(0,\sqrt{\lambda_{1}}-\sqrt{\lambda_{2}}-\sqrt{\lambda_{3}}-\sqrt{\lambda_{4}})
\end{eqnarray}
where	$\lambda_{i}'s$ are the eigenvalues of $\rho\widetilde{\rho}$, arranged in descending order. Here, $\widetilde{\rho}$=$(\sigma_{y}\otimes\sigma_{y})\rho(\sigma_{y}\otimes\sigma_{y})$.\\
Let $|\psi \rangle_{AB}^{{d_1\otimes d_2}}$ be any vector in $d_1\otimes d_2$ dimensional system. Then, the definition of concurrence for $d_1\otimes d_2$ dimensional bipartite pure quantum system may be generalized as \cite{rungta2},
\begin{eqnarray}
C(|\psi\rangle_{AB}^{d_1\otimes d_2})=\sqrt{2\nu_{d_1}\nu_{d_2}[1-Tr(\rho_{A}^{2})]}
\end{eqnarray}
where $\rho_{A}$ is the reduced state of $|\psi \rangle_{AB}^{d_1\otimes d_2}$ and $\nu_{d_1}$ and $\nu_{d_2}$ are positive constants. Except for the two-qubit system, we don't have any closed formula for the concurrence of bipartite $d_1\otimes d_2$ dimensional mixed state. For higher dimensional mixed states, we can only estimate the amount of entanglement through the lower bound of the concurrence\cite{albevario,mintert2,adhikari3}.

\noindent \textbf{Negativity:} Negativity is another measure of entanglement based on the negative eigenvalues of the partially transposed matrix. It was first introduced as an entanglement measure by Vidal and Werner \cite{vidal}. The negativity for $d\otimes d$ dimensional system described by the density operator $\rho_{AB}$ may be defined as \cite{lee},
\begin{eqnarray}
N(\rho_{AB})=\frac{||\rho_{AB}^{T_B}||_1-1}{d-1}=\frac{2}{d-1}\sum_{\lambda_i<0}{|\lambda_i(\rho_{AB}^{T_B})|}
\label{negativity}
\end{eqnarray}
where $||.||_1$ denotes trace norm and $\rho^{T_B}$ is the partial transposition of the density matrix $\rho_{AB}$ with respect to the subsystem $B$. In the entanglement measure, negativity is useful in comparison to concurrence because unlike concurrence, it depends on the negative eigenvalues of the partially transposed state and thus the exact value of negativity can be calculated very easily even for higher dimensional quantum systems.  Although negativity can be calculated exactly in theory for any arbitrary dimensional system it cannot be implemented in the laboratory. The reason behind this is that the partial transposition operation represents a positive but not a completely positive map. Recently, Bartkiewicz et.al. have studied the problem of experimental implementation of the entanglement measure negativity for a two-qubit system and found a way to calculate the negativity in an experiment by using three experimentally accessible moments of the partially transposed density matrix\cite{norek}. For a two-qubit system, a feasible scheme for experimental detection and quantification of entanglement was introduced using PPT criteria \cite{chimczak}. \\
%Quantification of a two-qubit system has also been defined using partial moments and they have extended their result to characterize entanglement in the case of a multi-qubit system\cite{shen2}.
To get rid of the difficulty of implementing negativity in an experiment, we have defined a new measure of entanglement and named it "structured negativity". We called the introduced measure "structured negativity" because this measure is based on the method of SPA-PT.
\section{Structured Negativity}
\noindent For the sake of completeness, let us recapitulate a few important points about SPA-PT. We start with the partial transposition (PT) map denoted by $I\otimes T$, which corresponds to a positive but not completely positive map, and thus it cannot be implemented in the laboratory. Therefore, we apply SPA-PT operation to implement the PT map in the laboratory.\\
For $d\otimes d$ dimensional system described by the density operator $\rho$, the SPA-PT of the state $\rho$ denoted as $\widetilde{\rho}$ and it may be expressed as \cite{horodecki6},
\begin{equation}
\widetilde{\rho}=\frac{d}{d^3+1}I\otimes I +\frac{1}{d^3+1}[I\otimes T](\rho)
\label{spapt}
\end{equation}
where $I\otimes I$ denote the identity matrix in $d\otimes d$ dimensional system.\\
The state $\rho$ is separable if and only if \cite{horodecki6}
\begin{eqnarray}
\lambda_{min}(\widetilde{\rho}) \geq \frac{d}{d^3+1}
\label{separability}
\end{eqnarray}
where $\lambda_{min}(\widetilde{\rho})$ denote the minimum eigenvalue of $\widetilde{\rho}$. Otherwise, the state $\rho$ is entangled.\\
Now we are in a position to define a new measure of entanglement using the separability criteria given in (\ref{separability}). We may term this new measure of entanglement as structured negativity and it is denoted by $N_S({\rho})$. Therefore, for $d\otimes d$ system, the structured negativity may be defined as,
\begin{eqnarray}
N_S({\rho})=K.max\{\frac{d}{d^3+1}-\lambda_{min}(\widetilde{\rho}),0\}
\label{structured negativity}
\end{eqnarray}
where $K=d(d^3+1)$.\\
\textbf{Lemma 1:} Any $d\otimes d$ system which is described by the density operator $\rho=\sum_k{p_k\rho_k}$, it's SPA-PT is given by,
\begin{eqnarray}
\widetilde{\rho}=\sum_k{p_k\widetilde{\rho_k}}
\end{eqnarray}
\textbf{Proof:} Consider a bipartite state $\rho=\sum_k{p_k\rho_k}$ in $d\otimes d$ dimensional system, then its SPA-PT may be written as
\begin{eqnarray}
\widetilde{\rho}&=&\frac{d}{d^3+1}(I\otimes I)+\frac{1}{d^3+1}(I\otimes T)\rho\nonumber\\
&=&\frac{d}{d^3+1}(I\otimes I)+\frac{1}{d^3+1}(I\otimes T)\sum_k{p_k\rho_k}\nonumber\\
&=&\frac{d}{d^3+1}(I\otimes I)+\sum_{k}{p_k}\frac{1}{d^3+1}(I\otimes T)\rho_k\nonumber\\
&=&\sum_{k}{p_k\frac{d}{d^3+1}}(I\otimes I)+\sum_{k}{p_k}\frac{1}{d^3+1}(I\otimes T)\rho_k\nonumber\\
&=&\sum_{k}{p_k[}\frac{d}{d^3+1}(I\otimes I)+\frac{1}{d^3+1}(I\otimes T)\rho_k]\nonumber\\
&=& \sum_{k}{p_k\widetilde{\rho_k}}
\end{eqnarray}\\
To show $N_S({\rho})$, a valid measure of entanglement, we need to show that it satisfies a few properties.\\ \ \\
\textbf{P1:} $N_S({\rho})$ vanishes if $\rho$ is separable.\\
\textbf{Proof}: In $d\otimes d$ dimensional quantum system, if $\rho$ is separable then $\lambda_{min}(\widetilde{\rho}) \geq \frac{d}{d^3+1}$\cite{horodecki6}. Thus, max$\{\frac{d}{d^3+1}-\lambda_{min}(\widetilde{\rho}),0\}$=0. Hence, $N_S({\rho})=0$.\\ \ \\
\textbf{P2:} $N_S({\rho})$ is invariant under local unitary transformation.\\
\textbf{Proof:}	$N_S({\rho})$ is invariant under a local change of basis since eigenvalues does not change under local change of basis \cite{ziman,anuma}. Thus, $N_S({\rho})=N_S(U_A \otimes U_B {\rho} U_A^{\dagger} \otimes U_B^{\dagger})$ where $U_A$ and $U_B$ denotes the unitaries acting on the subsystem A and B respectively.\\ \ \\
\textbf{P3:} $N_S({\rho})$ satisfies convexity property i.e.
\begin{eqnarray}
N_S(\sum_k{p_k{\rho_k}}) \leq \sum_k{p_k}N_S({\rho_k})
\end{eqnarray}
\textbf{Proof:} Let us consider a quantum state described by the density operator $\rho=\sum_k{p_k{\rho_k}}$, $(0\leq p_k \leq 1)$. The structured negativity of $\rho$ is given by
\begin{eqnarray}
N_S(\rho)=N_S(\sum_k{p_k{\rho_k}})=K[\frac{d}{d^3+1}-\lambda_{min}(\widetilde{\rho})]
\label{convexity}
\end{eqnarray}
Using Lemma 1, RHS of equation (\ref{convexity}) may be re-expressed as
\begin{eqnarray}
N_S(\sum_k{p_k{\rho_k}})=K[\frac{d}{d^3+1}-\lambda_{min}(p_1\widetilde{\rho_1}+\sum_{k \neq 1}{p_k\widetilde{\rho_k}})]
\label{weyl}
\end{eqnarray}
Using Weyl's inequality in (\ref{weyl}), we get
\begin{eqnarray}
N_S(\sum_k{p_k{\rho_k}}) \leq K [\frac{d}{d^3+1}-(\lambda_{min}(p_1\widetilde{\rho_1})+\lambda_{min}(\sum_{k \neq 1}{p_k\widetilde{\rho_k}})))]
\label{eq14}
\end{eqnarray}
Using Weyl's inequality repeatedly (k-1) times, equation (\ref{eq14}) may be re-written as,
\begin{eqnarray}
N_S(\sum_k{p_k{\rho_k}}) &\leq& K [\frac{d}{d^3+1}-\sum_{k}\lambda_{min}({p_k\widetilde{\rho_k}}))]\nonumber\\
&=& K[\sum_{k}{p_k}(\frac{d}{d^3+1}-\lambda_{min}({\widetilde{\rho_k}}))]\nonumber\\
&=& \sum_k{p_k}.K.(\frac{d}{d^3+1}-\lambda_{min}({\widetilde{\rho_k}}))\nonumber\\
&=& \sum_k{p_k}{N_S({\rho_k})}
\end{eqnarray}\\ \ \\
\textbf{P4: }$N_S({\rho})$ does not increase on average under LOCC\cite{plenio} i.e.
\begin{eqnarray}
N_S(\rho) \geq \sum_i{p_i}N_S[(K_i\otimes I)\rho (K_i^{\dagger}\otimes I)]
\label{locc}
\end{eqnarray}
where $K_i$ are the Kraus operators.\\
\textbf{Proof:} Let us consider an entangled state described by a density operator $\rho$. Now, consider the right-hand side of (\ref{locc}) that can be expressed as
\begin{eqnarray}
\sum_i{p_i}N_S[(K_i\otimes I)\rho (K_i^{\dagger}\otimes I)]&=&K\sum_ip_i[\frac{d}{d^3+1}-\lambda_{min}(\widetilde{(K_i\otimes I)\rho (K_i^{\dagger}\otimes I))}]\nonumber\\&=&K[\frac{d}{d^3+1}-\sum_i{p_i\lambda_{min}(\widetilde{(K_i\otimes I)\rho (K_i^{\dagger}\otimes I))}}]\nonumber \\&\geq& K[\frac{d}{d^3+1}-\sum_i{\lambda_{min}(\widetilde{(K_i\otimes I)\rho (K_i^{\dagger}\otimes I))}}]
\label{eq1}
\end{eqnarray}
The first step follows from the definition of structured negativity given in (\ref{structured negativity}). In the third step, the inequality follows from the fact $0\leq p_i\leq 1$ and $\sum_i{p_i}=1$.\\ Using Weyl's inequality, (\ref{eq1}) may be expressed as
\begin{eqnarray}
\sum_i{p_i}N_S[(K_i\otimes I)\rho (K_i^{\dagger}\otimes I)] &\geq&K[\frac{d}{d^3+1}-\lambda_{min}[\sum_i{(\widetilde{(K_i\otimes I)\rho (K_i^{\dagger}\otimes I))}}]\nonumber\\&=& K[\frac{d}{d^3+1}-\lambda_{min}[\sum_i(\frac{d}{d^3+1}I\otimes I\nonumber\\&+&\frac{1}{d^3+1}(I\otimes T)((K_i\otimes I)\rho(K_i^{\dagger}\otimes I)))]]
\label{eq17}
\end{eqnarray}
In (\ref{eq17}), the second step follows from the definition (\ref{spapt}) of SPA-PT  of $(K_i\otimes I)\rho(K_i^{\dagger}\otimes I)$.
Let us assume that the entangled state described by the density operator $\rho$ may be evolved as $\rho^{'}=\sum_{i=1}^{m}((K_i\otimes I)\rho (K_i^{\dagger}\otimes I))$, where $m$ denote the number of Kraus operators.\\
Then, (\ref{eq17}) can be re-expressed as			\begin{eqnarray}
\sum_i{p_i}N_S[(K_i\otimes I)\rho (K_i^{\dagger}\otimes I)]&\geq&K[\frac{d}{d^3+1}-\lambda_{min}[\frac{md}{d^3+1}I\otimes I+\frac{1}{d^3+1}(I\otimes T)\rho^{'}]]\nonumber\\&=&K[\frac{d}{d^3+1}-\lambda_{min}[\frac{md}{d^3+1}I\otimes I+\frac{1}{d^3+1}({\rho^{'}}^{T_B})]]
\label{eq2}
\nonumber \\&\geq&K[\frac{d}{d^3+1}-\lambda_{min}[(\widetilde{\rho^{'}})+\frac{(m-1)d}{d^3+1}I\otimes I] ]
\label{eq3}
\end{eqnarray}
Using the upper bound of Weyl's inequality, (\ref{eq3}) may be expressed as
\begin{eqnarray}
&&\sum_i{p_i}N_S[(K_i\otimes I)\rho (K_i^{\dagger}\otimes I)]\nonumber \\&\geq&K[\frac{d}{d^3+1}-\lambda_{min}(\widetilde{\rho^{'}})-\frac{(m-1)d}{d^3+1}]
\label{eq4}
\end{eqnarray}
To show	$N_S(\rho) \geq \sum_i{p_i}N_S[(K_i\otimes I)\rho (K_i^{\dagger}\otimes I)]$, it is sufficient to show that $N_S(\rho)-\sum_i{p_i}N_S[(K_i\otimes I)\rho (K_i^{\dagger}\otimes I)]\geq0$.\\
Using (\ref{structured negativity}) and(\ref{eq4}), we have
\begin{eqnarray}
&&N_S(\rho)-\sum_i{p_i}N_S[(K_i\otimes I)\rho (K_i^{\dagger}\otimes I)]\nonumber\\&\geq& K[\lambda_{min}(\widetilde{\rho^{'}})+\frac{(m-1)d}{d^3+1}-\lambda_{min}(\widetilde{\rho})]
\label{eq5}
\end{eqnarray}
For an entangled state $\rho$, we have
\begin{eqnarray}
\lambda_{min}(\widetilde{\rho})<\frac{d}{d^3+1}
\label{entangled}
\end{eqnarray}
where $\widetilde{\rho}$ denote the SPA-PT of $\rho$.\\
Using (\ref{entangled}), the inequality (\ref{eq5}) reduces to
\begin{eqnarray}
&&N_S(\rho)-\sum_i{p_i}N_S[(K_i\otimes I)\rho (K_i^{\dagger}\otimes I)]\nonumber\\&\geq& K[\lambda_{min}(\widetilde{\rho^{'}})+\frac{(m-2)d}{d^3+1}]
\end{eqnarray}
Since $m\geq 2$ so, $\lambda_{min}(\widetilde{\rho^{'}})+\frac{(m-2)d}{d^3+1} \geq 0$. Thus
\begin{eqnarray}
&&N_S(\rho)-\sum_i{p_i}N_S[(K_i\otimes I)\rho (K_i^{\dagger}\otimes I)]\geq 0\nonumber\\&\implies&N_S(\rho)\geq \sum_i{p_i}N_S[(K_i\otimes I)\rho (K_i^{\dagger}\otimes I)]
\end{eqnarray}
Hence proved.

\section{Relation between negativity and structured negativity}
In this section, we derive the relationship between the negativity and structured negativity of a given quantum state $\rho$.\\
\textbf{Result-1:} For any quantum state $\rho$ in $d\otimes d$ dimensional system, the relation between negativity and structured negativity is given by
\begin{eqnarray}
N(\rho)\leq 2(1-\frac{1}{d})N_S({\rho})
\label{finalnrho}
\end{eqnarray}
\textbf{Proof:}  For any two Hermitian matrices $A, B \in M_n$, Weyl's inequality may be defined as \cite{horn,tao,adhikari5}
\begin{eqnarray}
\lambda_k(A+B)\leq \lambda_{k+j}(A)+\lambda_{n-j}(B),~~k=\{1,2,..n\},~~ j=\{0,1,...n-1\}
\label{eq18}
\end{eqnarray}
where eigenvalues of the matrices $A$, $B$, and $A+B$ are arranged in increasing order.\\
For k=1, Weyl's inequality reduces to
\begin{eqnarray}
\lambda_{min}(A+B)\leq \lambda_{1+j}(A)+\lambda_{n-j}(B)
\label{min}
\end{eqnarray}
for $j=0,1...n-1$.\\
If $\widetilde{\rho}$ denote the SPA-PT of $\rho \in M_{d^2}$, then by taking $A=\frac{1}{d^3+1}({I\otimes T)}\rho \equiv\frac{1}{d^3+1}\rho^{T_B}$ and $B=\frac{d}{d^3+1}I \otimes I$, (\ref{min}) reduces to
\begin{eqnarray}
\lambda_{min}({\widetilde{\rho}})\leq \lambda_{1+j}(\frac{1}{d^3+1}\rho^{T_B})+\frac{d}{d^3+1}\lambda_{d^2-j}(I\otimes I)
\label{k1}
\end{eqnarray}
for $j=0,1...n-1$.\\
Let us suppose that $(I\otimes T)\rho$ has $q (\leq (d-1)^2)$ number of negative eigenvalues, if $\rho$ is an entangled state. \\
Putting $j=0,1...q-1$ in (\ref{k1}) and adding, we get
\begin{eqnarray}
q.\lambda_{min}(\widetilde{\rho}) \leq \frac{1}{d^3+1}\sum_{i=1,\lambda_i <0}^{q}{\lambda_i(\rho^{T_B})}+\frac{dq}{d^3+1}
\label{eq20}
\end{eqnarray}
Exploiting the definition of negativity given in (\ref{negativity}), the above equation (\ref{eq20}) reduces to
\begin{eqnarray}
\lambda_{min}(\widetilde{\rho}) \leq -\frac{(d-1)N(\rho)}{2q(d^3+1)}+\frac{d}{d^3+1}
\label{nrho}
\end{eqnarray}
If $\rho$ is an entangled state, then the minimum eigenvalue of $\widetilde{\rho}$ (given in (\ref{structured negativity})) may be expressed as,
\begin{eqnarray}
\lambda_{min}({\widetilde{\rho}})=\frac{d^2-N_S({{\rho}})}{d(d^3+1)}
\label{n1rho}
\end{eqnarray}
Substituting value of $\lambda_{min}({\widetilde{\rho}})$ in (\ref{nrho}), we get
\begin{eqnarray}
\frac{d^2-N_S({{\rho}})}{d(d^3+1)} \leq -\frac{(d-1)N(\rho)}{2q(d^3+1)}+\frac{d}{d^3+1}
\end{eqnarray}
After simplification, we get the required relation,
\begin{eqnarray}
N_S({\rho}) \geq \frac{d(d-1)}{2q}N(\rho)
\label{nrhon1rho}
\end{eqnarray}
Thus, the structured negativity of a given quantum state is always greater than or equal to $\frac{d(d-1)}{2q}$ times to its negativity.
The number of negative eigenvalues of $\rho^{T_B}$ for a $d\otimes d$ dimensional system are at most $(d-1)^2$. Substituting $q\leq (d-1)^2$, (\ref{nrhon1rho}) becomes,
\begin{eqnarray}
N_S({\rho})\geq \frac{d}{2(d-1)}N(\rho)
\end{eqnarray}
This implies,
\begin{eqnarray}
N(\rho)\leq 2(1-\frac{1}{d})N_S({\rho})
\label{finalnrho}
\end{eqnarray}
Hence proved.\\
\textbf{Remark:} For sufficiently large d, we have $1-\frac{1}{d} \approx 1$. Thus for higher dimensional system, the inequality (\ref{finalnrho}) reduces to $N(\rho) \leq 2N_S({\rho})$.%We observe that equality achieved in (\ref{finalnrho}) for two-qubit system.\\
\subsection{Examples}
\noindent As we have mentioned in Sec 6.1, there does not exist any closed formula for concurrence of arbitrary dimensional bipartite mixed state, nevertheless, there exists a lower bound of the concurrence for arbitrary $d\otimes d$ dimensional system \cite{mintert2,albevario,adhikari3}. One such lower bound of concurrence obtained in \cite{albevario} which may be given by
\begin{eqnarray}
C(\rho)&\geq& \sqrt{\frac{2}{d(d-1)}}(max(||\rho^{T_B}||,||R(\rho)||)-1)\nonumber\\
&\equiv& C_{lb}(\rho)
\end{eqnarray}
where $C(\rho)$ and $C_{lb}$ respectively denote the concurrence and lower bound of the concurrence of the state $\rho$ in an arbitrary $d\otimes d$ dimensional system. $||.||$ denote the trace norm of $(.)$. $R(\rho)$ and $\rho^{T_B}$ respectively denote the realignment operation and partial transposition operation with respect to subsystem $B$.\\
In this subsection, we will provide a few examples by which we can show that Result-1 is indeed true. For the given state $\rho$, we have compared three measures of entanglement such as negativity $(N(\rho))$, structured negativity $(N_S(\rho))$ and the lower bound of the concurrence $(C_{lb}(\rho))$ and found that structured negativity is always greater than or equals negativity. In a few cases, structured negativity is greater than the lower bound of concurrence. Also, we have observed that for $d\otimes d$ dimensional system, negativity and structured negativity coincides when the number of negative eigenvalues of the partially transposed matrix are $\frac{d(d-1)}{2}$ i.e equality holds in (\ref{nrhon1rho}), when $q=\frac{d(d-1)}{2}$, and strict inequality holds when $q\neq \frac{d(d-1)}{2}$.\\ \ \\
\textbf{Example 1: Two-qubit Werner state}\\
Let us consider a two-qubit Werner state defined as \cite{adhikari5,horodecki3}
\begin{eqnarray}
\rho_W=F|\psi^{-}\rangle \langle \psi^{-}|+(1-F)\frac{I}{4}
\end{eqnarray}
where
\begin{eqnarray}
|\psi^{-}\rangle=\frac{1}{\sqrt{2}}|01\rangle -|10\rangle
\end{eqnarray}
The family of Werner states is the only states invariant under the transformation \cite{horodecki3}
\begin{eqnarray}
\rho_W \longrightarrow U\otimes U \rho_W U^{\dagger} \otimes U^{\dagger}
\end{eqnarray}
where $U$ is a unitary transformation.
The state is entangled for $\frac{1}{3} <F\leq1$. Negativity of $\rho_W$ is $\frac{3F-1}{2}$  and structured negativity, $N_S(\rho_W)=18[\frac{2}{9}+\frac{1}{12}(-3+F)]$. From Fig-1, Result-1 is verified.
\begin{figure}[h]
	\centering
	\includegraphics[scale=0.35]{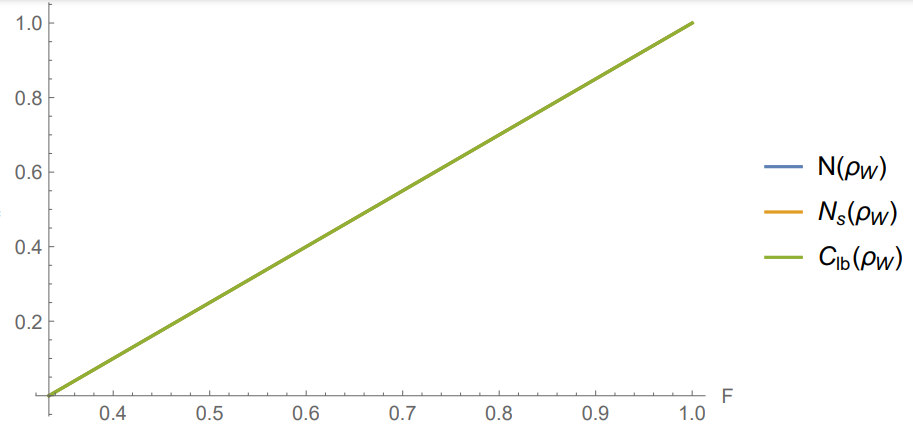}
	\caption{The negativity, structured negativity, and lower bound of concurrence for the two-qubit Werner state coincide with each other. Therefore, equality in (\ref{finalnrho}) has been achieved for a family of Werner state.}
\end{figure}\\ \ \\
\textbf{Example 2: Two-qubit MEMS state}\\
Consider a maximally entangled mixed state (MEMS) introduced by Munro et.al. \cite{munro,adhikari6}
\begin{eqnarray}
\rho_{MEMS}=
\begin{pmatrix}
h(C) & 0 & 0 & \frac{C}{2}\\
0 & 1-2h(C) & 0 & 0\\
0 & 0 & 0 & 0\\
\frac{C}{2} & 0 & 0 & h(C)
\end{pmatrix}
\end{eqnarray}
where,
\begin{eqnarray}
h(C)=
\begin{cases}
\frac{C}{2} & C\geq \frac{2}{3}\\
\frac{1}{3} & C<\frac{2}{3}
\end{cases}
\end{eqnarray}	
where C denotes the concurrence of $\rho_{MEMS}$.\\
For $C\geq \frac{2}{3}$, $N(\rho_{MEMS})=-1+C+\sqrt{1-2C+2C^2}$,  $N_S(\rho_{MEMS})=18(\frac{2}{9}+\frac{1}{18}(-5+C+\sqrt{1-2C+2C^2}))$. From Fig-2, it can be seen that negativity, structured negativity and $C_{lb}(\rho_{MEMS})$ of the state $\rho_{MEMS}$ coincide for $C\geq \frac{2}{3}$.
\begin{figure}[h]
	\centering
	\includegraphics[scale=0.35]{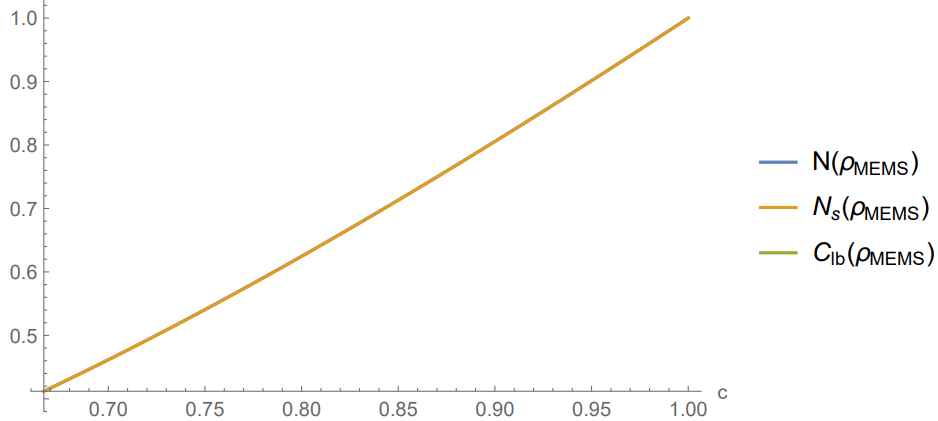}
	\caption{The negativity, structured negativity and lower bound of concurrence coincide for the state  $\rho_{MEMS}$ for $C\geq\frac{2}{3}$.}
	\label{MEMS1}
\end{figure}
Also for $C<\frac{2}{3}$ negativity and structured negativity of $\rho_{MEMS}$ are given by: $N(\rho_{MEMS})=\frac{1}{3}(-1+\sqrt{1+9C^2})$ and $N_S(\rho_{MEMS})=18(\frac{2}{9}+\frac{1}{54}(-13+\sqrt{1+9C^2}))$. From Fig-3, it can be seen that negativity is the same as the structured negativity for $\rho_{MEMS}$($C<\frac{2}{3}$).
\begin{figure}[h]
	\centering
	\includegraphics[scale=0.35]{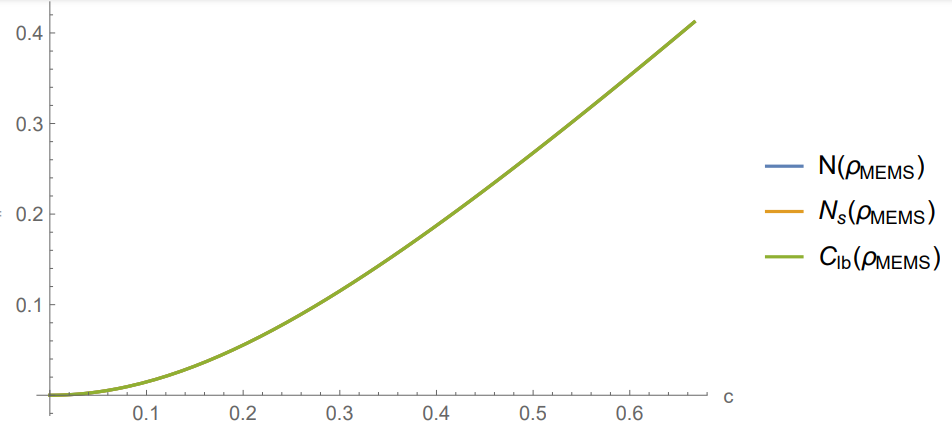}
	\caption{The negativity, structured negativity and lower bound of concurrence coincides for the state  $\rho_{MEMS}$ for $C< \frac{2}{3}$.}
	\label{MEMS1}
\end{figure}\\
\textbf{Example 3: Two-qutrit state described by the density operator $\rho_{a}$}\\
Consider a two-qutrit state defined in \cite{rana}, which is described by the density operator
\begin{eqnarray}
\rho_a=\frac{1}{5+2a^2}\sum_{i=1}^{3}{|\psi_i\rangle \langle \psi_i|},~~\frac{1}{\sqrt{2}}\leq a \leq 1
\end{eqnarray}
where, $|\psi_i\rangle=|0i\rangle-a|i0\rangle$, for $i=\{1,2\}$ and\\ $|\psi_3\rangle=\sum_{i=0}^{n-1}{|ii\rangle}$. \\
For the state $\rho_a$ the negativity $N(\rho_a)$ and the structured negativity $N_S({\rho_a})$ can be calculated as
\begin{eqnarray}
N(\rho_a)&=&\frac{1}{5+2a^2}-\frac{1-2\sqrt{a}}{5+2a^2}\nonumber\\&-&\frac{1+a^2-\sqrt{5-2a^2+a^4}}{5+2a^2}\\
N_S({\rho_a})&=&84[\frac{3}{28}-\frac{7+3a^2}{14(5+2a^2)}]
\end{eqnarray} The comparison between $N(\rho_a)$, $N_S(\rho_a)$, $C_{lb}(\rho_{a})$ for the two-qutrit state $\rho_a$ has been studied in Fig-4.
\begin{figure}[h]
	\centering
	\includegraphics[scale=0.25]{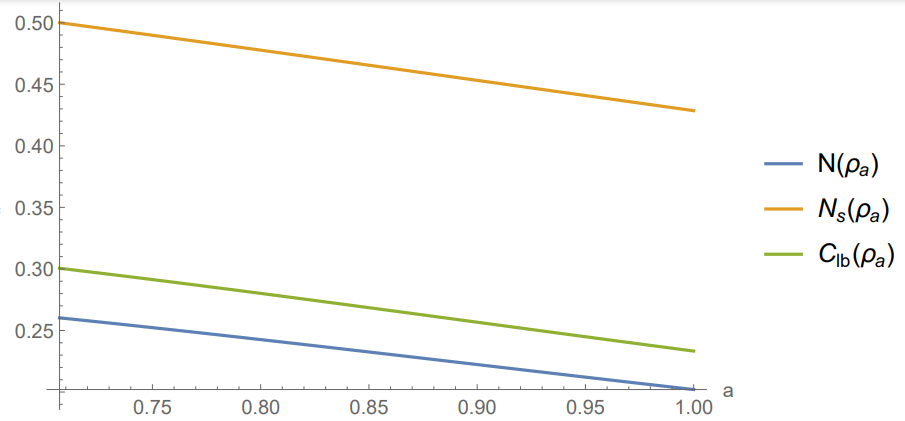}
	\caption{The structured negativity of the state $\rho_a$ is greater than the negativity as well as the lower bound of concurrence.}
	\label{MEMS2}
\end{figure}\\ \ \\
\textbf{Example 4: Two-qutrit $\alpha$ state}\\
Consider a two-qutrit state defined by\cite{horodecki11},
\begin{eqnarray}
\rho_{\alpha}=\frac{2}{7}|\psi^{+}\rangle \langle \psi^{+}|+\frac{\alpha}{7}\sigma_{+}+\frac{5-\alpha}{7}{\sigma_{-}}, ~~2\leq \alpha\leq 5
\end{eqnarray}
where,
\begin{eqnarray}
|\psi^{+}\rangle&=&\frac{1}{\sqrt{3}}[|00\rangle+|11\rangle+|22\rangle]\nonumber\\
\sigma_{+}&=&\frac{1}{3}(|01\rangle \langle 01|+|12\rangle \langle 12|+|20\rangle \langle 20|)\nonumber\\
\sigma_{-}&=&\frac{1}{3}(|10\rangle \langle 10|+|21\rangle \langle 21|+|02\rangle \langle 02|)
\end{eqnarray}
The given state $\rho_{\alpha}$ is NPTES for $4\leq \alpha \leq5$.\\
For the state $\rho_{\alpha}$, 
In this example, first consider $\rho_{\alpha}$ for $4\leq \alpha \leq 5$. The SPA-PT of $\rho_{\alpha}$ is given by, $N(\rho_a)$ and $N_S({\rho_a})$ may be calculated as
\begin{eqnarray}
N(\rho_{\alpha})&=&-\frac{1}{14}[-5+\sqrt{41-20\alpha+4\alpha^{2}}]\\
N_S({\rho_{\alpha}})&=&84[\frac{3}{28}-\frac{-131+\sqrt{41-20\alpha+4\alpha^{2}}}{1176}]
\end{eqnarray}
The comparison between $N(\rho_{\alpha})$, $N_S(\rho_{\alpha})$ and $C_{lb}(\rho_{\alpha})$ for the  state $\rho_{\alpha}$ has been studied in fig-5.
\begin{figure}[h]
	\centering
	\includegraphics[scale=0.25]{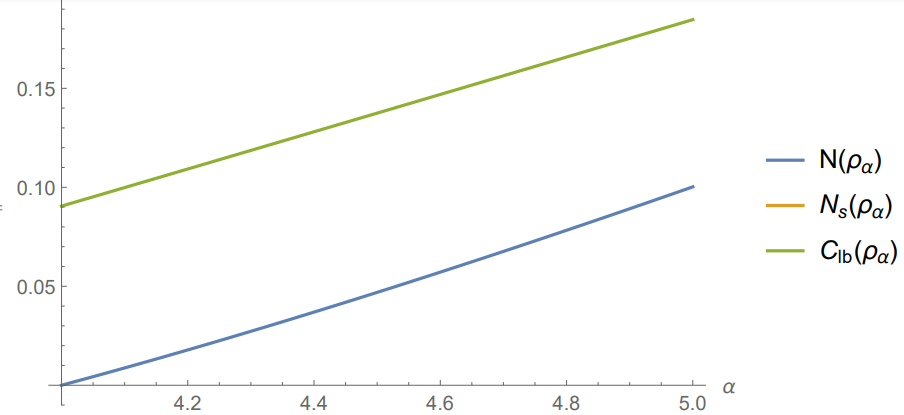}
	\caption{The negativity and structured negativity coincide for the state $\rho_{\alpha}$. In this case, $C_{lb}(\rho_{\alpha})$ is greater than the structured negativity. }
	\label{ex4}
\end{figure}
\section{Conclusion}
\noindent In this chapter, we have defined an entanglement measure based on the minimum eigenvalue of the SPA-PT of the arbitrary dimensional bipartite quantum system. We named the introduced measure structured negativity because it is based on the SPA-PT method and proved that the defined measure satisfies the properties of a valid entanglement measure. Since SPA-PT is a completely positive map, so, the proposed measure of entanglement may be realized in an experiment. The introduced measure of entanglement provides an advantage over the existing measure, negativity as negativity depends on the sum of the negative eigenvalues of the non-physical partial transposition operation whereas structured negativity depends on the minimum eigenvalue of SPA of the partially transposed matrix. We have established a relation between negativity and structured negativity and found that the negativity and structured negativity coincides for a large number of two-qubit systems. Thus, we conjecture that the negativity and structured negativity coincides when $q=\frac{d(d-1)}{2}$. In the end, we have compared negativity, structured negativity, and the lower bound of concurrence $C_{lb}$ and shown that the structured negativity is always greater than or equal to the negativity.

% Thus, our entanglement measure is physically realizable and can be implemented in a laboratory. Moreover, we have obtained an inequality between negativity and the structured negativity of a state. We have observed that in the obtained relation, equality holds for two-qubit states. We have illustrated our results with a few examples.

\begin{center}
	****************
\end{center}

%\input{Summary/Anu_scope}
%\bibliographystyle{amsplain}
%\bibliographystyle{plain}
%\bibliographystyle{unsrt}
%\bibliography{thesis}
%\renewcommand{\bibname}{Bibliography}
%\renewcommand{\baselinestretch}{}
%{\singlespacing}

\begin{center}
	****************
\end{center}

\newpage
\addcontentsline{toc}{chapter}{List of Publications}
\chapter*{List of Publications}
\begin{enumerate}
	
\item  {\bf\Student} and Satyabrata Adhikari; \emph{Detection of mixed bipartite entangled state in arbitrary dimension via structural physical approximation of partial transposition}, Physical Review A {\bf 100}, (2019),  052323(1-8). \textbf{Impact Factor (2.971)}
\item {\bf\Student} and Satyabrata Adhikari; \emph{Classification witness operator for the classification of different subclasses of three-qubit GHZ class}, Quantum Information Processing {\bf20} (2021)  316(1-25). \textbf{Impact Factor (1.965)}
\item {\bf\Student} and Satyabrata Adhikari; \emph{Structural physical approximation of partial transposition makes possible to distinguish SLOCC inequivalent classes of three-qubit system}, European Physical Journal D {\bf76} (2022) 73(1-9). \textbf{Impact Factor (1.611)}
\item {\bf\Student} and Satyabrata Adhikari; \emph{Structured Negativity: A physically realizable measure of entanglement based on structural physical approximation}, Annals of Physics \textbf{446} (2022) 169113(1-12). \textbf{Impact Factor (3.036)}
\item {\bf\Student} and Satyabrata Adhikari; \emph{Detection and Classification of Three-qubit States Using $l_{1}$ Norm of Coherence}\\
Status: Communicated
\end{enumerate}

\begin{center}
	****************
\end{center}

%\backmatter

%for index
%\markboth{Index}{Index}
%\renewcommand{\chaptermark}[1]{%
%\markboth{\thechapter.\ #1}{}}
%\thispagestyle{plain}
%\addcontentsline{toc}{chapter}{Index}
%\balance
%\printindex
%

\end{document}